\documentclass[twocolumn]{aastex631}

\usepackage{psfig}
\usepackage{graphicx}	

\usepackage{subfigure}

\shorttitle{Signatures of mass segregation}
\shortauthors{R. J. Parker \emph{et al.}}

\begin{document}

\title{Signatures of mass segregation from competitive accretion and monolithic collapse}

\correspondingauthor{Richard Parker}
\email{R.Parker@sheffield.ac.uk}

\author[0000-0002-1474-7848]{Richard J. Parker}
\altaffiliation{Royal Society Dorothy Hodgkin Fellow}
\affiliation{Department of Physics and Astronomy, The University of Sheffield, Hicks Building, Hounsfield Road, Sheffield, S3 7RH, UK}

\author{Emily J. Pinson}
\affiliation{Department of Physics and Astronomy, The University of Sheffield, Hicks Building, Hounsfield Road, Sheffield, S3 7RH, UK}

\author{Hayley L. Alcock}
\affiliation{Department of Physics and Astronomy, The University of Sheffield, Hicks Building, Hounsfield Road, Sheffield, S3 7RH, UK}

\author{James E. Dale}
\affiliation{Department of Peace and Conflict Research, Universitet Uppsala, Box 514, 751 20 Uppsala, Sweden}

\label{firstpage}

\begin{abstract}
The two main competing theories proposed to explain the formation of massive ($>10$\,M$_\odot$) stars -- competitive accretion and monolithic core collapse -- make different observable predictions for the environment of the massive stars during, and immediately after, their formation. Proponents of competitive accretion have long predicted that the most massive stars should have a different spatial distribution to lower-mass stars, either through the stars being mass segregated, or being in areas of higher relative densities, or sitting deeper in gravitational potential wells. We test these predictions by analysing a suite of SPH simulations where star clusters form massive stars via competitive accretion with and without feedback. We find that the most massive stars have higher relative densities, and sit in deeper potential wells, only in simulations in which feedback is not present. When feedback is included, only half of the simulations have the massive stars residing in deeper potential wells, and there are no other distinguishing signals in their spatial distributions. Intriguingly, in our simple models for monolithic core collapse, the massive stars may also end up in deeper potential wells, because if massive cores fragment the stars are still massive, and dominate their local environs. We find no robust diagnostic test in the spatial distributions of massive stars that can distinguish their formation mechanisms, and so other predictions for distinguishing between competitive accretion and monolithic collapse are required. 
\end{abstract}

\keywords{star forming regions (1565), massive stars (732), star formation (1569)}

\section{Introduction}

The formation of massive stars is one of the least understood aspects of star formation \citep{Zinnecker07}. It is unclear how massive stars can assemble their mass without breaching the Eddington limit \citep{Humphreys79,Sanyal15}, unclear whether they undergo a pre-main sequence phase in the same way that low-/intermediate-mass stars do \citep{Palla91,Behrend01}, unclear how massive they can be \citep{Figer05,Crowther10}, and unclear how exactly they form with such high masses \citep{Hosokawa09,Rosen16}.

There are two contrasting theories for how massive stars form. The first, referred to as `competitive accretion', posits that all stars initially start with similar masses, and it is the lucky ones that grow to high masses by virtue of forming in the most gas-rich areas of the star-forming region \citep{Zinnecker82,Bonnell97,Bonnell01}. Once they grow to substantial masses, they dominate their local environs and grow to even larger masses.

The second theory, sometimes referred to as `monolithic collapse', posits that stars inherit their masses from their pre-/proto-stellar cores \citep{McKee03,Krumholz05}, and it is the core mass that dictates the final mass of the star \citep{Alves07}. Here, the assumption is that the massive core does not fragment into lots of much less massive stars \citep{Krumholz06a}, and that the core does not accrete any significant amount of leftover gas from the surrounding envelope (in this scenario it would become even more difficult to distinguish between  this and competitive accretion).

In addition to these two models, \citet{Vazquez19} propose the `Global Hierarchical Collapse' (GHC) model, whereby the masses of stars are set by the collapse of the giant molecular cloud and subsequent Bondi--Hoyle--Lyttleton accretion \citep{Bondi44,Bondi52,Hoyle39}.  \citet{Padoan20} propose the `inertial--inflow' (I2) model, where massive stars gain their mass from the convergence of gas flows driven by turbulence in star-forming clouds. Recent observational studies \citep[e.g.][]{Liu23,Pillai23} have used both the GHC and the I2 models in the interpretation of their data.

  Whilst distinct from the competitive accretion and monolithic collapse scenarios, the GHC and I2 models do share similarities with both the more established theories, and from hereonin we only discuss competitive accretion and monolithic collapse as these are arguably the most different (and hence likely to be distinguishable in the spatial distributions of stars). 

There are very few diagnostic observational tests that could identify stars that formed via monolithic core collapse versus stars that formed via competitive accretion. A `smoking gun' for monolithic collapse would be the discovery of isolated massive stars \citep{Lamb10,Bressert12}. Isolated massive stars are those that could not have formed in a clustered environment (as is required by competitive accretion, \citealp{Smith09}, though see \citet{Hsu10}), but that have not been ejected from a star-forming region \citep{Schoettler19}.  Ruling out ejection is particularly challenging \citep{Gvaramadze08,Pflamm10,Oh15}, even in the era of multi-epoch and multi-dimensional data from \emph{Gaia} \citep{Farias20,Schoettler20}.

A prediction of the competitive accretion theory is that the most massive stars should sit in the deepest gravitational potential wells in a star-forming region, in the locations where the most gas was available to them as they were forming \citep{Bonnell06}. This has led to the idea that mass segregation -- where the most massive stars are more centrally concentrated than the average mass stars -- would be observed in star-forming regions where stars had formed via competitive accretion \citep{Maschberger11,TieLiu13,Wright14}.

In previous work \citep{Parker17b} we have shown that hydrodynamic simulations in which massive stars form via competitive accretion usually lead to the most massive stars residing in deeper potential wells, although this does not always translate into mass segregation, or the most massive stars residing in areas of relatively high stellar densities \citep{Maschberger11}.

In  this paper, we extend this work to look for signatures in the spatial distribution of massive stars in a larger suite of simulations in which stars form via competitive accretion. We compare the analysis of these simulations to idealised, or `synthetic' star-forming regions, where we move the massive stars to be more centrally concentrated, and to locations of high relative surface densities, in addition to random distributions. Finally, we analyse the spatial distribution of massive stars that form in a simple model of monolithic core collapse.

The paper is organised as follows. In Section~\ref{methods} we describe our methods to analyse the spatial distribution of massive stars, the SPH simulations from \citet{Dale14} and \citet{Dale17}, the synthetic star-forming regions, as well as our simple prescription for monolithic collapse. In Section~\ref{results} we present our results. We provide a discussion in Section~\ref{discuss} and we conclude in Section~\ref{conclude}.

\section{Method}
\label{methods}

\subsection{Quantifying the spatial distributions of massive stars}
\label{metrics}

\subsubsection{Gravitational potential}

We follow \citet{Parker17b} and calculate the gravitational potential for each individual star, $\Phi_j$:
\begin{equation}
\Phi_j = -\sum \frac{m_i}{r_{ij}},
\end{equation}
where $m_i$ is the mass of the $i^{\rm th}$ star, and $r_{ij}$ is the distance between the $j^{\rm th}$ star and the $i^{\rm th}$ star. When we calculate the gravitational potential for each star in the SPH simulations, we just consider the sink particles (stars), and not any gas remaining from star formation. We discuss this assumption in subsequent sections.

We calculate the median potential for all the stars in the star-forming region, $\tilde{\Phi}_{\rm all}$ and compare this to the median potential of a subset of the most massive stars, $\tilde{\Phi}_{\rm subset}$. We compare these quantities via the ratio, $\Phi_{\rm PDR}$,
\begin{equation}
\Phi_{\rm PDR} = \frac{\tilde{\Phi}_{\rm subset}}{\tilde{\Phi}_{\rm all}},
\end{equation}
and a Kolmogorov-Smirnov test between the two distributions. The significance of any differences between the distribution of all of the individual stellar potentials, compared to the distribution of the stellar potentials  of the subset, is quantified using the Kolmogorov-Smirnov test $d$ statistic, and we reject the hypothesis that the two distributions share the same underlying parent distribution is the associated $p-$value is less than 0.1.

\subsubsection{Mass segregation}

We quantify the amount of mass segregation, which we define as the most massive stars being closer to each other than average stars, using the \citet{Allison09a} mass segregation ratio, $\Lambda_{\rm MSR}$. This method constructs a minimum spanning tree (MST) -- a graph between a series of points where there are no closed loops -- for the subset of the most massive stars, and compares this to the MSTs of randomly chosen sets of stars.

$\Lambda_{\rm MSR}$ is the ratio of the average length of randomly chosen MSTs $\langle l_{\rm average} \rangle$, divided by the length of the MST of the chosen subset, ${l_{\rm subset}}$:
\begin{equation}
\Lambda_{\rm MSR} = {\frac{\langle l_{\rm average} \rangle}{l_{\rm subset}}} ^{+ {\sigma_{\rm 5/6}}/{l_{\rm subset}}}_{- {\sigma_{\rm 1/6}}/{l_{\rm subset}}}.
\label{lambda_eqn}
\end{equation}
We conservatively estimate the uncertainty on $\Lambda_{\rm MSR}$ by making an ordered list of the random MST lengths and taking the values that lie 1/6 and 5/6 of the way through the list.

Instances where $\Lambda_{\rm MSR} > 1$ indicate mass segregation \citep{Allison09a}, but subsequent work \citep{Parker15b} suggests that values of $\Lambda_{\rm MSR} > 2$ (with the lower uncertainty above unity) are significant.

\subsubsection{Relative densities}

\citet{Maschberger11} developed a method to quantify the relative local densities of massive stars compared to the local densities around all stars in a star-forming region. In order to make direct comparisons with observations, \citet{Maschberger11} used the local surface densities. For each star, the local surface density out to the ten nearest neighbours is calculated
\begin{equation}
\Sigma = \frac{N - 1} {\pi r_{N}^2},
\end{equation}
where $r_N$ is the distance to the $N^{\rm th}$ nearest neighbouring star (we adopt $N = 10$ throughout this work).

\citet{Kupper11} and \citet{Parker14b} introduced a ratio between the median surface density for all the stars in the region, $\tilde{\Sigma}_{\rm all}$, and the median surface density for a subset of the ten most massive stars, $\tilde{\Sigma}_{\rm subset}$:
\begin{equation}
\Sigma_{\rm LDR} = \frac{\tilde{\Sigma}_\mathrm{subset}}{\tilde{\Sigma}_\mathrm{all}}.
\end{equation} 
The significance of any differences between the distribution of all of the surface densities, compared to the distribution of surface densities of the subset, is quantified using a Kolmogorov-Smirnov test in a similar manner to the difference between the stellar potentials.

\subsection{SPH simulations}

We use simulations from \citet{Dale14} and \citet{Dale17}. These simulations are an evolutionary extension of the first simulations to model competitive accretion by \citet{Bonnell97,Bonnell01}, but in addition to modelling the formation of massive stars, they also implement photoionising and stellar wind feedback from the most massive stars. Once three massive stars exceeding 20\,M$_\odot$ have formed, each simulation is split into two versions; a control run without feedback, and a run that implements the feedback.

The two versions of each simulation are run until 3\,Myr. At this time the most massive stars begin to leave the main sequence, and dynamical interactions can be significant -- neither process is modelled in SPH simulations of this type. It is at this point that we extract the sink particle distribution (their masses, and positions) and then analyse them with the metrics described in the previous subsection. The simulations are described in detail in \citet{Dale14} and \citet{Dale17}, but we summarise them in Table~\ref{simulations}.

\begin{table*}
\caption{A summary of the ten different pairs of smoothed particle hydrodynamics (SPH) simulations. The values in the columns are: the corresponding Run ID from \citet{Dale14} or  \citet{Dale17}, the type of feedback in the SPH simulation (none, or photoionisation and stellar winds), the paper reference,  the initial virial ratio of the original clouds $\alpha_{\rm init}^{\rm SPH}$ (to distinguish bound from unbound clouds), the initial radius of the cloud in the SPH simulation ($R_{\rm cloud}$), the initial mass of the cloud ($M_{\rm cloud}$),   the number of stars that have formed at the end of the SPH simulation ($N_{\rm stars}$) and the total stellar mass of this star-forming region ($M_{\rm region}$).}
\begin{center}
\begin{tabular}{cccccccccc}
\hline 
 Simulation ID & Feedback & Ref.  & $\alpha_{\rm init}^{\rm SPH}$ &$R_{\rm cloud}$ & $M_{\rm cloud}$ & $N_{\rm stars}$ & $M_{\rm region}$ \\
\hline
J & None & \citet{Dale14} & 0.7 & 5\,pc & 10 000\,M$_\odot$ & 578 & 3207\,M$_\odot$ \\ 
J & Photoionisation + wind & \citet{Dale14} & 0.7 & 5\,pc & 10 000\,M$_\odot$ & 564 & 2186\,M$_\odot$ \\
\hline
I & None & \citet{Dale14} & 0.7 & 10\,pc & 10 000\,M$_\odot$ & 186 & 1270\,M$_\odot$ \\
I & Photoionisation + wind & \citet{Dale14} & 0.7 & 10\,pc & 10 000\,M$_\odot$ & 132 & 766\,M$_\odot$ \\ 
\hline 
UF & None & \citet{Dale14} & 2.3 & 10\,pc & 30 000\,M$_\odot$ & 66 & 1392\,M$_\odot$ \\ 
UF & Photoionisation + wind & \citet{Dale14} & 2.3 & 10\,pc & 30 000\,M$_\odot$ & 93 & 841\,M$_\odot$ \\
\hline
UP & None & \citet{Dale14} & 2.3 & 2.5\,pc & 10 000\,M$_\odot$ & 340 & 2718\,M$_\odot$ \\ 
UP & Photoionisation + wind & \citet{Dale14} & 2.3 & 2.5\,pc & 10 000\,M$_\odot$ & 343 & 1926\,M$_\odot$ \\ 
\hline
UQ & None & \citet{Dale14} & 2.3 & 5\,pc & 10 000\,M$_\odot$ & 48 & 723\,M$_\odot$ \\ 
UQ & Photoionisation + wind & \citet{Dale14} & 2.3 & 5\,pc & 10 000\,M$_\odot$ & 77 & 594\,M$_\odot$ \\
\hline
R11O & None & \citet{Dale17} & 1.1 & 5\,pc & 10 000\,M$_\odot$ & 239 & 2679\,M$_\odot$\\
R11O & Photoionisation + wind & \citet{Dale17} & 1.1 & 5\,pc & 10 000\,M$_\odot$ & 372 & 1853\,M$_\odot$ \\
\hline
R15L & None & \citet{Dale17} & 1.5 & 5\,pc & 10 000\,M$_\odot$ & 170 & 2084\,M$_\odot$\\
R15L & Photoionisation + wind & \citet{Dale17} & 1.5 & 5\,pc & 10 000\,M$_\odot$ & 282 & 1456\,M$_\odot$\\
\hline 
R15M & None & \citet{Dale17} & 1.5 & 2.5\,pc & 10 000\,M$_\odot$ & 543 & 4747\,M$_\odot$ \\
R15M & Photoionisation + wind & \citet{Dale17} & 1.5 & 2.5\,pc & 10 000\,M$_\odot$ & 777 & 3177\,M$_\odot$ \\
\hline
R19S & None & \citet{Dale17} & 1.9 & 5\,pc & 10 000\,M$_\odot$ & 80 & 1281\,M$_\odot$ \\
R19S & Photoionisation + wind &  \citet{Dale17} & 1.9 & 5\,pc & 10 000\,M$_\odot$ & 160 & 1095\,M$_\odot$\\
\hline
R19T & None & \citet{Dale17} & 1.9 & 2.5\,pc & 10 000\,M$_\odot$ & 377 & 3530\,M$_\odot$\\
R19T & Photoionisation + wind & \citet{Dale17} & 1.9 & 2.5\,pc & 10 000\,M$_\odot$ & 544 & 2492\,M$_\odot$ \\
\hline
\end{tabular}
\end{center}
\label{simulations}
\end{table*}

\subsection{Synthetic star-forming regions}

In order to demonstrate the effects of moving massive stars on the average gravitational potential, degree of mass segregation, and relative surface densities of the most massive stars, we construct various synthetic star-forming regions with different morphologies.

A significant amount of confusion abounds in the literature \citep[e.g.][]{Maschberger11,Parker15b,Guszejnov22} on whether the $\Lambda_{\rm MSR}$ or $\Sigma_{\rm LDR}$ methods measure the same properties (we contend that they do not), or whether $\Lambda_{\rm MSR}$ actually measures mass segregation (we contend that it does). By including these synthetic morphologies we hope to negate some of this confusion. These idealised morphologies will be used to interpret the outcome of both the SPH simulations, and our simulations of cores undergoing fragmentation and monolithic collapse, which we describe in the next subsection (Sect.~\ref{methods:core_frag}).

Each star-forming region contains $N_\star = 300$ stars. We adopt three different spatial distributions. The first is a \citet{Plummer11} sphere, in which we assign positions to stars based on the method in \citet{Aarseth74}.

Second, we create associations following the method in \citet{Parker12d}. Here, we randomly produce 10 subgroups within a cubic volume (and then project in two-dimensions), and then randomly populate each of the subgroups with $N_\star/10$ of our stars, distributed in spheres where the number density of stars $n$ at position $r$ follows the relation
  \begin{equation}
n \propto r^{-2.5}.
    \end{equation}

For completeness, we also adopt the box-fractal generating method from \citet{Goodwin04a}. We refer the interested reader to \citet{Goodwin04a} for details on how the box fractals are generated. For one set of simulations we adopt a fractal dimension $D = 1.6$, which results in a highly substructured distribution (in three dimensions -- again we project into two dimensions) and for another set of simulations we adopt $D = 2.0$, which results in a moderate degree of substructure. 

In all regions, we select stellar masses from the \citet{Maschberger13} IMF, which has a probability distribution function of the form
\begin{equation}
p(m) \propto \left(\frac{m}{\mu}\right)^{-\alpha}\left(1 + \left(\frac{m}{\mu}\right)^{1 - \alpha}\right)^{-\beta}.
\label{maschberger_imf}
\end{equation}
In Eqn.~\ref{maschberger_imf}  $\mu = 0.2$\,M$_\odot$ is the scale parameter, or `peak' of the IMF \citep{Bastian10,Maschberger13}, $\alpha = 2.3$ is the \citet{Salpeter55} power-law exponent for higher mass stars, and $\beta = 1.4$ describes the slope of the IMF for low-mass objects. We randomly sample this distribution in the mass range 0.1 -- 50\,M$_\odot$, such that brown dwarfs are not included in the simulations.

We create five realisations of each of the four morphologies to check that none of the realisations we analyse are outliers -- the realisations we plot in Section~\ref{results} are the most representative of each set of five. For each of the five realisations, we have three versions, one where the massive stars are distributed randomly, one where the massive stars are placed in the most central locations in the realisation, and one where the massive stars are placed in the locations with the highest local densities.

\subsection{Core fragmentation}
\label{methods:core_frag}

We produce a further set of model star-forming regions in which we set up a distribution of pre-stellar cores, and fragment them into stars according to a simple prescription. The full method is detailed in \citet{Alcock19}, but we briefly restate it here.

We assume the core mass function is the precursor to the stellar initial mass function, but shifted to higher masses by  a factor equal to the inverse of the star formation efficiency, $\epsilon$. We set $\epsilon = 0.333$ and draw masses from the \citet{Maschberger13} IMF (Eqn.~\ref{maschberger_imf}), but we now adopt $\mu_{\rm core} = 0.6$\,M$_\odot$, and sample masses in the range $m_{\rm core} = 0.3$ -- 300\,M$_\odot$.

We distribute $N_{\rm core} = 300$ cores in a fractal distribution using the box-fractal method described in \citet{Goodwin04a}, \citet{Lomax11} and \citet{DaffernPowell20}, with a high degree of substructure (fractal dimension $D = 1.6$). For completeness, we also checked the results using a smoother distribution (fractal dimension $D = 2.6$). In two sets of simulations, we randomly allow the cores to fragment into between 1 and 5 pieces \citep{Alcock19}. This produces a fragment mass function that is consistent with the initial mass function (likely to be because the mean number of fragments in this scenario will be $\sim$3, roughly the inverse of our assumed star formation efficiency). In the first set of simulations, we apply a random offset to the positions of the fragments of 0.05\,pc. In the second set of simulations we increase the size of the random offset to 0.25\,pc, to mimic a small degree of dynamical evolution. 

In the third set of simulations, we restrict the fragmentation just to the lower-mass cores, such that the most massive cores form just one massive star but a core with mass $\leq$10\,M$_\odot$ fragments into 5 pieces. We apply an offset of 0.05\,pc to the fragments.

In a final set of simulations, we restrict the fragmentation to just the high mass cores, such that the least massive cores form just one star, but a core with mass $>$10\,M$_\odot$ fragments into 5 pieces. We apply an offset of 0.05\,pc to the fragments.
  
We then calculate the potentials, degree of mass segregation and relative surface densities on the fragments, as detailed in Section~\ref{metrics}.

\section{Results}
\label{results}

\subsection{Competitive accretion in SPH simulations}

We first describe the results from two versions of the same Smoothed Particle Hydrodynamics simulation of star formation. The first version is a control run, where the gas is converted into stars (sink particles) but there is no feedback from the most massive stars. The second version is a run where photoionising feedback, and stellar wind feedback, from the most massive stars, acts upon the gas that has not yet formed stars.

This particular simulation \citep[R15M from][]{Dale17} was chosen because the results are fairly representative of the behaviour of both the control runs, and the runs with feedback, in the simulations. However, we list the results for all pairs of simulations in Table~\ref{simulation_results}, and present the figures for the remaining pairs in Figs.~\ref{J113_sim_results_control}--\ref{R19T_sim_results_dual} in the Appendix.

\subsubsection{SPH control run}

\begin{figure*}
  \begin{center}
\setlength{\subfigcapskip}{10pt}
\hspace*{-1.5cm}\subfigure[SPH control $x-y$]{\label{SPH_sim_results_control-a}\rotatebox{270}{\includegraphics[scale=0.25]{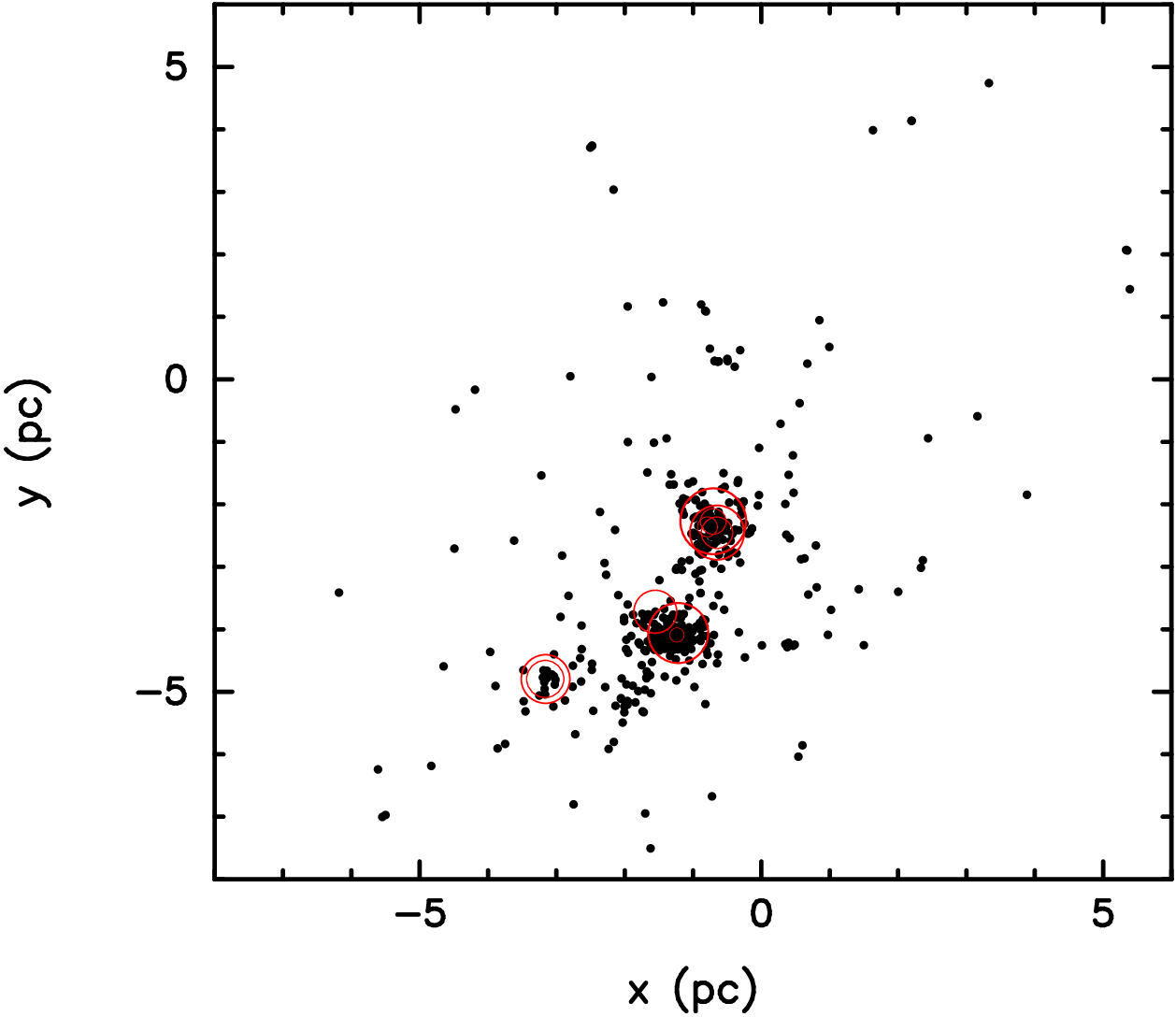}}}
\hspace*{0.3cm}
\subfigure[SPH control potential]{\label{SPH_sim_results_control-b}\rotatebox{270}{\includegraphics[scale=0.25]{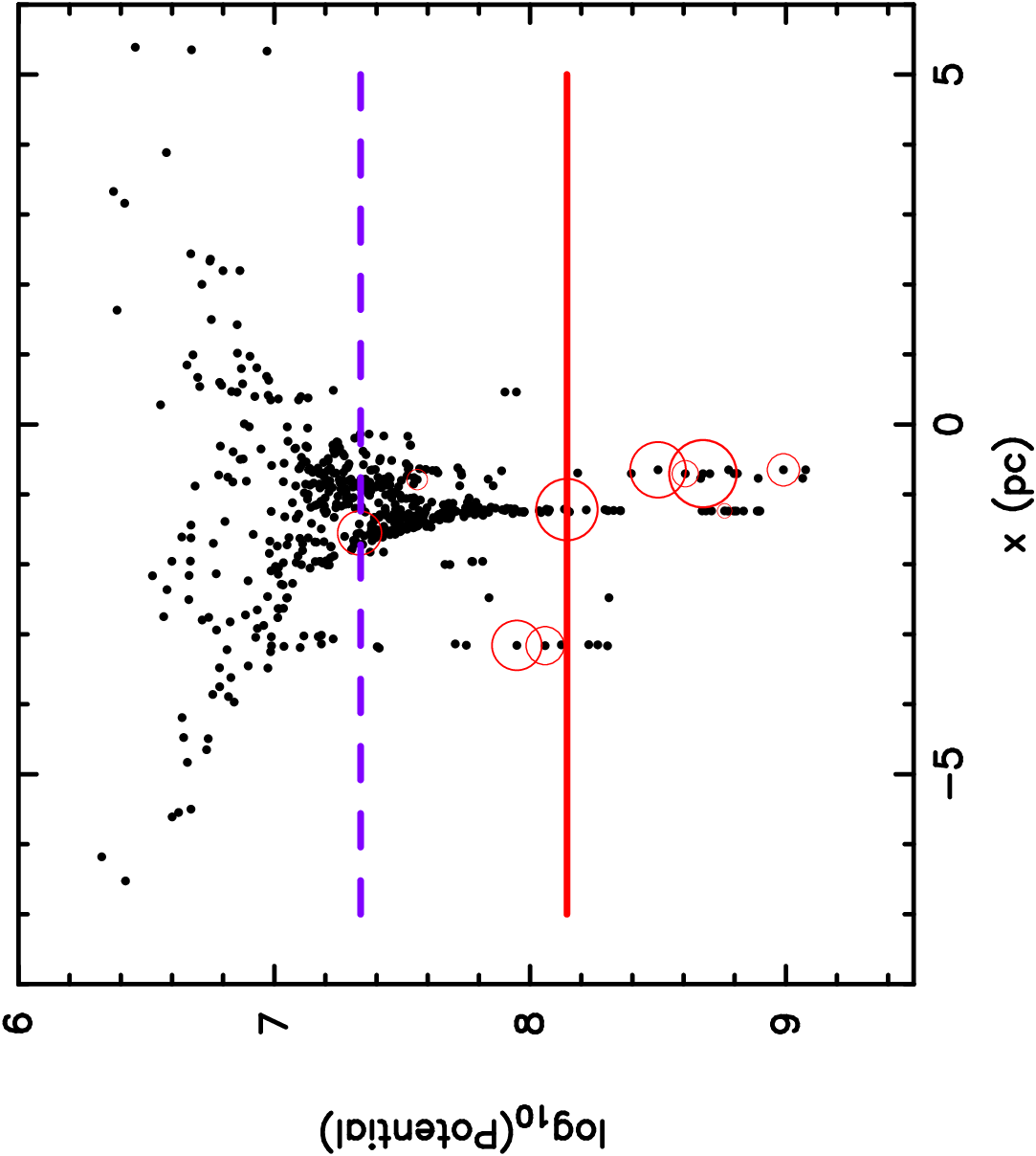}}}
\hspace*{0.3cm}
\subfigure[SPH control $\Phi - m$]{\label{SPH_sim_results_control-c}\rotatebox{270}{\includegraphics[scale=0.25]{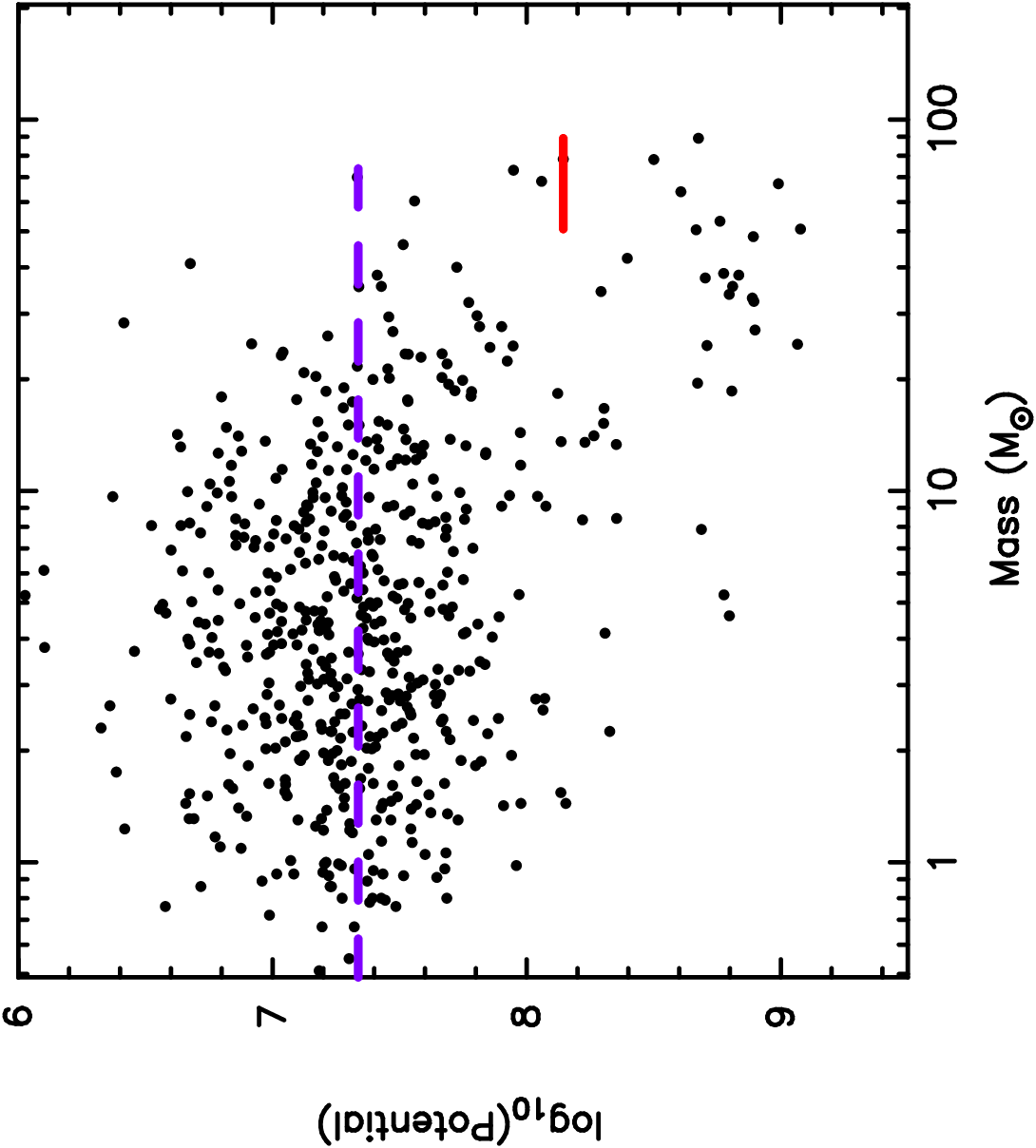}}}
\hspace*{-1.5cm}\subfigure[SPH control $\Lambda_{\rm MSR}$]{\label{SPH_sim_results_control-d}\rotatebox{270}{\includegraphics[scale=0.35]{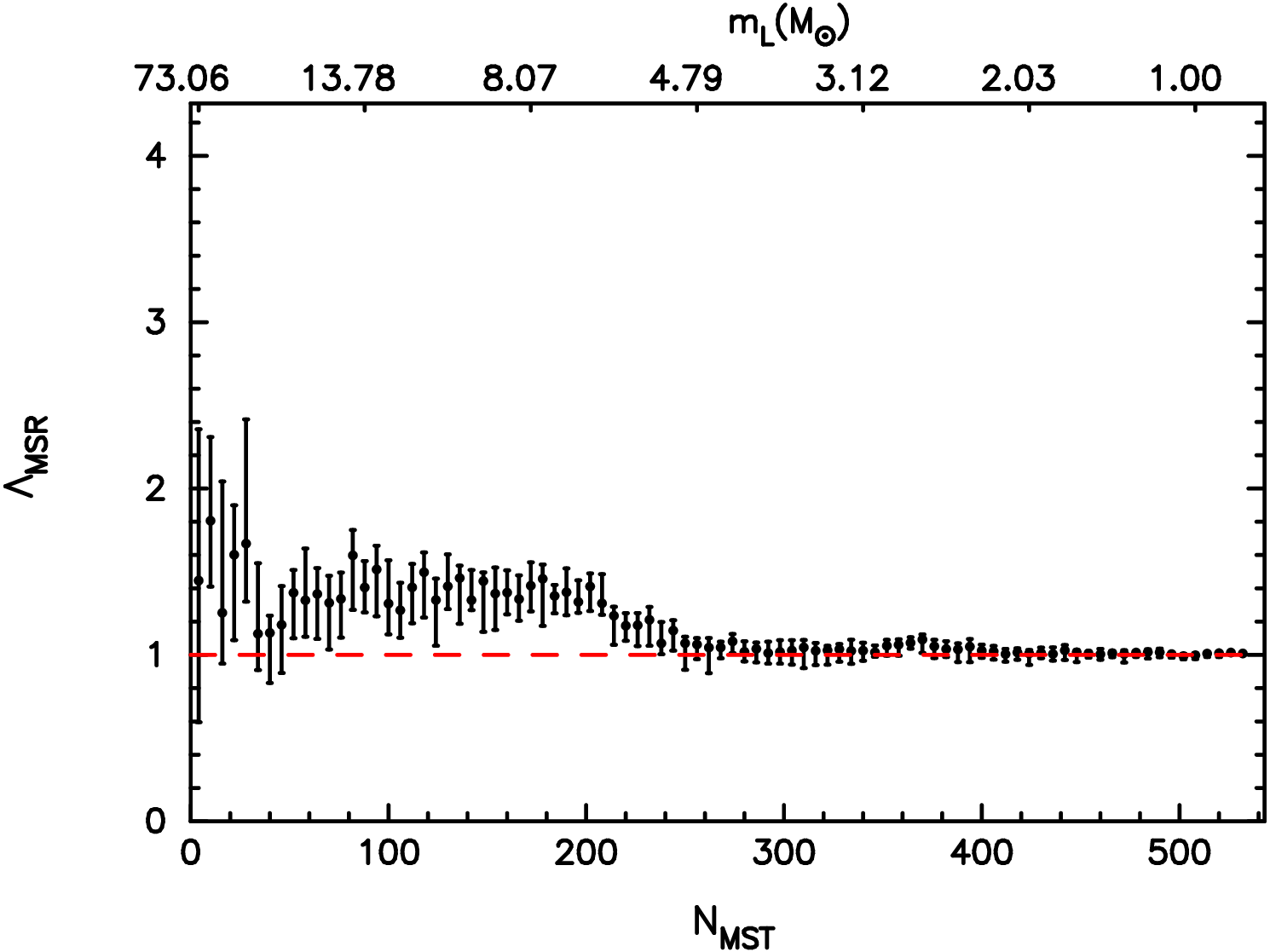}}}
\hspace*{0.3cm}
\subfigure[SPH control $\Sigma - m$]{\label{SPH_sim_results_control-e}\rotatebox{270}{\includegraphics[scale=0.4]{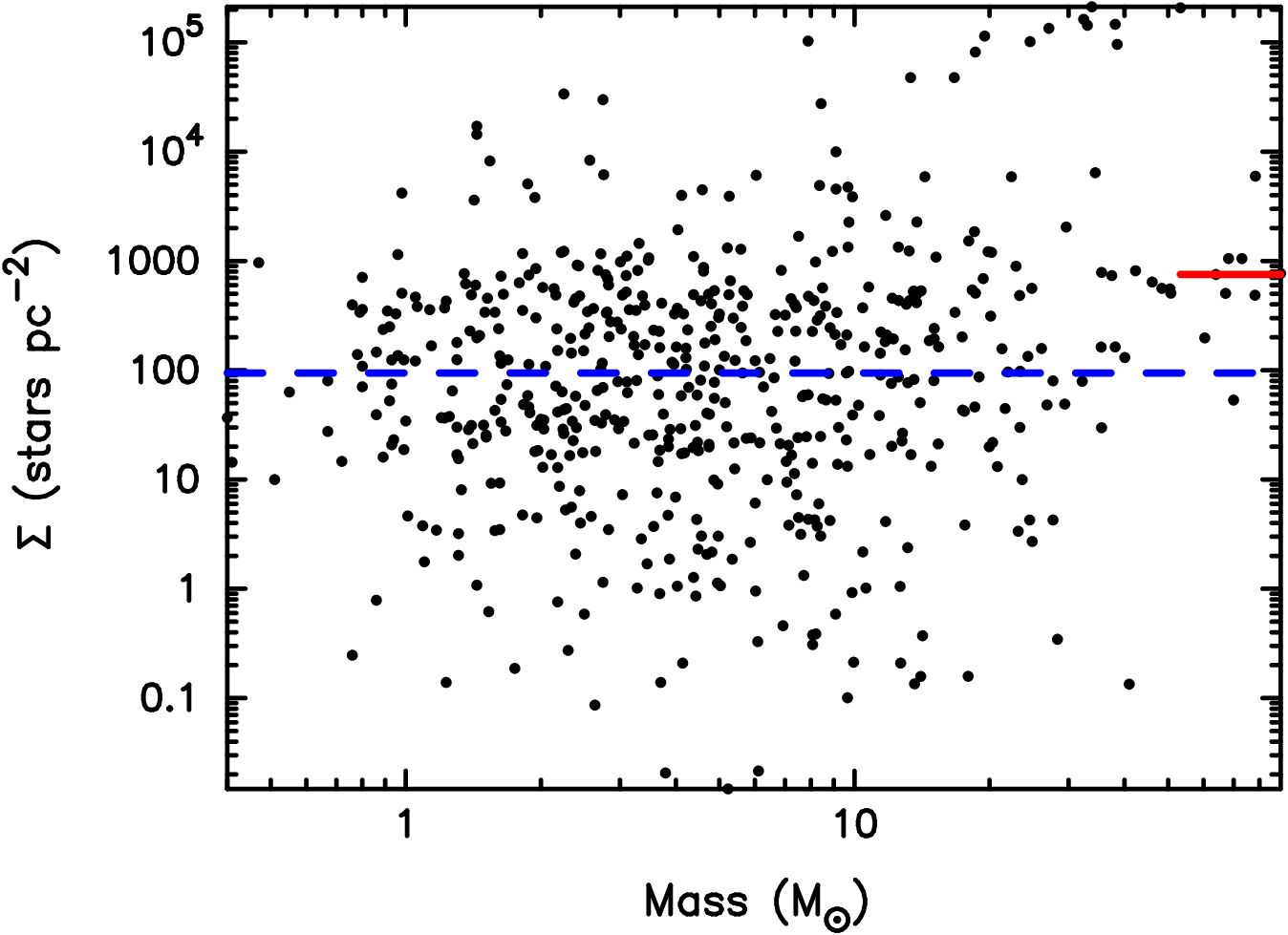}}}

\caption{Results from a representative SPH control run (i.e.\,\,where there is no feedback from the most massive stars), simulation R15M from \citet{Dale17}. In all panels we show the end-point of the simulation.  In panel (a) we show a projection of the simulation in the $x - y$ plane. In panel (b) we show the gravitational potential of each star  in the $x$-axis. In panels (a) and (b) the ten most massive stars are shown by the red points. The relative size of the red point indicates where it resides in the list of the most massive stars, with the largest circle indicating the most massive star, and the smallest circle indicating the tenth most massive star. In panel (c) we show the potential of each star plotted against the individual stellar masses. The median potential for all of the stars is shown by the dashed horizontal purple line, and the median potential for the ten most massive stars is shown by the solid red line. In panel (d) we show the $\Lambda_{\rm MSR}$ mass segregation ratio as a function of the $N_{\rm MST}$ most massive stars. The mass corresponding to this $N_{\rm MST}$ value is shown on the top ordinate. The red horizonal dashed line represents $\Lambda_{\rm MSR} = 1$, i.e.\,\,no mass segregation. In panel (e) we show the local surface density of each star plotted against the individual stellar masses. The horizontal blue dashed line shows the median surface density of all stars, and the solid red line shows the median surface density of the ten most massive stars.    }
\label{SPH_sim_results_control}
  \end{center}
\end{figure*}

\begin{figure*}
  \begin{center}
\setlength{\subfigcapskip}{10pt}
\hspace*{-1.5cm}\subfigure[SPH dual feedback $x-y$]{\label{SPH_sim_results_dual-a}\rotatebox{270}{\includegraphics[scale=0.25]{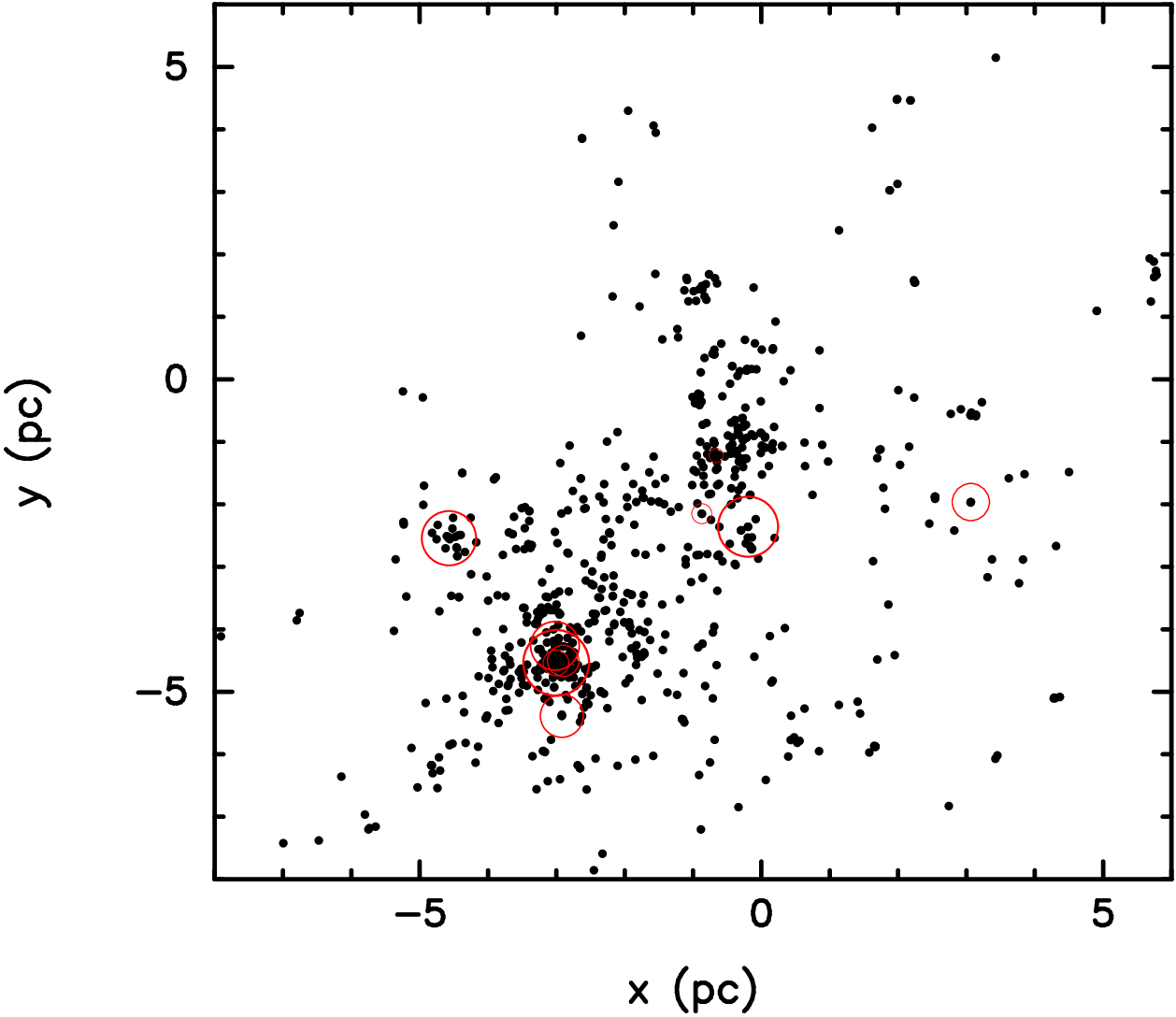}}}
\hspace*{0.3cm}
\subfigure[SPH dual feedback potential]{\label{SPH_sim_results_dual-b}\rotatebox{270}{\includegraphics[scale=0.25]{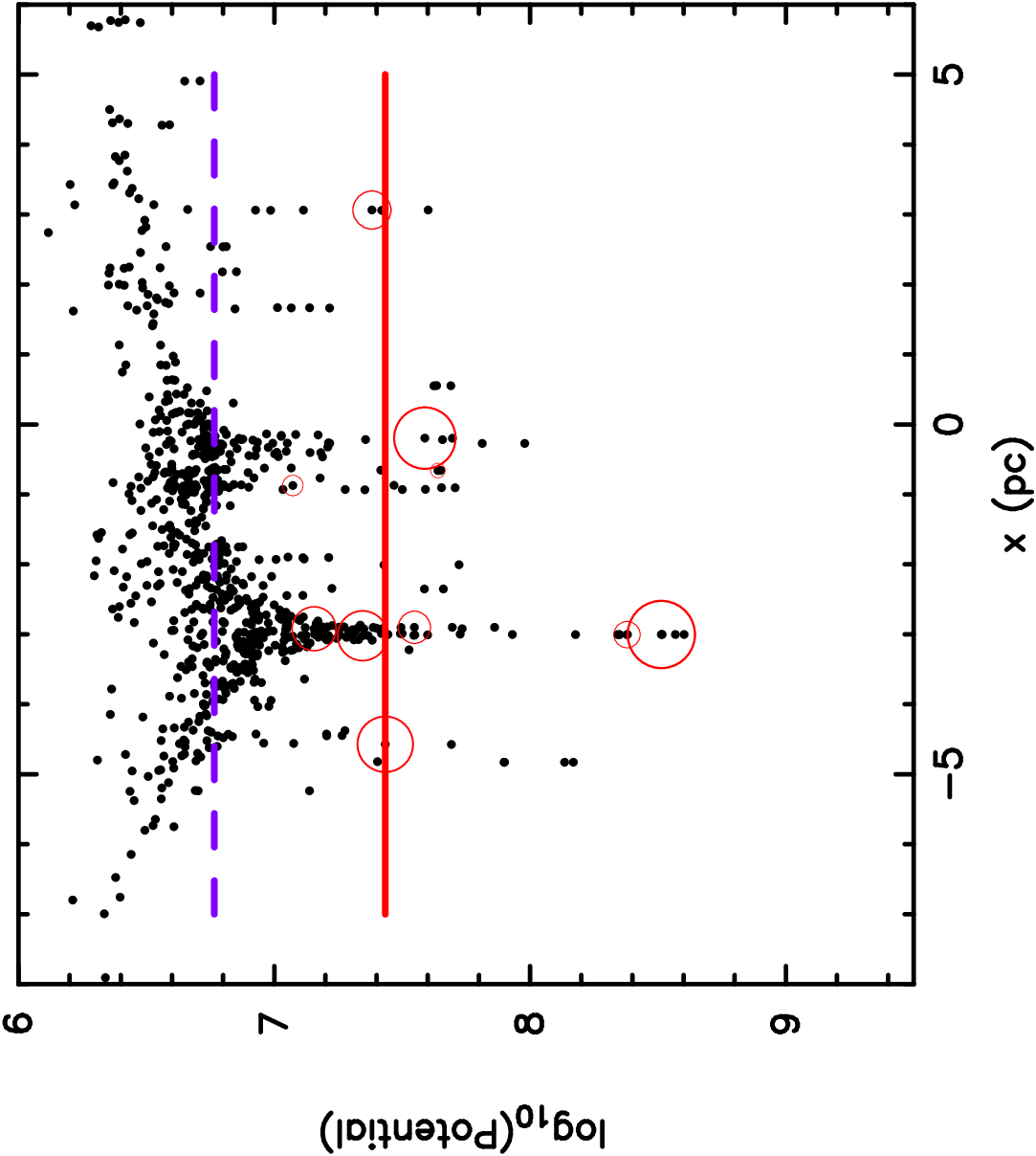}}}
\hspace*{0.3cm}
\subfigure[SPH dual feedback $\Phi -m$]{\label{SPH_sim_results_dual-c}\rotatebox{270}{\includegraphics[scale=0.25]{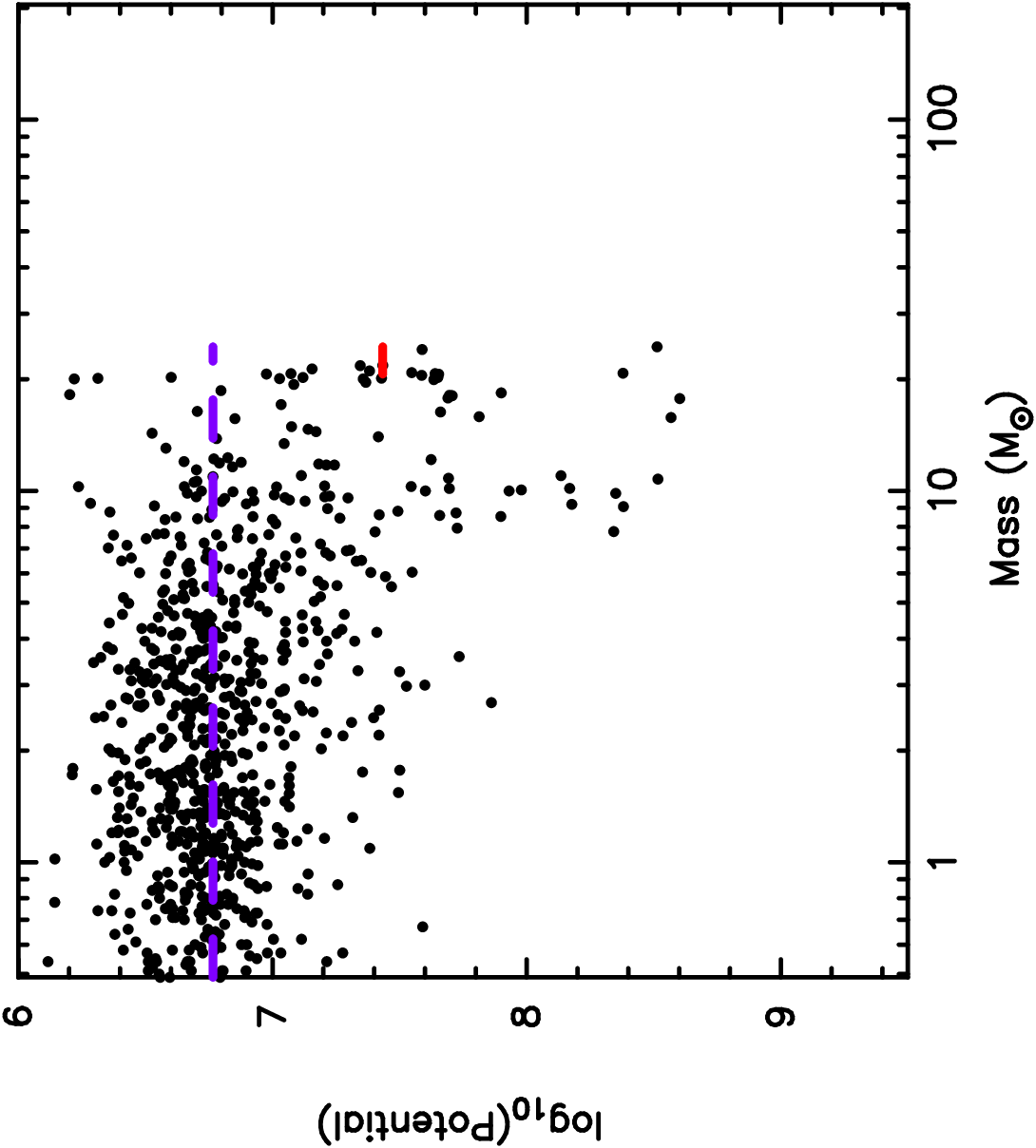}}}
\hspace*{-1.5cm}\subfigure[SPH dual feedback $\Lambda_{\rm MSR}$]{\label{SPH_sim_results_dual-d}\rotatebox{270}{\includegraphics[scale=0.35]{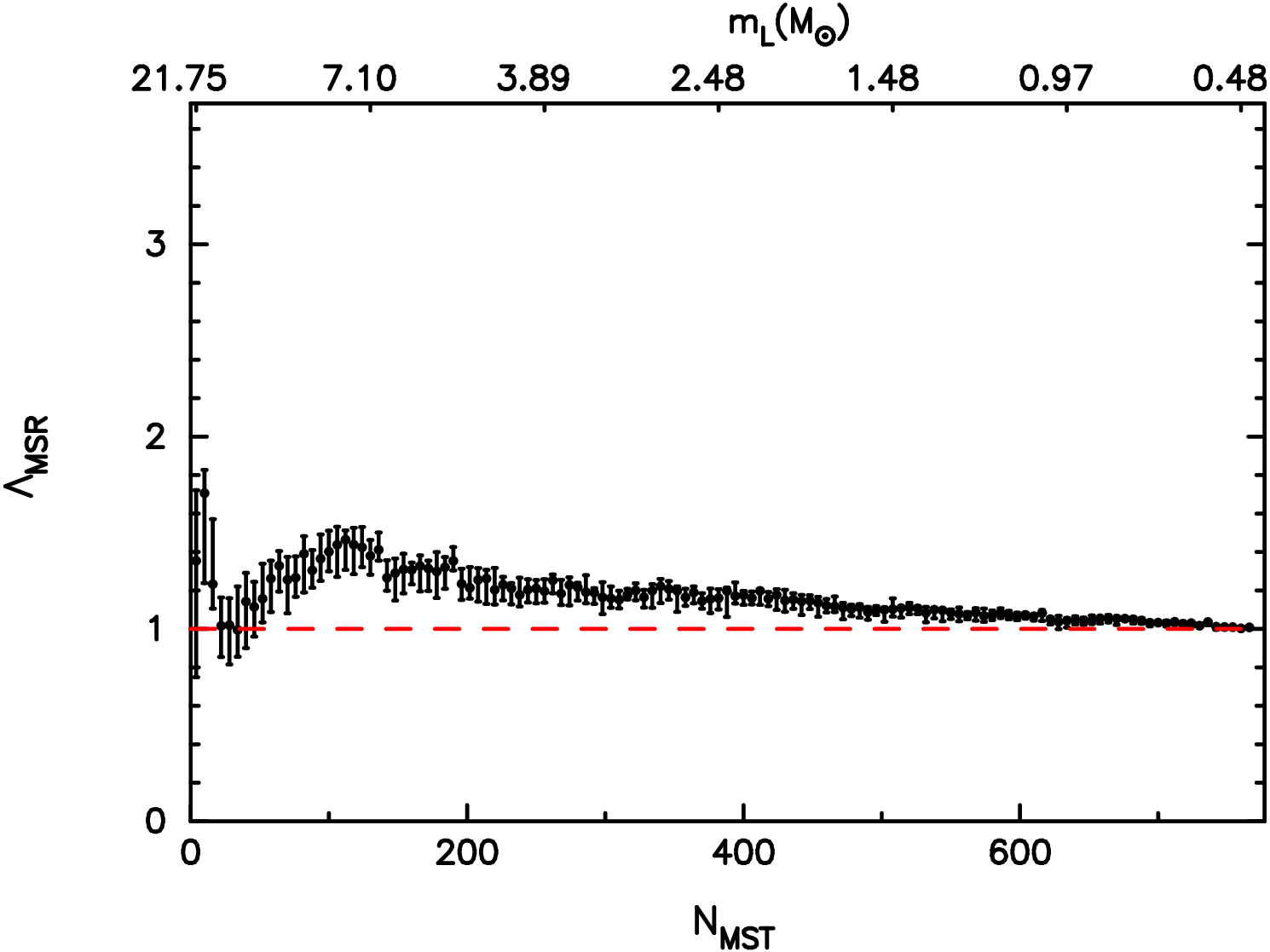}}}
\hspace*{0.3cm}
\subfigure[SPH dual feedback $\Sigma - m$]{\label{SPH_sim_results_dual-e}\rotatebox{270}{\includegraphics[scale=0.4]{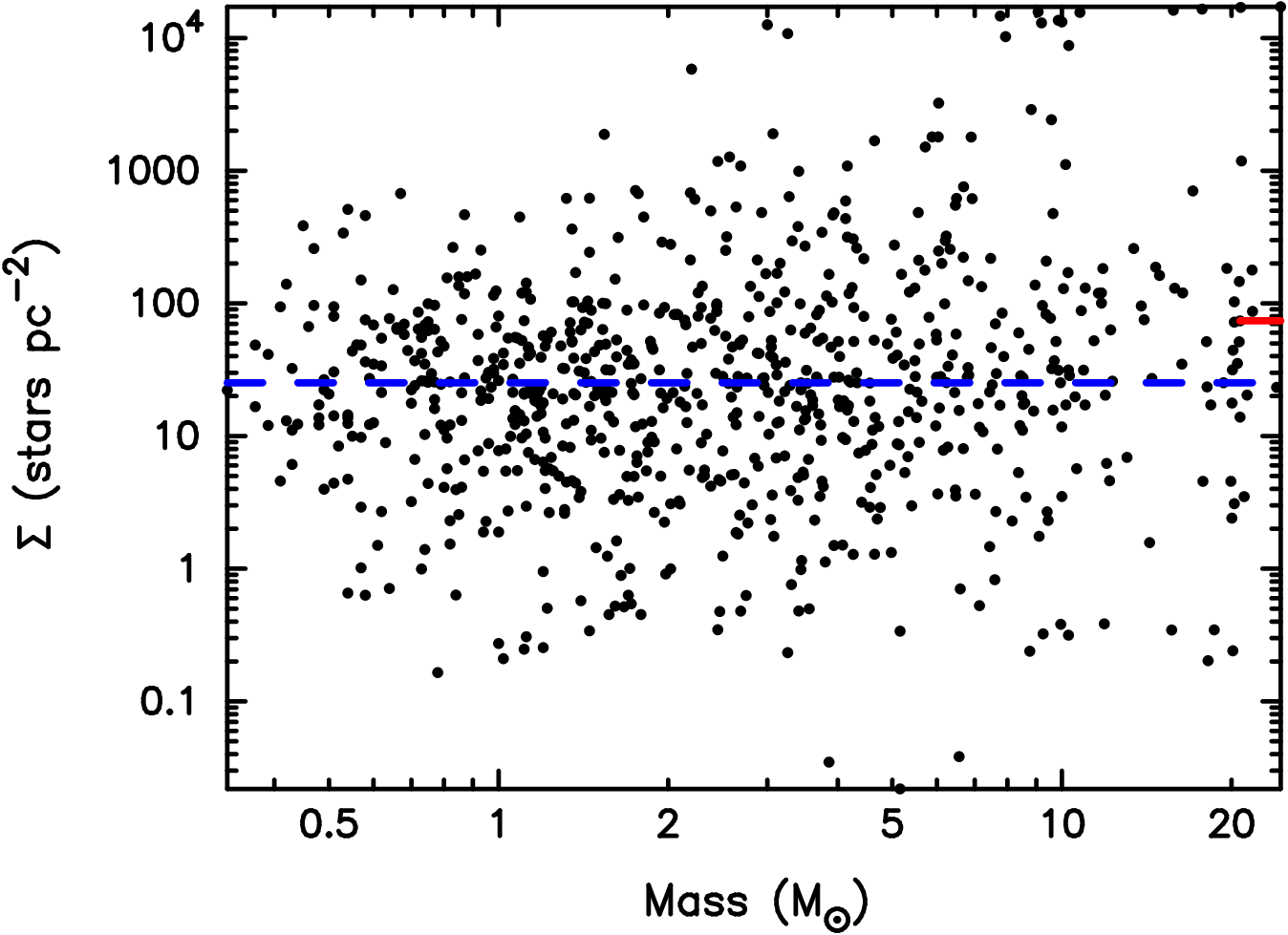}}}

\caption{Results from a representative SPH dual feedback run (i.e.\,\,where there is photionising feedback, and feedback from the stellar winds, from the most massive stars), simulation R15M from \citet{Dale17}. In all panels we show the end-point of the simulation.  In panel (a) we show a projection of the simulation in the $x - y$ plane. In panel (b) we show the gravitational potential of each star  in the $x$-axis. In panels (a) and (b) the ten  most massive stars are shown by the red points. The relative size of the red point indicates where it resides in the list of the most massive stars, with the largest circle indicating the most massive star, and the smallest circle indicating the tenth most massive star.  In panel (c) we show the potential of each star plotted against the individual stellar masses. The median potential for all of the stars is shown by the dashed horizontal purple line, and the median potential for the ten most massive stars is shown by the solid red line. In panel (d) we show the $\Lambda_{\rm MSR}$ mass segregation ratio as a function of the $N_{\rm MST}$ most massive stars. The mass corresponding to this $N_{\rm MST}$ value is shown on the top ordinate. The red horizonal dashed line represents $\Lambda_{\rm MSR} = 1$, i.e.\,\,no mass segregation. In panel (e) we show the local surface density of each star plotted against the individual stellar masses. The horizontal blue dashed line shows the median surface density of all stars, and the solid red line shows the median surface density of the ten most massive stars. }
\label{SPH_sim_results_dual}
  \end{center}
\end{figure*}

\begin{table*}
\caption{A summary of the results from the ten different pairs of smoothed particle hydrodynamics (SPH) simulations. The values in the columns are: the corresponding Run ID from \citet{Dale14} or  \citet{Dale17}, the type of feedback in the SPH simulation (none, or photoionisation and stellar winds), the mass segregation ratio (and uncertainties) for the four most massive stars, $\Lambda_{{\rm MSR},4}$, the mass segregation ratio (and uncertainties) for the ten most massive stars, $\Lambda_{{\rm MSR},10}$, the relative surface density ratio for the ten most massive stars, $\Sigma_{\rm LDR}$, and its associated $p$-value, and the relative potential difference ratio of the ten most massive stars, $\Phi_{\rm PDR}$, and its associated $p$-value.}
\begin{center}
\begin{tabular}{cccccccccc}
\hline 
 Simulation ID & Feedback & $\Lambda_{{\rm MSR},4}$  & $\Lambda_{{\rm MSR},10}$ & $\Sigma_{\rm LDR}$ & $p$-value & $\Phi_{\rm PDR}$ & $p$-value \\
\hline
J & None  & $0.41^{+0.54}_{-0.22}$ & $0.76^{+1.02}_{-0.64}$ & 15.3 & $5 \times 10^{-5} $ & 1.03 & $3 \times 10^{-2}$ \\
J & Photoionisation + wind & $1.69^{+2.01}_{-0.96}$ & $1.16^{+1.61}_{-1.10}$ & 0.60 & 0.79  & 1.01  & 0.24 \\
\hline
I & None & $125^{+165}_{-49}$ & $1.55^{+2.31}_{-1.28}$ & 5.22 & $4 \times 10^{-2}$ & 1.11 & $8 \times 10^{-4}$ \\
I & Photoionisation + wind & $1.85^{+2.24}_{-1.39}$ & $0.85^{+0.94}_{-0.68}$ & 0.25 & 0.39 & 1.03 & 0.40 \\
\hline
UF & None & $0.33^{+0.63}_{-0.07}$ & $1.08^{+1.39}_{-0.65}$ & 0.03 & 0.19 & 1.05 & $8 \times 10^{-3}$ \\
UF &  Photoionisation + wind & $11.1^{+13.6}_{-7.6}$ & $1.00^{+1.13}_{-0.85}$ & 1.04 & 0.60 & 1.06 & 0.26 \\
\hline
UP & None & $0.46^{+0.63}_{-0.09}$ & $0.88^{+1.12}_{-0.72}$ & 0.72 & 0.43 & 1.09 & $8 \times 10^{-4}$ \\
UP & Photoionisation + wind & $0.53^{+0.68}_{-0.20}$ & $0.84^{+1.12}_{-0.69}$ & 0.57 & 0.78 & 1.07 & $6 \times 10^{-2}$ \\
\hline
UQ & None & $0.36^{+0.59}_{-0.06}$ & $0.67^{+0.90}_{-0.54}$ & 11.1 & $2 \times 10^{-2}$ & 1.21 & $2 \times 10^{-2}$ \\
UQ & Photoionisation + wind & $2.10^{+2.38}_{-1.33}$ & $0.86^{+0.93}_{-0.67}$  & 3.92 & 0.32 & 1.11 & $1 \times 10^{-2}$ \\
\hline
R11O & None & $0.69^{+1.08}_{-0.37}$ & $0.93^{+1.24}_{-0.81}$ & 0.52 &  0.96 & 1.08 & $1 \times 10^{-4}$ \\
R11O & Photoionisation + wind & $1.33^{+1.62}_{-0.66}$ & $1.29^{+1.61}_{-0.99}$ & 0.89 & 0.91 & 1.00 & 0.77 \\
\hline
R15L & None & $0.79^{+1.09}_{-0.64}$ & $1.17^{+1.41}_{-0.64}$ & 8.11 & $5 \times 10^{-2}$ & 1.10 & $1 \times 10^{-6}$ \\
R15L & Photoionisation + wind & $2.04^{+2.78}_{-1.53}$ & $1.16^{+1.31}_{-1.00}$ & 1.06 & 0.41 & 1.13 & $1 \times 10^{-2}$ \\
\hline
R15M & None & $1.45^{+2.36}_{-0.59}$ & $1.81^{+2.31}_{-1.41}$ & 7.97 & $2 \times 10^{-3}$ & 1.11 & $4 \times 10^{-5}$ \\
R15M & Photoionisation + wind & $1.35^{+1.72}_{-0.75}$ & $1.71^{+1.83}_{-1.24}$ & 2.93 & 0.15 & 1.10 & $7 \times 10^{-7}$ \\
\hline
R19S & None & $1.14^{+2.06}_{-0.31}$  & $1.18^{+1.59}_{-0.72}$ & 4.18 & $9 \times 10^{-2}$ & 1.10 & $1 \times 10^{-2}$ \\
R19S & Photoionisation + wind & $1.00^{+1.25}_{-0.61}$ & $1.00^{+1.16}_{-0.82}$  &  1.82 & 0.60 & 1.08 & $3 \times 10^{-2}$ \\
\hline
R19T & None & $3.18^{+4.79}_{-0.78}$  & $1.46^{+1.68}_{-0.79}$  & 2.55  & 0.25  &  1.11 & $6 \times 10^{-5}$ \\
R19T & Photoionisation + wind &  $1.46^{+2.00}_{-0.62}$  & $1.15^{+1.38}_{-0.86}$  & 0.97  & 0.56  & 1.10  & $2 \times 10^{-4}$ \\
\hline
\end{tabular}
\end{center}
\label{simulation_results}
\end{table*}

\begin{figure*}
  \begin{center}
\setlength{\subfigcapskip}{10pt}
\hspace*{-1.5cm}\subfigure[SPH simulations without feedback]{\label{SPH_sim_results-a}\rotatebox{270}{\includegraphics[scale=0.35]{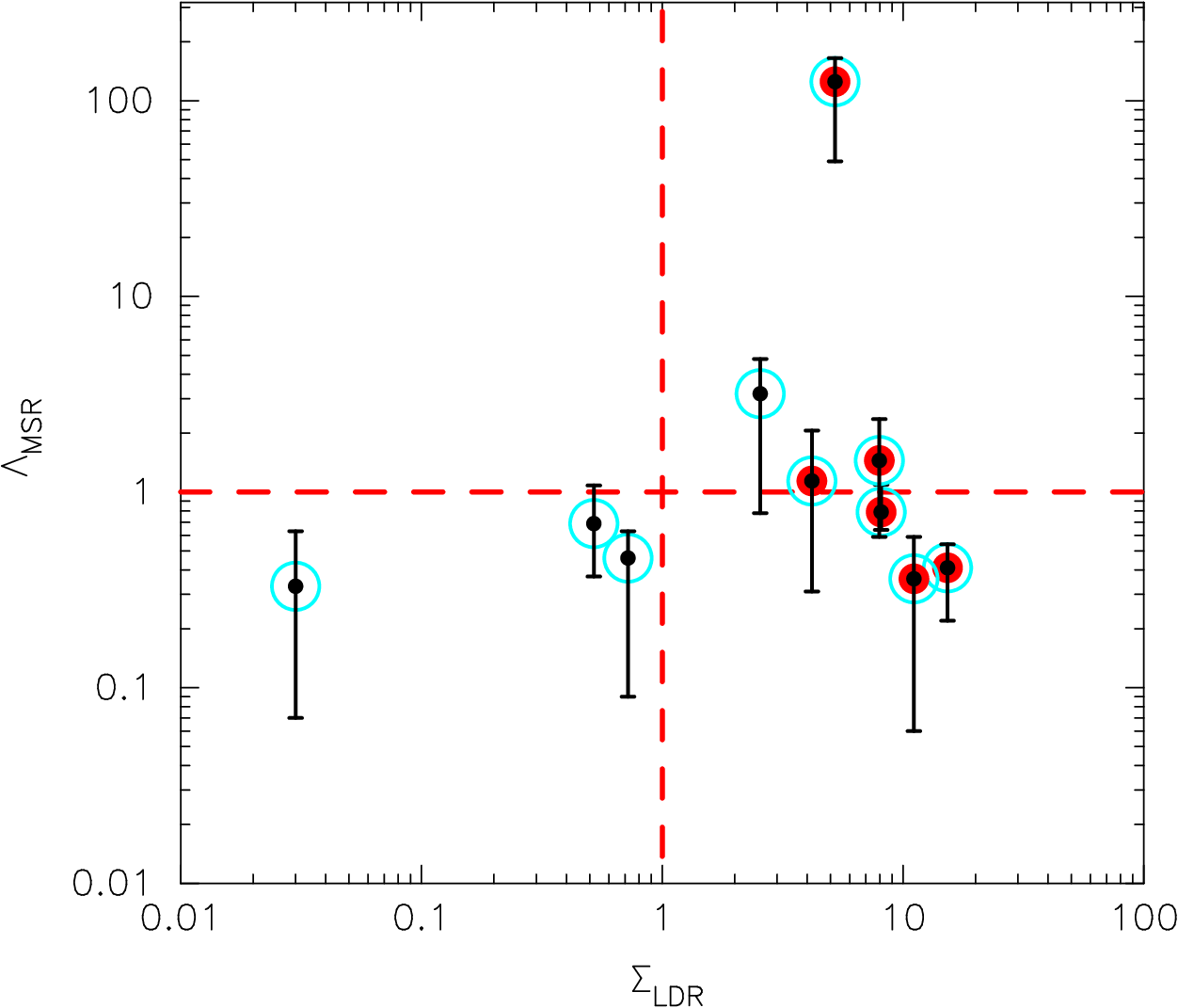}}}
\hspace*{0.3cm} 
\subfigure[SPH simulations with feedback]{\label{SPH_sim_results-b}\rotatebox{270}{\includegraphics[scale=0.35]{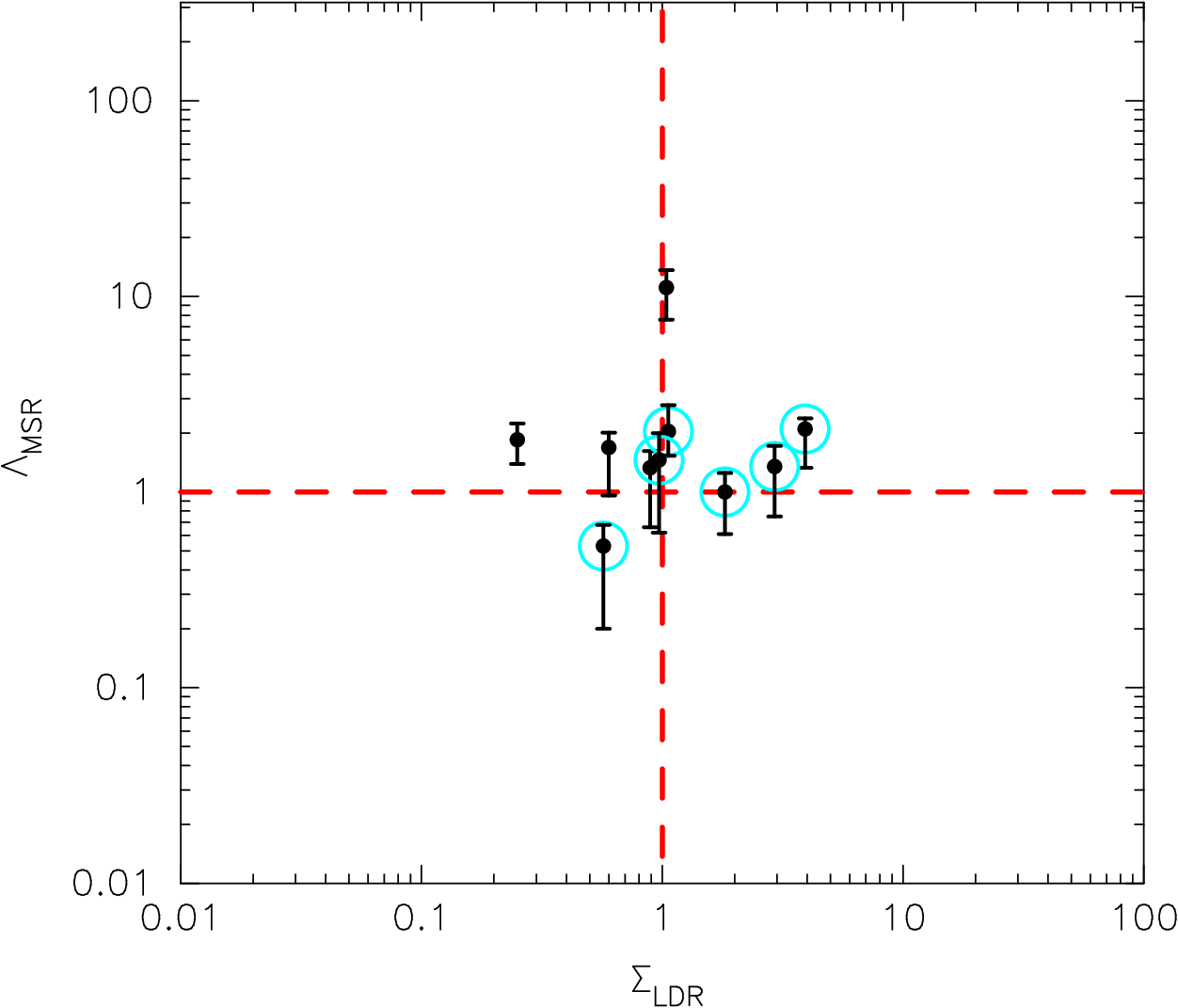}}}
\caption{A summary of the results from our full suite of SPH simulations. In panel (a) we show the results from the control simulations, and in panel (b) we show the results for the SPH simulations with dual feedback (photoionising radiation and winds from massive stars). We show the mass segregation ratio for the four most massive stars, $\Lambda_{\rm MSR, 4}$, with uncertainties as defined in Eqn.~\ref{lambda_eqn}, against the local surface density ratio $\Sigma_{\rm LDR}$. When $\Sigma_{\rm LDR}$ significantly exceeds unity (i.e.\,\,the most massive stars are in areas of higher than average local densities) we plot a larger filled circle. Where the most massive stars sit in significantly deeper potential wells than the average mass stars, we also plot a larger open circle. The horizontal dashed line indicates $\Lambda_{\rm MSR} = 1$ (no mass segregation) and the vertical dashed line indicates $\Sigma_{\rm LDR} = 1$ (the massive stars are not in areas of higher or lower surface density). }
\label{SPH_sim_results}
  \end{center}
\end{figure*}

In Fig.~\ref{SPH_sim_results_control} we show the results for the SPH control simulation. Panel (a)  shows the $x-y$ projection, with the ten most massive stars shown by the red points. The relative size of the red point indicates where it resides in the list of the most massive stars, with the largest circle indicating the most massive star, and the smallest circle indicating the tenth most massive star.

In panel (b) we show the gravitational potential of the individual stars, as a function of their distance from the origin along the $x$-axis. Again, the red points show the locations of the most massive stars.

The most massive stars appear to sit in deeper potentials than most of the lower mass stars, and this is confirmed in panel (c), which shows the gravitational potential as a function of stellar mass. The purple dashed line indicates the median potential for all stars in the simulation, and the solid red line indicates the median potential of the ten most massive stars. A KS test between the entire sample of stars and the subset of the ten most massive stars returns a $p$-value of $4 \times 10^{-5}$, meaning we can reject the null hypothesis that they share the same underlying parent distribution.

The $\Lambda_{\rm MSR}$ mass segregation ratio is shown in Fig.~\ref{SPH_sim_results_control-d}. We plot the evolution of  $\Lambda_{\rm MSR}$ for the $N_{\rm MST}$ most massive stars, starting with $N_{\rm MST} = 4$ and increasing the numbers of stars in the most massive subset. In no subsets do we detect a significant deviation from unity, i.e.\,\,this simulation is not mass segregated. In our ten sets of SPH control simulations, only one (Run I) displays significant mass segregation, and then only for the four most massive stars. Inspection of Fig.~\ref{I180_sim_results_dual} suggests the reason for this is that seven of the most massive stars sit within the same subcluster in the simulation.

The most massive stars do, however, reside in areas of significantly higher stellar densities than the average stars in the simulation. This is shown in Fig.~\ref{SPH_sim_results_control-e} where we plot the local surface density around each star as a function of its mass. The median surface density for the cluster is shown by the black dashed line, and the median surface density for the ten most massive stars is shown by the solid red line. The surface density ratio, $\Sigma_{\rm LDR} = 7.97$, and a KS test between the densities for all stars and the densities of the most massive subset returns a $p$-value of $2 \times 10^{-3}$, suggesting we can reject the hypothsesis that they share the same underlying parent distribution.

The simulation shown in Fig.~\ref{SPH_sim_results_control} is broadly representative of all of the SPH control run simulations; in all of the control run simulations the most massive stars sit in deeper gravitational potentials, in 6/10 simulations they have significantly higher than average surface densities, but are not mass-segregated (i.e.\,\,the massive stars are not closer together than the average mass stars -- the exception being Run I). In one simulation (UF) the relative surface density of the most massive stars is very low ($\Sigma_{\rm LDR}$ = 0.03), although this is not statistically significant due to the very small total number of stars in this simulation ($N_{\rm stars} = 66$).

\subsubsection{SPH dual feedback run}

In Fig.~\ref{SPH_sim_results_dual} we present the same SPH simulation (R15M), but the version in which feedback from photoionising radiation, and the stellar winds,  from massive stars, is implemented. The visual differences between simulations with feedback, compared to those without, are only obvious when looking at the column densities of gas, and we refer the reader to \citet{Dale14} and \citet{Dale17} for examples of these.

We show the projection in the $x-y$ plane in Fig.~\ref{SPH_sim_results_dual-a}, where the ten most massive stars are shown by the red points. We then show the gravitational potential as a function of the distance from the centre of the star-forming region along the $x$-axis in Fig.~\ref{SPH_sim_results_dual-b}. We quantify the depth of the gravitational potential for all stars compared to the most massive stars in Fig.~\ref{SPH_sim_results_dual-c}, where we show the gravitational potential for each star as a function of the individual mass of the stars. The median potential for all stars is shown by the purple dashed line, and the median potential for the ten most massive stars is shown by the solid red line. As with the control run, the median potential of the most massive stars is larger than the median for all stars, and a KS test between the two samples returns a $p$-value of $7 \times 10^{-7}$ that they share the same underlying parent distribution. The massive stars sit in significantly deeper potentials in 6/10 simulations with feedback, in contrast to the control runs, where the massive stars sit in deeper potentials in all runs.

The mass segregation ratio, $\Lambda_{\rm MSR}$ is shown in Fig.~\ref{SPH_sim_results_dual-d}. There is some deviation from unity; however \citet{Parker15b} show in synthetic data that such deviations could be random, and recommend a threshold of $\Lambda_{\rm MSR} > 2$ for significantly high mass segregation ratios, which is not met in this simulation.

In contrast to the control run without feedback, in this simulation with feedback the most massive stars do not reside in areas of significantly higher than average stellar surface densities (Fig.~\ref{SPH_sim_results_dual-e}). The dashed black line indicates the median surface density for all stars, and the solid red line indicates the median surface density for the subset of the ten most massive stars. The surface density ratio is $\Sigma_{\rm LDR} = 2.93$, but a KS test between all of the stars, and the most massive, returns a $p$-value of 0.15, meaning we cannot reject the hypothesis that the two datasets share the same underlying parent distribution. The massive stars do not attain significantly higher surface densities in \emph{any} of the simulations with feedback, compared to 6/10 of the control runs. \\

In total we have ten pairs of SPH simulations (one set of control runs, and one set of runs with feedback). In Fig.~\ref{SPH_sim_results} we plot the mass segregation ratio for the four most massive stars $\Lambda_{\rm MSR, 4}$ and its associated uncertainties with the error bars, against the surface density ratio, $\Sigma_{\rm LDR}$, for all ten pairs. Where the surface density ratio is significantly above unity (as defined by a KS test between the most massive stars and all stars in the simulation, where the $p$-value is less than 0.1), we plot a filled red circle. If the most massive stars sit significantly deeper in the gravitational potential (again, as defined by the KS test $p$-value $p<0.1$) then we plot a larger cyan circle.

In the control run simulations (Fig.~\ref{SPH_sim_results-a}), the massive stars \emph{always} sit in deeper potentials, which is as predicted by the earliest incarnations of the competitive accretion theory \citep{Zinnecker82}. However, in only half of these simulations are the massive stars in areas of higher than average stellar density, and only in one region are the most massive stars mass segregated, i.e.\,\,closer to each other than the average stars.

When feedback is implemented in the simulations (Fig.~\ref{SPH_sim_results-b}), the massive stars may or may not sit in deeper potentials (they do in half of these simulations). The main action of feedback in the context of the spatial distribution of massive stars is that they no longer reside in areas of high stellar surface densities, and only in one simulation (UF) do we see  a high degree of mass segregation.

\subsection{Synthetic datasets}

We now use `synthetic' star-forming regions to determine the spatial distributions of massive stars, including their gravitational potentials with respect to lower-mass stars, their relative surface densities, and the degree of mass segregation, for three different scenarios.

In the first, we assume the stars (including the most massive) are distributed randomly. In the second, we assume that the stars form in the most central locations, and we switch the positions of the most massive stars with the stars that are closest to the centre of the star-forming region. In the third scenario, we assume that the massive stars form in areas of high relative surface density, and we switch the positions of the most massive stars with the stars of highest surface density.

\begin{figure*}
  \begin{center}
    \setlength{\subfigcapskip}{10pt}
        \vspace*{-0.5cm}
\hspace*{-1.5cm}\subfigure[Plummer, random, $x-y$]{\label{synthetic_plummer-a}\rotatebox{270}{\includegraphics[scale=0.25]{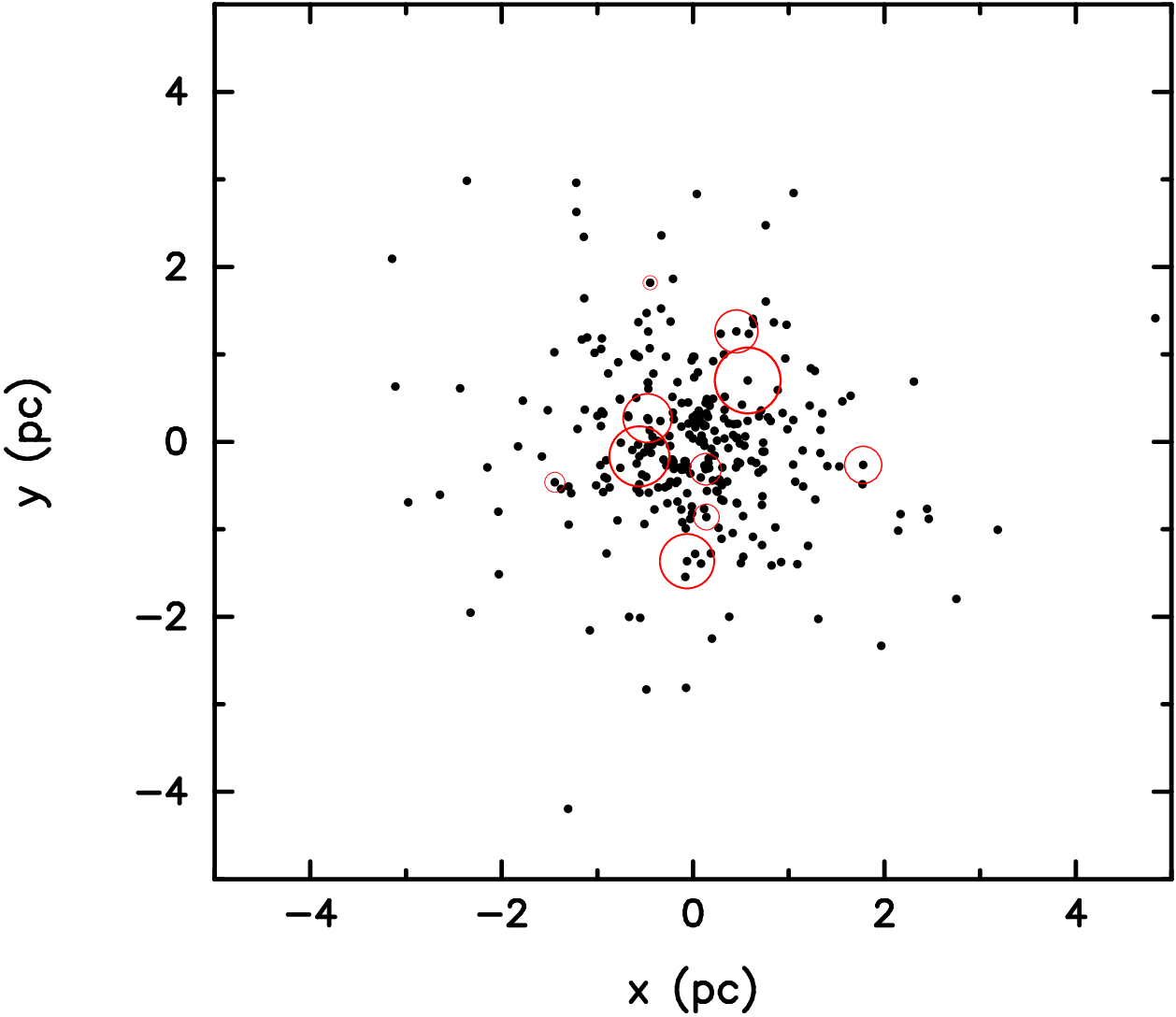}}}
\hspace*{0.3cm} 
\subfigure[Plummer, cent. conc., $x-y$]{\label{synthetic_plummer-b}\rotatebox{270}{\includegraphics[scale=0.25]{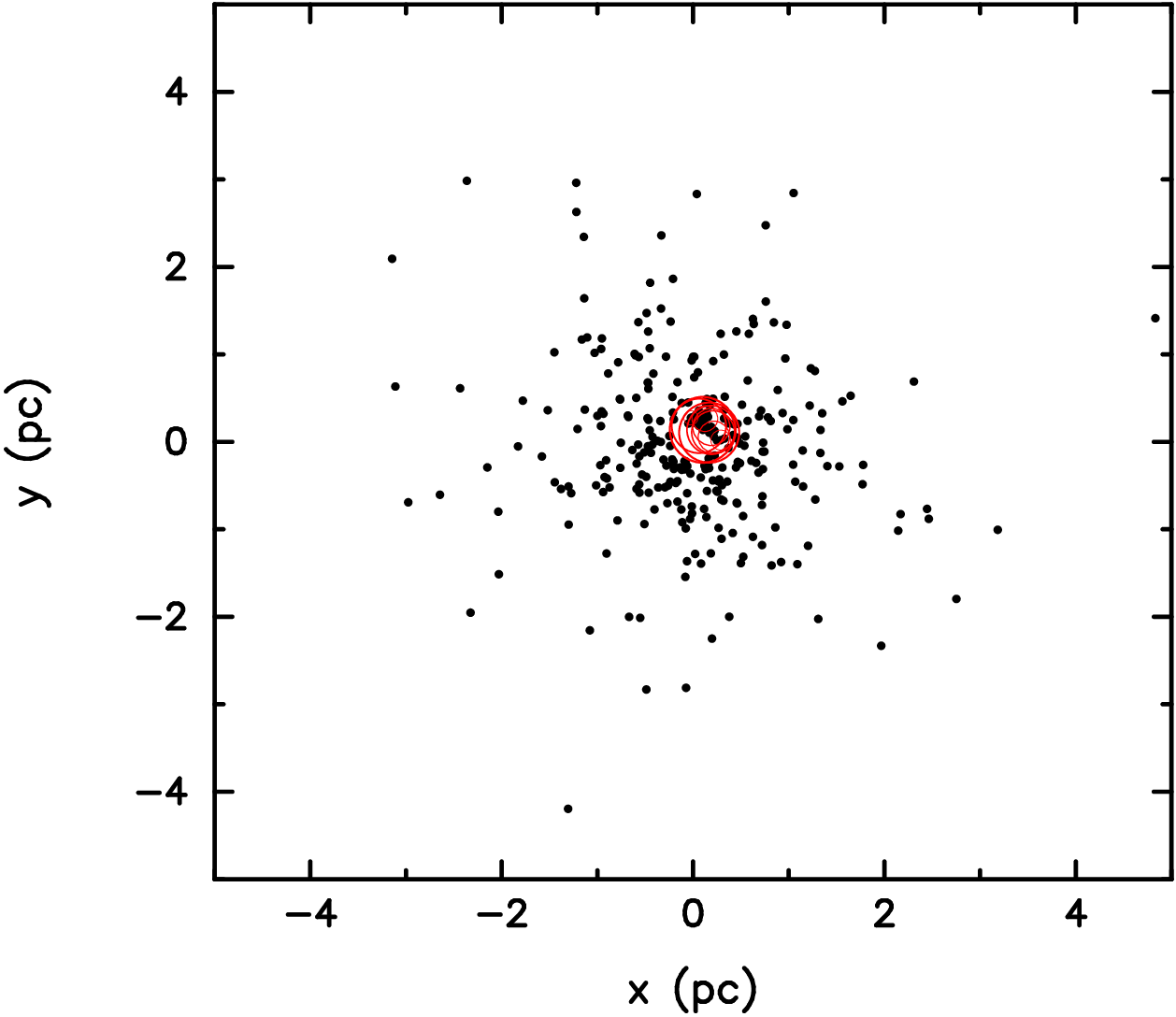}}}
\hspace*{0.3cm}
\subfigure[Plummer, high dens., $x-y$]{\label{synthetic_plummer-c}\rotatebox{270}{\includegraphics[scale=0.25]{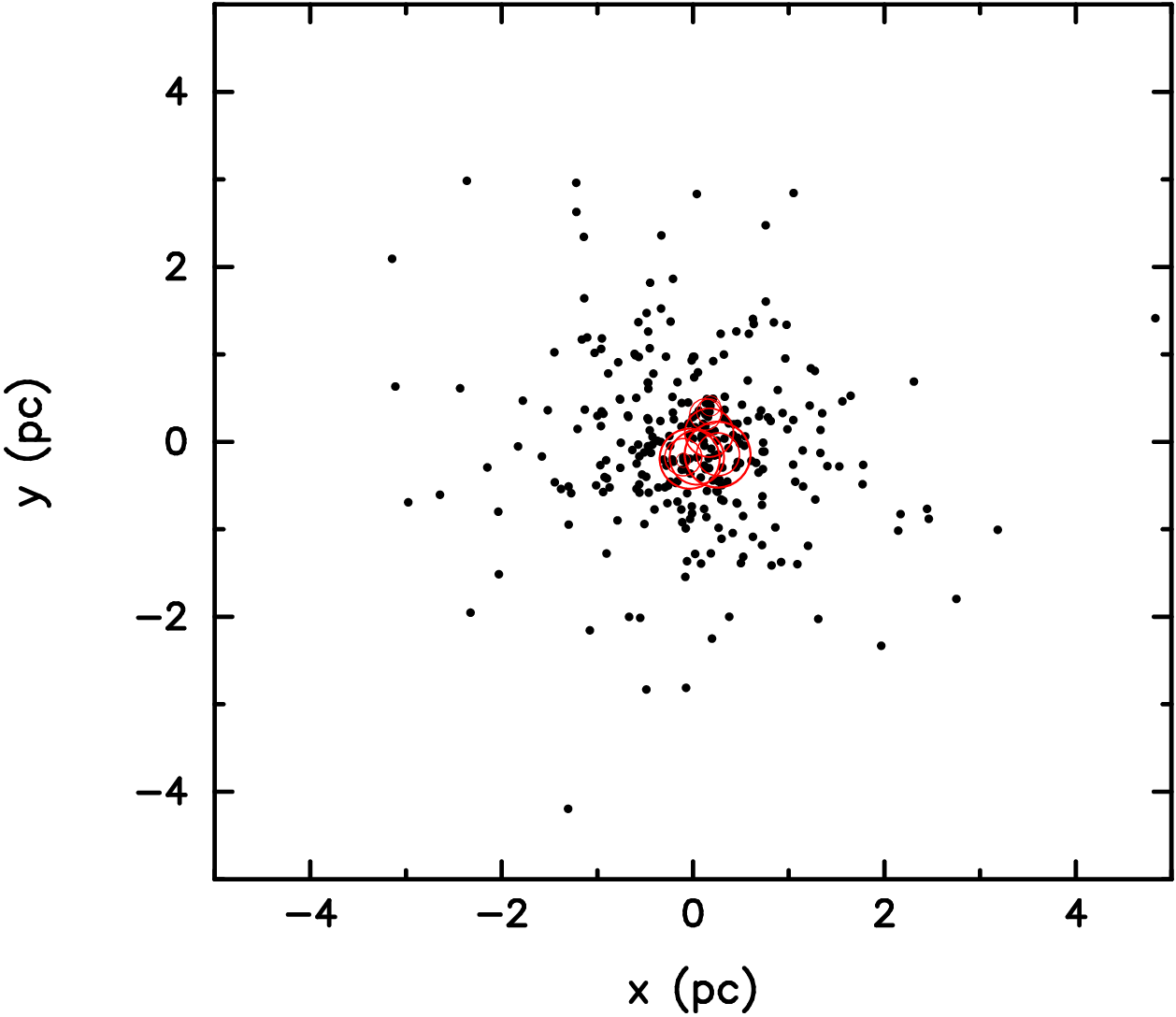}}}

\hspace*{-1.5cm}\subfigure[Plummer, random, potential]{\label{synthetic_plummer-d}\rotatebox{270}{\includegraphics[scale=0.25]{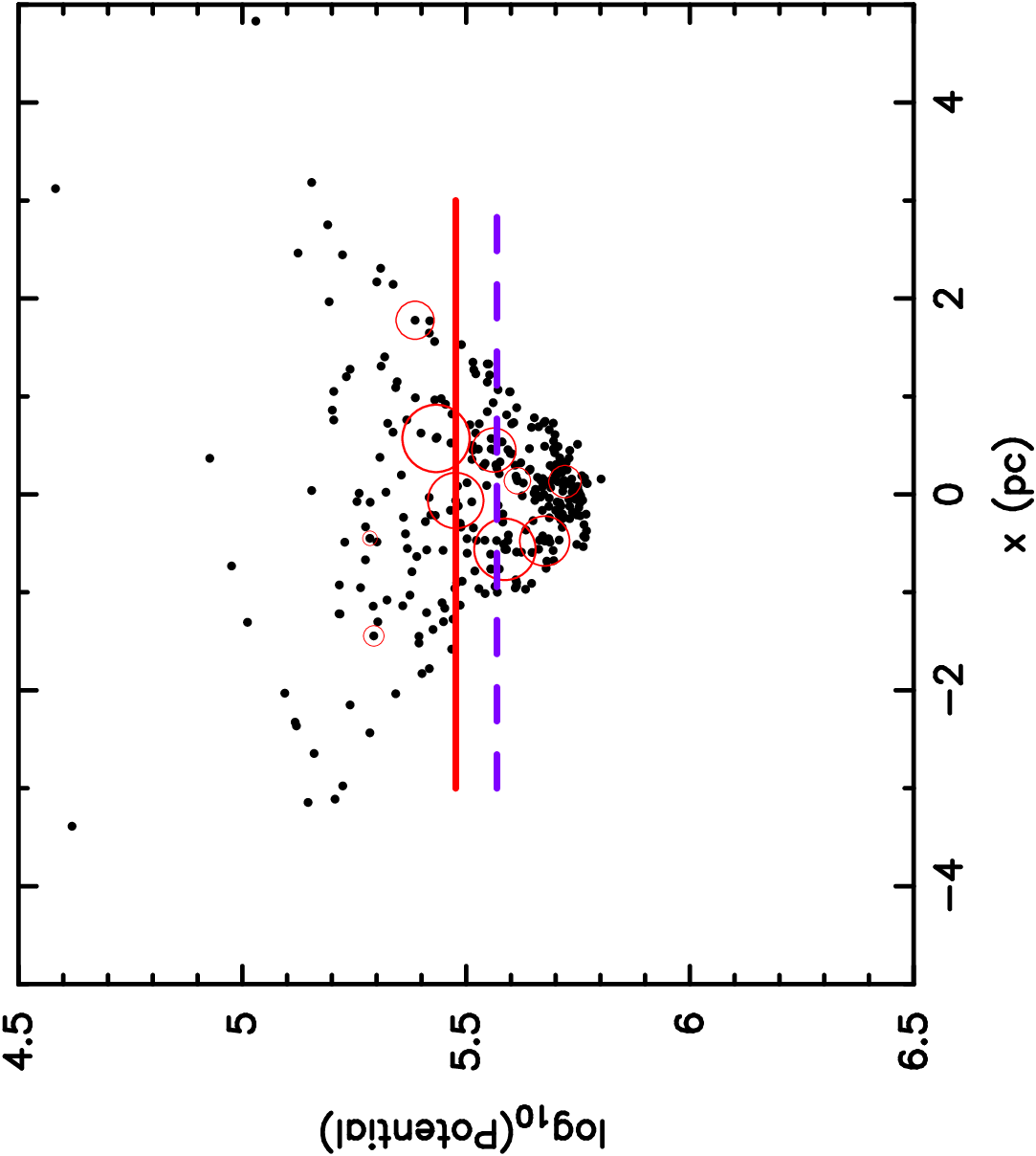}}}
\hspace*{0.3cm} 
\subfigure[Plummer, cent. conc., potential]{\label{synthetic_plummer-e}\rotatebox{270}{\includegraphics[scale=0.25]{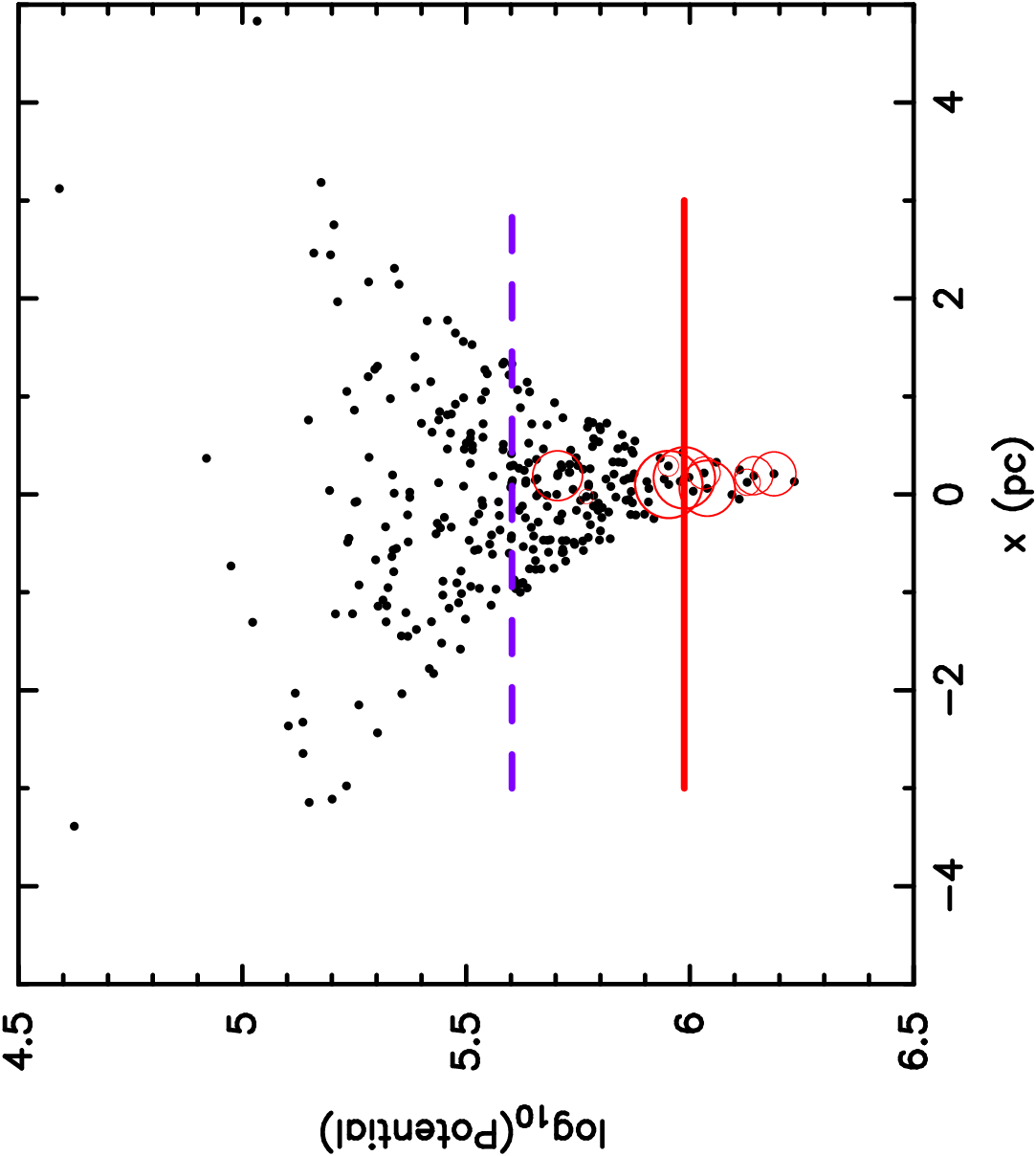}}}
\hspace*{0.3cm}
\subfigure[Plummer, high dens., potential]{\label{synthetic_plummer-f}\rotatebox{270}{\includegraphics[scale=0.25]{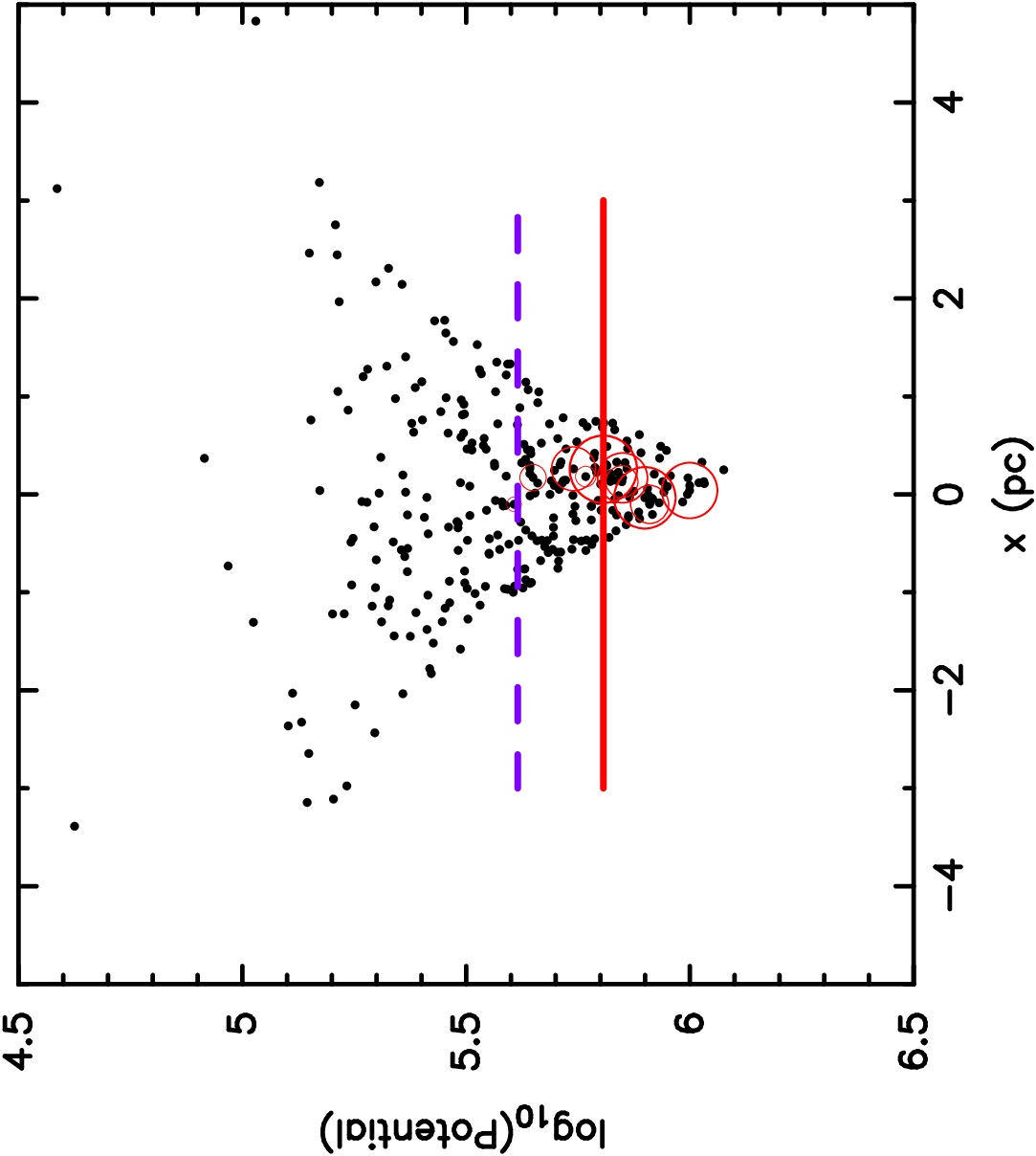}}}

\hspace*{-1.5cm}\subfigure[Plummer, random, $\Lambda_{\rm MSR}$]{\label{synthetic_plummer-g}\rotatebox{270}{\includegraphics[scale=0.23]{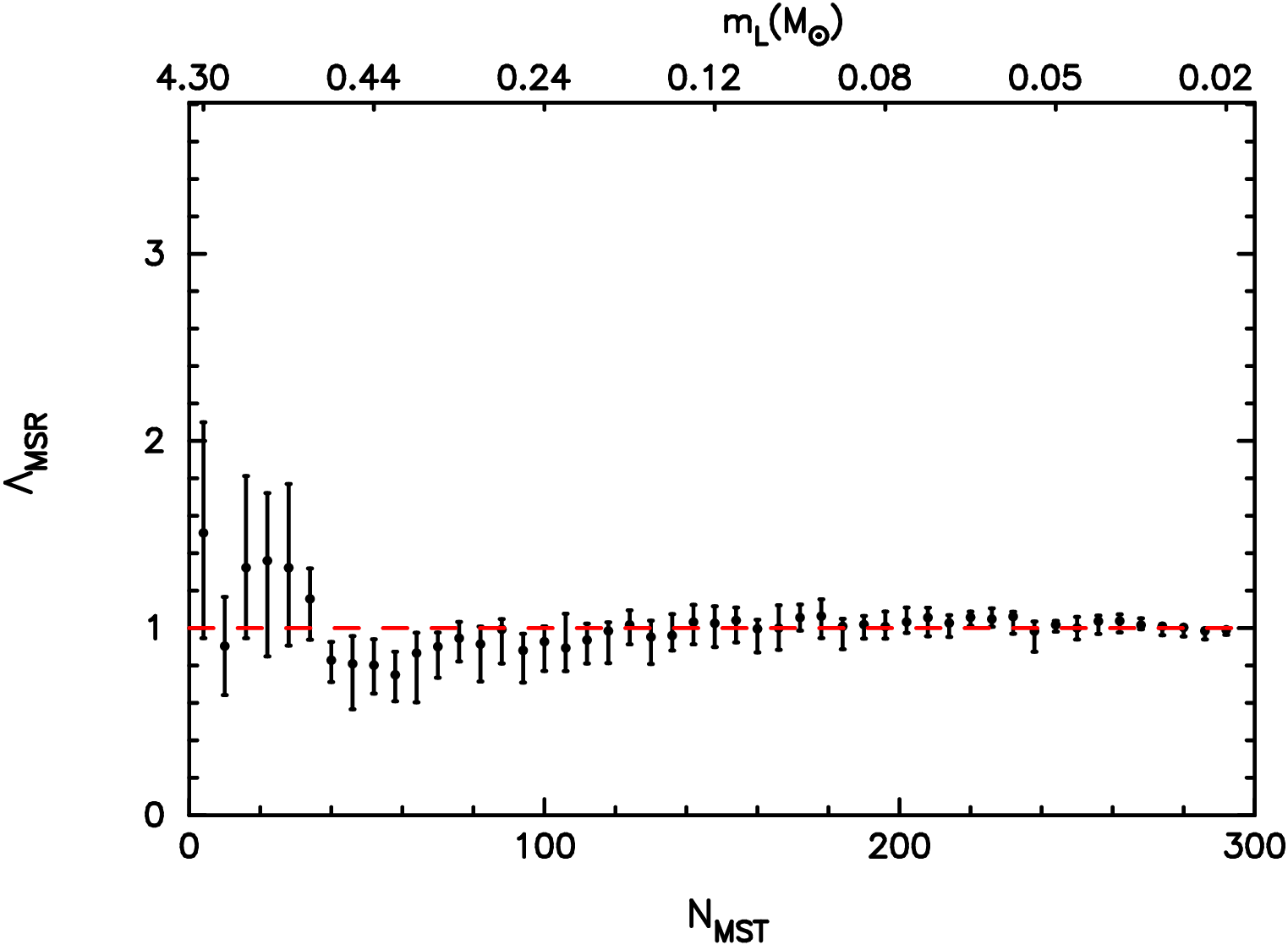}}}
\hspace*{0.1cm} 
\subfigure[Plummer, cent. conc.,  $\Lambda_{\rm MSR}$]{\label{synthetic_plummer-h}\rotatebox{270}{\includegraphics[scale=0.23]{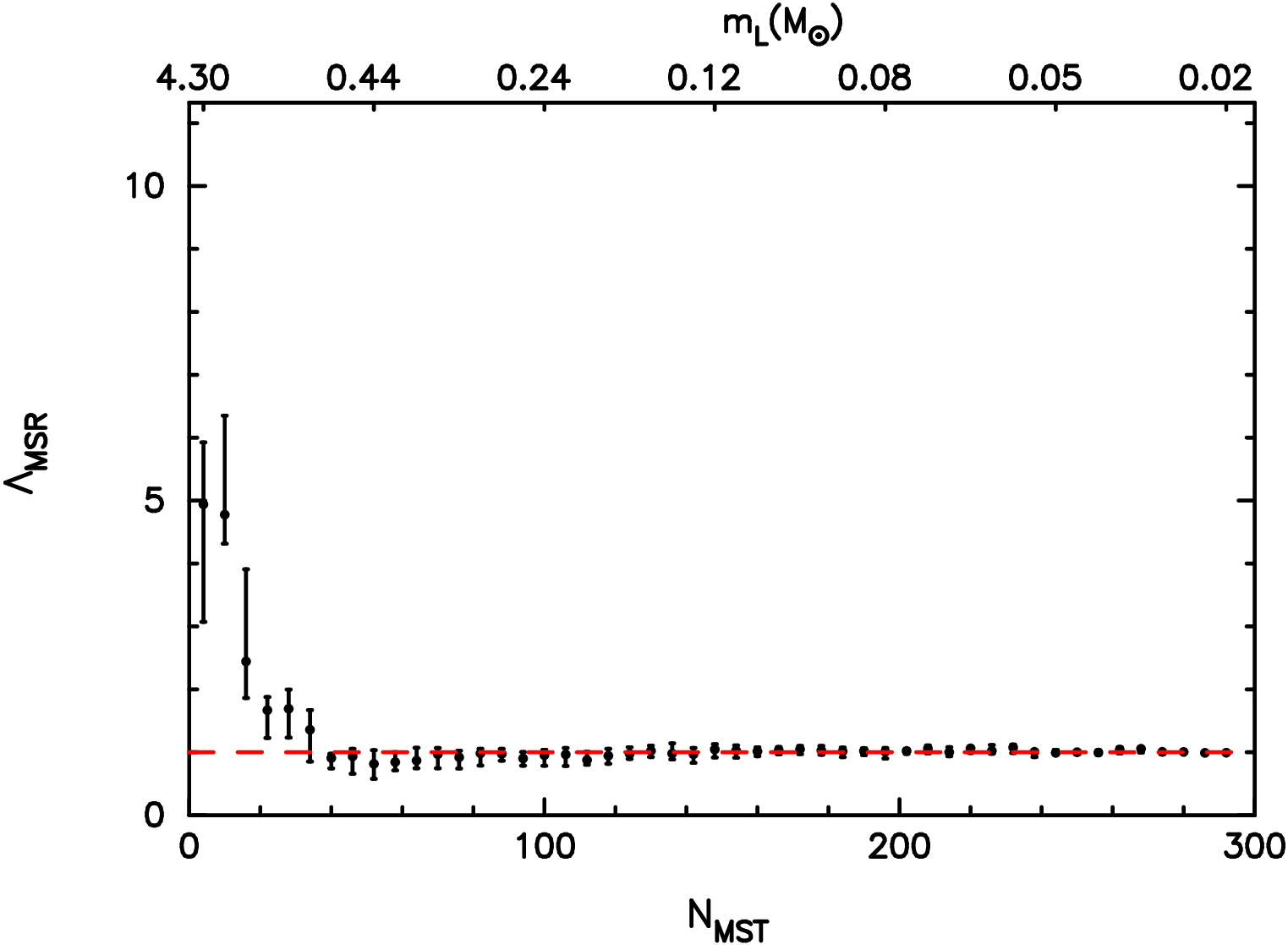}}}
\hspace*{0.1cm}
\subfigure[Plummer, high dens.,  $\Lambda_{\rm MSR}$]{\label{synthetic_plummer-i}\rotatebox{270}{\includegraphics[scale=0.23]{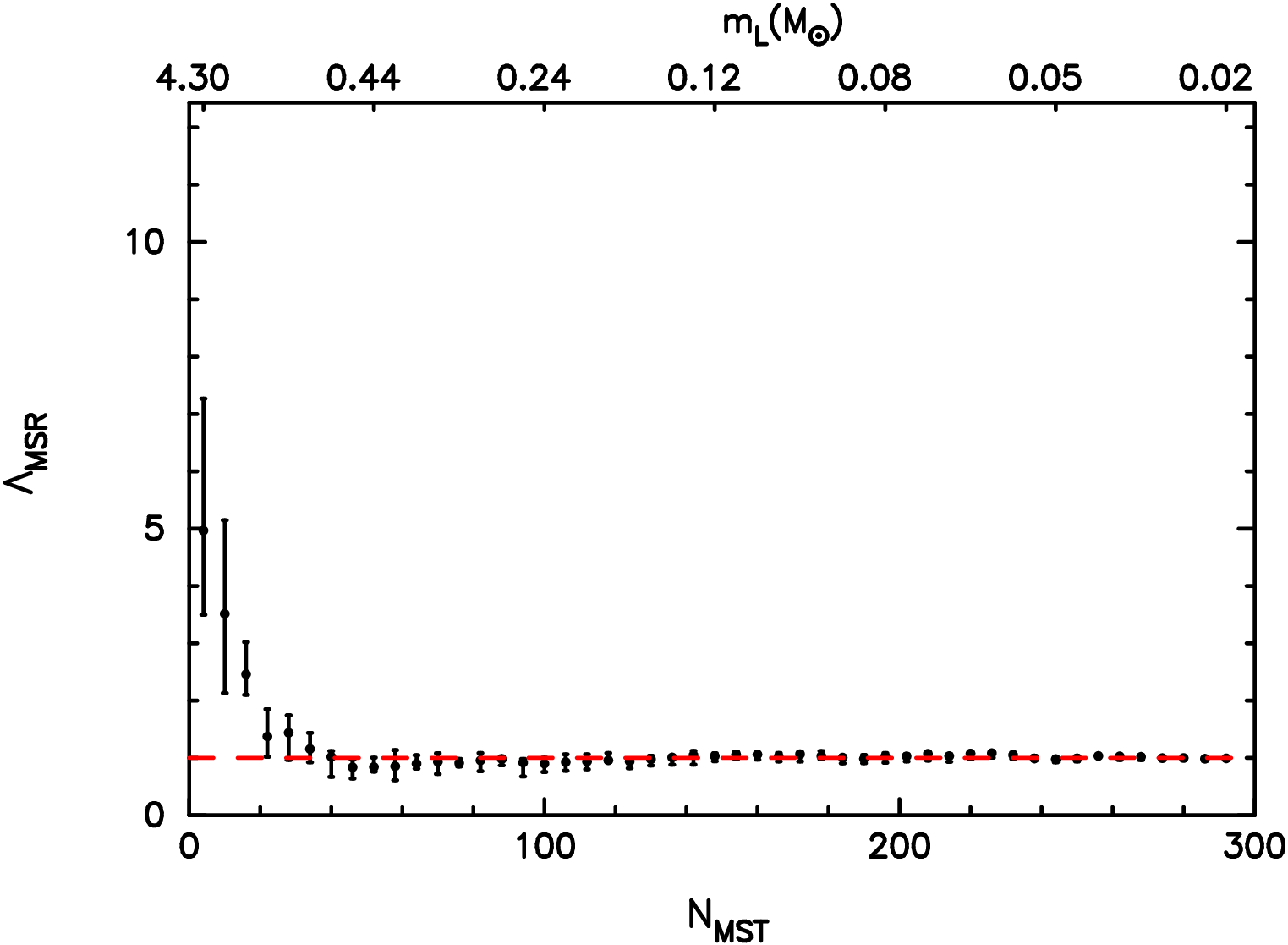}}}

\hspace*{-1.5cm}\subfigure[Plummer, random, $\Sigma_{\rm LDR}$]{\label{synthetic_plummer-j}\rotatebox{270}{\includegraphics[scale=0.25]{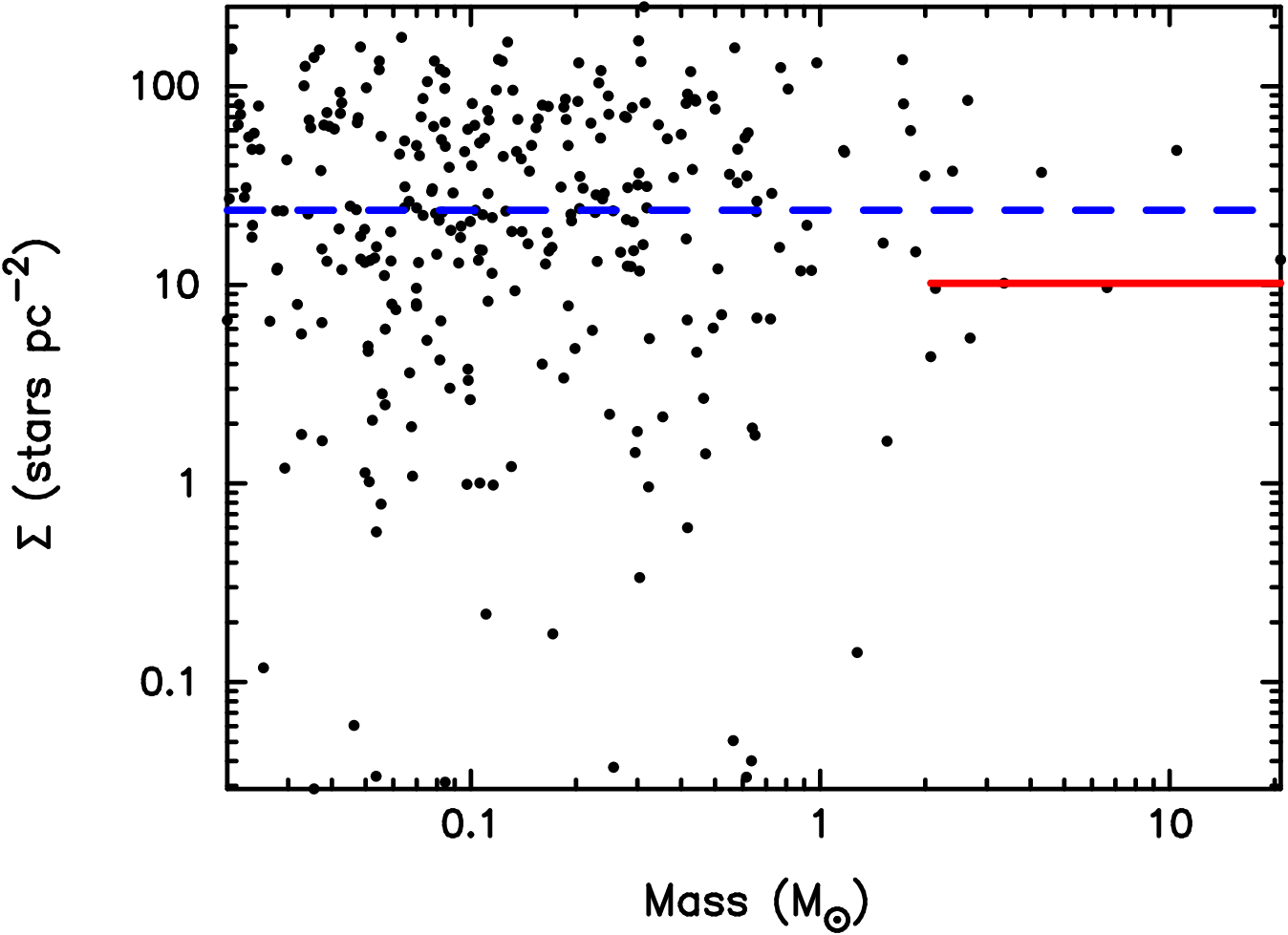}}}
\hspace*{0.3cm} 
\subfigure[Plummer, cent. conc.,  $\Sigma_{\rm LDR}$]{\label{synthetic_plummer-k}\rotatebox{270}{\includegraphics[scale=0.25]{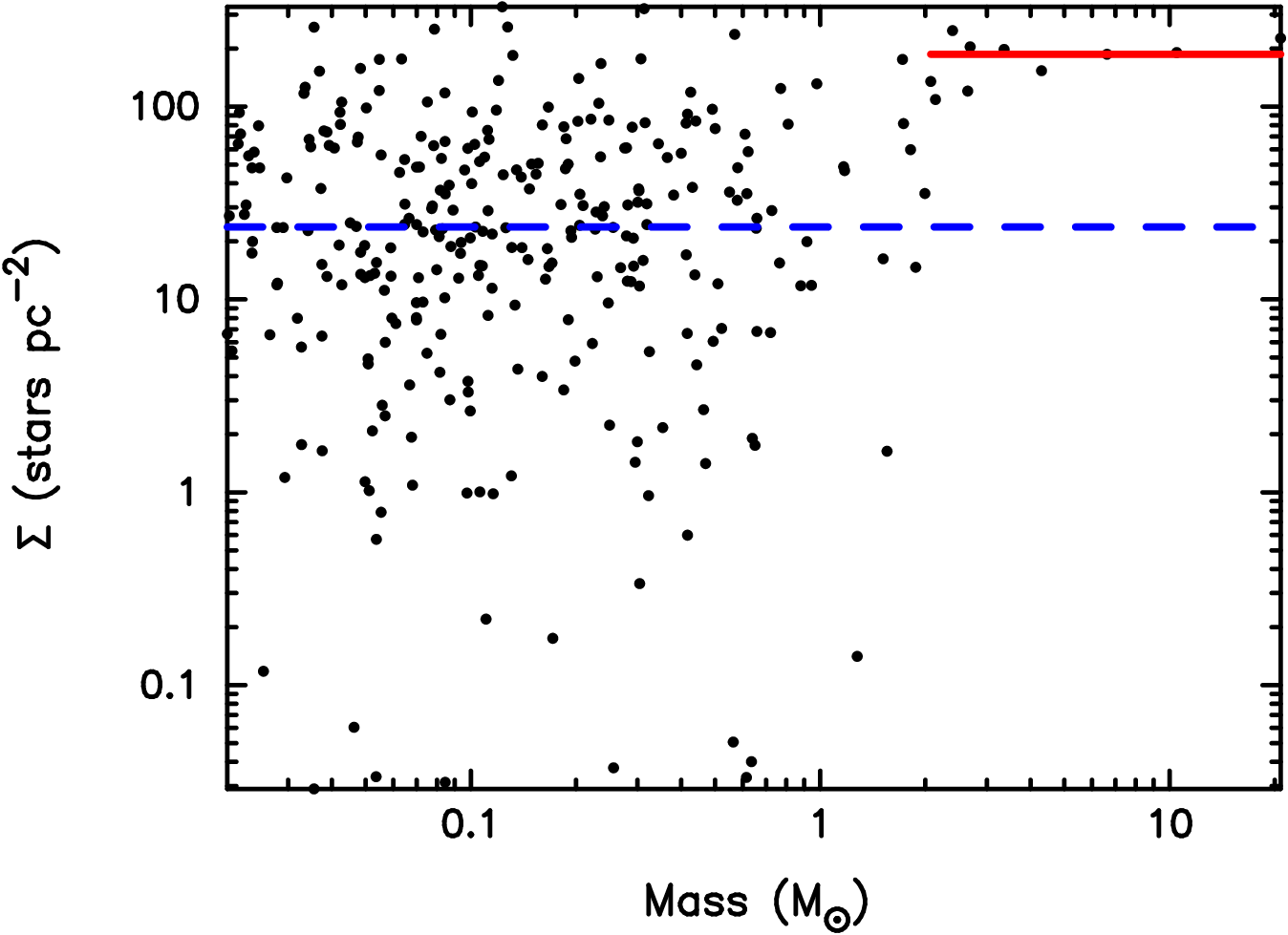}}}
\hspace*{0.3cm}
\subfigure[Plummer, high dens.,  $\Sigma_{\rm LDR}$]{\label{synthetic_plummer-l}\rotatebox{270}{\includegraphics[scale=0.25]{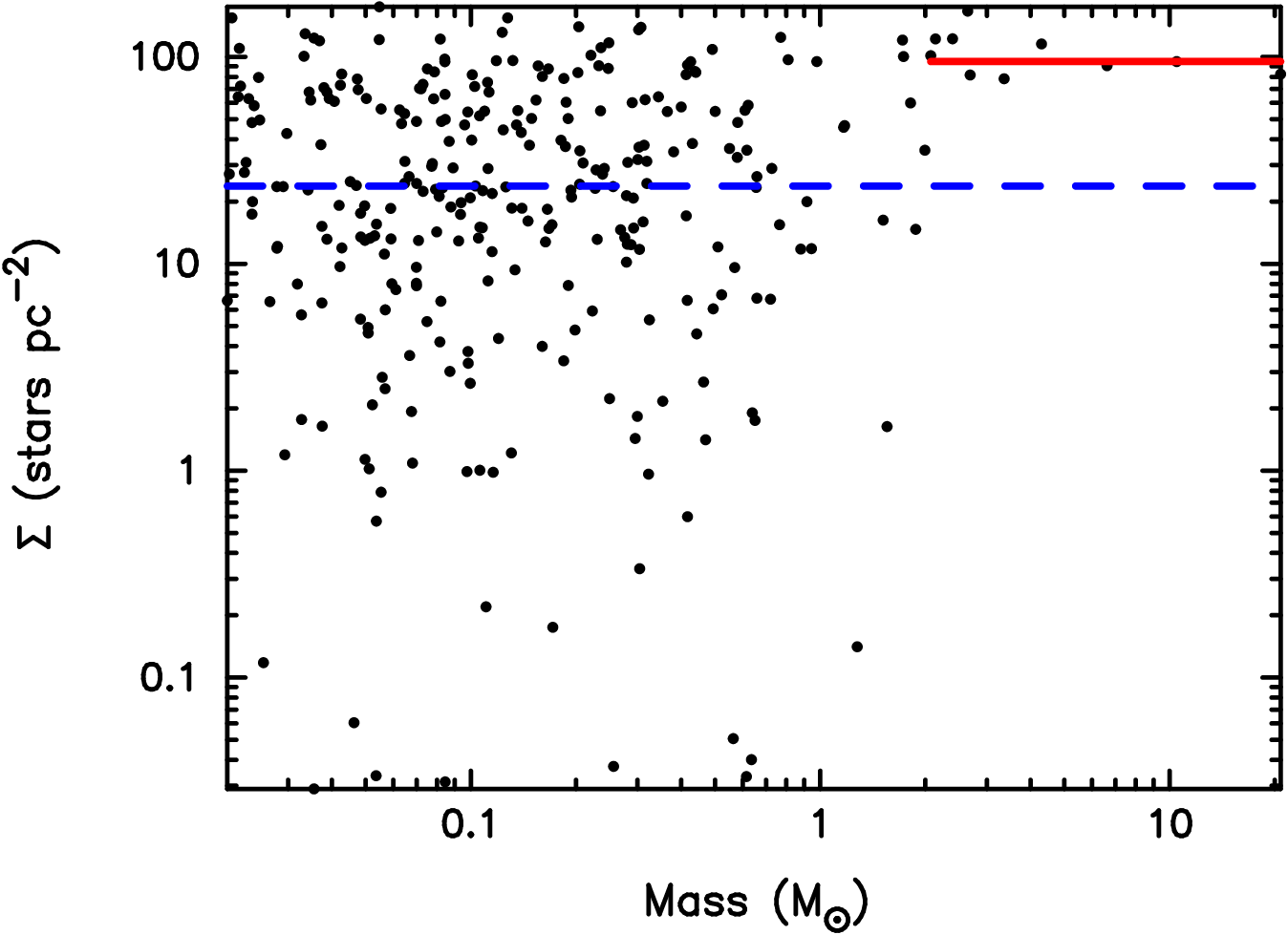}}}

\caption{A typical realisation of a synthetic Plummer sphere. The lefthand column shows the results when the ten most massive stars are distributed randomly, the middle column shows the results when the most massive stars are moved to the most central locations, and the righthand column shows the results when the massive stars are moved to the locations with the highest relative densities. The top row shows the $x-y$ positions, with the most massive stars shown in red. The relative size of the red point indicates where it resides in the list of the most massive stars, with the largest circle indicating the most massive star, and the smallest circle indicating the tenth most massive star. The second row shows the gravitational potential for each star, where the solid red line is the median potential for the most massive stars and the dashed purple line is the median potential for all stars. The fourth row shows the $\Lambda_{\rm MSR}$ mass segregation ratio, where $\Lambda_{\rm MSR} = 1$ is shown by the red dashed line. The bottom row shows the surface density for each star as a function of its mass; the solid red line indicates the median for the most massive stars and the blue dashed line shows the median for all stars.}
\label{synthetic_plummer}
  \end{center}
\end{figure*}

We do this for four different types of morphology; we start by adopting a Plummer sphere which has a smooth, centrally concentrated distribution. We then show the results for a substructured association, and then the results for a highly substructured box fractal with fractal dimension $D = 1.6$. We also analysed the results for a more moderately substructured box fractal with fractal dimension $D = 2.0$, but did not find any significant differences compared to the more substructured fractal, so we omit these results from the paper.

The results for a typical realisation of a Plummer sphere are shown in Fig.~\ref{synthetic_plummer}. In this figure, the first column shows the results for the random distribution of stars, the middle column shows the results for centrally concentrated massive stars, and the righthand column shows the results for massive stars in the locations of highest surface density.

Due to the morphology of the Plummer sphere, the spatial distributions of the massive stars are very similar when they are centrally concentrated as they are when the massive stars are in areas of highest surface density (as in effect the massive stars reside at the centre in both cases). The first row shows the $x-y$ projections, the second row shows the gravitional potential along the $x$-axis, the third row shows the $\Lambda_{\rm MSR}$ mass segregation ratio and the final row shows the $\Sigma-m$ plot, used to quantify differences in the relative densities of the most massive stars.

\begin{figure*}
  \begin{center}
    \setlength{\subfigcapskip}{10pt}
    \vspace*{-0.5cm}
\hspace*{-1.5cm}\subfigure[Association, random, $x-y$]{\label{synthetic_morphologies-a}\rotatebox{270}{\includegraphics[scale=0.25]{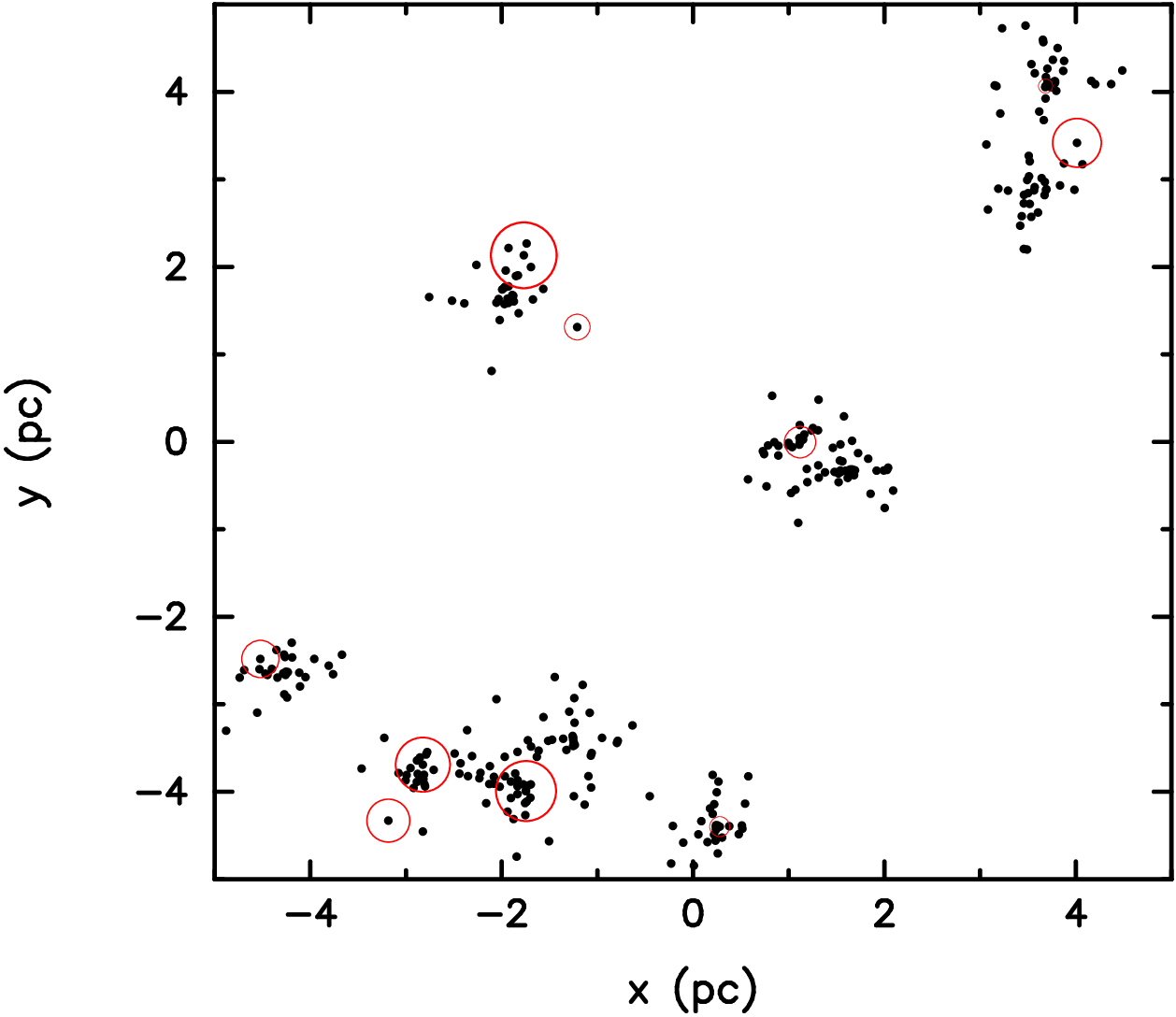}}}
\hspace*{0.3cm} 
\subfigure[Association, cent. conc., $x-y$]{\label{synthetic_morphologies-b}\rotatebox{270}{\includegraphics[scale=0.25]{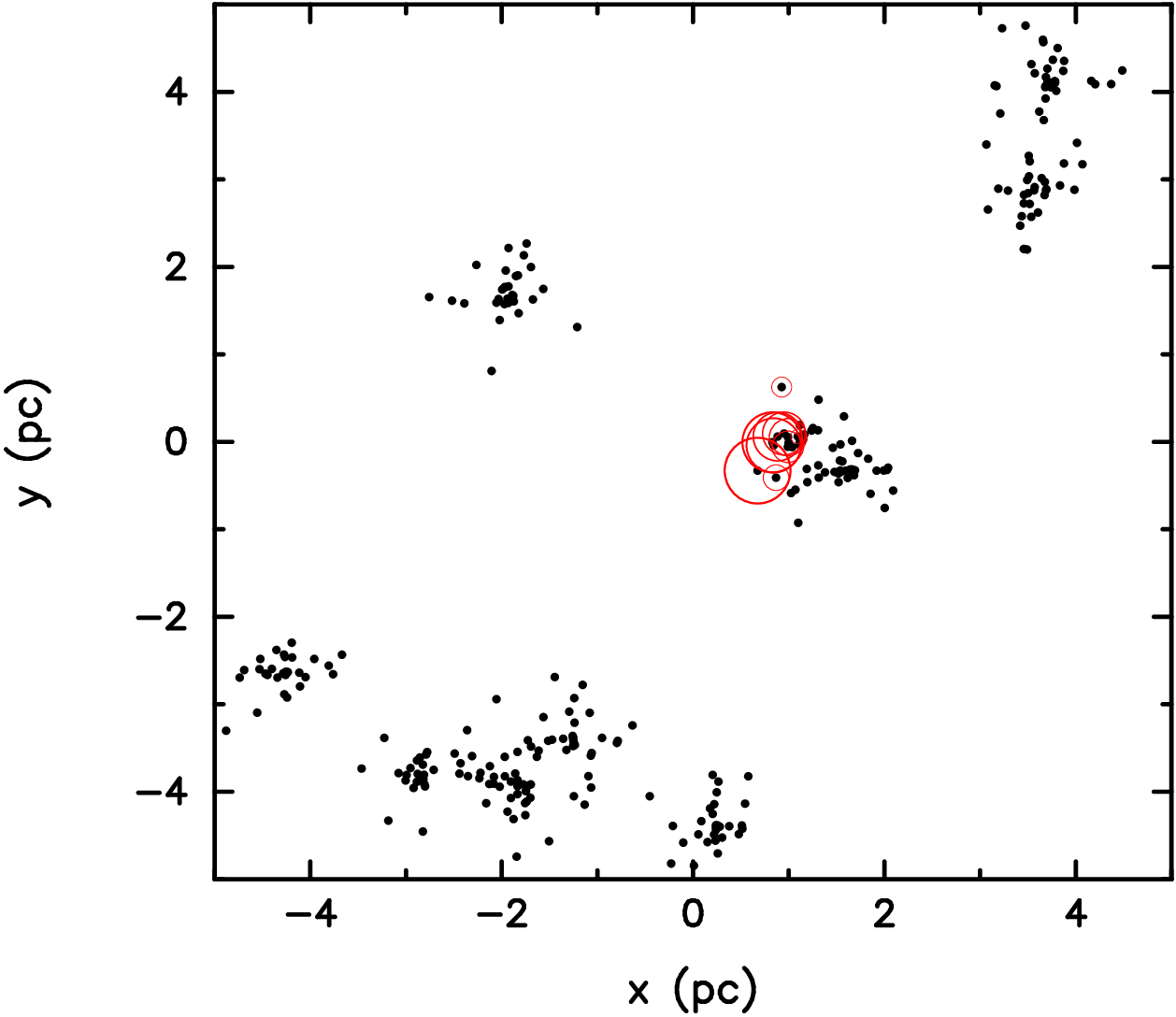}}}
\hspace*{0.3cm}
\subfigure[Association, high dens., $x-y$]{\label{synthetic_morphologies-c}\rotatebox{270}{\includegraphics[scale=0.25]{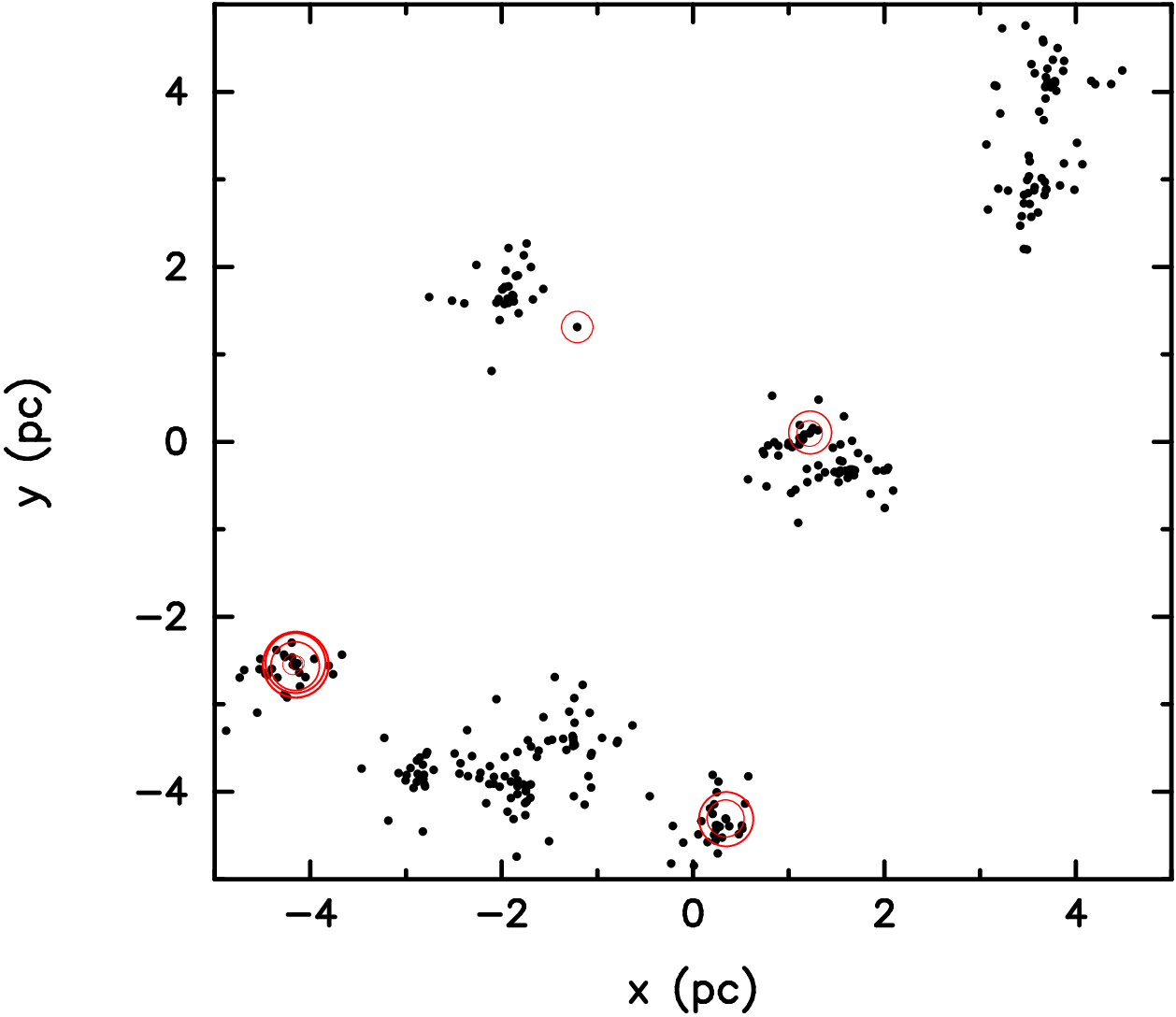}}}

\hspace*{-1.5cm}\subfigure[Association, random, potential]{\label{synthetic_morphologies-d}\rotatebox{270}{\includegraphics[scale=0.25]{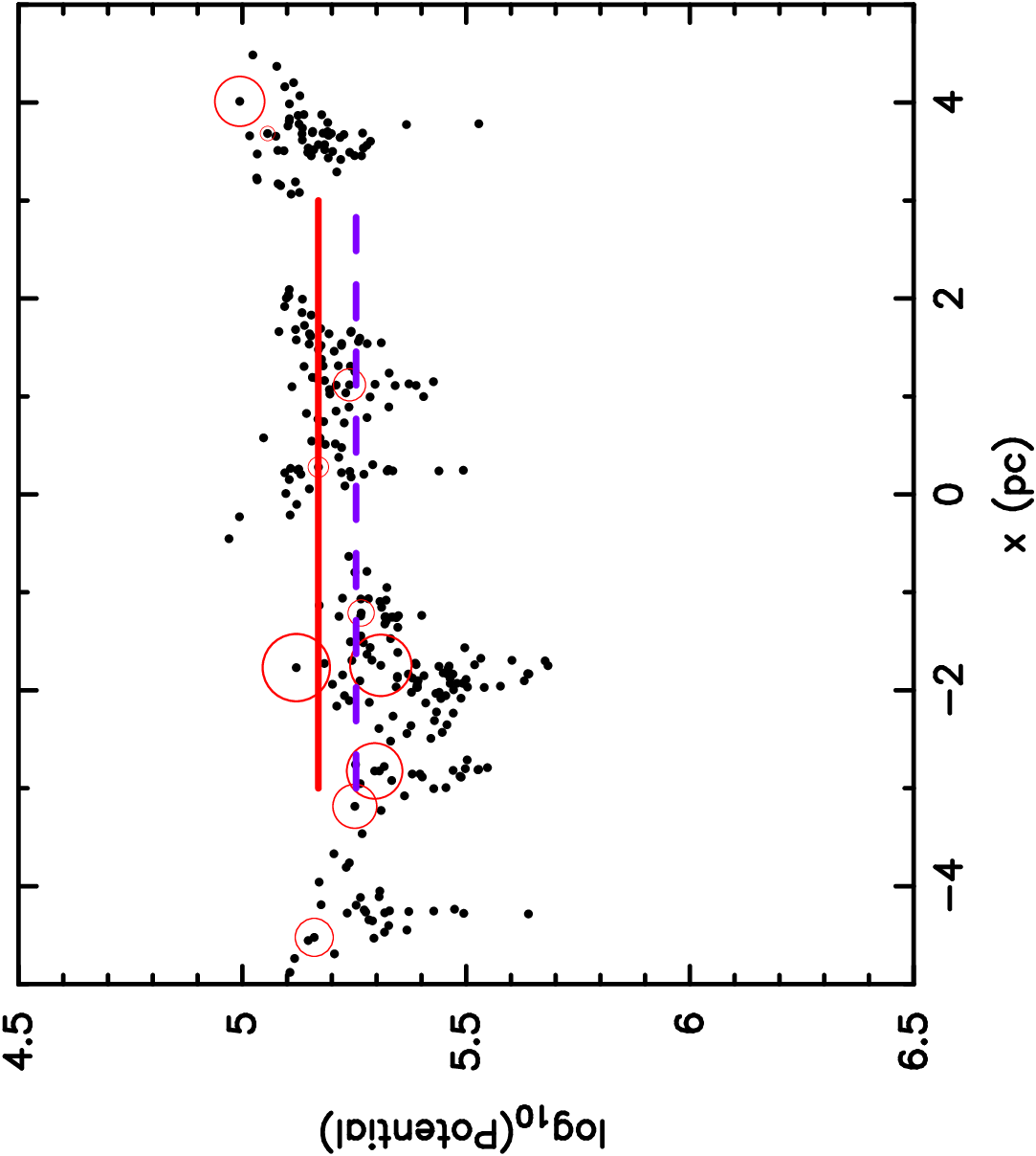}}}
\hspace*{0.3cm} 
\subfigure[Association, cent. conc., potential]{\label{synthetic_morphologies-e}\rotatebox{270}{\includegraphics[scale=0.25]{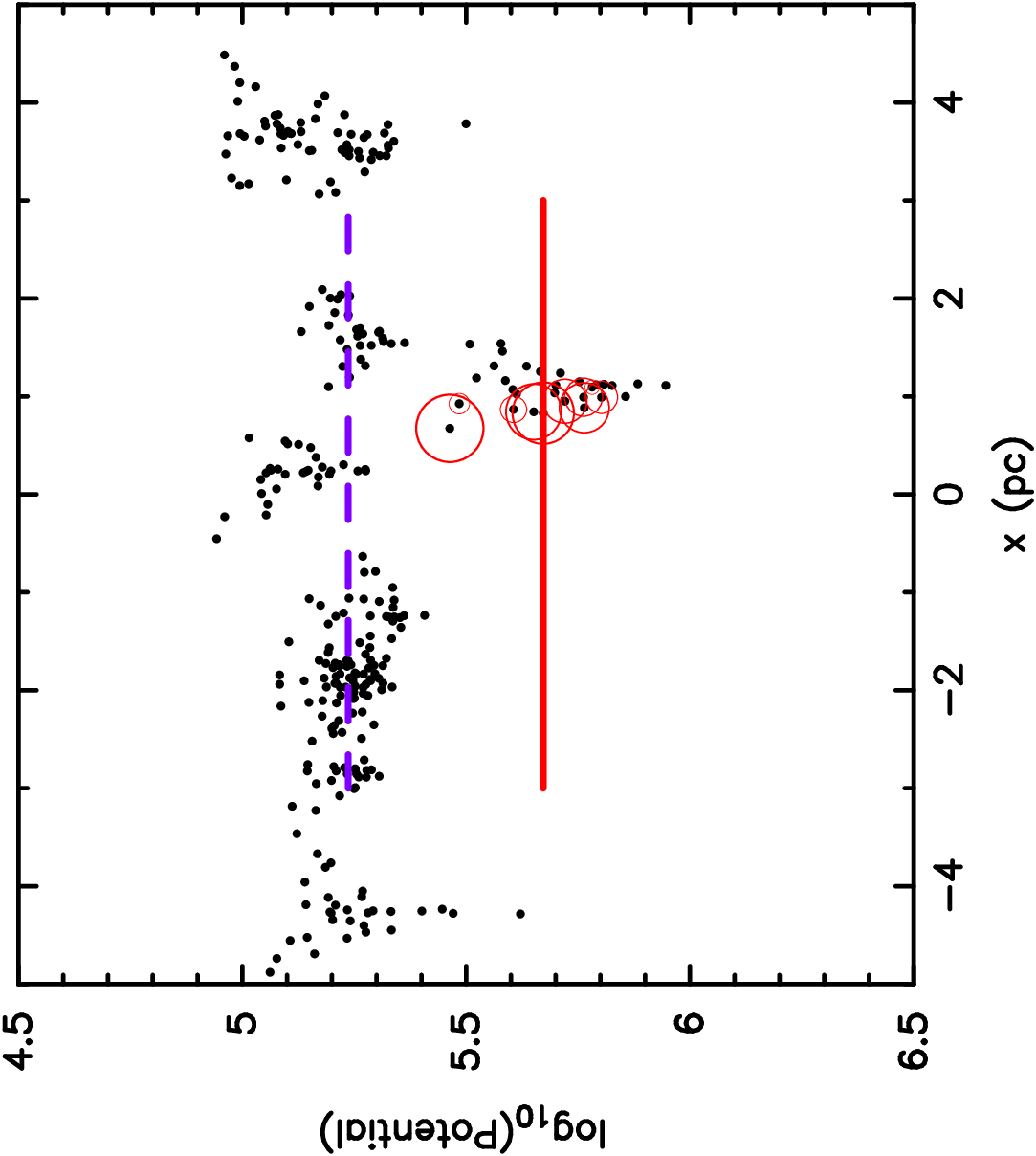}}}
\hspace*{0.3cm}
\subfigure[Association, high dens., potential]{\label{synthetic_morphologies-f}\rotatebox{270}{\includegraphics[scale=0.25]{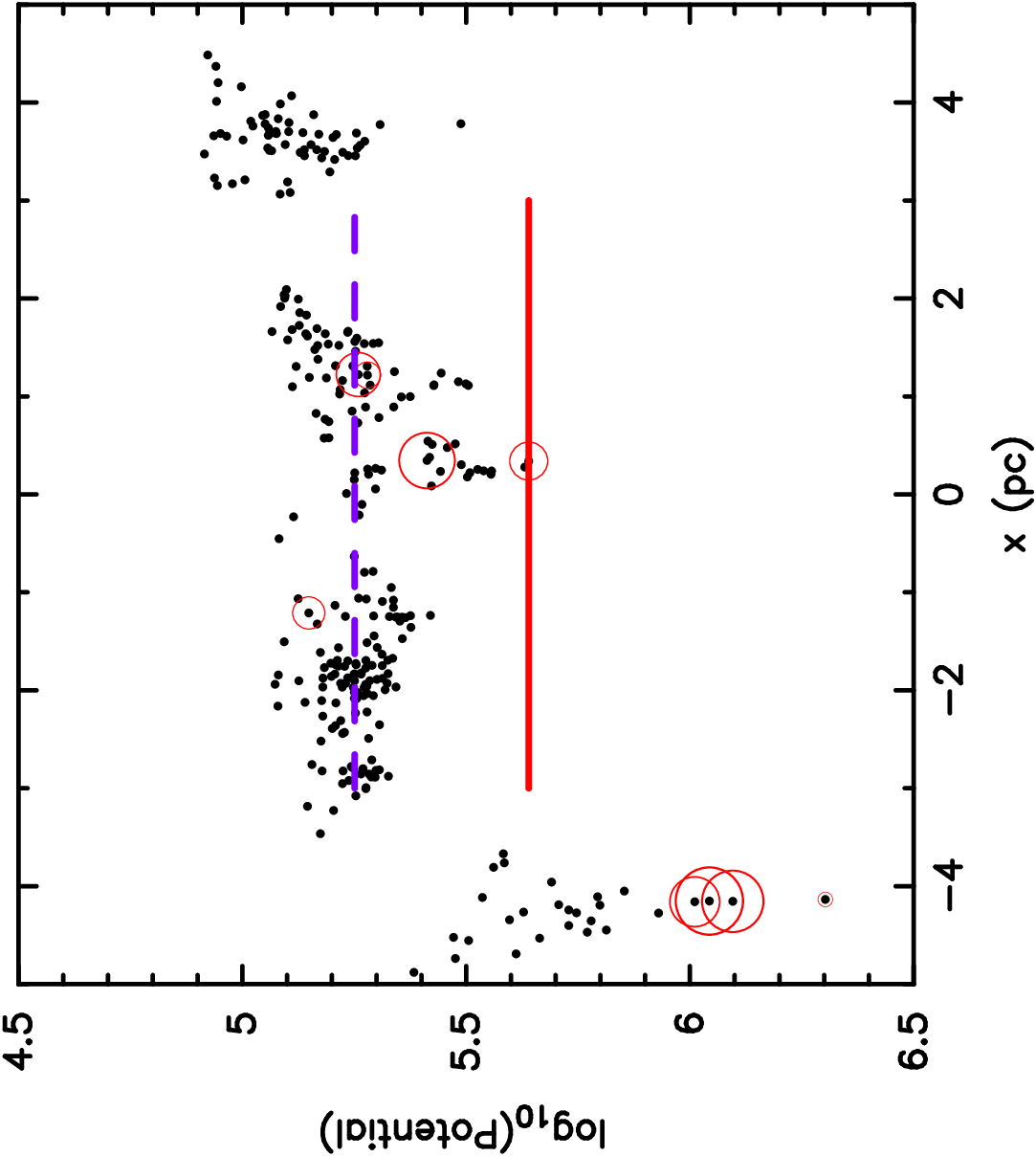}}}

\hspace*{-1.5cm}\subfigure[Association, random, $\Lambda_{\rm MSR}$]{\label{synthetic_morphologies-g}\rotatebox{270}{\includegraphics[scale=0.23]{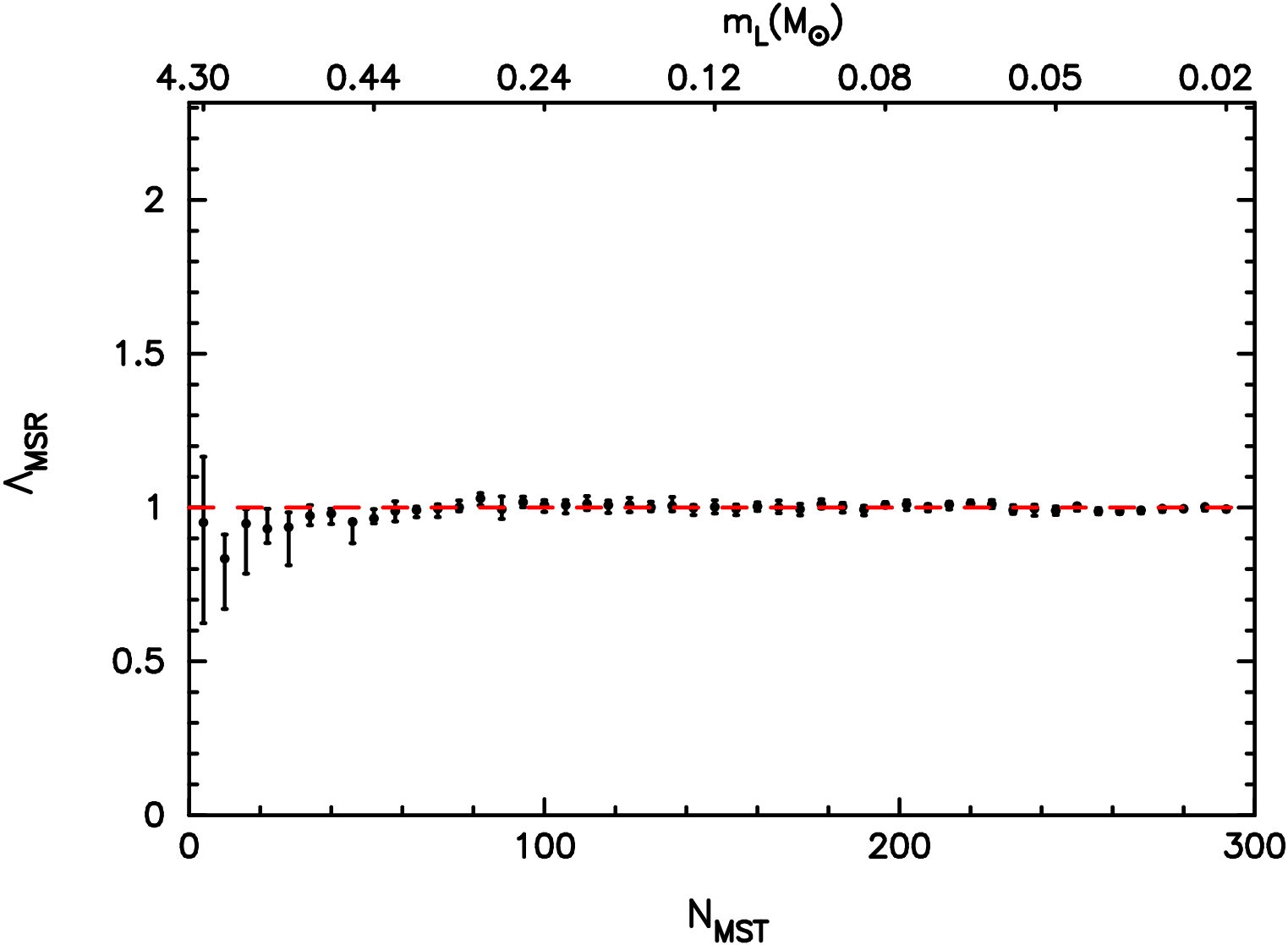}}}
\hspace*{0.1cm} 
\subfigure[Association, cent. conc., $\Lambda_{\rm MSR}$]{\label{synthetic_morphologies-h}\rotatebox{270}{\includegraphics[scale=0.23]{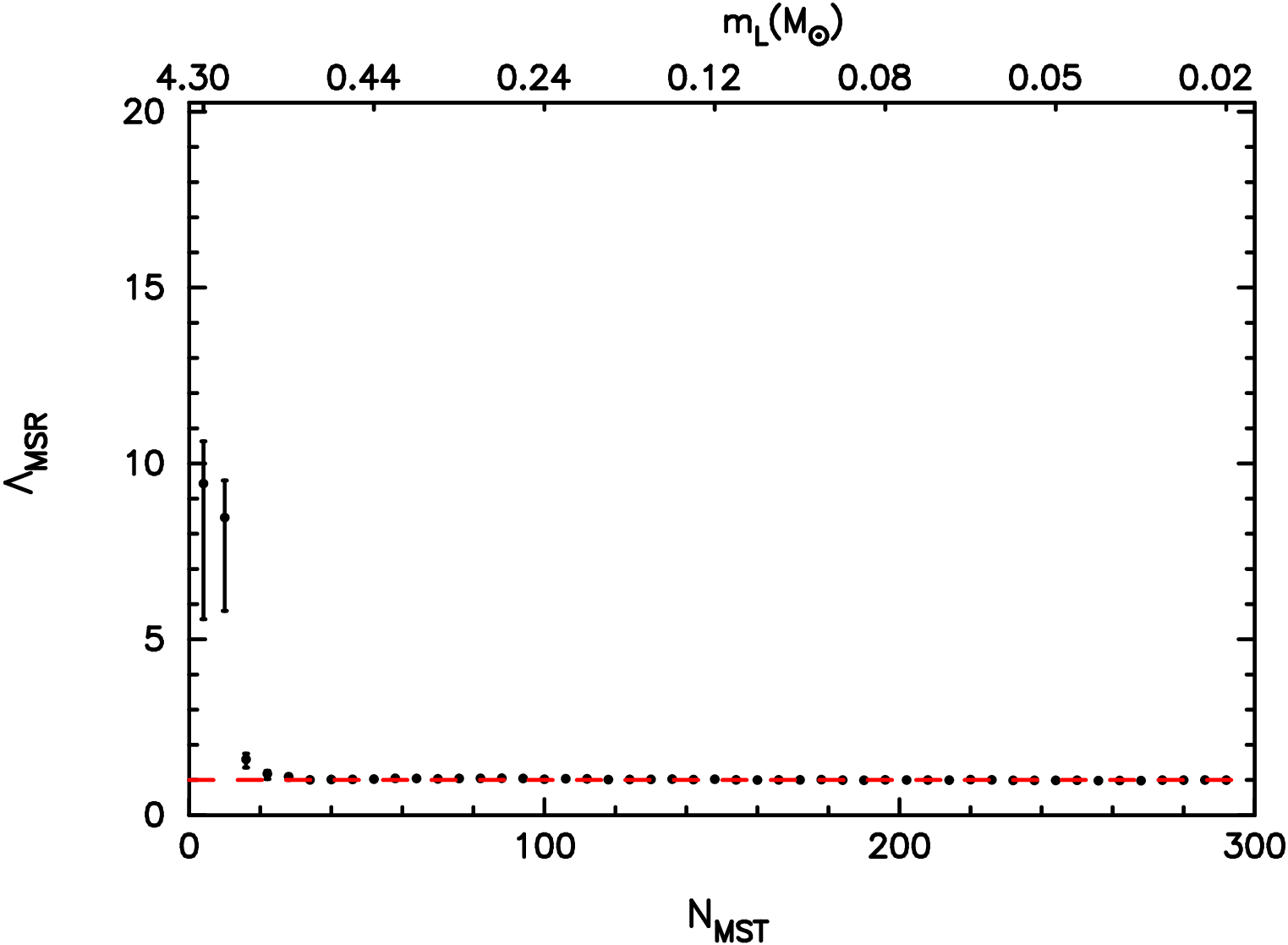}}}
\hspace*{0.1cm}
\subfigure[Association, high dens., $\Lambda_{\rm MSR}$]{\label{synthetic_morphologies-i}\rotatebox{270}{\includegraphics[scale=0.23]{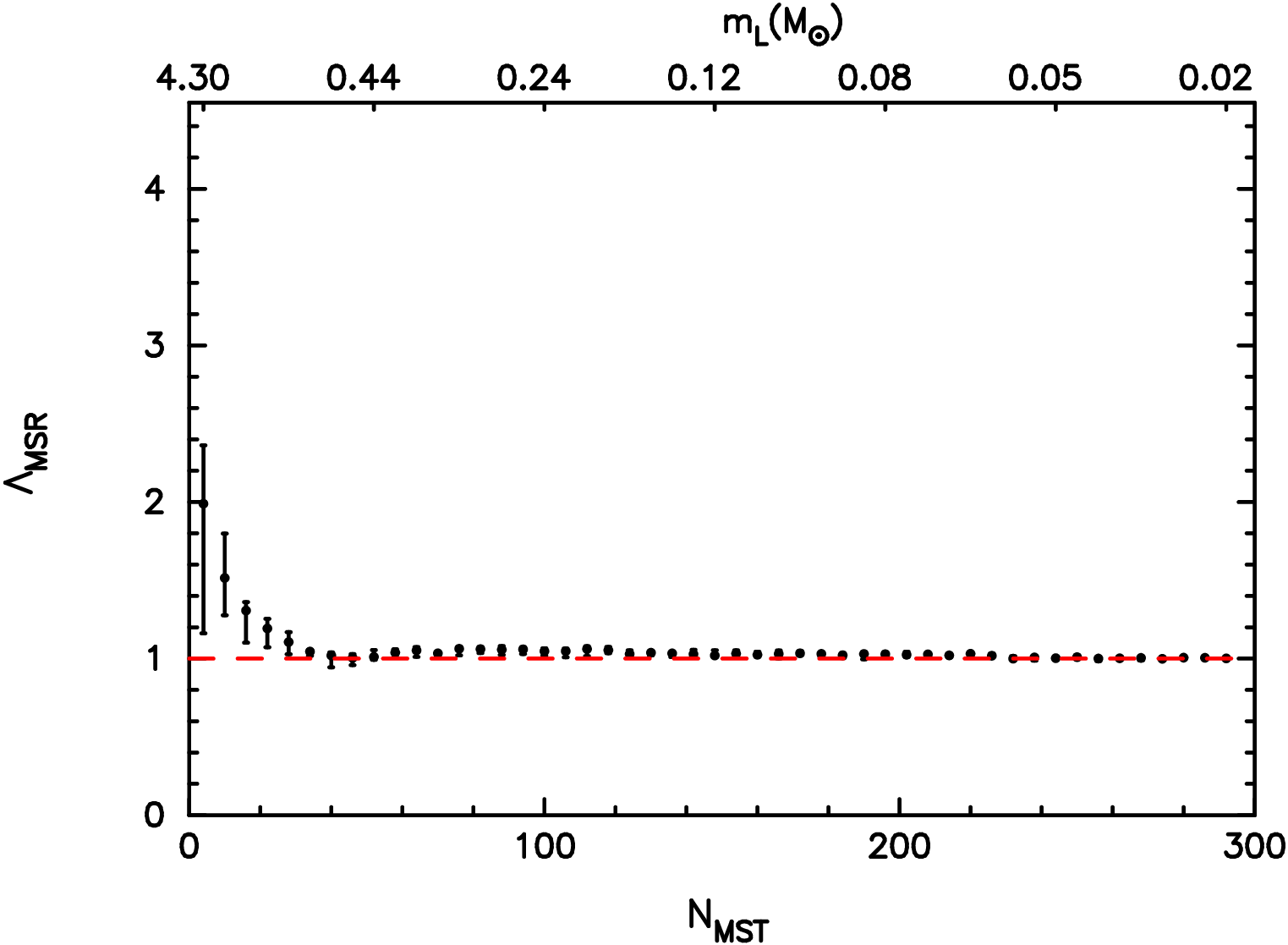}}}

\hspace*{-1.5cm}\subfigure[Association, random, $\Sigma - m$]{\label{synthetic_morphologies-j}\rotatebox{270}{\includegraphics[scale=0.25]{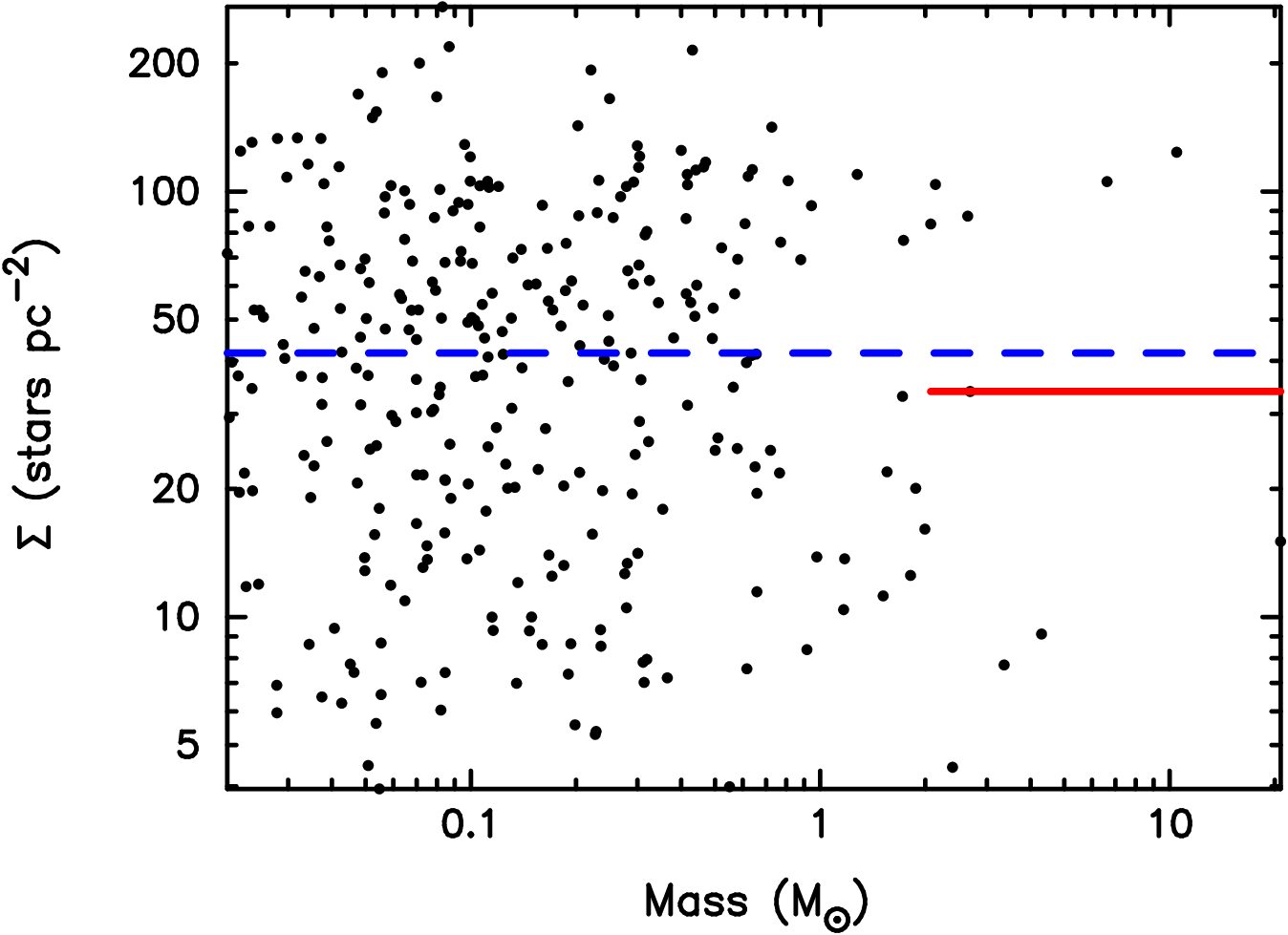}}}
\hspace*{0.3cm} 
\subfigure[Association, cent. conc., $\Sigma - m$]{\label{synthetic_morphologies-k}\rotatebox{270}{\includegraphics[scale=0.25]{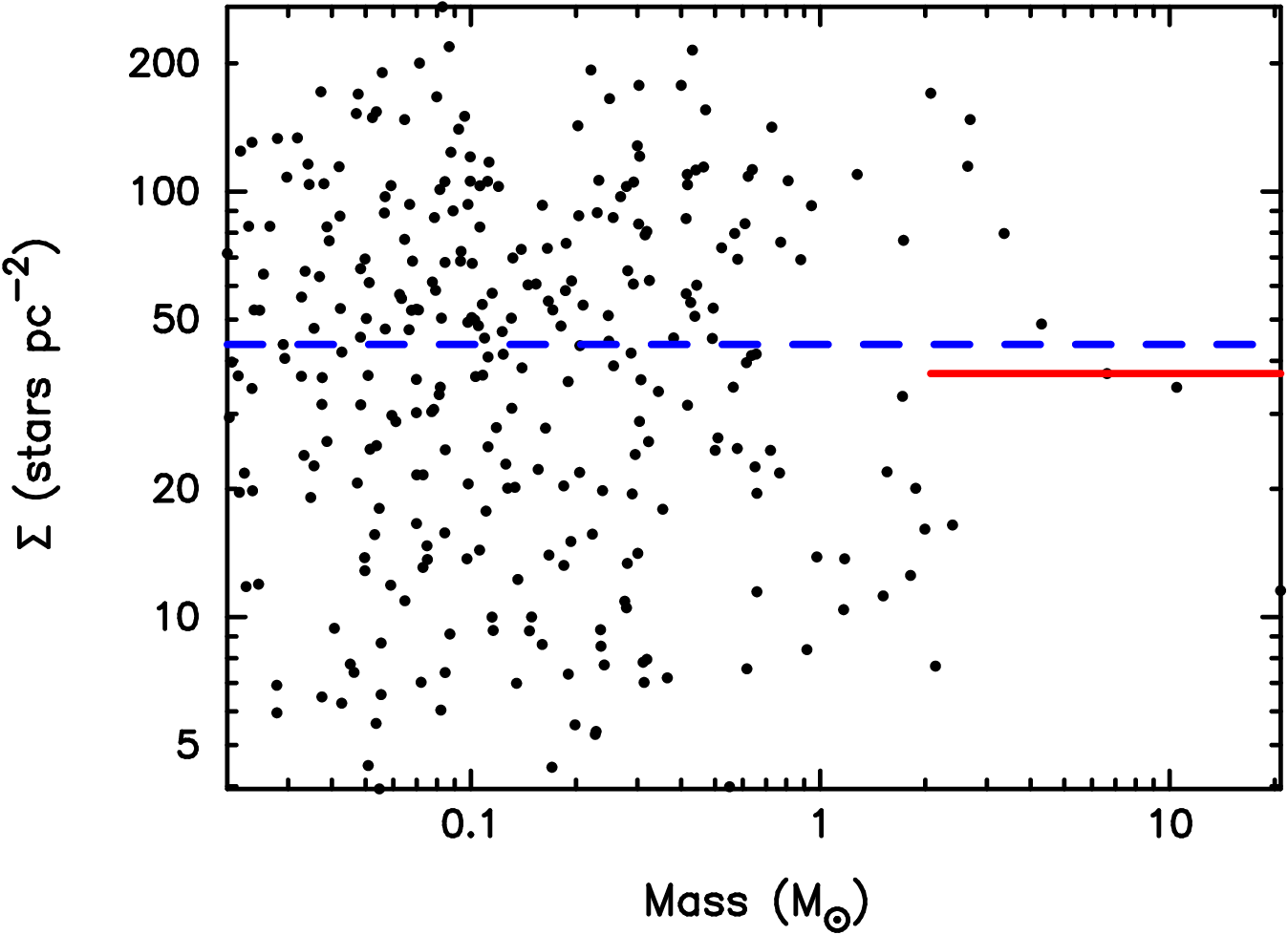}}}
\hspace*{0.3cm}
\subfigure[Association, high dens., $\Sigma - m$]{\label{synthetic_morphologies-l}\rotatebox{270}{\includegraphics[scale=0.25]{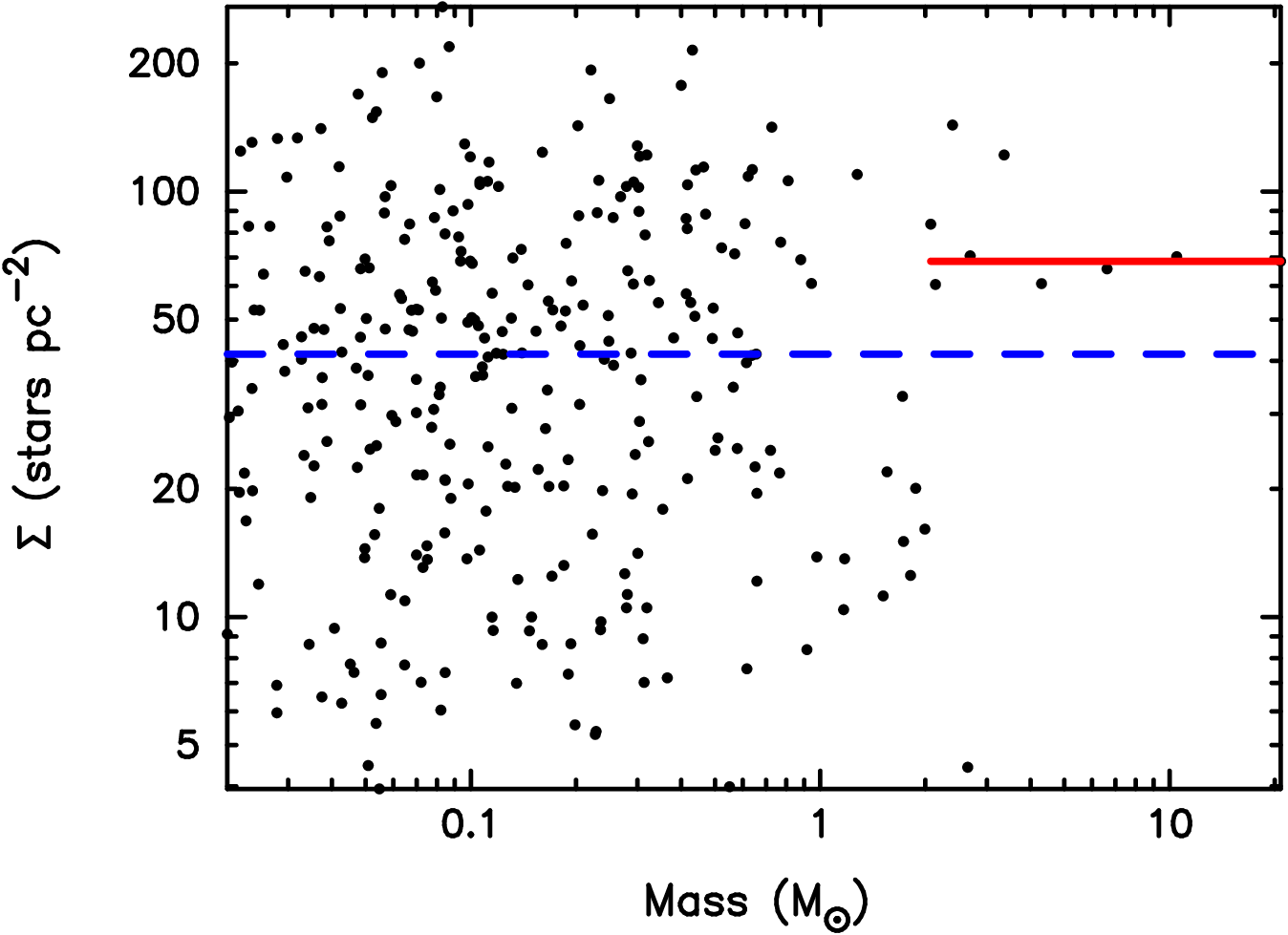}}}

\caption{A typical realisation of a synthetic association. The lefthand column shows the results when the massive stars are distributed randomly in the association, the middle column shows the results when the most massive stars are moved to the most central locations, and the righthand column shows the results when the ten most massive stars are moved to the locations with the highest relative densities. The top row shows the $x-y$ positions, with the most massive stars shown in red. The relative size of the red point indicates where it resides in the list of the most massive stars, with the largest circle indicating the most massive star, and the smallest circle indicating the tenth most massive star. The second row shows the gravitational potential for each star, where the solid red line is the median potential for the most massive stars and the dashed purple line is the median potential for all stars. The fourth row shows the $\Lambda_{\rm MSR}$ mass segregation ratio, where $\Lambda_{\rm MSR} = 1$ is shown by the red dashed line. The bottom row shows the surface density for each star as a function of its mass; the solid red line indicates the median for the most massive stars and the blue dashed line shows the median for all stars.}
\label{synthetic_morphologies}
  \end{center}
\end{figure*}

When the massive stars are randomly distributed (left column), the massive stars do not sit in deeper potentials than the average stars, nor are they mass segregated, nor are they in areas of higher than average surface density. However, when they are centrally concentrated \emph{or} in locations of higher than average density, they sit deeper in the potential, are mass segregated and display high surface densities.

\begin{figure*}
  \begin{center}
    \setlength{\subfigcapskip}{10pt}
    \vspace*{-0.5cm}
\hspace*{-1.5cm}\subfigure[Fractal, random, $x-y$]{\label{synthetic_fractal-a}\rotatebox{270}{\includegraphics[scale=0.25]{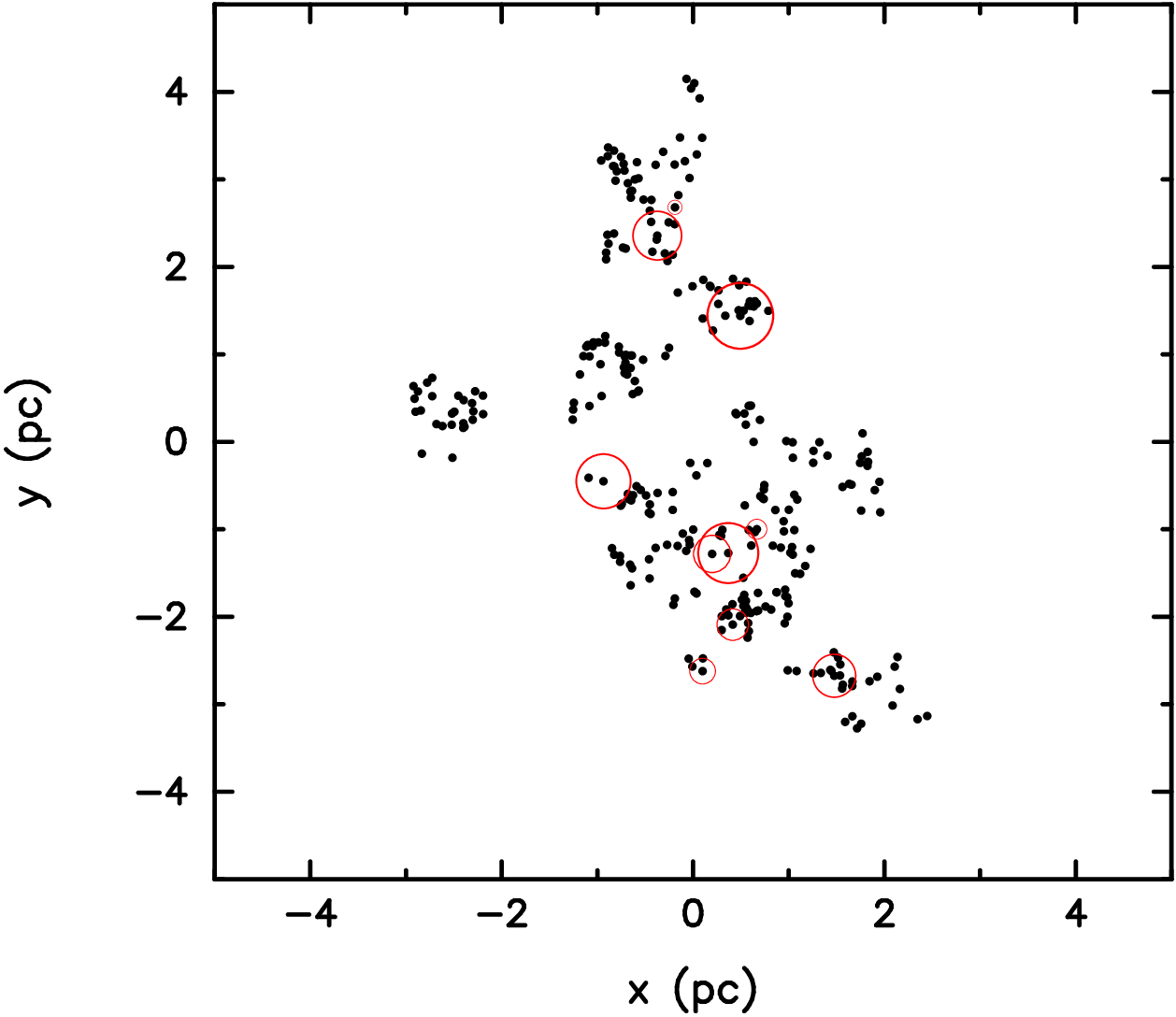}}}
\hspace*{0.3cm} 
\subfigure[Fractal, cent. conc., $x-y$]{\label{synthetic_fractal-b}\rotatebox{270}{\includegraphics[scale=0.25]{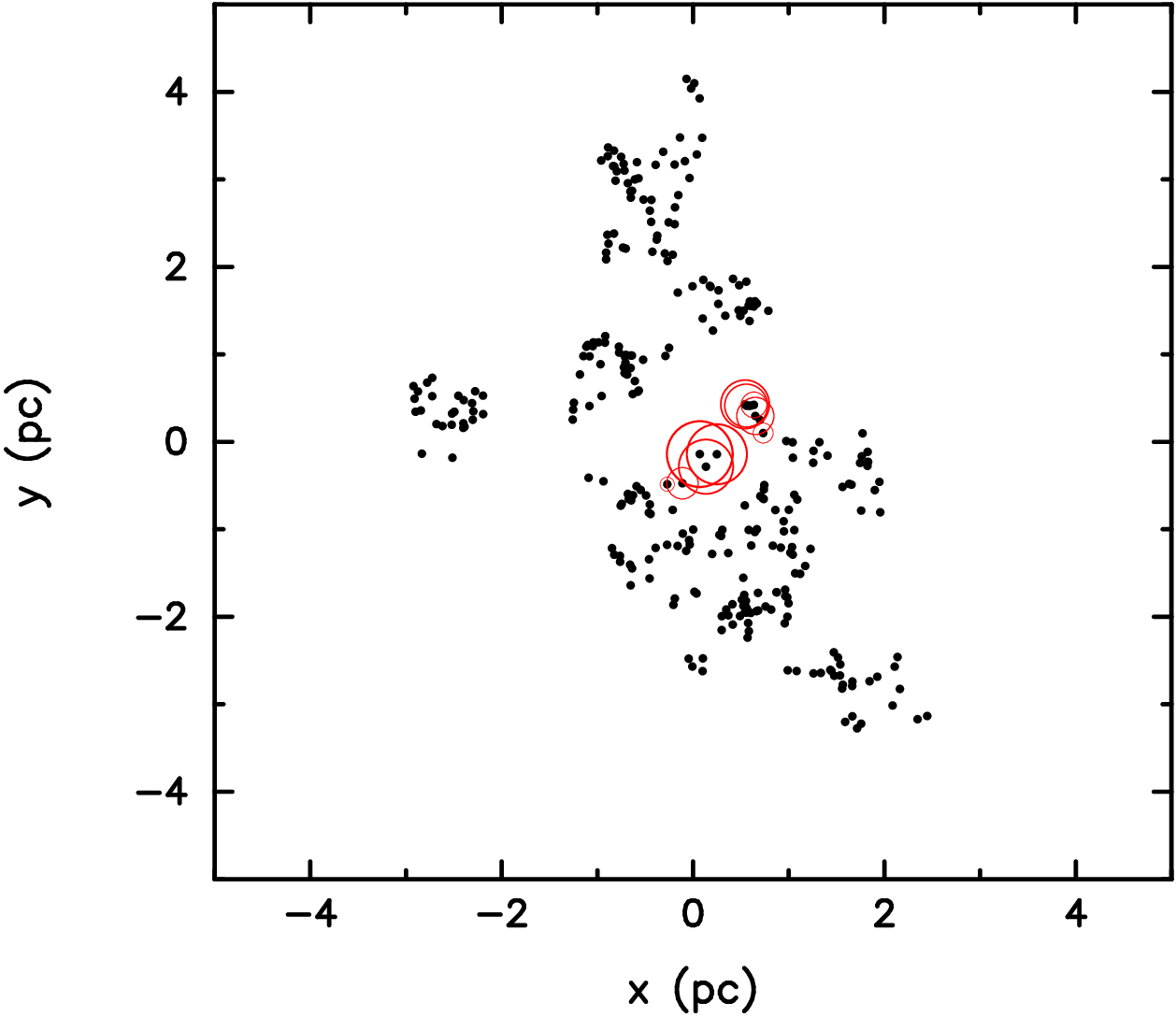}}}
\hspace*{0.3cm}
\subfigure[Fractal, high dens., $x-y$]{\label{synthetic_fractal-c}\rotatebox{270}{\includegraphics[scale=0.25]{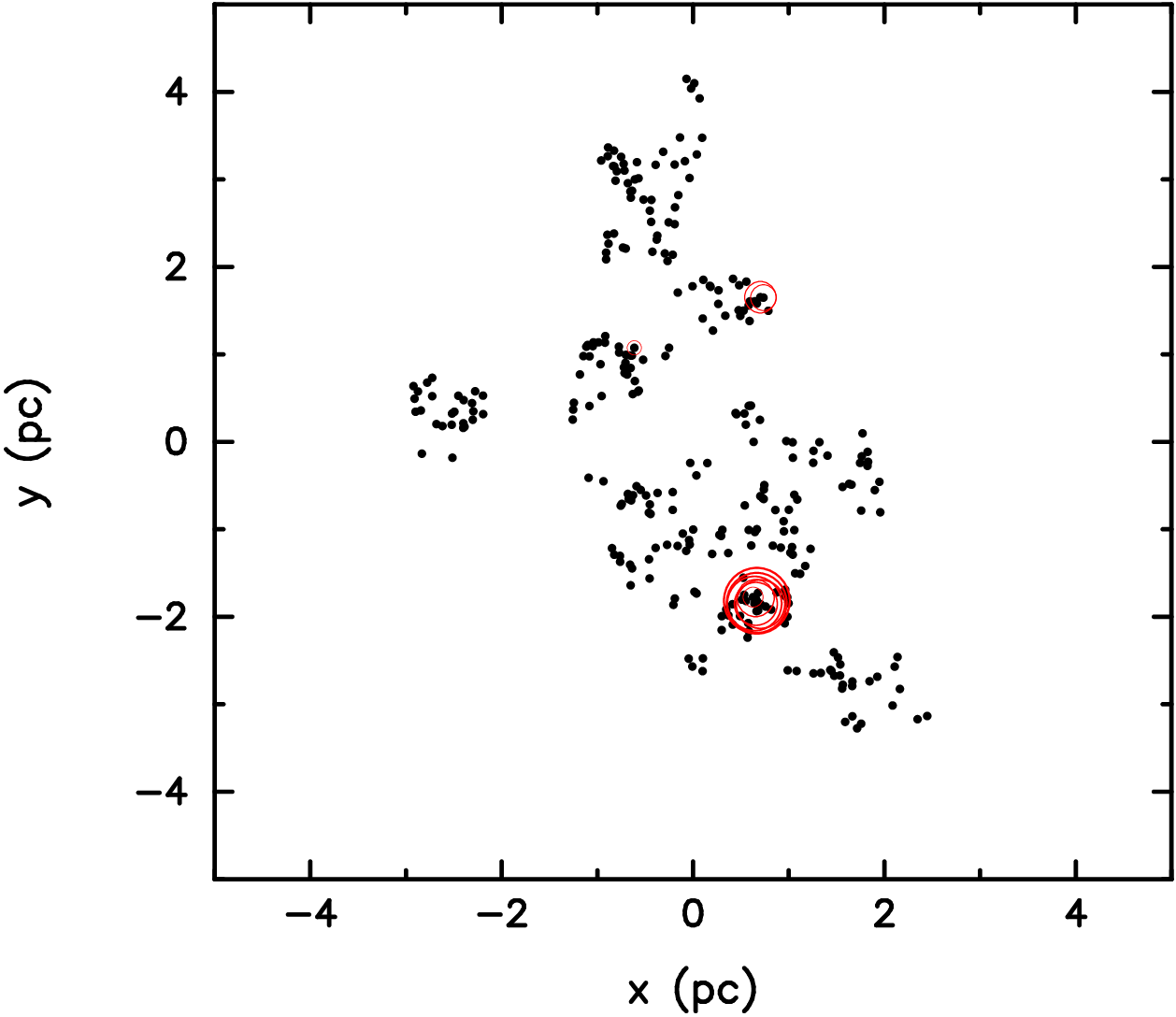}}}

\hspace*{-1.5cm}\subfigure[Fractal, random, potential]{\label{synthetic_fractal-d}\rotatebox{270}{\includegraphics[scale=0.25]{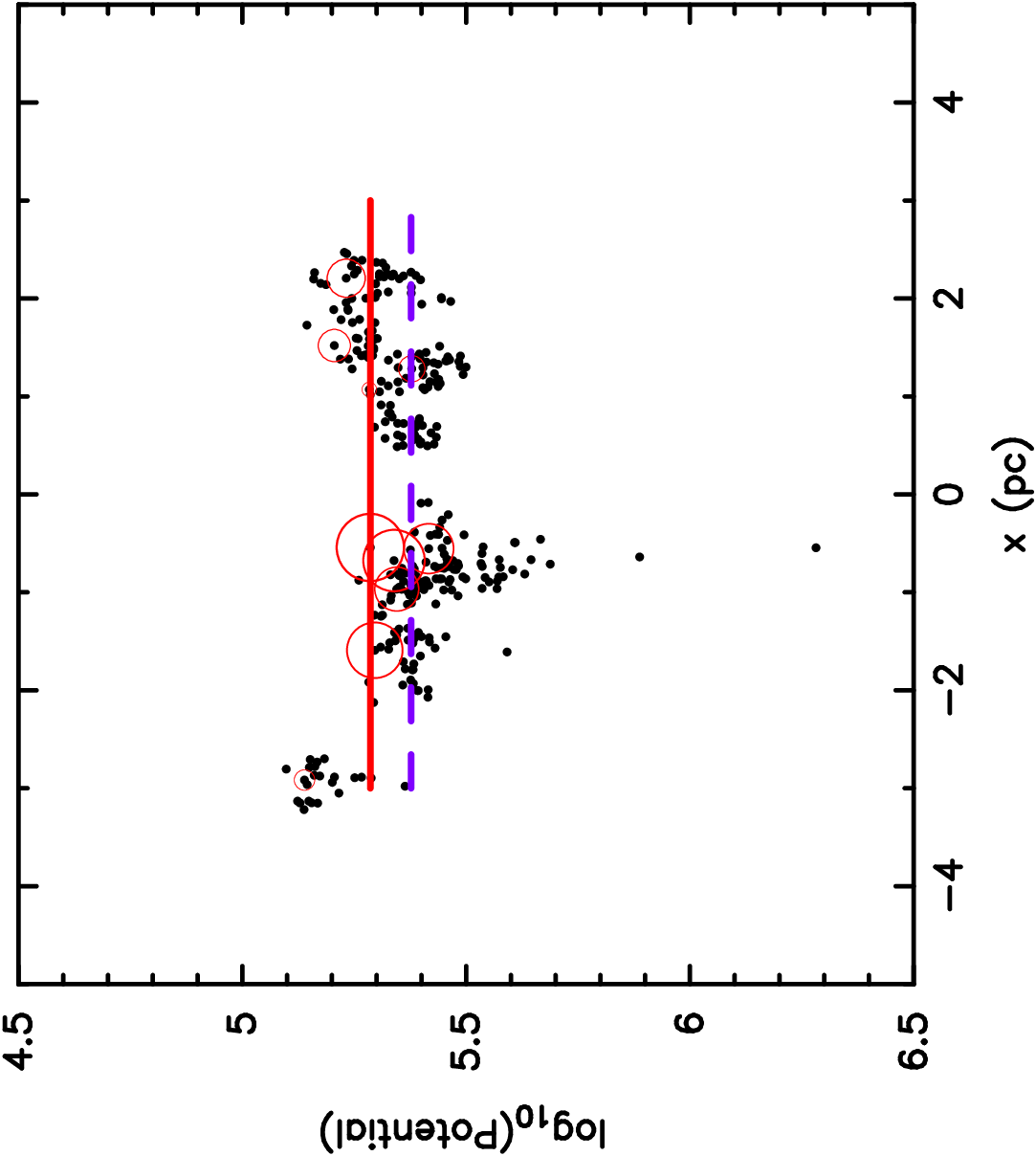}}}
\hspace*{0.3cm} 
\subfigure[Fractal, cent. conc., potential]{\label{synthetic_fractal-e}\rotatebox{270}{\includegraphics[scale=0.25]{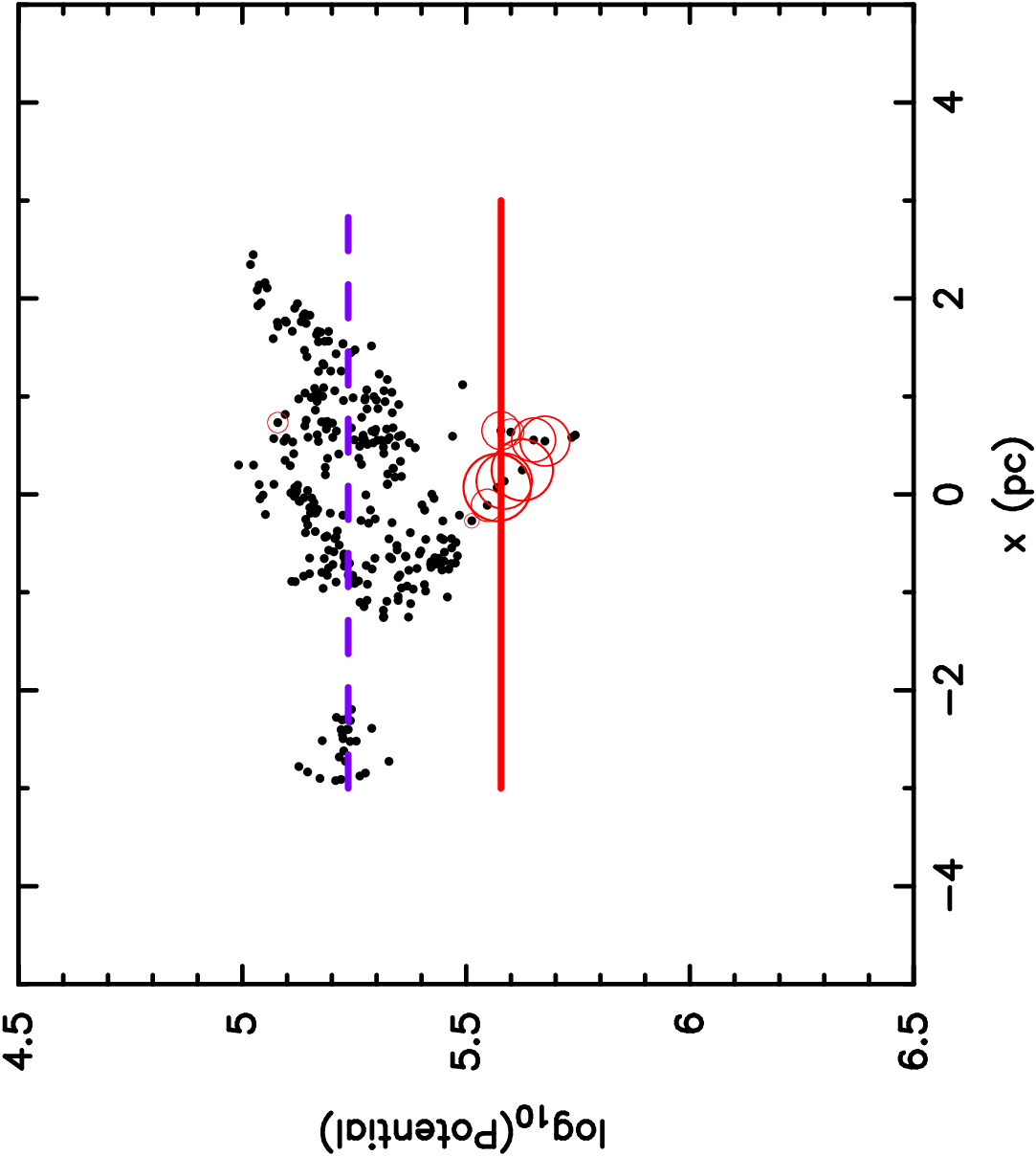}}}
\hspace*{0.3cm}
\subfigure[Fractal, high dens., potential]{\label{synthetic_fractal-f}\rotatebox{270}{\includegraphics[scale=0.25]{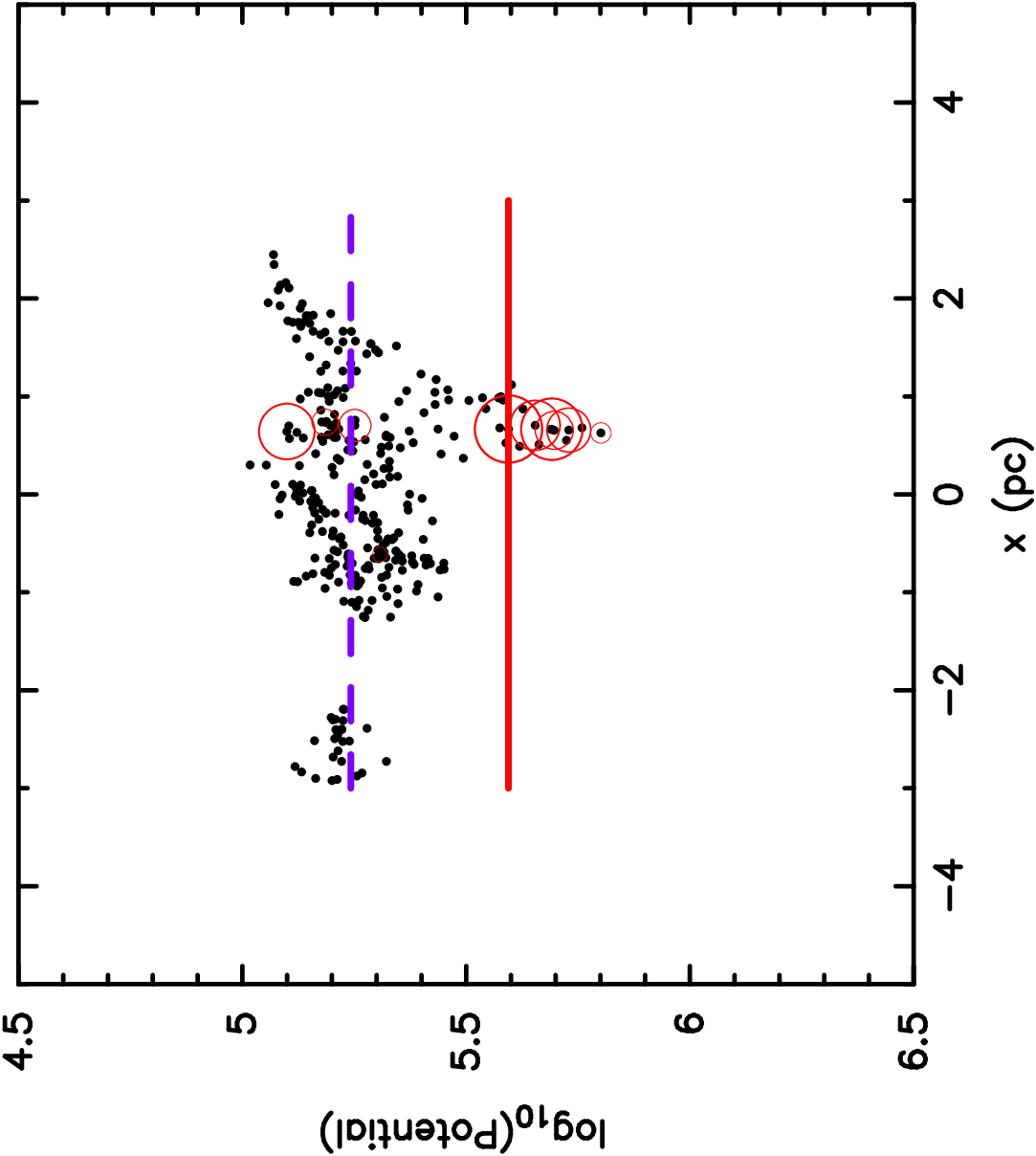}}}

\hspace*{-1.5cm}\subfigure[Fractal, random, $\Lambda_{\rm MSR}$]{\label{synthetic_fractal-g}\rotatebox{270}{\includegraphics[scale=0.23]{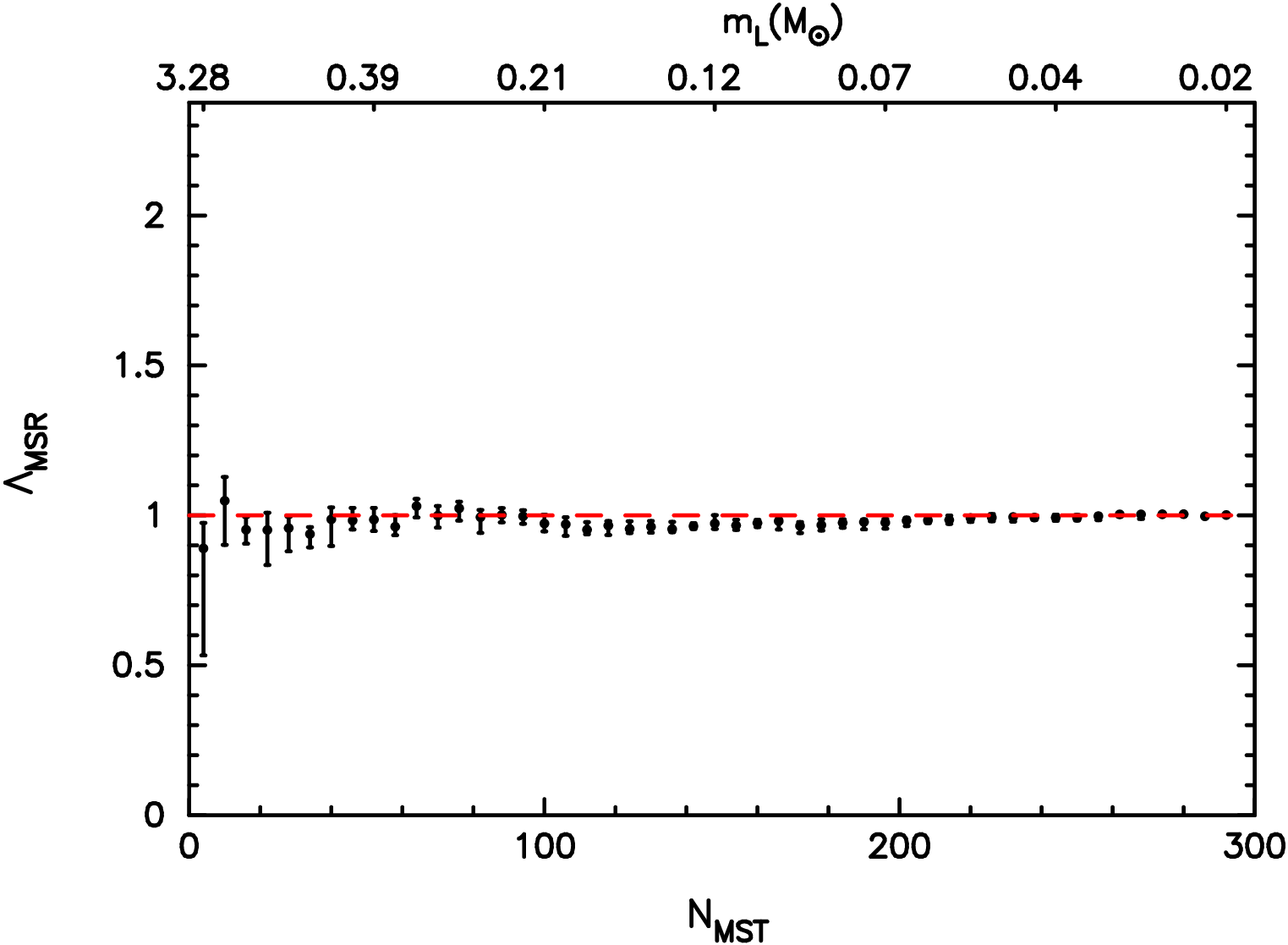}}}
\hspace*{0.1cm} 
\subfigure[Fractal, cent. conc., $\Lambda_{\rm MSR}$]{\label{synthetic_fractal-h}\rotatebox{270}{\includegraphics[scale=0.23]{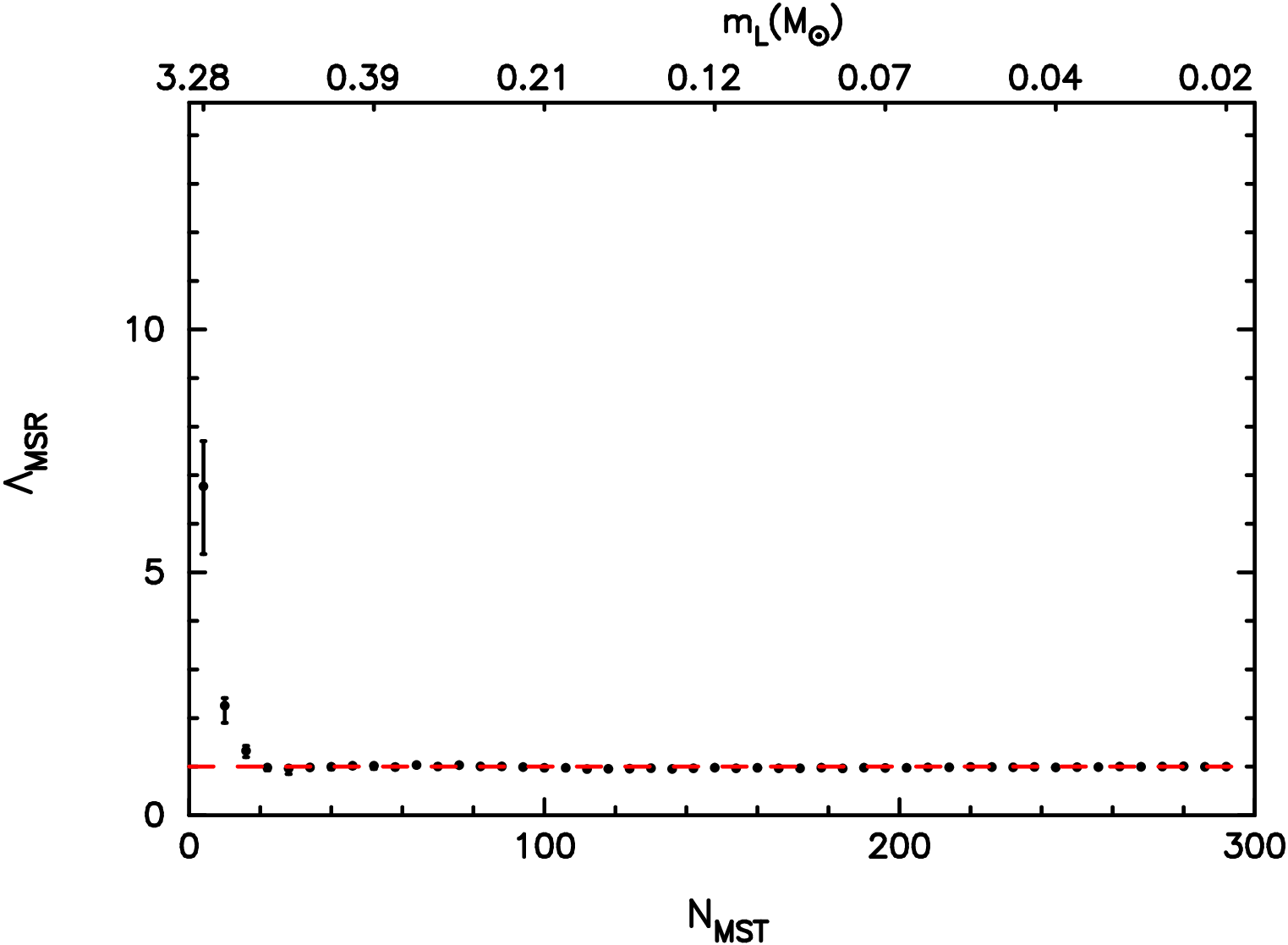}}}
\hspace*{0.1cm}
\subfigure[Fractal, high dens., $\Lambda_{\rm MSR}$]{\label{synthetic_fractal-i}\rotatebox{270}{\includegraphics[scale=0.23]{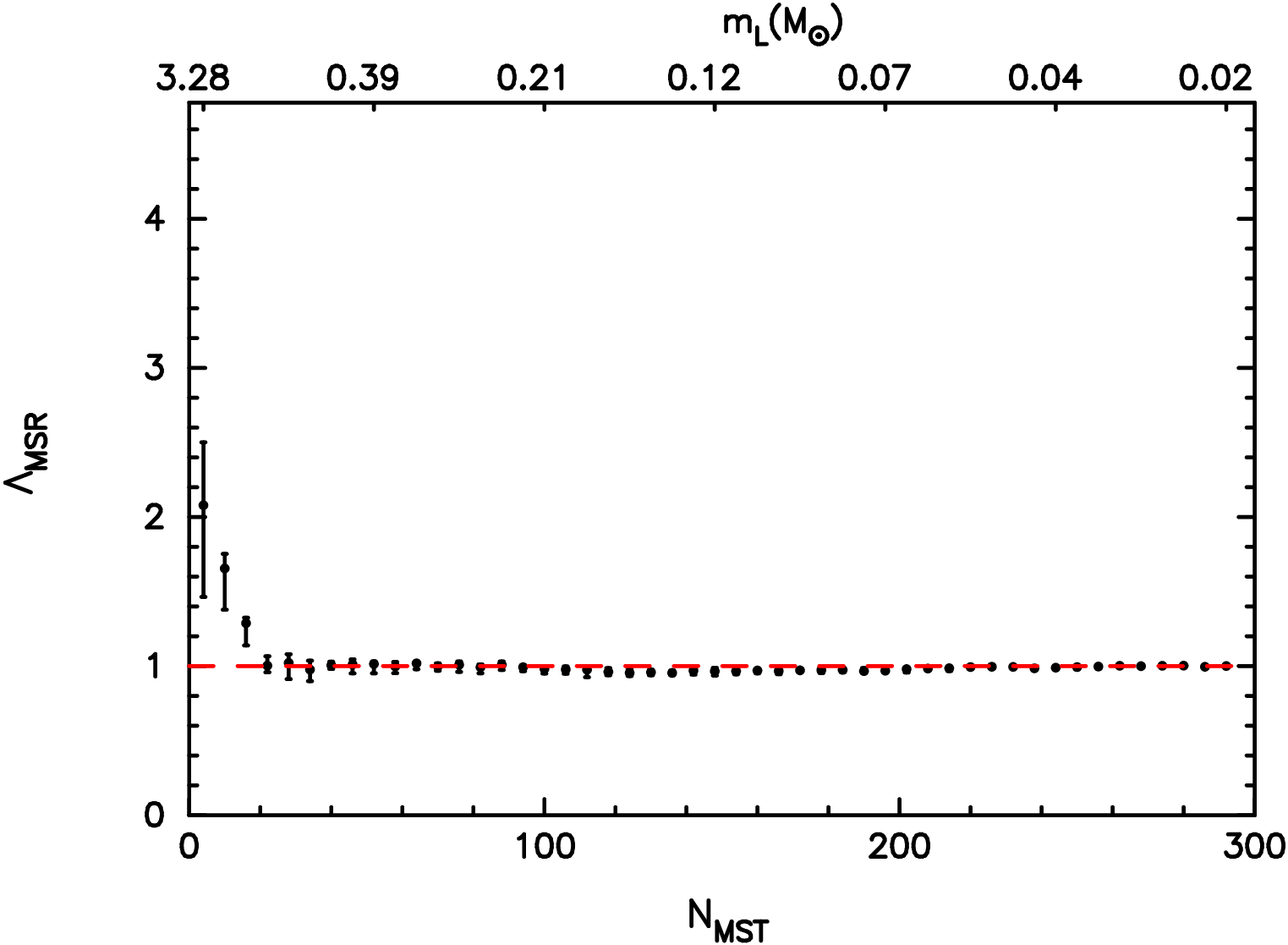}}}

\hspace*{-1.5cm}\subfigure[Fractal, random, $\Sigma - m$]{\label{synthetic_fractal-j}\rotatebox{270}{\includegraphics[scale=0.25]{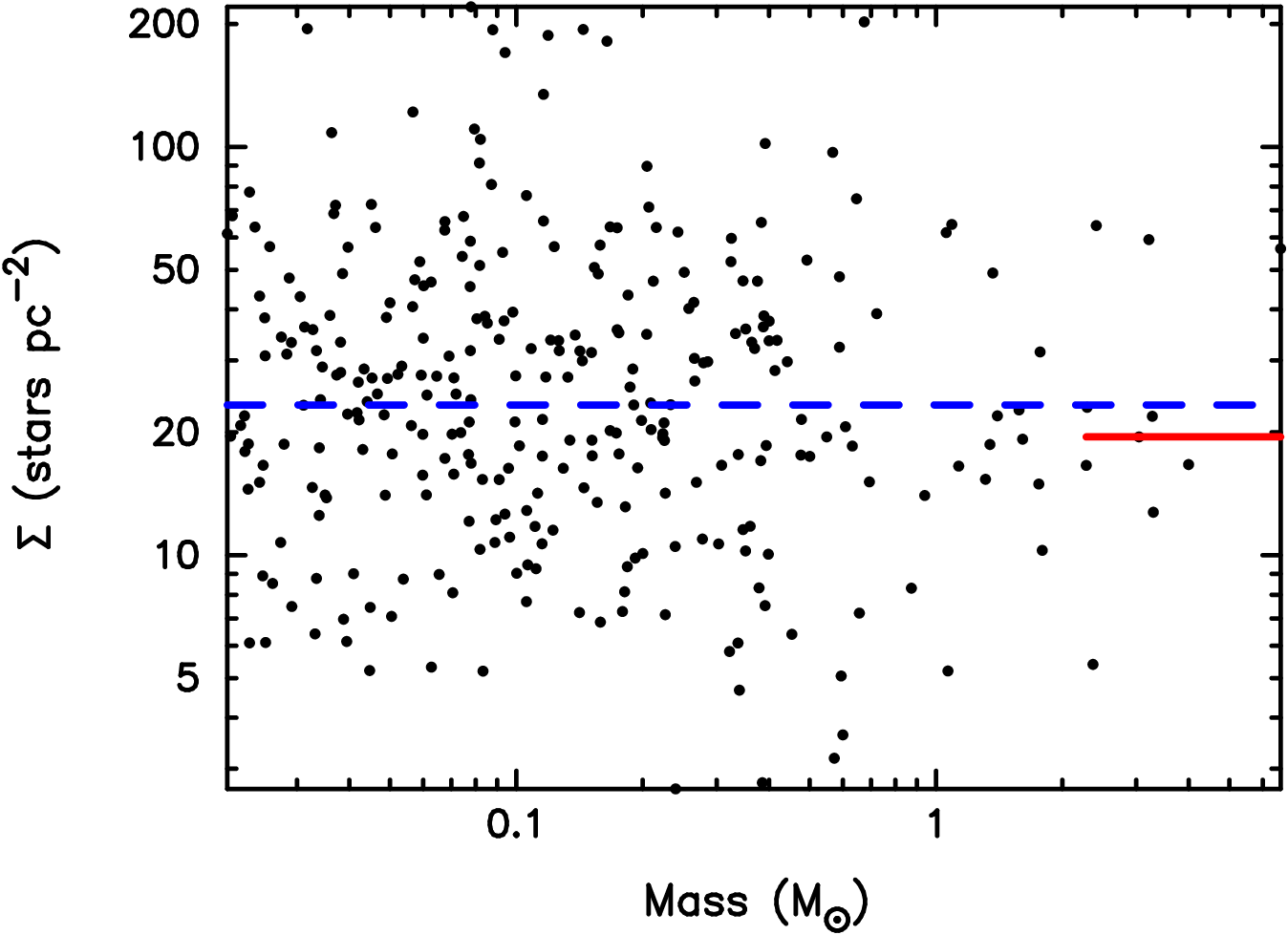}}}
\hspace*{0.3cm} 
\subfigure[Fractal, cent. conc., $\Sigma - m$]{\label{synthetic_fractal-k}\rotatebox{270}{\includegraphics[scale=0.25]{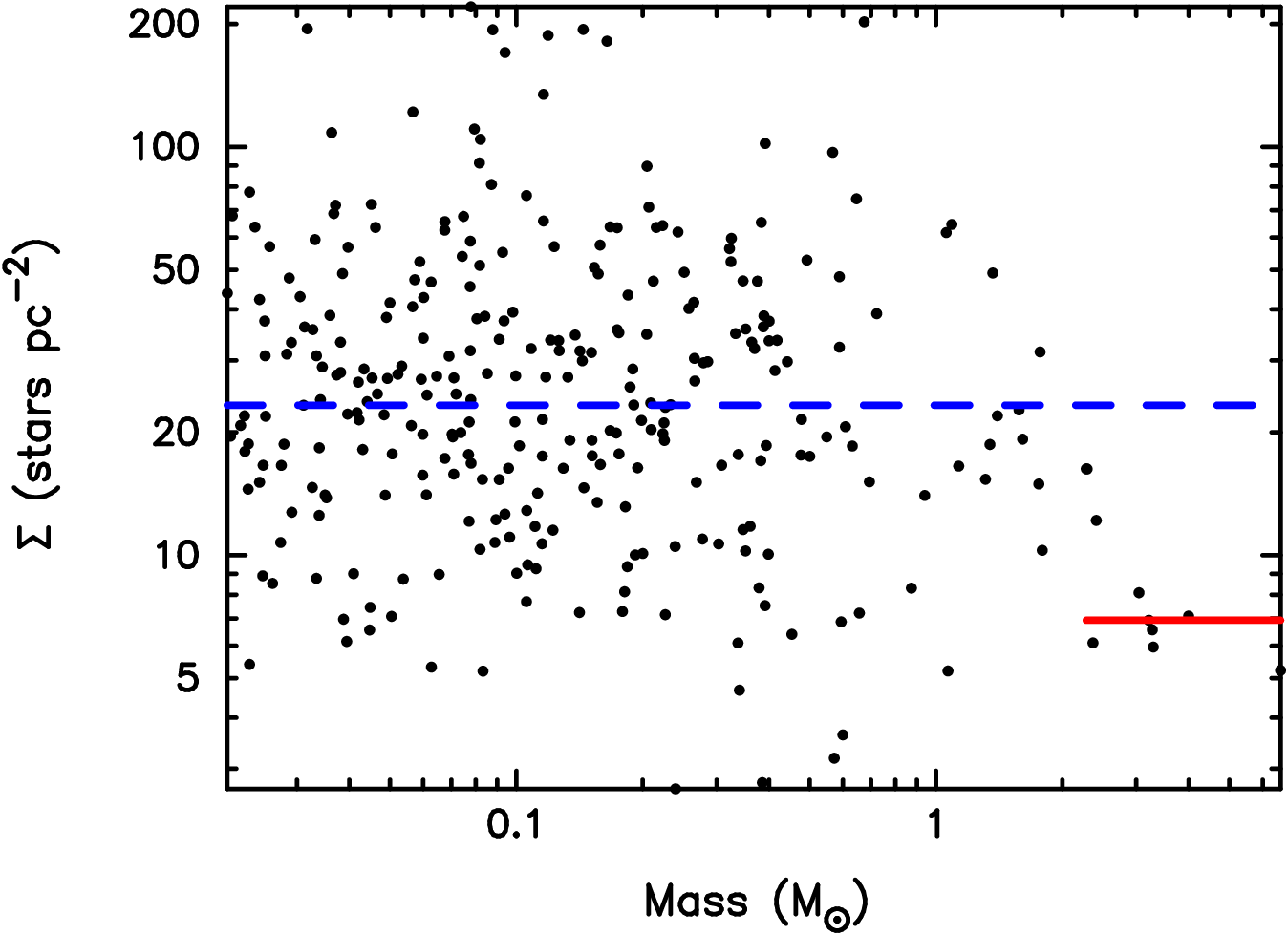}}}
\hspace*{0.3cm}
\subfigure[Fractal, high dens., $\Sigma - m$]{\label{synthetic_fractal-l}\rotatebox{270}{\includegraphics[scale=0.25]{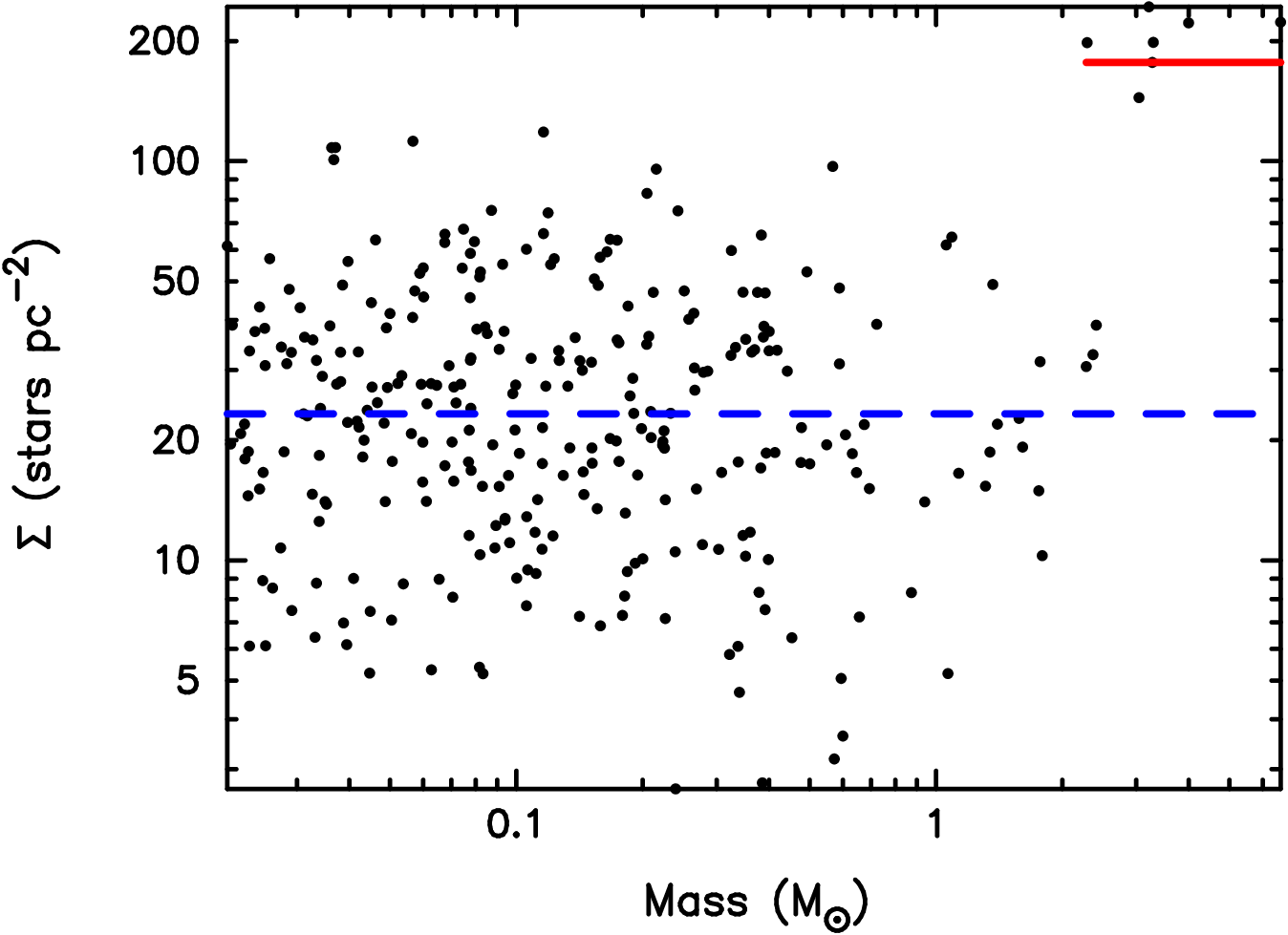}}}

\caption{A typical realisation of a synthetic fractal.The lefthand column shows the results when the massive stars are distributed randomly in the fractal, the middle column shows the results when the most massive stars are moved to the most central locations, and the righthand column shows the results when the ten most massive stars are moved to the locations with the highest relative densities. The top row shows the $x-y$ positions, with the most massive stars shown in red. The relative size of the red point indicates where it resides in the list of the most massive stars, with the largest circle indicating the most massive star, and the smallest circle indicating the tenth most massive star. The second row shows the gravitational potential for each star, where the solid red line is the median potential for the most massive stars and the dashed purple line is the median potential for all stars. The fourth row shows the $\Lambda_{\rm MSR}$ mass segregation ratio, where $\Lambda_{\rm MSR} = 1$ is shown by the red dashed line. The bottom row shows the surface density for each star as a function of its mass; the solid red line indicates the median for the most massive stars and the blue dashed line shows the median for all stars.}
\label{synthetic_fractal}
  \end{center}
\end{figure*}

Whilst a Plummer sphere is a reasonable spatial distribution for an old, evolved star cluster, observations \citep{Gomez93,Cartwright04,Sanchez09,Hacar16,Buckner19} and simulations \citep{Schmeja06,Dale12c,Dale13a,Dale14} suggest that most star-forming regions form with spatial and kinematic substructure.

Our next synthetic morphology is designed to mimic the substructure in a young star-forming region, or an older association, and we show an example in Fig.~\ref{synthetic_morphologies}. As with the Plummer sphere, the lefthand column shows the association when the massive stars are distributed randomly, the middle column shows the association when the most massive stars are centrally concentrated, and the righthand column shows the association when the most massive stars have been moved to the positions where the stellar surface density is highest.

In the random distribution, the most massive stars do not sit deeper in the gravitional potential, nor are they mass segregated, nor do the most massive stars reside in significantly higher surface densities. When the most massive stars are moved to the most central locations (middle column of Fig.~\ref{synthetic_morphologies}), they do sit deeper within the potential, and they are mass segregated. However, the massive stars are not located in area of high surface densities (Fig.~\ref{synthetic_morphologies-k}). When we move the most massive stars to the locations of highest surface densities, they again sit deeper in the gravitational potential, but are not mass segregated (Fig.~\ref{synthetic_morphologies-i}).

We have repeated this experiment for two other types of substructured regions, namely a box fractal with a high degree of substructure ($D = 1.6$) and a box fractal with a moderate degree of substructure ($D = 2.0$). We show an example of the highly substructured ($D = 1.6$) fractal in Fig.~\ref{synthetic_fractal}. The results are qualitatively similar to the those for the association shown in Fig.~\ref{synthetic_morphologies}, save for the scenario where the most massive stars are moved to the most central locations of the fractal; this causes the massive stars to reside in locations of significantly lower surface densities than the average stars in the region (compare the association in Fig.~\ref{synthetic_morphologies-k} to the fractal in Fig.~\ref{synthetic_fractal-k}).

In reality, massive stars would be unlikely to reside in a centrally concentrated configuration without also being surrounded by many low-mass stars (as is the case in the centrally concentrated Plummer spheres). This would occur either because low mass stars would also congregate in the main gravitational potential in the event of dynamical mass segregation, or if the massive stars formed in a centrally concentrated configuration they would likely be surrounded by a significant number of low-mass stars \citep{Weidner06}.

  Of all our synthetic models, the most likely scenario might be the occurrence of massive stars in regions of higher than average surface density (like those shown in the righthand column of Fig.~\ref{synthetic_morphologies}). However, as we have seen in the SPH simulations with stellar feedback (Fig.~\ref{SPH_sim_results-b}), we might not expect high surface densities around the most massive stars from formation.

  \subsection{Core fragmentation and monolithic collapse}

\begin{figure*}
  \begin{center}
\setlength{\subfigcapskip}{10pt}
\hspace*{-1.5cm}\subfigure[Random fragmentation]{\label{fragmentation-a}\rotatebox{270}{\includegraphics[scale=0.39]{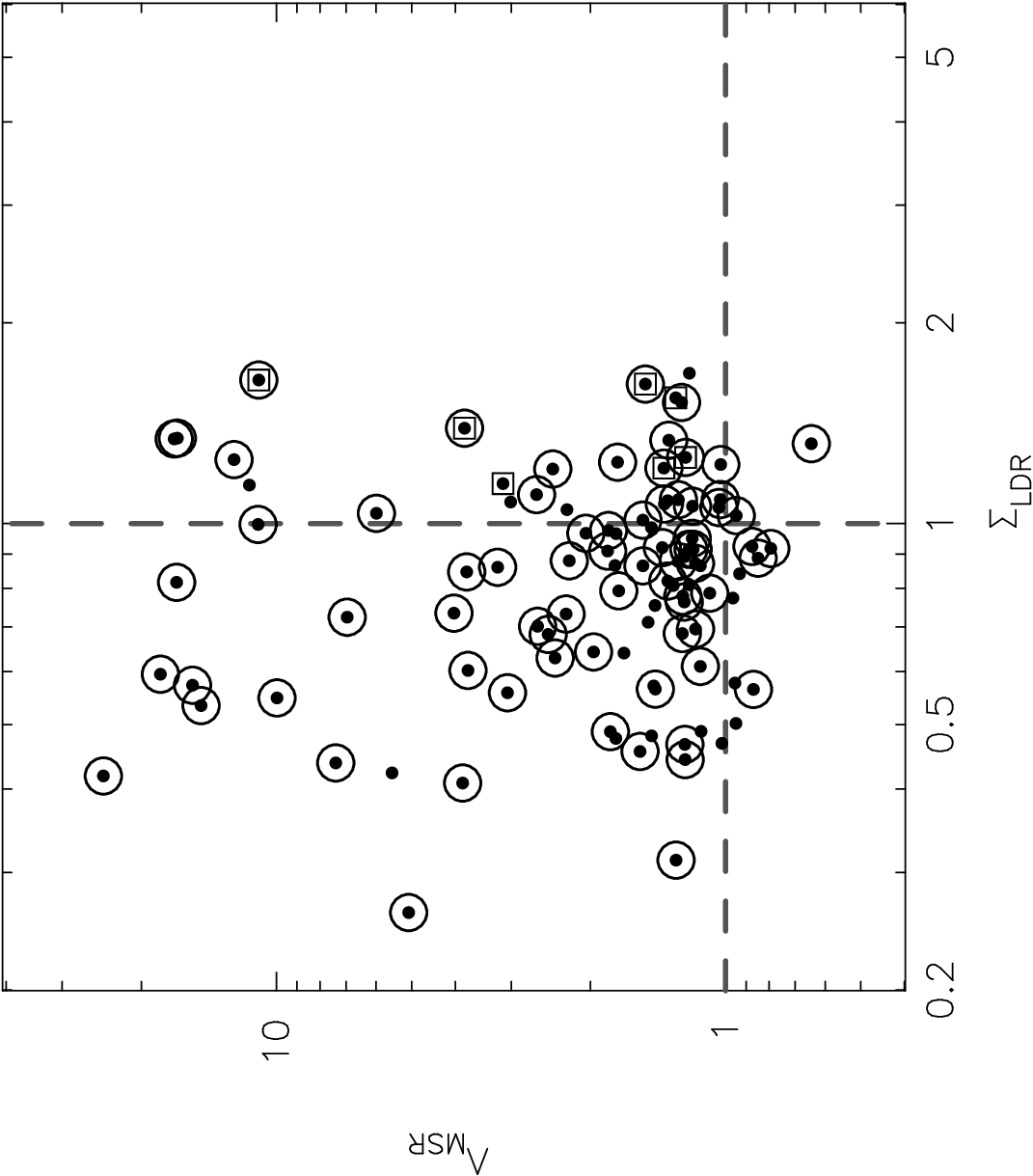}}}
\hspace*{0.3cm} 
\subfigure[Random fragmentation, dynamical evolution]{\label{fragmentation-b}\rotatebox{270}{\includegraphics[scale=0.39]{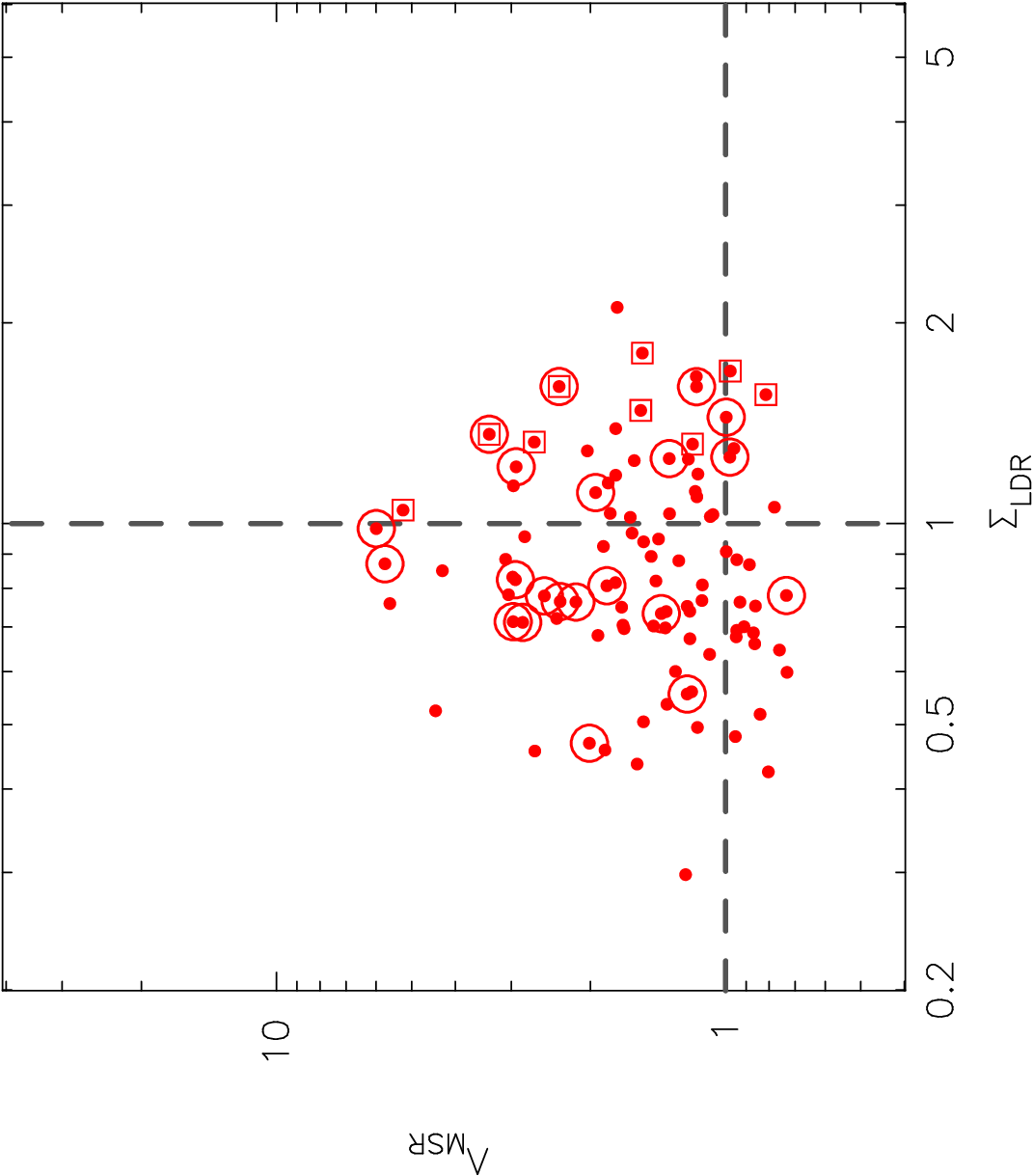}}}
\hspace*{-1.5cm}
\subfigure[Only low-mass ($\leq$10\,M$_\odot$) cores fragment]{\label{fragmentation-c}\rotatebox{270}{\includegraphics[scale=0.39]{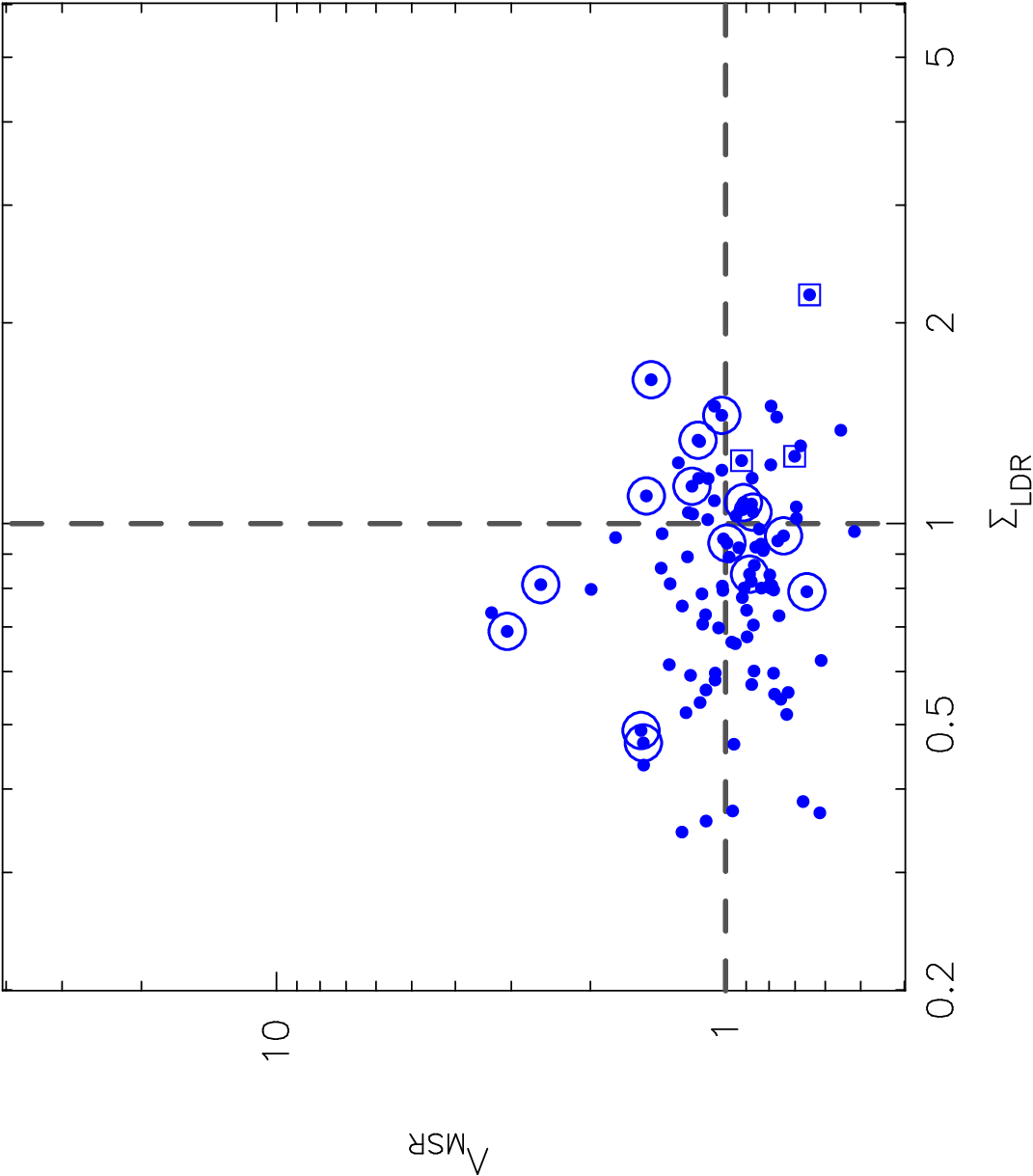}}}
\hspace*{0.3cm}
\subfigure[Only high-mass ($>$10\,M$_\odot$) cores fragment]{\label{fragmentation-d}\rotatebox{270}{\includegraphics[scale=0.39]{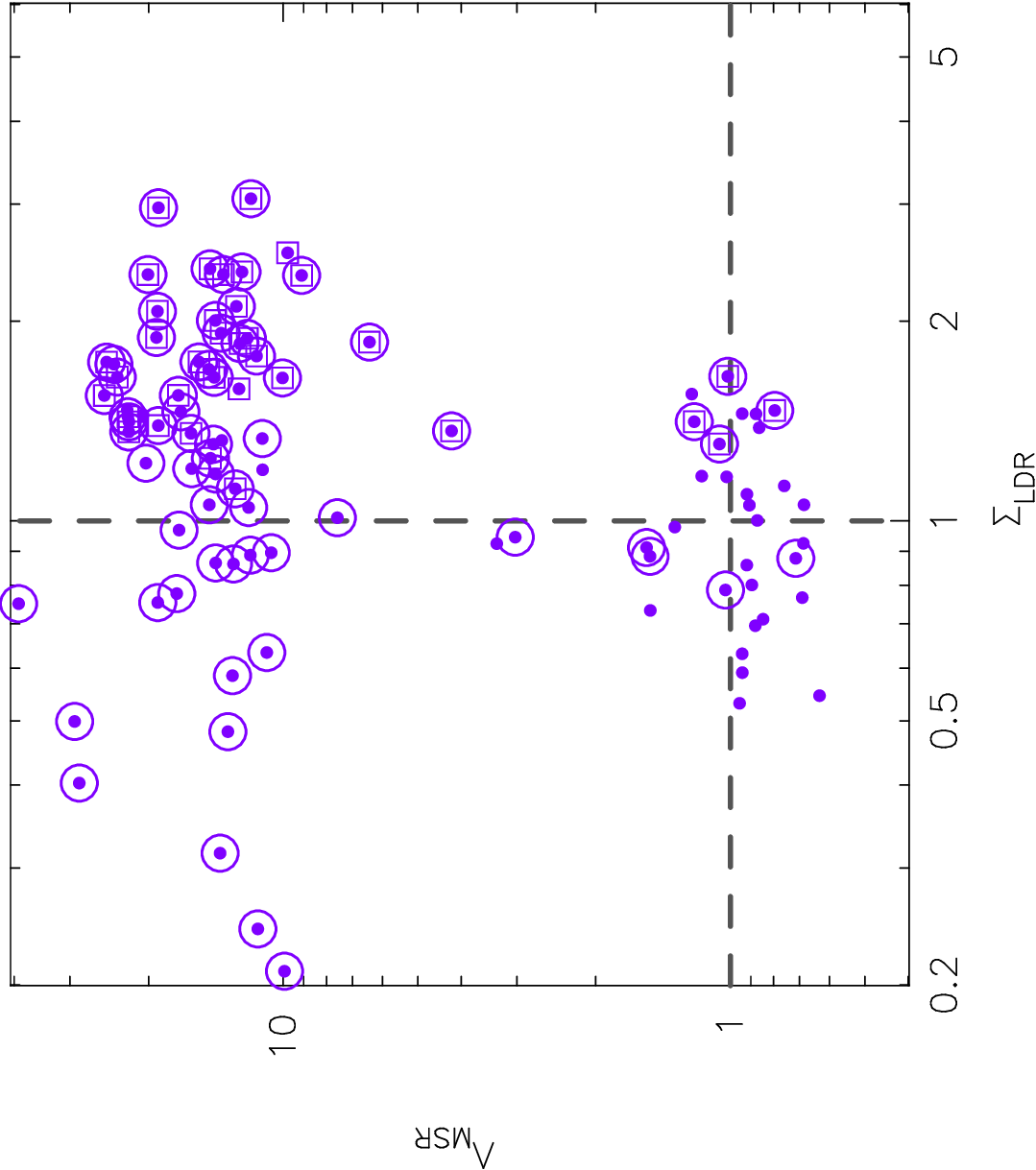}}}
\caption{Simulations in which a population of pre/proto-stellar cores are allowed to fragment into stars and then we measure the mass segregation ratio, $\Lambda_{\rm MSR}$, relative surface density ratio $\Sigma_{\rm LDR}$ and relative gravitational potential, of the most massive stars, compared to low mass stars. Each point represents one simulation (we repeat the experiment 100 times). Where the massive stars have significantly higher surface densities compared to the average stars, we add an open square symbol. Where the most massive stars sit in significantly deeper gravitational potentials, we plot a larger open circle. Panel (a) shows the results for cores that are allowed to randomly fragment into between 1 and 5 stars; panel (b) shows the same experimeent but the fragments are allowed to move up to 0.25\,pc from their birth core, panel (c) shows the results where the most massive cores do not fragment, but low-mass ($\leq 10$M$_\odot$) cores fragment into five stars, and panel (d) shows the results where the most massive cores fragment, but low-mass cores do not. In all panels the horizontal dashed line represents $\Lambda_{\rm MSR} = 1$ (no mass segregation) and the dashed vertical lines represents  $\Sigma_{\rm LDR} = 1$ (the massive stars do not reside in areas of relatively higher or lower surface densities).  }
\label{fragmentation}
  \end{center}
\end{figure*}

  We now describe a set of simulations similar to those in \citet{Alcock19} in which we allow a distribution of pre/proto-stellar cores (a fractal distribution like that shown in Fig.~\ref{synthetic_fractal}) to fragment into stars. The results are shown in Fig.~\ref{fragmentation} and in Tables~\ref{fragmentation_results}~and~\ref{fragmentation_results_d2p6}. In Fig.~\ref{fragmentation}, panel (a) shows the scenario where our cores randomly fragment into between one and five stars, but do not travel far from their fragmentation sites. Panel (b) shows the scenario where objects formed from random core fragmentation into between one and five stars, but the fragments have an offset of 0.25\,pc applied in a random direction, to mimic the effects of dynamical evolution. Panel (c) shows the scenario for fragments where the cores more massive than 10\,M$_\odot$ remain as a single object, but the lower mass cores fragment into between one and five stars. Panel (d) shows the scenario for fragments where the cores more massive than 10\,M$_\odot$ fragment into five stars, but the low-mass cores only form single stars. Each point represents a different realisation of the same simulation, and we run 100 of each case. Where the most massive stars have significantly higher densities than the average stars in a region, we plot an open square in addition to the point, and when the most massive stars sit in deeper potentials we plot a larger open circle. 

Similar to \citet{Alcock19}, we find that enabling cores to fragment often produces significant mass segregation in the entire distribution of fragments. This is seen in Fig.~\ref{fragmentation-a} in the number of black points with $\Lambda_{\rm MSR} > 2$. Interestingly, the majority of these reside in areas of relatively low surface density (more than half the points have $\Sigma_{\rm LDR} < 1$), but in most of the regions the massive stars reside in deeper gravitational potentials than the low-mass stars (the regions where the massive stars reside in deeper potentials are shown by a larger open circle). These distributions are most similar to the scenario shown in the middle columns of Figs.~\ref{synthetic_morphologies} and \ref{synthetic_fractal}, where the massive stars can be centrally concentrated but not in areas of high surface density. 

In the simulations where the fragments are allowed to move up to 0.25\,pc due to dynamical evolution (Fig.~\ref{fragmentation-b}), the numbers of mass segregated regions is lower, as are the numbers of regions where the most massive stars reside in deeper potentials than the low-mass stars. The numbers of regions where the most massive stars high higher relative surface densities is similar.

\begin{table*}
\caption{Summary of the main core fragmentation results for a highly substructured (fractal dimension $D = 1.6$) star-forming region. We list the results for the four different core fragmentation cases: (a) random, (b) random with some dynamical evolution of the fragments, (c) fragmentation of the low-mass ($\leq 10$\,M$_\odot$) cores only, and (d) fragmentation of the high-mass ($>10$\,M$_\odot$) cores only. Each of these cases is shown in the respective panel in Fig.~\ref{fragmentation}. We show the numbers of simulations which are significantly mass-segregated ($\Lambda_{\rm MSR} > 2$) \emph{only}, the numbers which have significantly high relative surface densities \emph{only}, the numbers where massive stars sit in significantly deeper potentials \emph{only}. We then show the numbers of simulations with various combinations of significantly high $\Lambda_{\rm MSR}$, $\Sigma_{\rm LDR}$, $\Phi_{\rm PDR}$, and finally show the numbers of simulations in which all three measures are significantly high. }
\begin{center}
\begin{tabular}{cccccccc}
\hline 
Type of fragmentation & $\Lambda_{\rm MSR}$ & $\Sigma_{\rm LDR}$ & $\Phi_{\rm PDR}$ &  $\Lambda_{\rm MSR}$ \& $\Sigma_{\rm LDR}$ & $\Lambda_{\rm MSR}$ \& $\Phi_{\rm PDR}$ & $\Phi_{\rm PDR}$ \& $\Sigma_{\rm LDR}$ &  $\Lambda_{\rm MSR}$ \& $\Sigma_{\rm LDR}$ \& $\Phi_{\rm PDR}$ \\
\hline
Random & 6 & 1 & 39 & 1 & 29 & 3 & 2 \\
Random + dynamics & 10 & 5 & 9 & 2 & 10 & 0 & 2 \\
Only low-mass fragmentation & 1 & 3 & 13 & 0 & 2 & 0 & 0 \\
Only high-mass fragmentation & 5 & 0 & 4 & 2 & 30 & 4 & 32 \\
\hline
\end{tabular}
\end{center}
\label{fragmentation_results}
\end{table*}

\begin{table*}
\caption{Summary of the main core fragmentation results for an almost smooth (fractal dimension $D = 2.6$) star-forming region. We list the results for the four different core fragmentation cases: (a) random, (b) random with some dynamical evolution of the fragments, (c) fragmentation of the low-mass ($\leq 10$\,M$_\odot$) cores only, and (d) fragmentation of the high-mass ($>10$\,M$_\odot$) cores only. We show the numbers of simulations which are significantly mass-segregated ($\Lambda_{\rm MSR} > 2$) \emph{only}, the numbers which have significantly high relative surface densities \emph{only}, the numbers where massive stars sit in significantly deeper potentials \emph{only}. We then show the numbers of simulations with various combinations of significantly high $\Lambda_{\rm MSR}$, $\Sigma_{\rm LDR}$, $\Phi_{\rm PDR}$, and finally show the numbers of simulations in which all three measures are significantly high.}
\begin{center}
\begin{tabular}{cccccccc}
\hline 
Type of fragmentation & $\Lambda_{\rm MSR}$ & $\Sigma_{\rm LDR}$ & $\Phi_{\rm PDR}$ &  $\Lambda_{\rm MSR}$ \& $\Sigma_{\rm LDR}$ & $\Lambda_{\rm MSR}$ \& $\Phi_{\rm PDR}$ & $\Phi_{\rm PDR}$ \& $\Sigma_{\rm LDR}$ &  $\Lambda_{\rm MSR}$ \& $\Sigma_{\rm LDR}$ \& $\Phi_{\rm PDR}$ \\
\hline
Random & 2 & 0 & 45 & 0 & 33 & 4 & 5 \\
Random + dynamics & 8  & 1 & 20 & 2 & 16 & 4 & 2 \\
Only low-mass fragmentation & 0 & 1 & 12 & 0 & 1 & 3 & 0 \\
Only high-mass fragmentation & 2 & 2 & 9 & 1 & 30 & 7 & 24 \\
\hline
\end{tabular}
\end{center}
\label{fragmentation_results_d2p6}
\end{table*}

Thirdly, we restrict the fragmentation of the cores such than the most massive cores ($>$10\,M$_\odot$) do not fragment, but the lower-mass cores fragment into five stars. This is to mimic a scenario where the most massive cores are prevented from fragmenting, but the low-mass cores are allowed to fragment, such that we recover the IMF \citep{Alcock19}.

The results are shown in Fig.~\ref{fragmentation-c}. In this scenario, the most massive stars do not have a different spatial distribution to the low-mass stars in 80\,per cent of the star-forming regions. However, we note that in 15/100 regions the most massive stars sit in deeper potentials, although in no region does this also correspond to the most massive stars residing in locations of high relative densities. Two of the regions where the most massive stars sit in deeper potentials display significant mass segregation  ($\Lambda_{\rm MSR} > 2$).

Finally, when we prevent the low-mass cores from fragmenting, but allow the high-mass cores to fragment into 5 objects, we see significant numbers of star-forming regions where the fragments are mass-segregated, and they lie in deeper potential wells (Fig.~\ref{fragmentation-d}). Some of these regions (though not all)  also have massive stars with high relative surface densities. The reason for the high numbers of mass-segregated fragments is due to an already massive core splitting into closely-associated massive objects \citep{Alcock19}.

If we compare the fragmentation scenarios in Fig.~\ref{fragmentation} with the SPH results summarised in Fig.~\ref{SPH_sim_results}, we see some overlap, but there appears to be no fragmentation scenario that is fully consistent with the competitive accretion simulations. The closest is the random fragmentation scenario with no dynamical evolution (Fig.~\ref{fragmentation-a}), where about 1/10 of the fragmentation simulations are mass-segregated, the majority have massive stars sitting in deeper potential wells, and several have massive stars residing in areas of high local density. This is consistent with the SPH control simulations (Fig.~\ref{SPH_sim_results-a}), though clearly these SPH simulations (by design) are missing important physics, namely feedback. 

The fragmentation simulations where only the low-mass cores fragment (Fig.~\ref{fragmentation-c}) is most similar to the SPH simulations that have feedback (Fig.~\ref{SPH_sim_results-b}), though these fragmentation simulations produce some regions where the massive stars reside in high area of local density, whereas none of the ten SPH simulations show this.

\section{Discussion}
\label{discuss}

One of the predictions of the competitive accretion theory is that the most massive stars might be expected to have a different spatial distribution to low-mass stars, if the reason for their growth is due to access to a larger reservoir of gas. This might be either deep in a gravitational potential well, and/or in the central region of the star forming region as it is forming stars, which may also translate into the most massive stars residing in areas of high relative densities.

We find in our SPH simulations where the massive stars form from competitive accretion, the massive stars do sit in deeper potential wells, and in the majority of these simulations have high relative surface densities, \emph{only if} the growth of massive stars procedes unchecked by feedback from the massive stars. When feedback is implemented, the runaway growth of massive stars is prevented, and the massive stars sit deeper in potentials in only 6/10 simulations.

We have attempted to map the relation between the massive stars sitting in deeper potentials, and other measures of the spatial distribution of massive stars, such as mass segregation. In only 1/10 simulations are the massive stars truly mass segregated (more centrally concentrated than the average stars), and there is no dependence on feedback.

Other simulations of massive star formation find similar results for star formation that occurs without feedback \citep{Maschberger11,Myers14}, namely that the most massive stars tend to be in areas of high relative densities. Our simulations do not include magnetic fields, and \citet{Myers14} and \citet{Guszejnov22} find that magnetic fields may significantly affect the final spatial distribution of the most massive stars.

In our synthetic simulations, the massive stars sitting in deeper potential wells corresponds to high relative surface densities \emph{and} mass segregation only if the region is smooth and centrally concentrated (like an old, evolved star cluster). When there is spatial substructure (as is the case in our SPH simulations, and in observed star-forming regions), then mass-segregated stars sit in deep potentials, but do not have high relative densities. Conversely, if the massive stars are in locations of high relative densities, they sit deeper in pontential wells but are not mass-segregated.

We find that massive stars can also end up in deeper potentials (and have high surface densities and/or be mass segregated) in models in which we assume they form via monolithic collapse and fragmentation of very massive cores. We see significant suppression of these signals if massive cores are not allowed to fragment, but the lower mass cores are. However, if we allow some fragmentation of the massive cores, we still form massive stars, and these massive stars still dominate their local gravitational potentials.

We note, however, that our analytic models of massive star formation via monolithic collapse contain no physics, beyond the assumptions that stars will form in a substructured distribution \citep{Cartwright04,Hacar16}, that they will form cores with a mass distribution similar to the CMF in \citet{Alves07}, and that these cores fragment into several stars. Ideally, the spatial distributions of the most massive stars in our SPH simulations of competitive accretion should be compared to the equivalent simulations that model monolithic collapse to determine what fraction of star-forming regions form massive stars that reside in deeper potential wells, and have high relative surface densities, and/or mass segregation.

The calculations of the gravitational potentials do not include the gravitational potential of the gas, as this was only present in the SPH simulations, and not in our synthetic datasets or core fragmentation simulations. We would expect that the gas might contribute to the potentials in the SPH control runs, but in the simulations with feedback, the massive stars clear out gas from their vicinities.

  \citet{Dale15b} calculate the potentials in some of the SPH simulations in this paper (J, I, UF, UP \& UQ) and include the gas particles, as well as the star particles. In their fig.~4 they present the results for Run I (see Figs.~\ref{I180_sim_results_control}~and~\ref{I180_sim_results_dual} in this paper), and whilst the inclusion of the gas particles  does produce different distributions, the main differences are similar -- the control run has only one main potential well, whereas the feedback run has several.

\section{Conclusions}
\label{conclude}

We present the results of SPH simulations of massive star formation via `competitive accretion', both with and without feedback, to search for signs that the most massive stars have a different spatial distribution to the low-mass, or average-mass stars. We also present synthetic models of star-forming regions in which the massive stars have a random spatial distribution, a centrally concentrated distribution, and where the massive stars are in areas of high surface densities. We also present analytic models of massive star formation via core collapse and fragmentation, to mimic the `monolithic collapse' scenario. Our conclusions are the following:

(i) Massive stars that form via competitive accretion sit in deeper gravitional potentials than the average mass stars, and usually have higher than average local densities, when forming in simulations with no subsequent feedback from the massive stars. They do not, however, exhibit mass segregation in the sense of being more centrally concentrated than the average mass stars.

(ii) When feedback is implemented, massive stars reside in deeper gravitational potentials in only half of simulations, and in none of these simulations are the massive stars in locations of relatively high surface densities. One simulation is mass-segregated, but the remaining nine are not.

(iii) Our synthetic data models suggest that any interpretation of the spatial distribution of massive stars must take into account the morphology of a star-forming region. Where the massive stars form in a smooth, centrally concentrated star-forming region, the most massive stars will sit in deeper potentials, be mass segregated and have high relative surface densities when the massive stars are centrally concentrated, or when they are in areas of high surface densities.

(iv) Most star-foming regions (both observed and simulated) seem to form with spatial substructure. In this scenario, the massive stars being centrally concentrated is a very different type of distribution compared to when they are in locations of high relative density. In the former scenario, the massive stars sit in deeper potential wells, but have low relative densities and are mass-segregated; in the latter scenario the massive stars again are in deeper potential wells but have  high relative densities and no mass segregation (like the SPH control runs).

(v) Analytic models that mimic massive star formation via core collapse and fragmentation can also lead to the most massive stars sitting in deeper potentials, even when the most massive stars fragment differently to lower-mass stars.

(vi) The most basic analytic fragmentation models produce spatial distributions of massive stars that are similar to those produced by SPH control runs with no feedback (which are most like the competitive accretion scenario). Whilst these SPH runs lack important physics compared to the runs with feedback, they demonstrate a significant degeneracy in distinguishing between signatures of competitive accretion and monolithic collapse.

In summary, we conclude that observations of massive stars residing in deep gravitational potentials (and higher relative surface densities, and/or mass segregation) may not be an indication of their formation by competitive accretion, and it is difficult to distinguish from their formation from monolithic collapse in an individual star-forming region.

\section*{Acknowledgements}

We thank the anonymous referee for their detailed and helpful reports. RJP acknowledges support from the Royal Society in the form of a Dorothy Hodgkin Fellowship. For the purpose of open access, the corresponding author has applied a Creative Commons Attribution (CC BY) licence to any Author Accepted Manuscript version arising.



\bibliographystyle{aasjournal}  
\bibliography{general_ref}

\appendix

In this Appendix we provide plots for the remaining nine pairs of SPH simulations that are not shown in Fig.~\ref{SPH_sim_results_control} and Fig.~\ref{SPH_sim_results_dual}. The symbols, lines, legends, etc., are the same as in Figs.~\ref{SPH_sim_results_control}~and~\ref{SPH_sim_results_dual}.

\begin{figure*}
  \begin{center}
\setlength{\subfigcapskip}{10pt}

\subfigure[]{\rotatebox{270}{\includegraphics[scale=0.21]{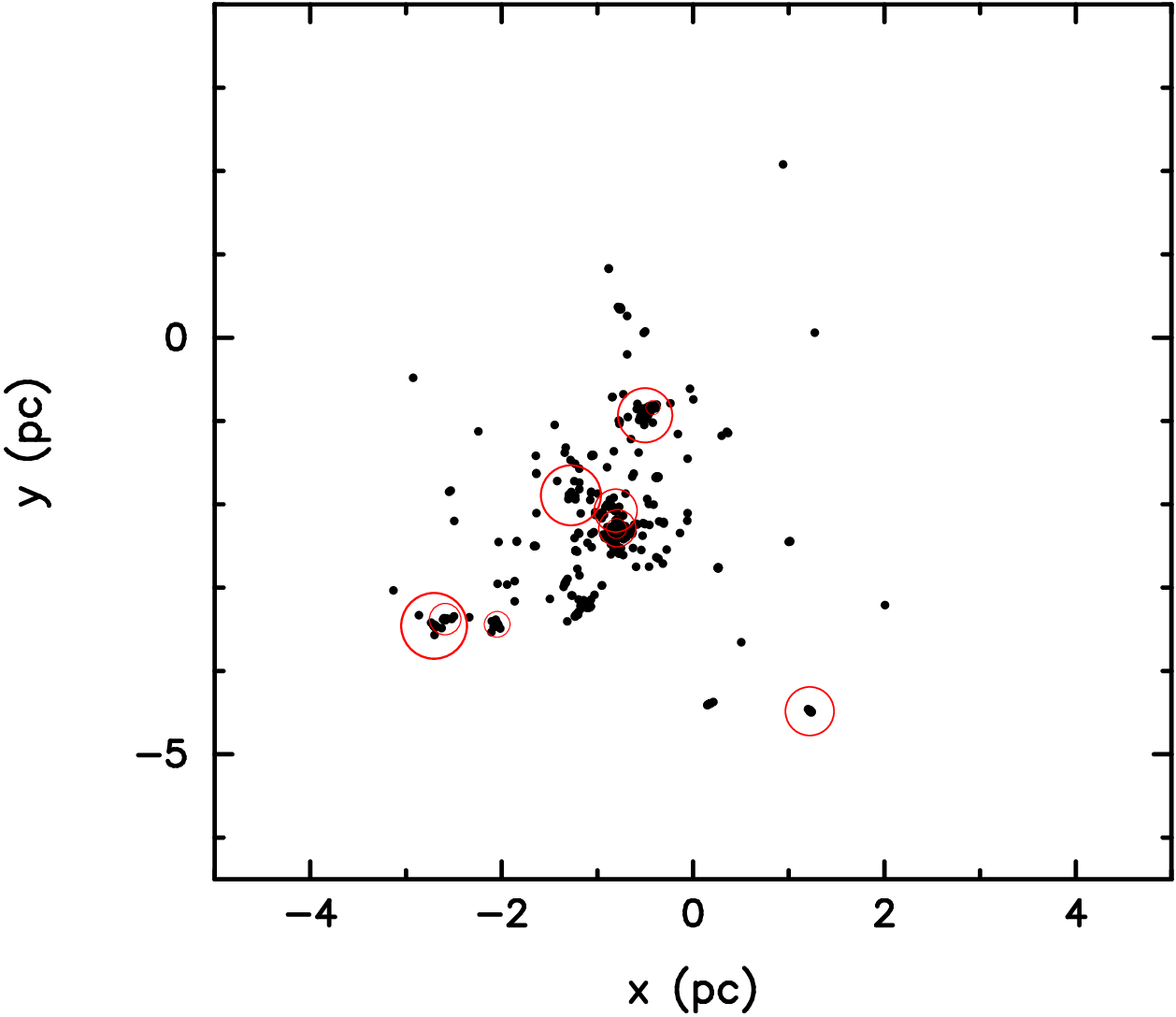}}}
\subfigure[]{\rotatebox{270}{\includegraphics[scale=0.21]{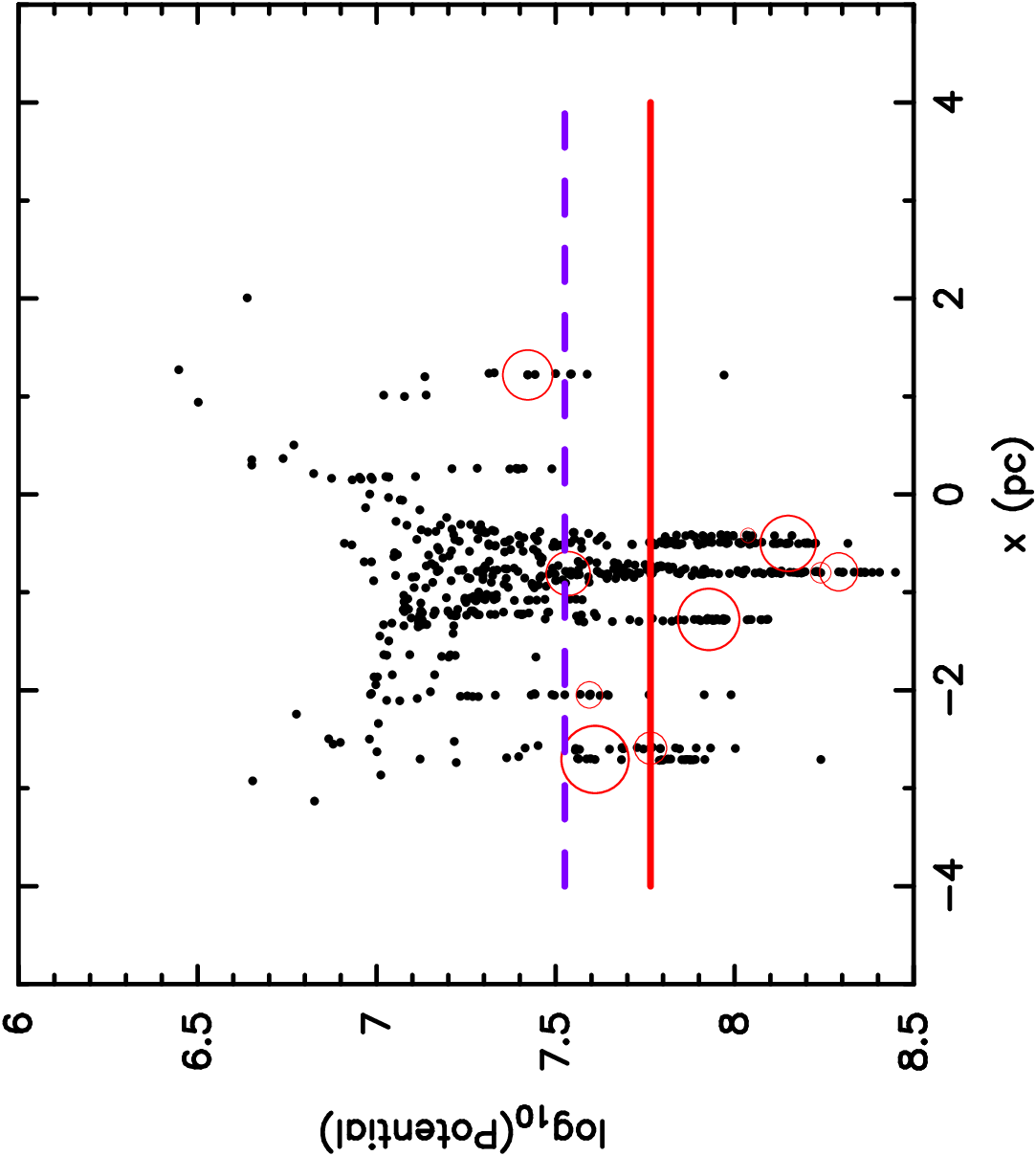}}}
\subfigure[]{\rotatebox{270}{\includegraphics[scale=0.21]{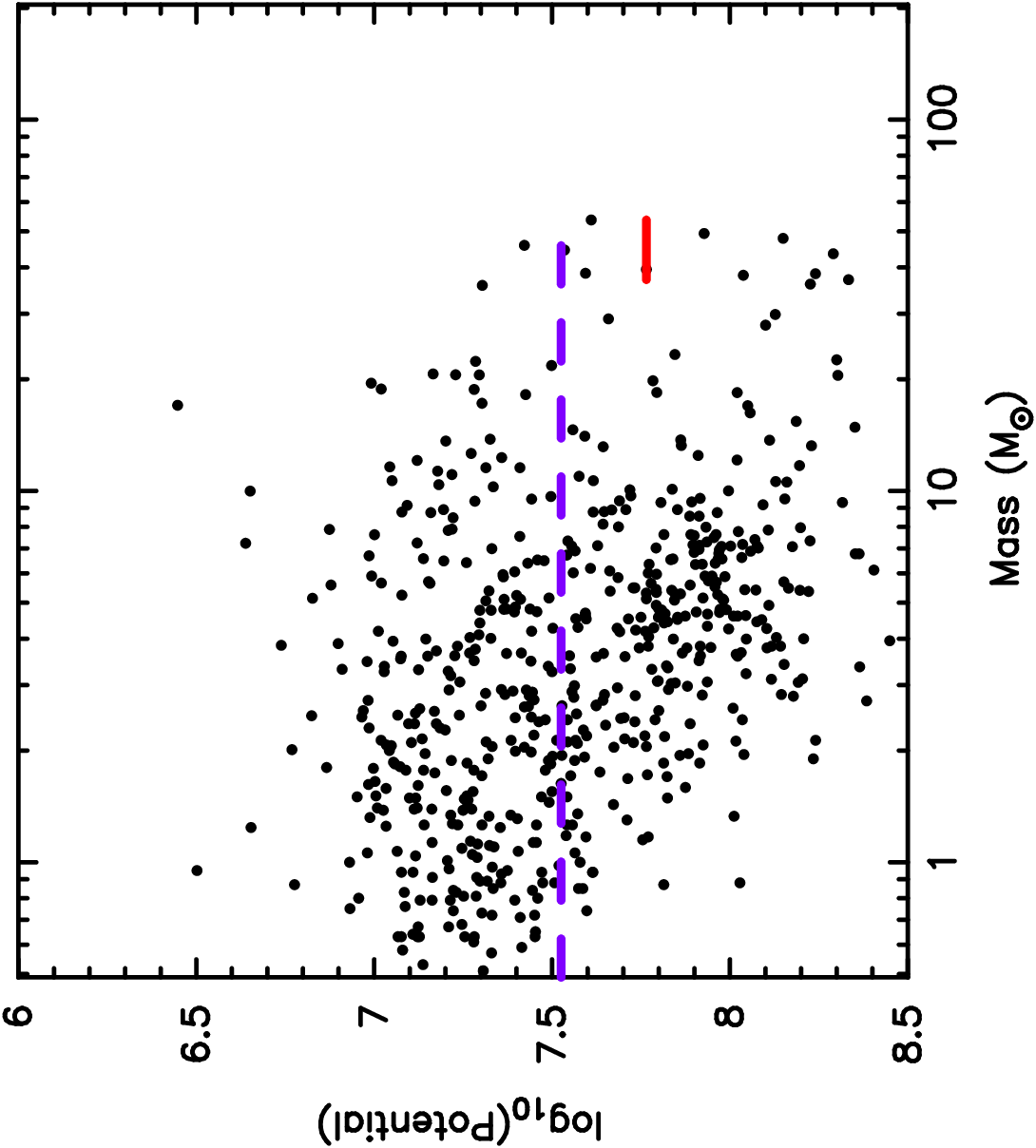}}}
\subfigure[]{\rotatebox{270}{\includegraphics[scale=0.21]{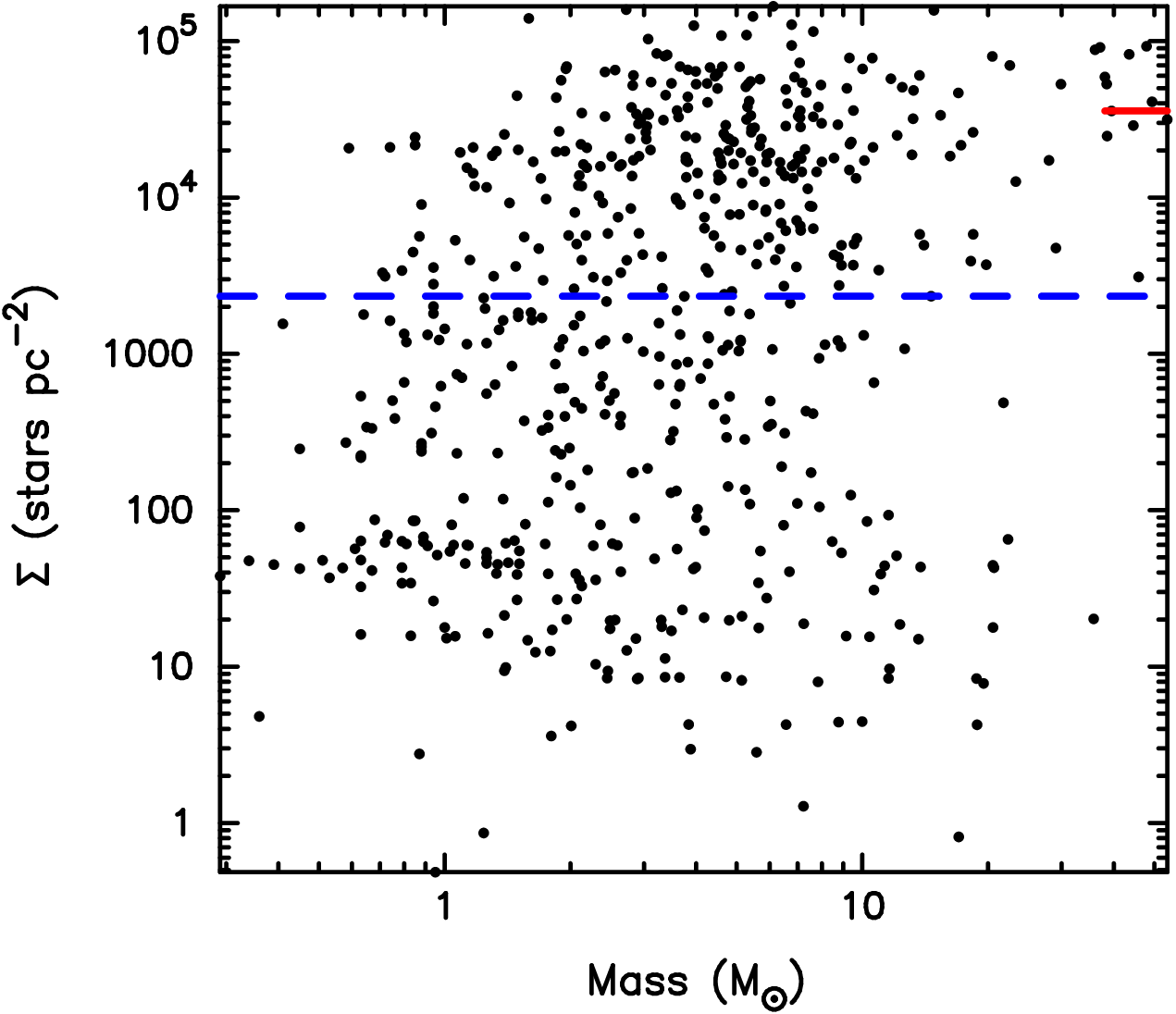}}}

\caption{J Control run.}
\label{J113_sim_results_control}
  \end{center}
\end{figure*}

\begin{figure*}
  \begin{center}
\setlength{\subfigcapskip}{10pt}

\subfigure[]{\rotatebox{270}{\includegraphics[scale=0.21]{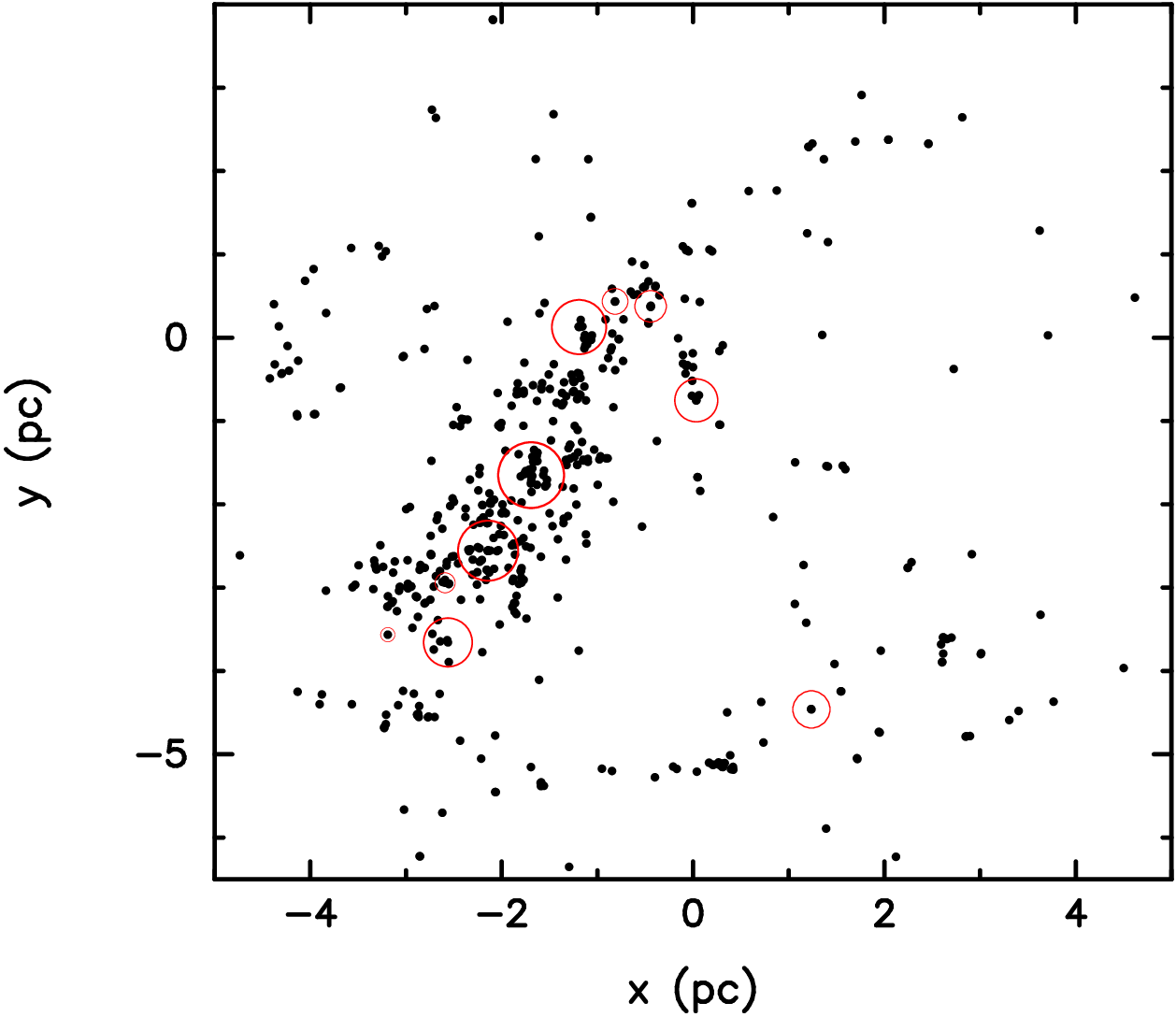}}}
\subfigure[]{\rotatebox{270}{\includegraphics[scale=0.21]{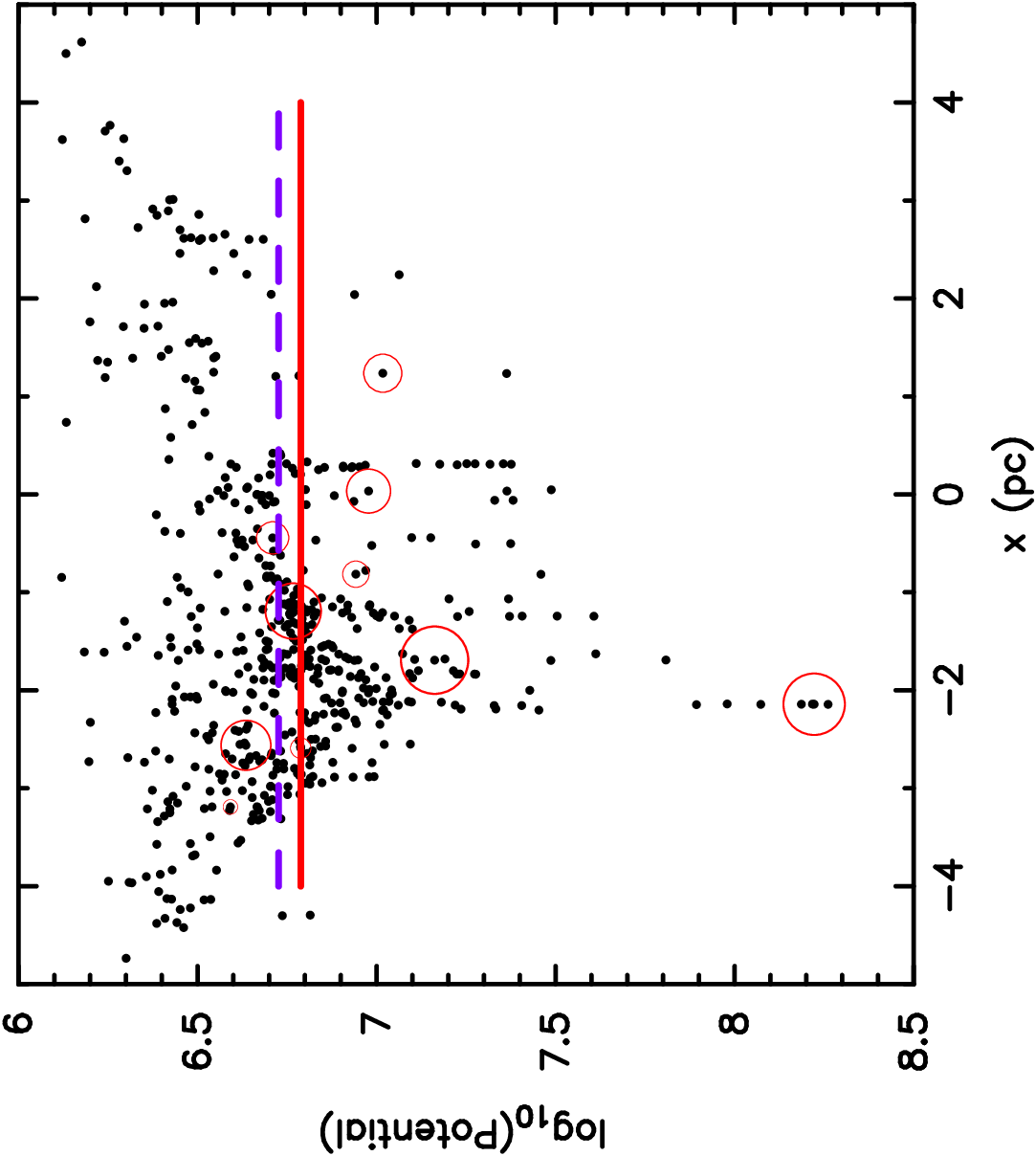}}}
\subfigure[]{\rotatebox{270}{\includegraphics[scale=0.21]{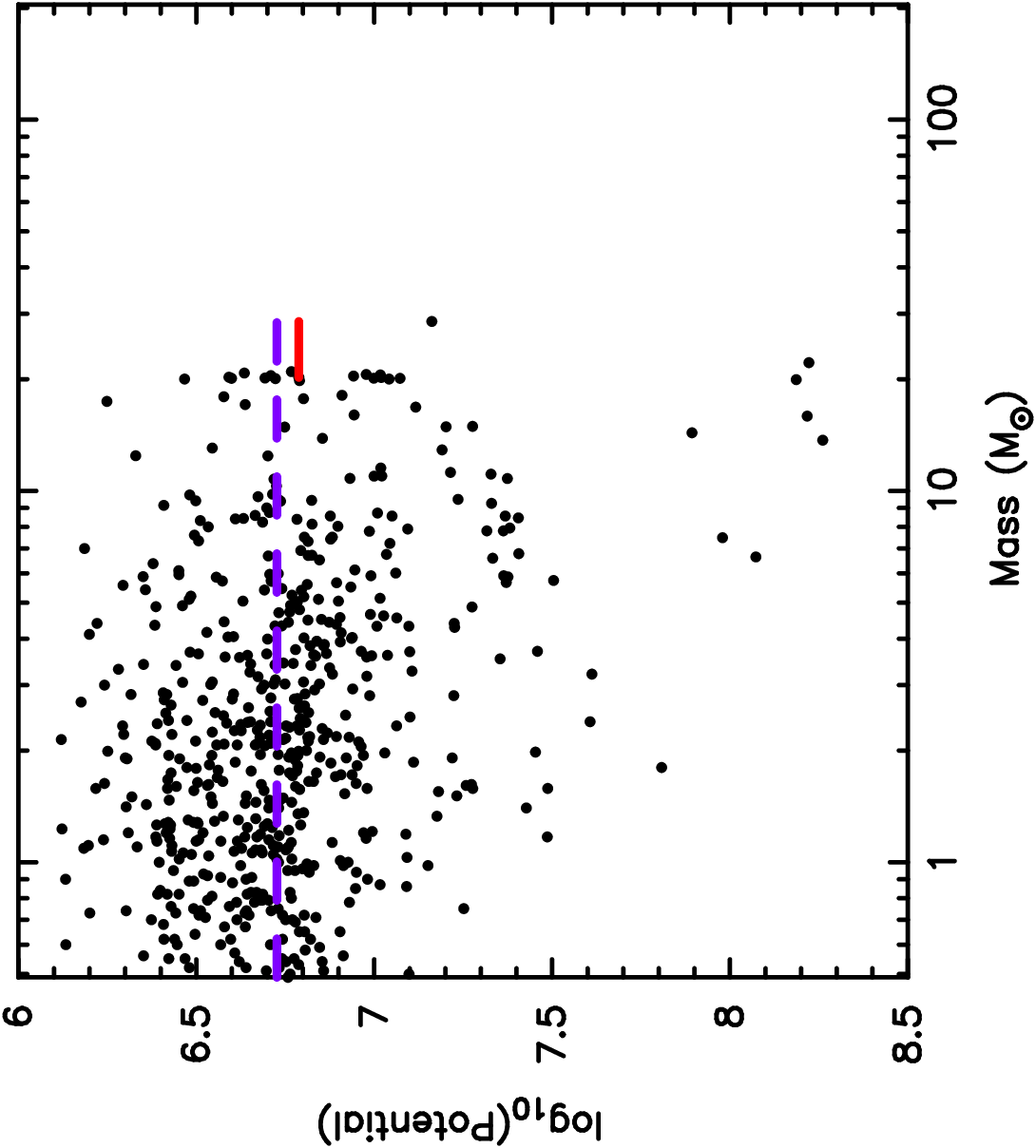}}}
\subfigure[]{\rotatebox{270}{\includegraphics[scale=0.21]{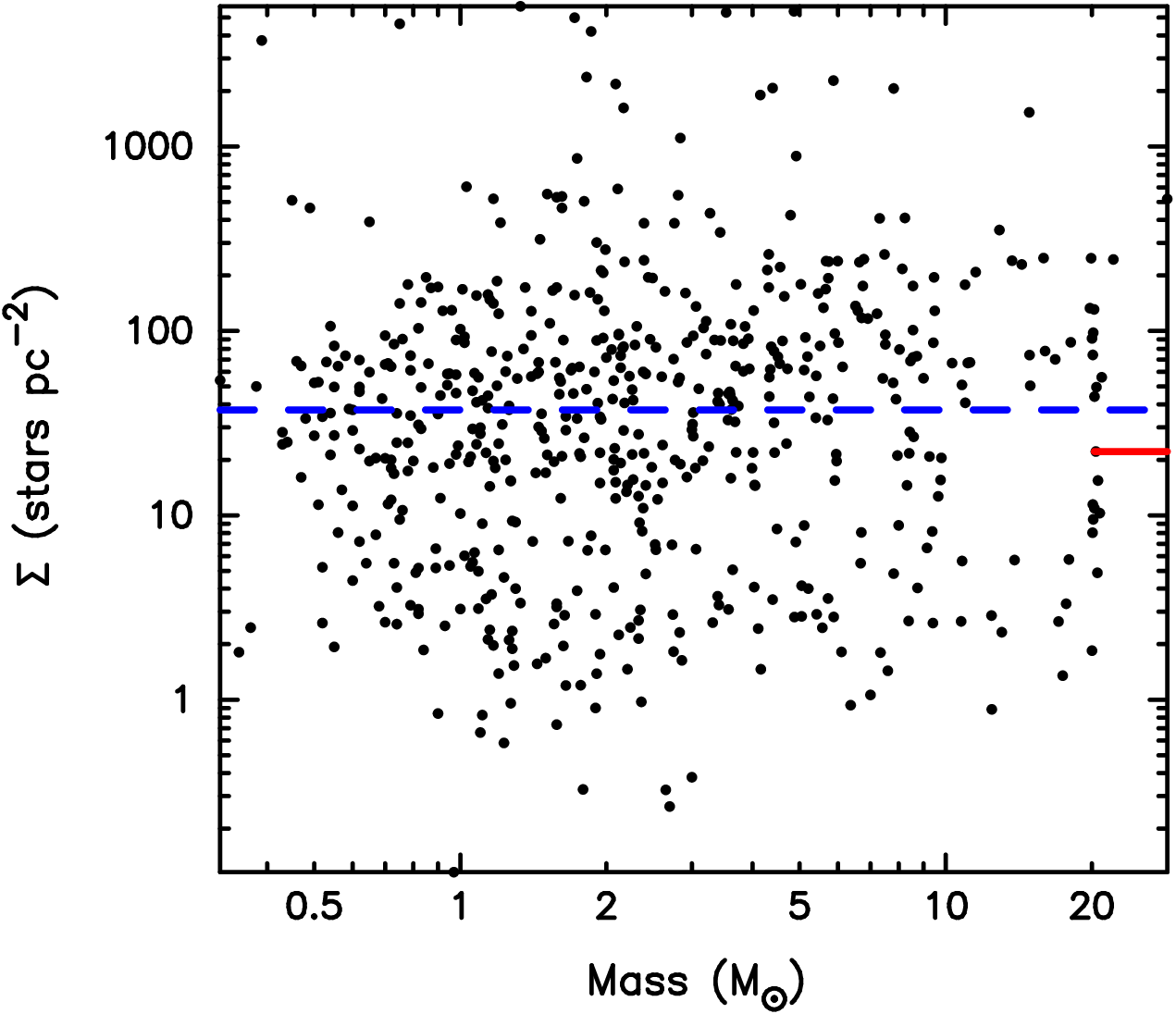}}}

\caption{J Dual feedback run.  }
\label{J113_sim_results_dual}
  \end{center}
\end{figure*}

\begin{figure*}
  \begin{center}
\setlength{\subfigcapskip}{10pt}

\subfigure[]{\rotatebox{270}{\includegraphics[scale=0.21]{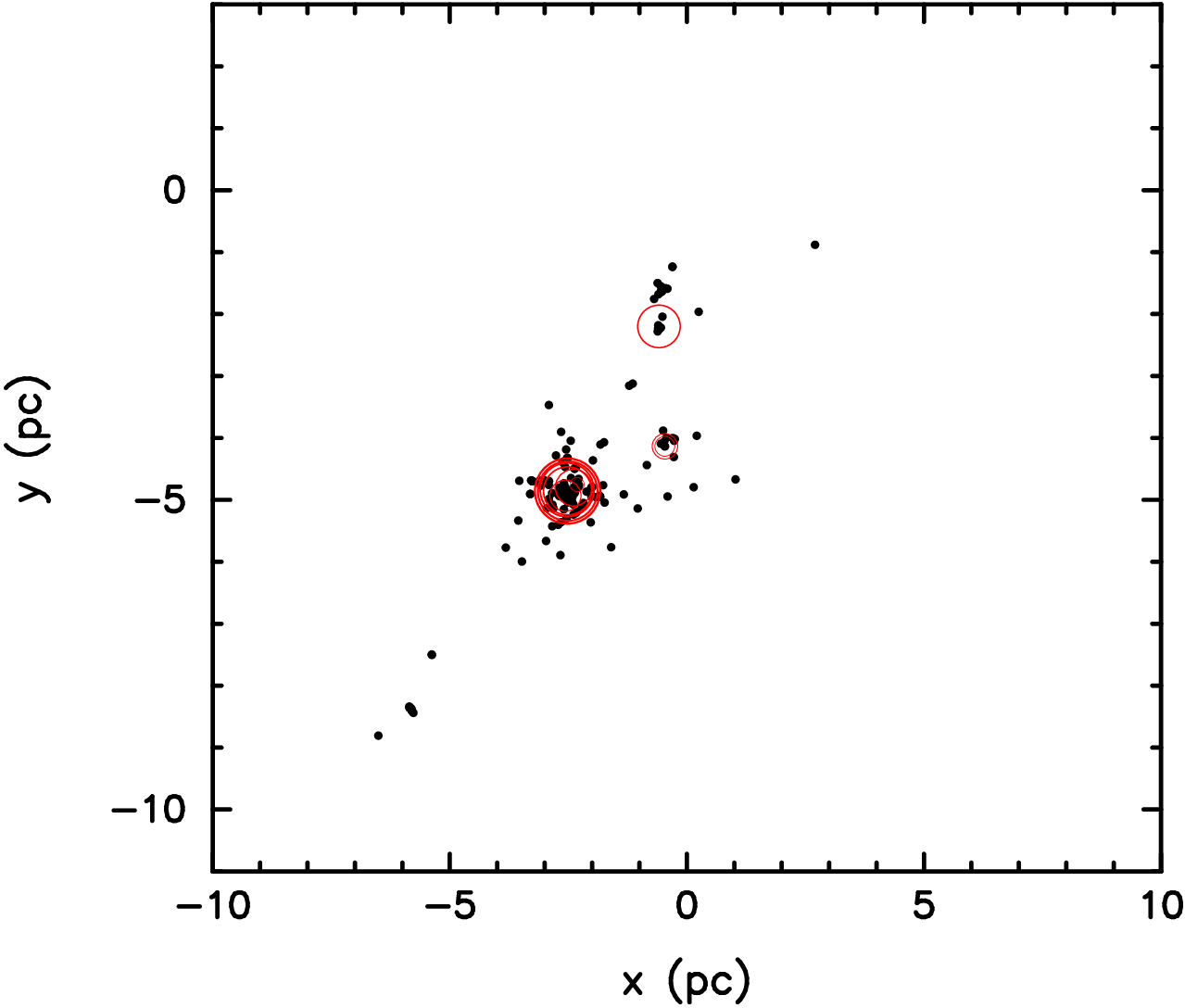}}}
\subfigure[]{\rotatebox{270}{\includegraphics[scale=0.21]{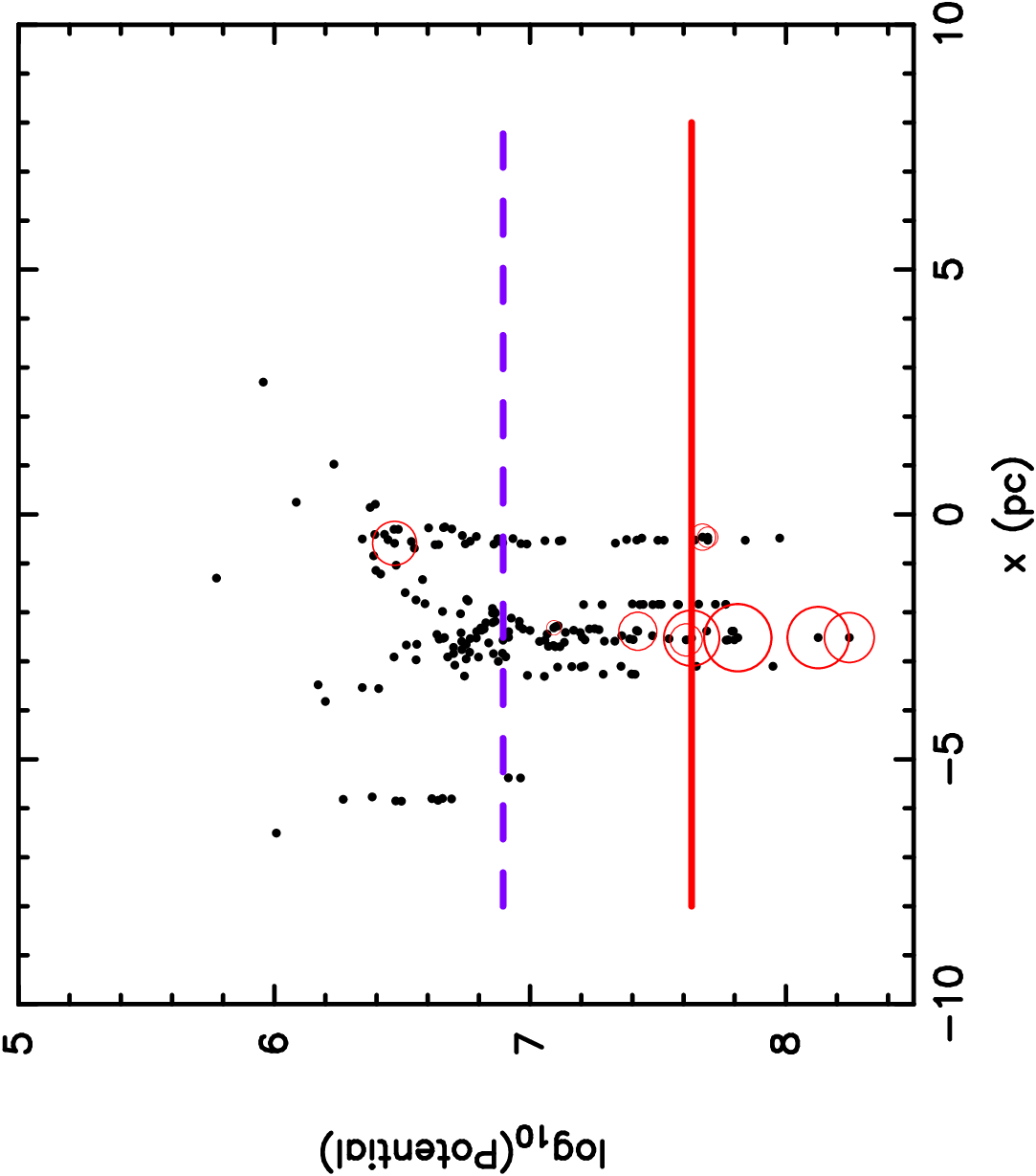}}}
\subfigure[]{\rotatebox{270}{\includegraphics[scale=0.21]{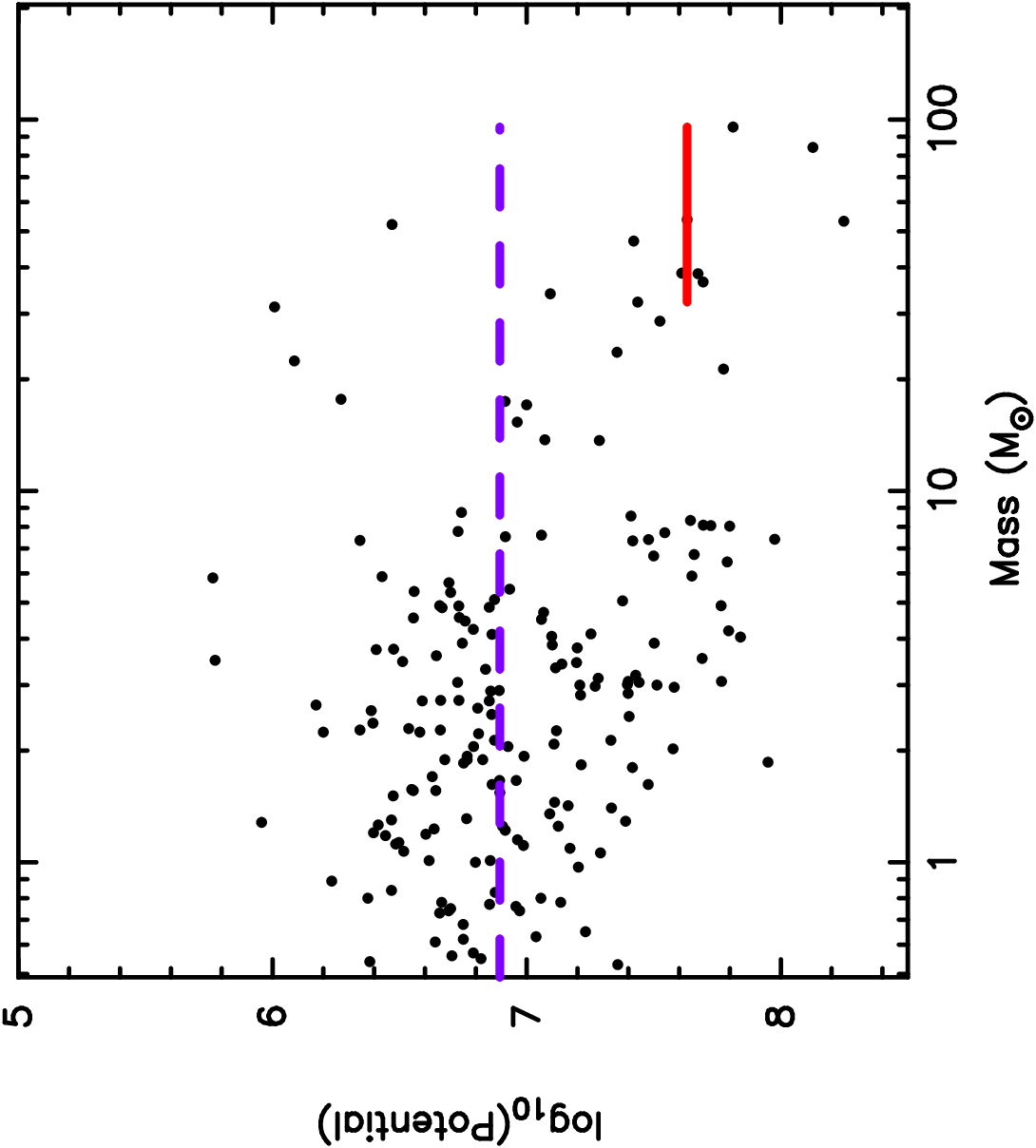}}}
\subfigure[]{\rotatebox{270}{\includegraphics[scale=0.21]{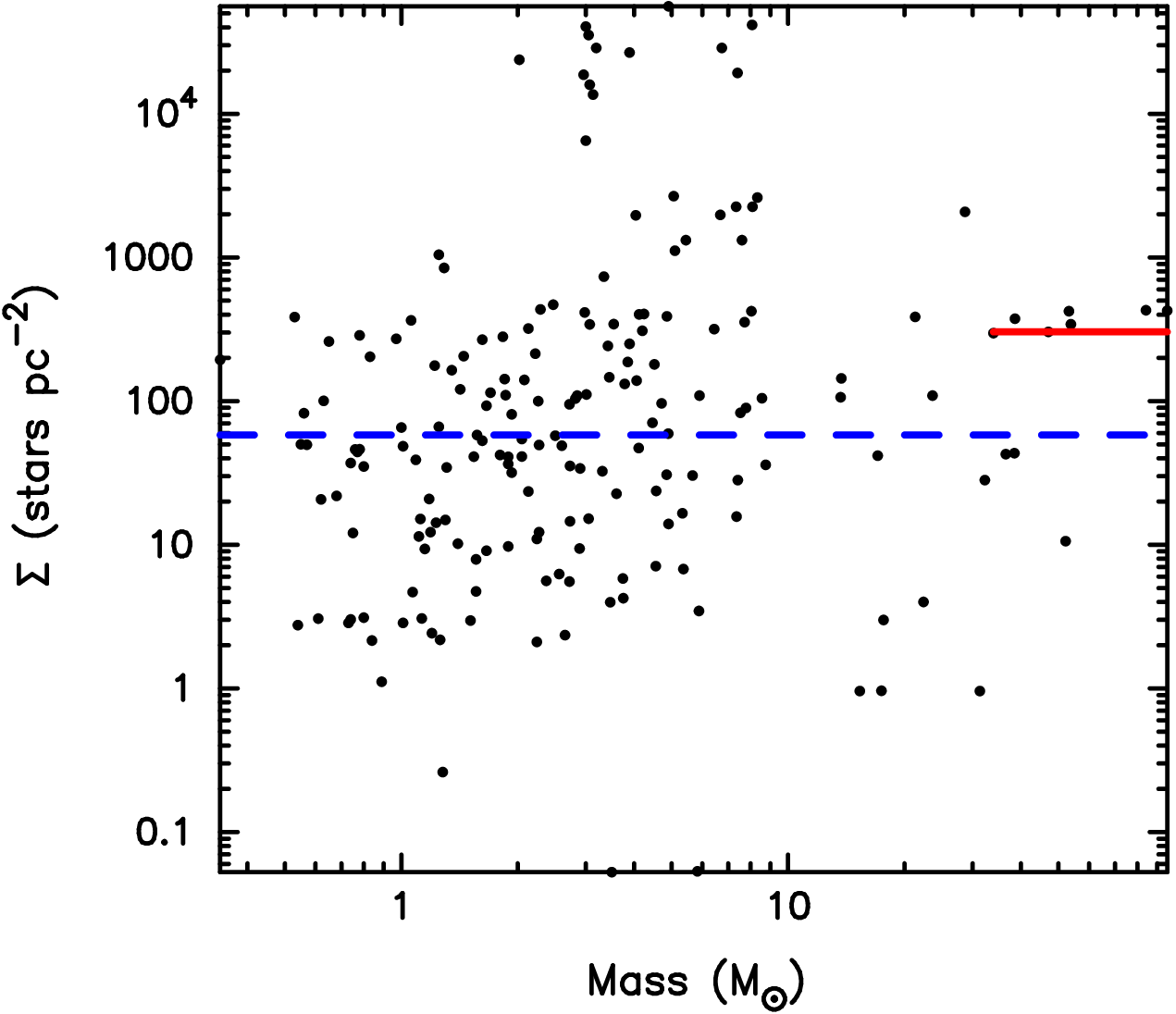}}}

\caption{I Control  run. }
\label{I180_sim_results_control}
  \end{center}
\end{figure*}

\begin{figure*}
  \begin{center}
\setlength{\subfigcapskip}{10pt}

\subfigure[]{\rotatebox{270}{\includegraphics[scale=0.21]{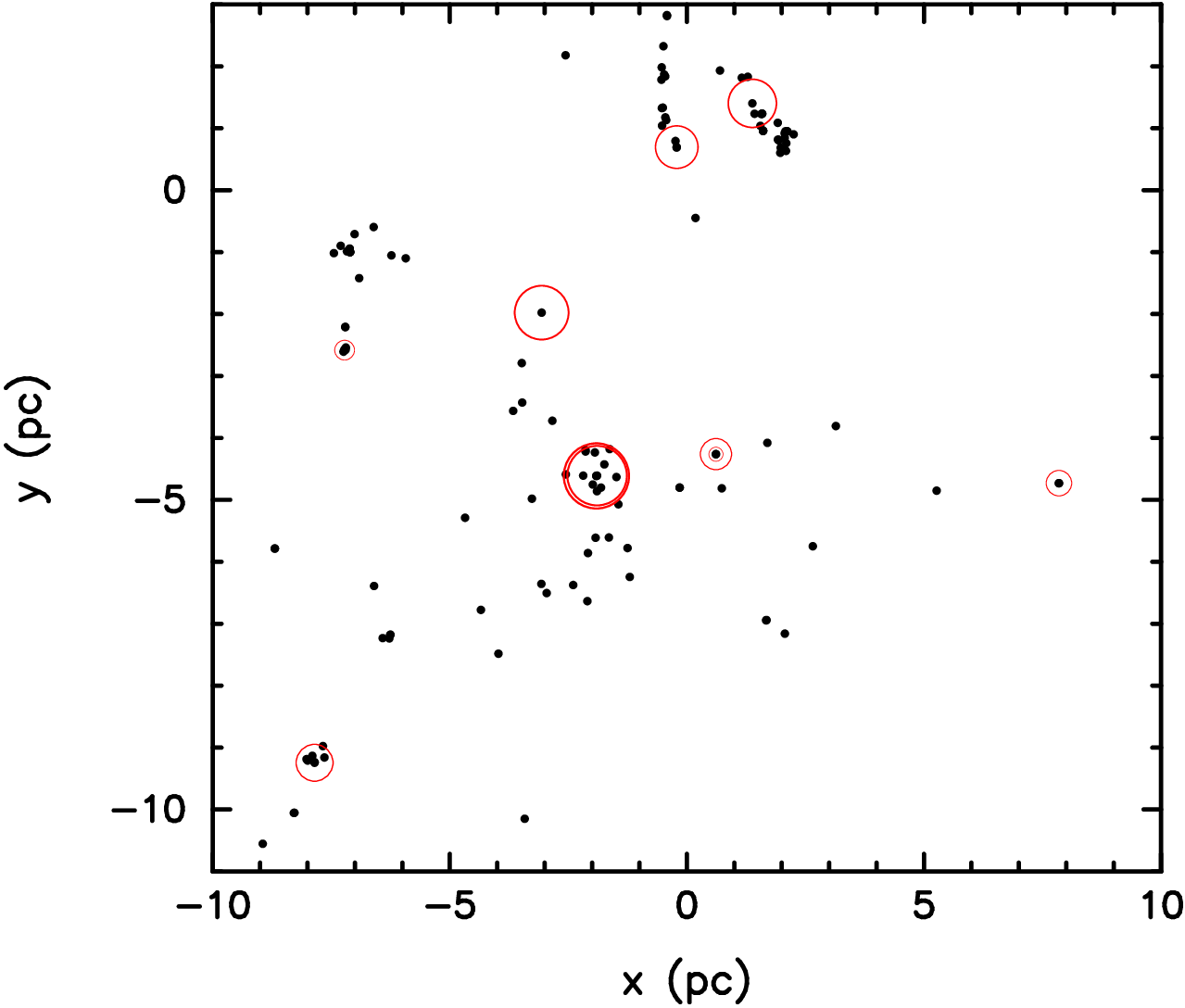}}}
\subfigure[]{\rotatebox{270}{\includegraphics[scale=0.21]{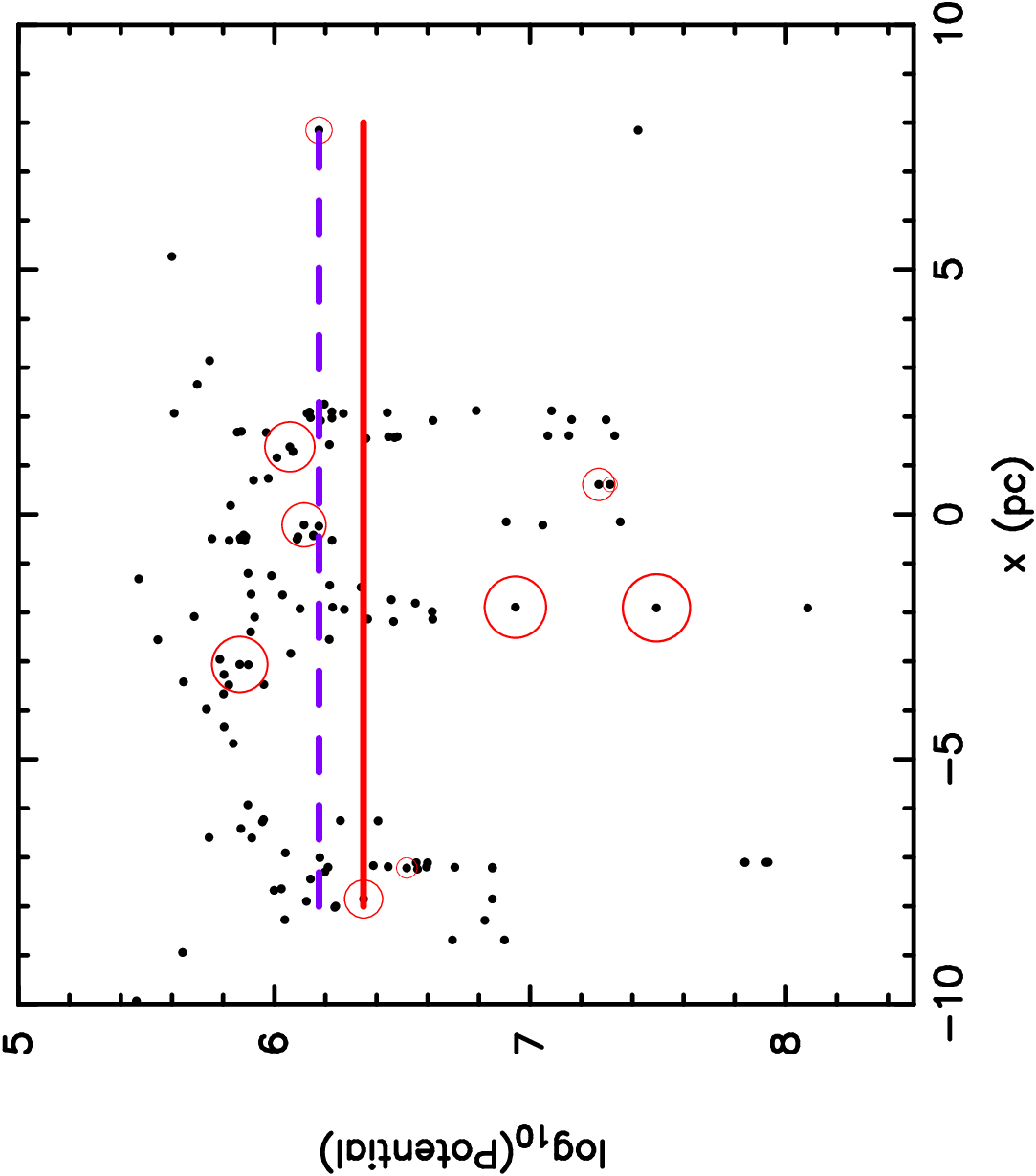}}}
\subfigure[]{\rotatebox{270}{\includegraphics[scale=0.21]{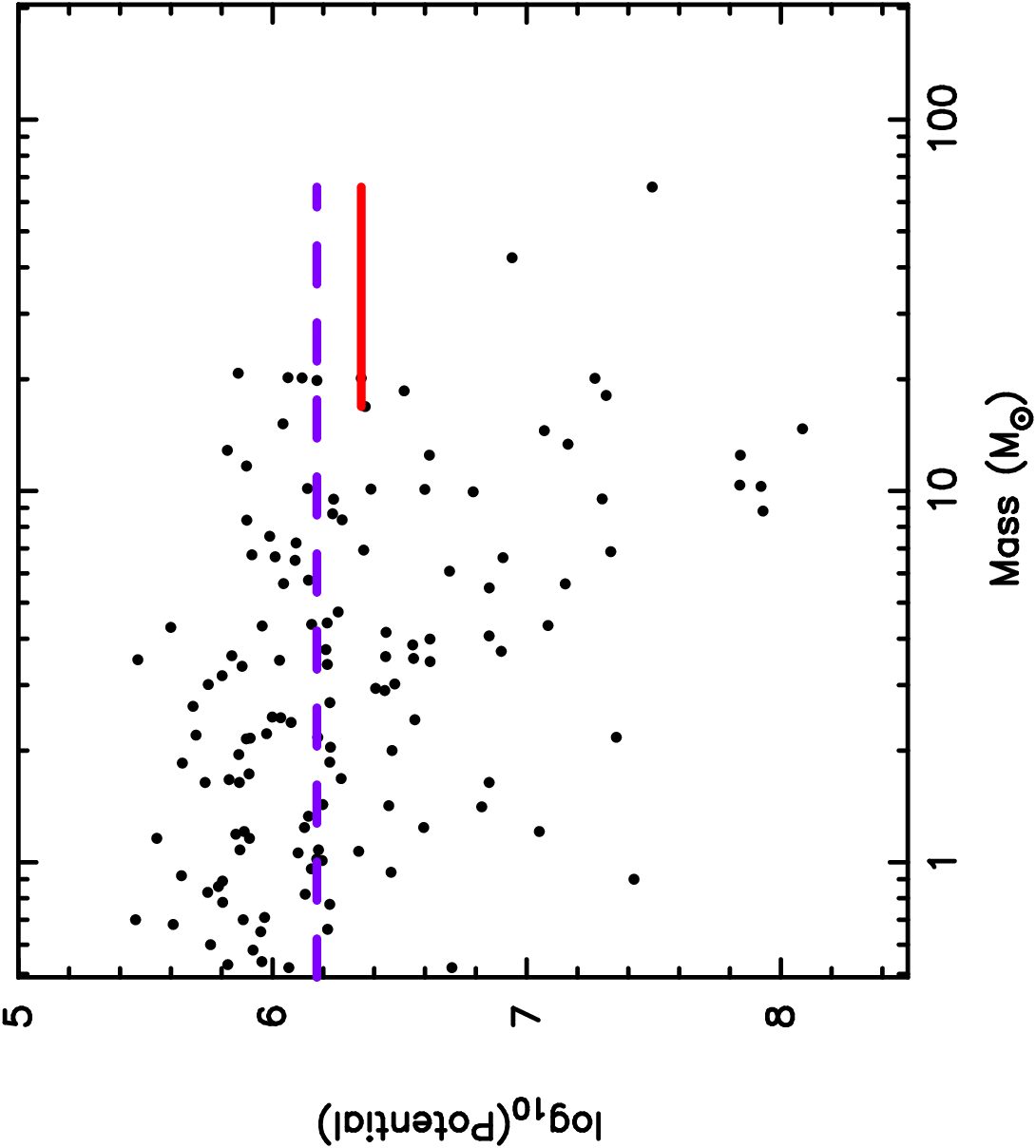}}}
\subfigure[]{\rotatebox{270}{\includegraphics[scale=0.21]{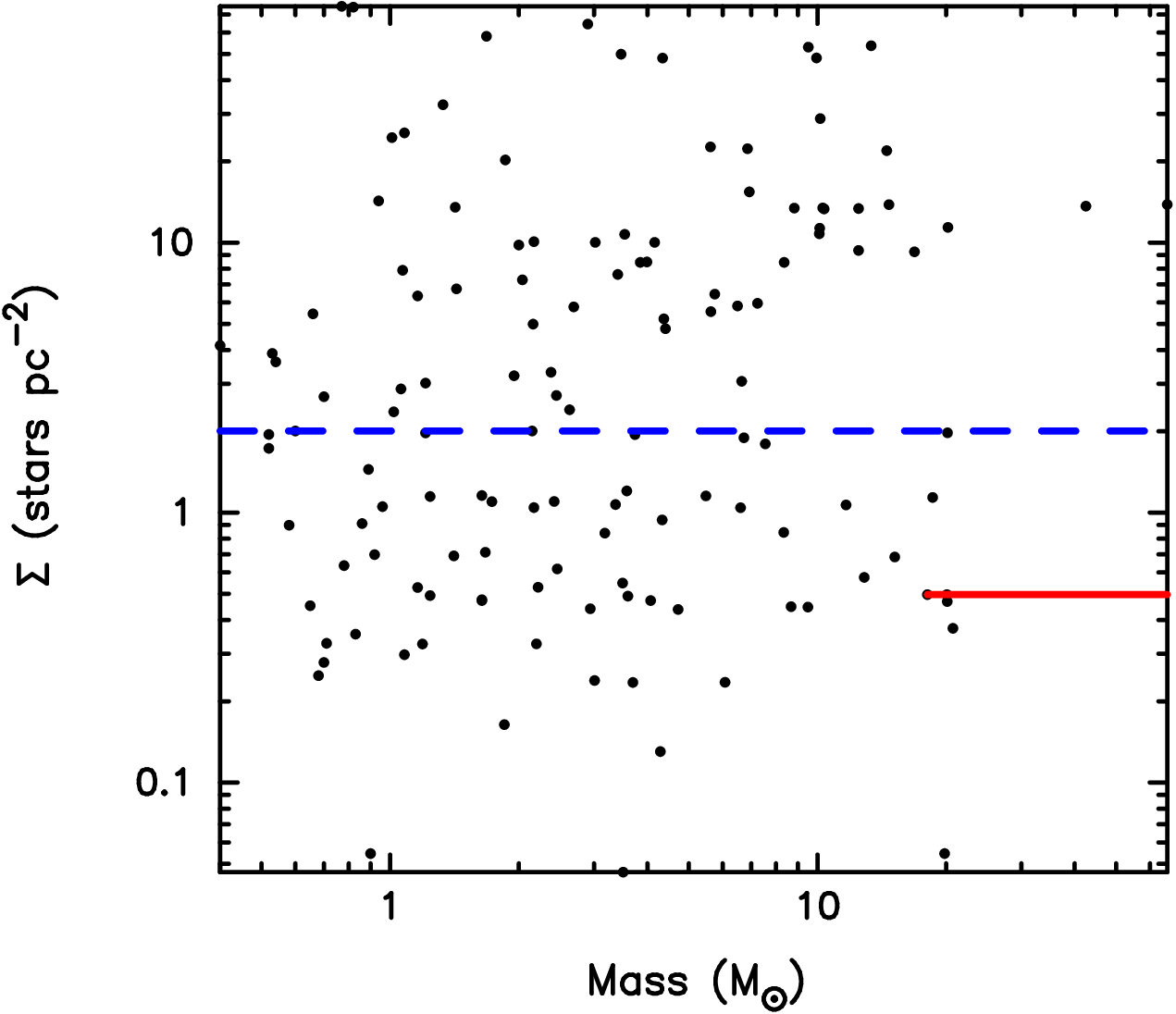}}}

\caption{I Dual feedback  run. }
\label{I180_sim_results_dual}
  \end{center}
\end{figure*}

\begin{figure*}
  \begin{center}
\setlength{\subfigcapskip}{10pt}

\subfigure[]{\rotatebox{270}{\includegraphics[scale=0.21]{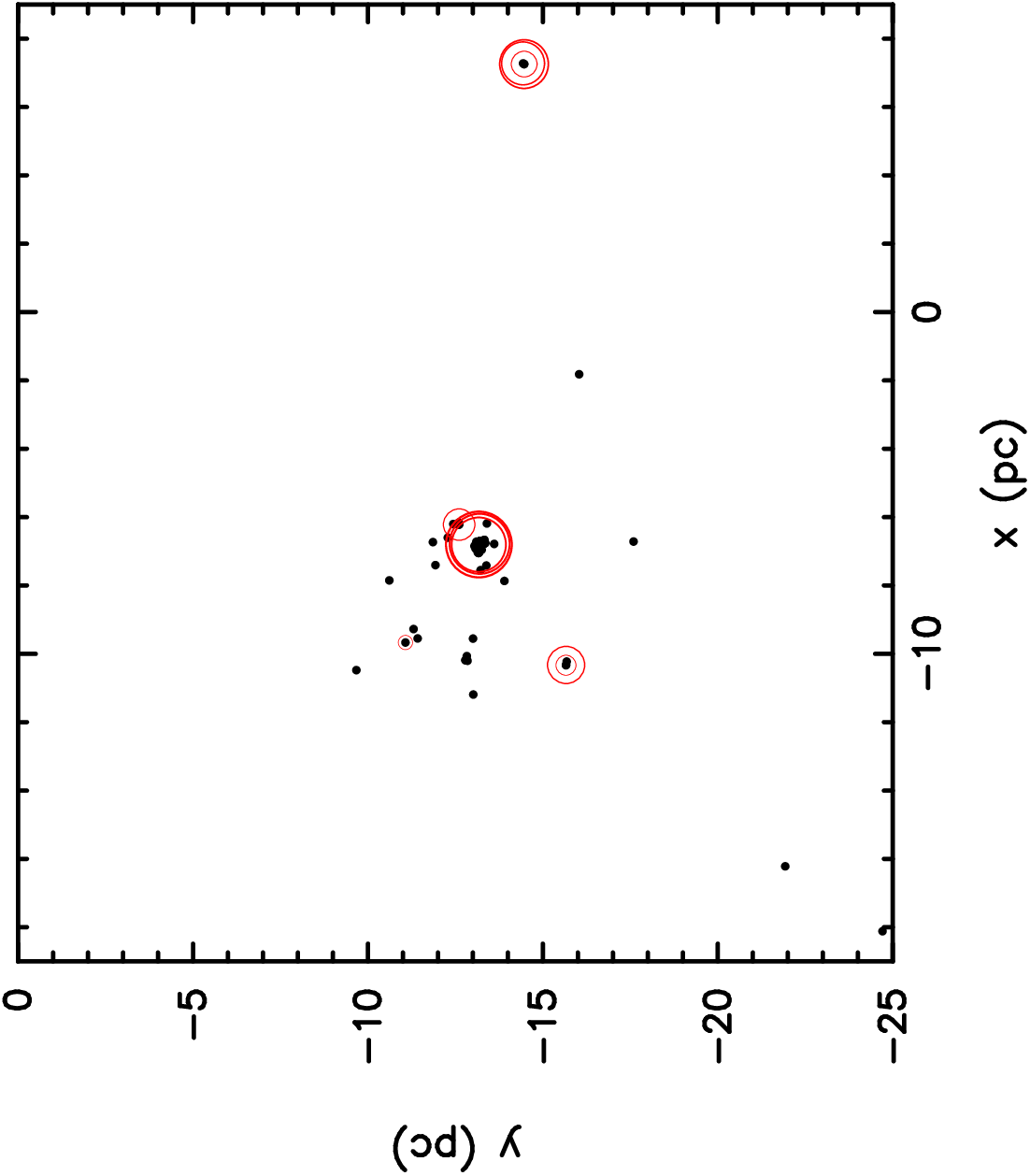}}}
\subfigure[]{\rotatebox{270}{\includegraphics[scale=0.21]{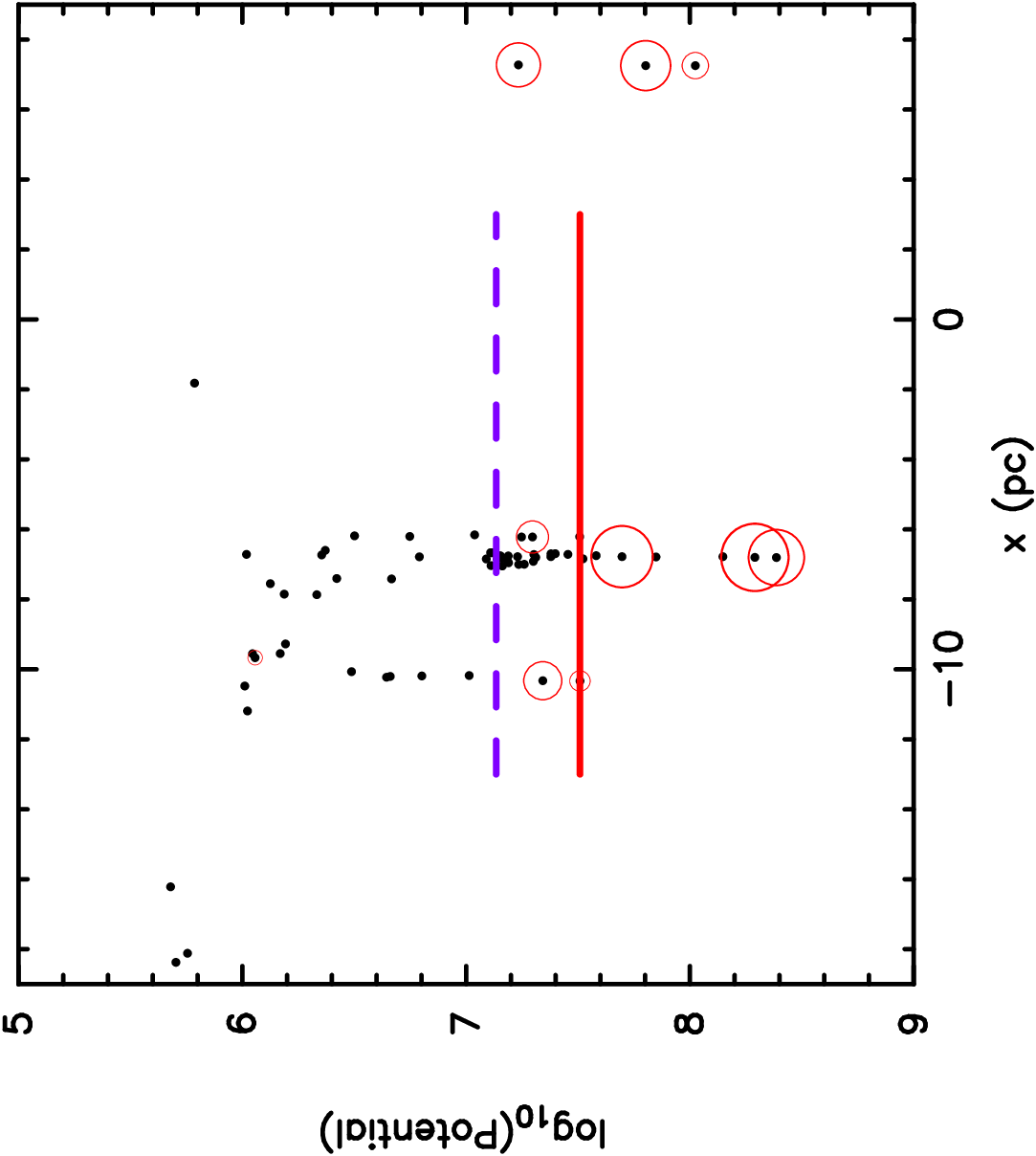}}}
\subfigure[]{\rotatebox{270}{\includegraphics[scale=0.21]{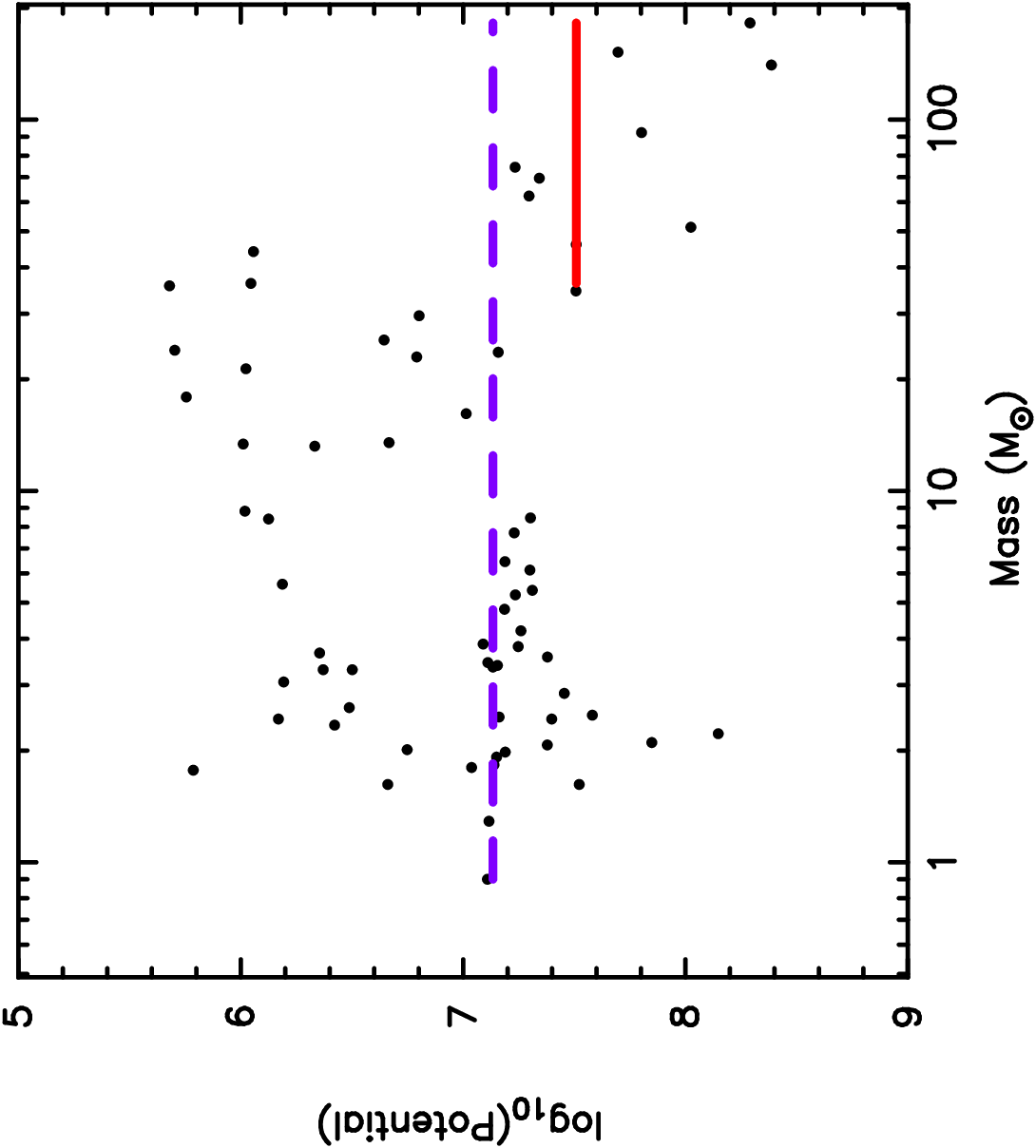}}}
\subfigure[]{\rotatebox{270}{\includegraphics[scale=0.21]{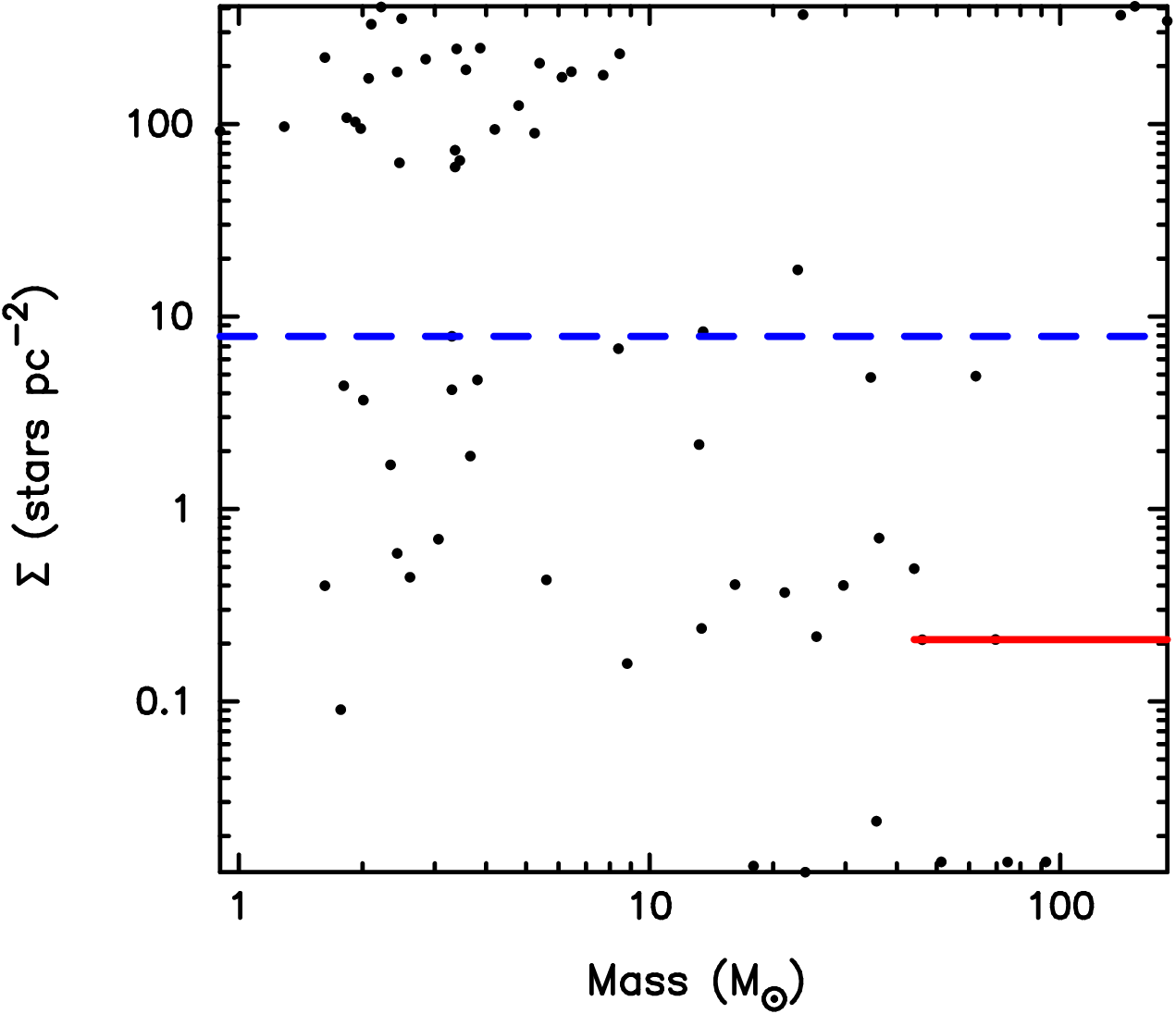}}}

\caption{UF Control   run.}
\label{UF216_sim_results_control}
  \end{center}
\end{figure*}

\begin{figure*}
  \begin{center}
\setlength{\subfigcapskip}{10pt}

\subfigure[]{\rotatebox{270}{\includegraphics[scale=0.21]{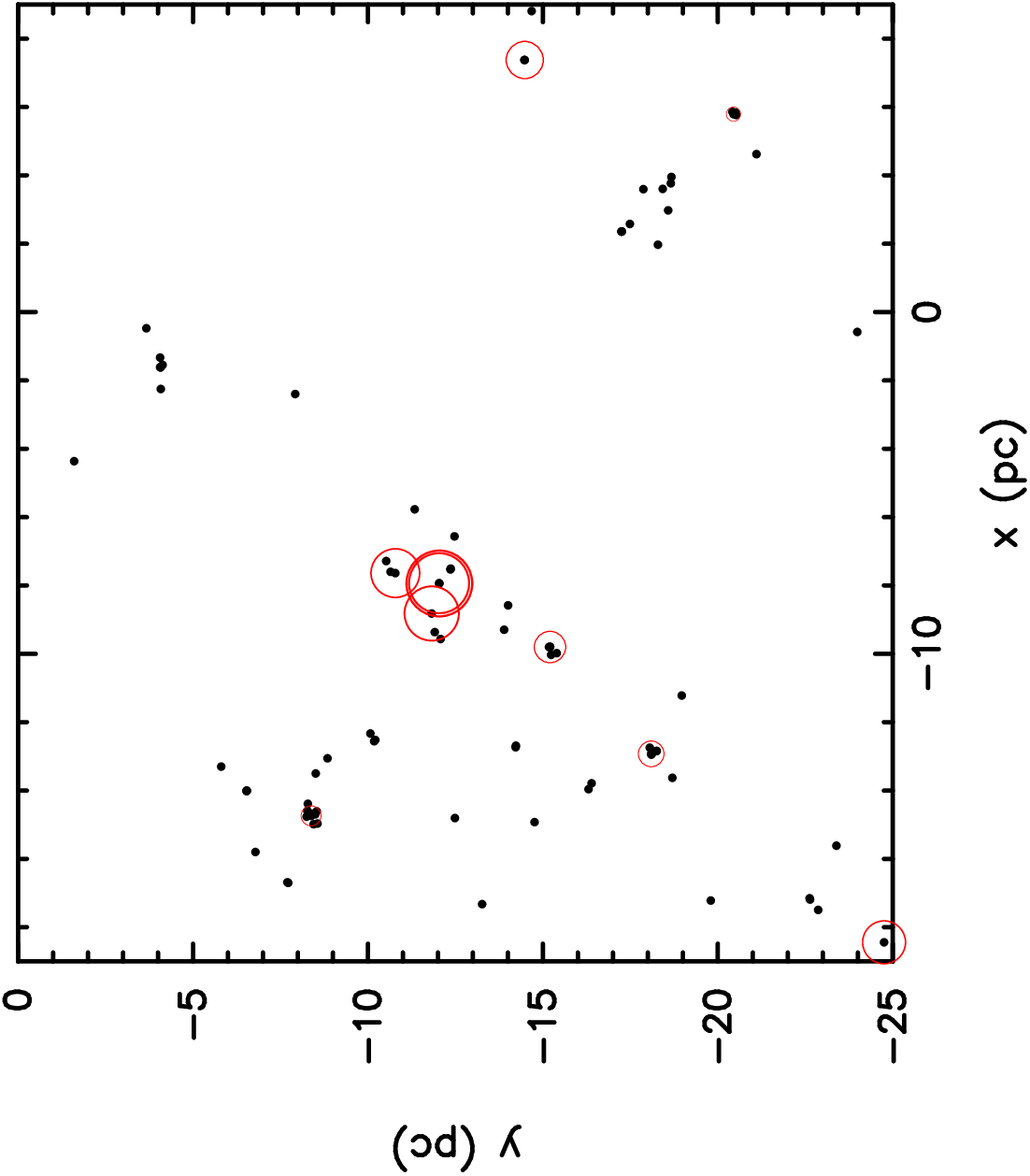}}}
\subfigure[]{\rotatebox{270}{\includegraphics[scale=0.21]{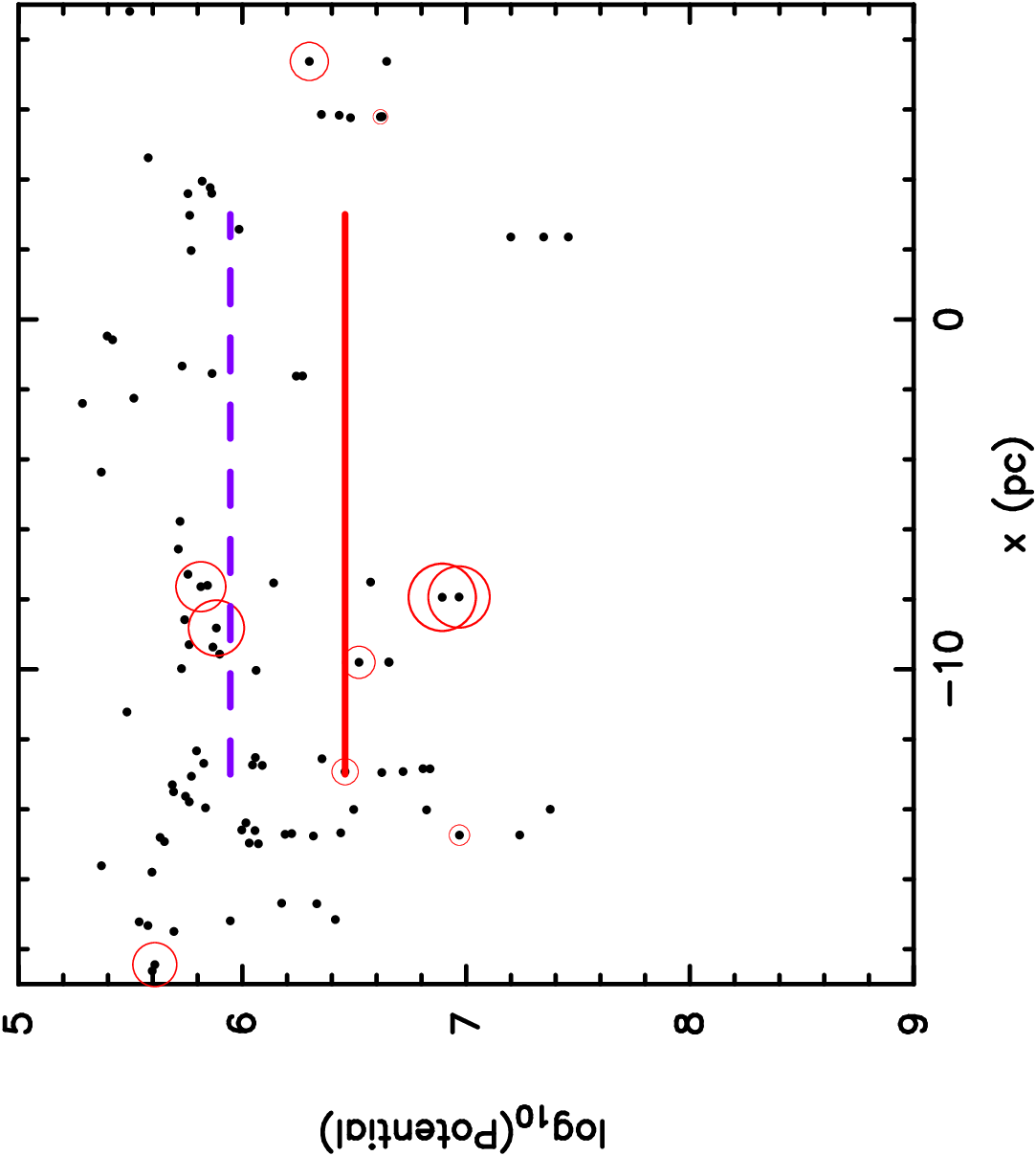}}}
\subfigure[]{\rotatebox{270}{\includegraphics[scale=0.21]{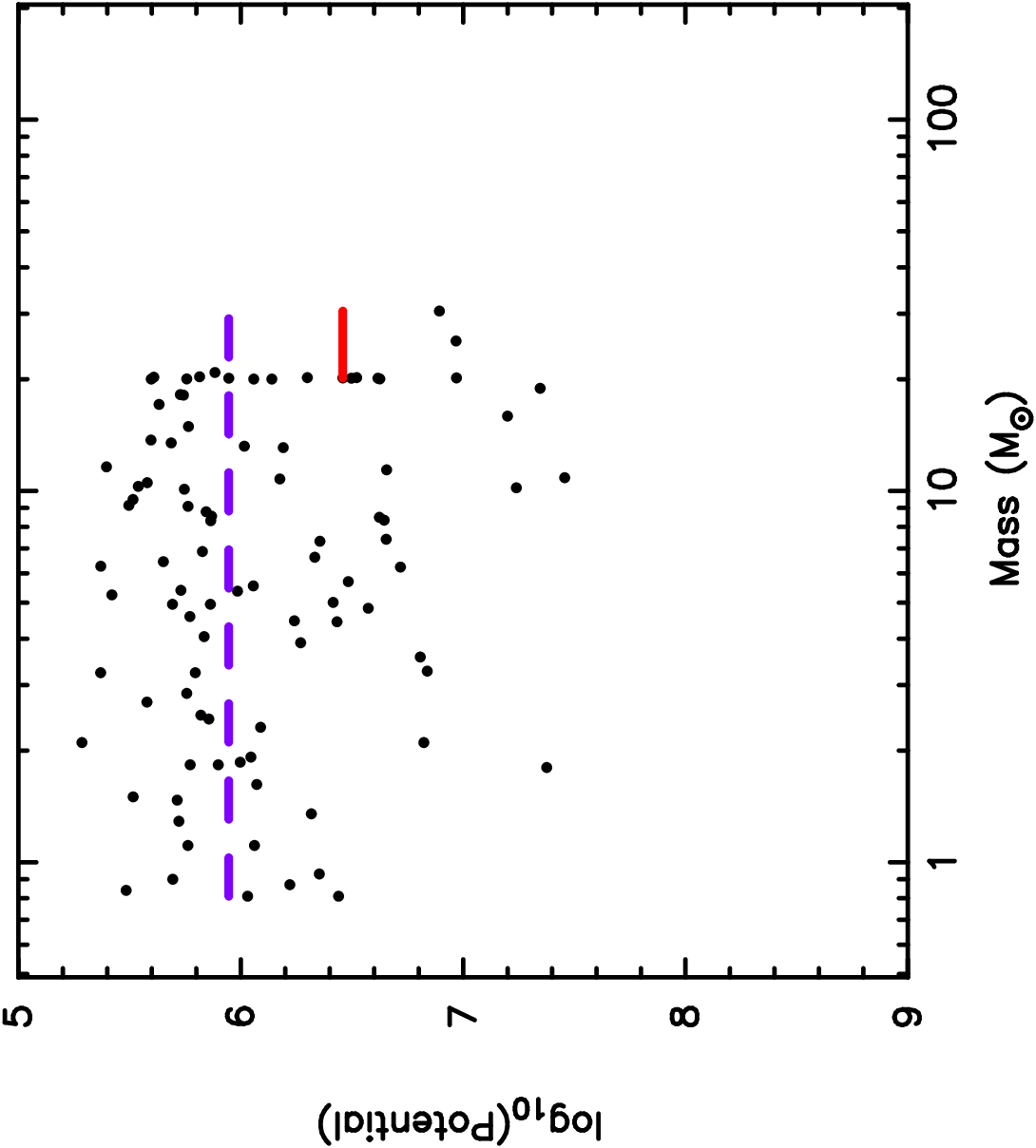}}}
\subfigure[]{\rotatebox{270}{\includegraphics[scale=0.21]{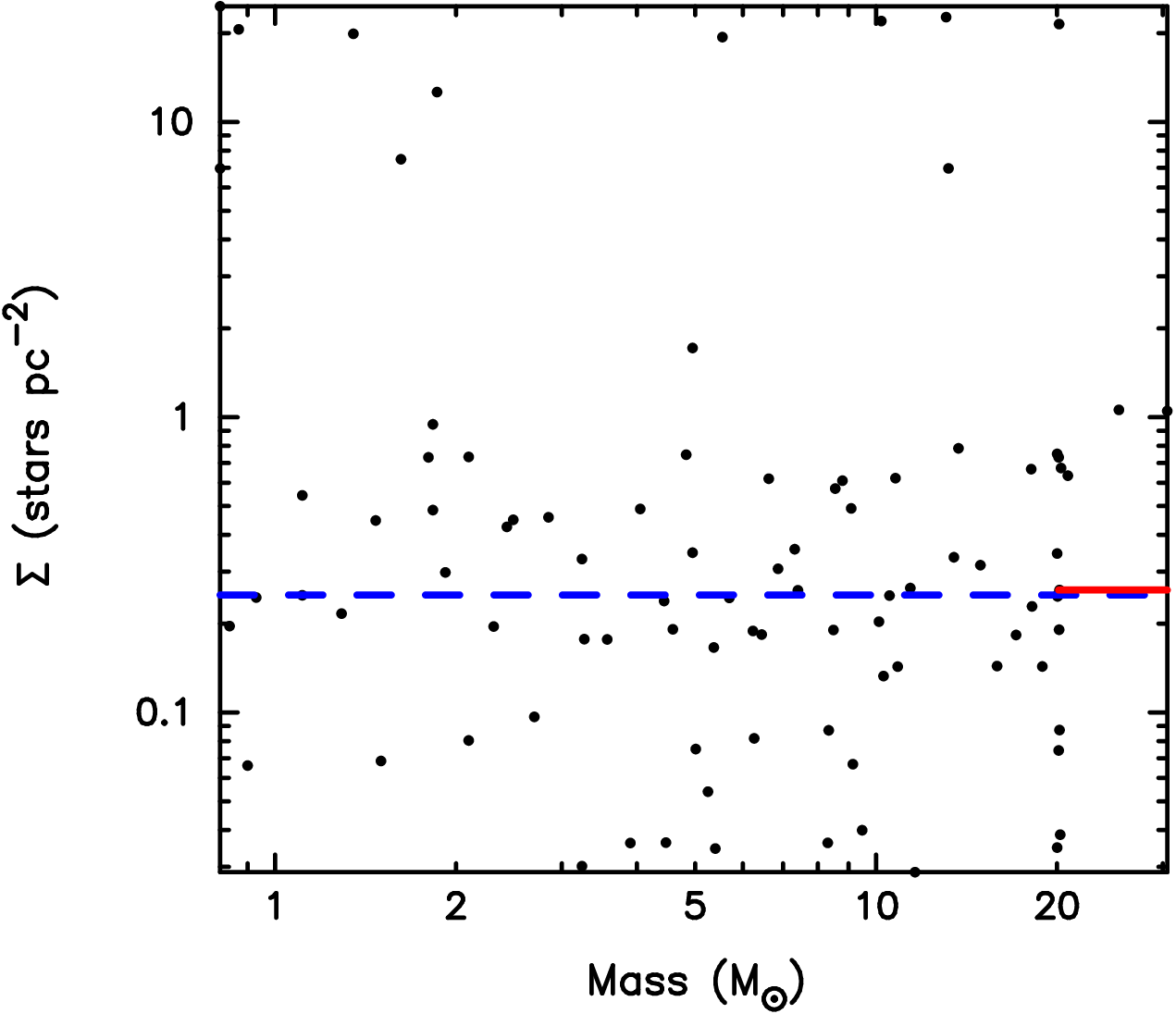}}}

\caption{UF Dual feedback   run.}
\label{UF216_sim_results_dual}
  \end{center}
\end{figure*}

\begin{figure*}
  \begin{center}
\setlength{\subfigcapskip}{10pt}
\hspace*{-0.1cm}\subfigure[]{\rotatebox{270}{\includegraphics[scale=0.21]{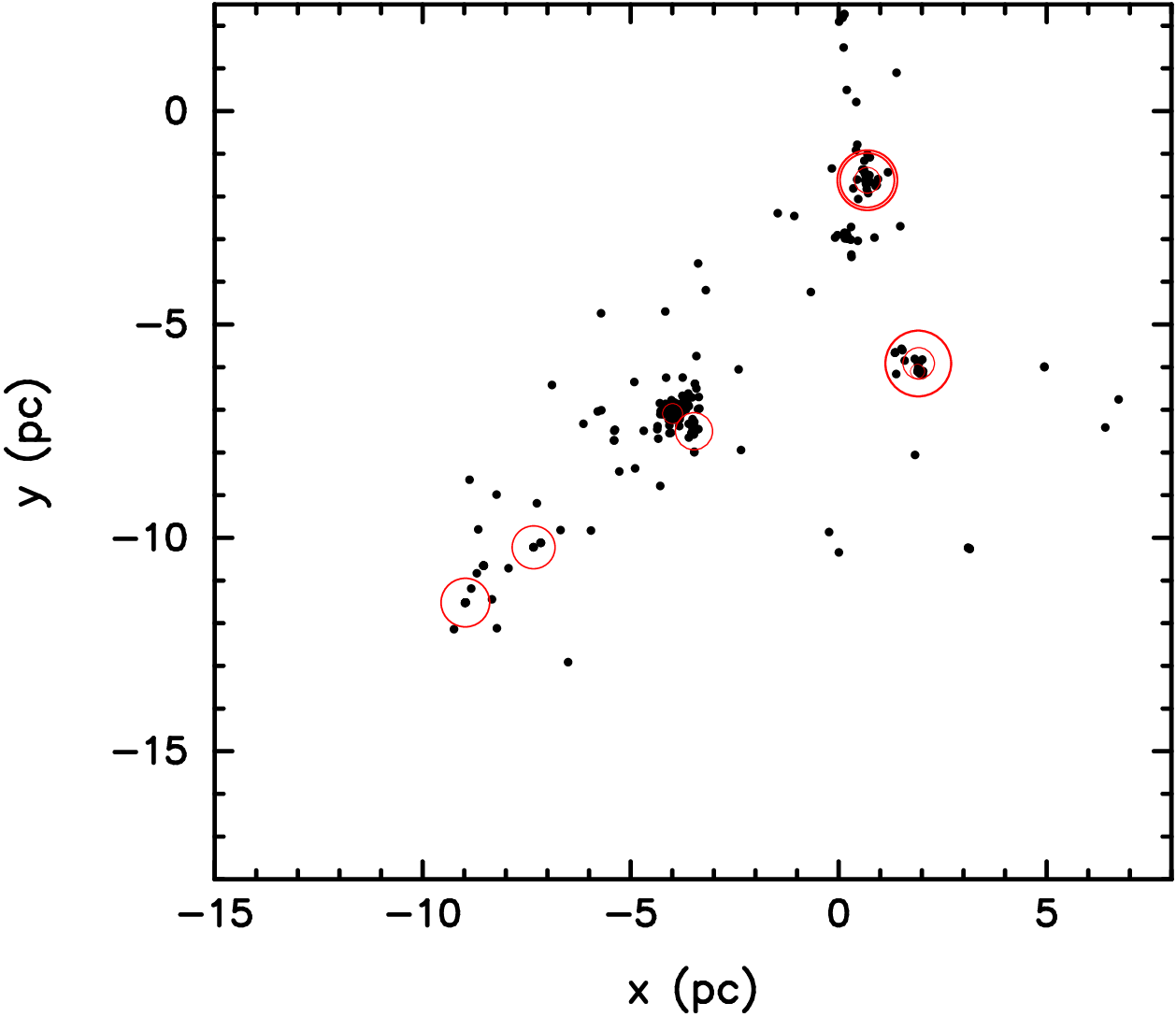}}}
\subfigure[]{\rotatebox{270}{\includegraphics[scale=0.21]{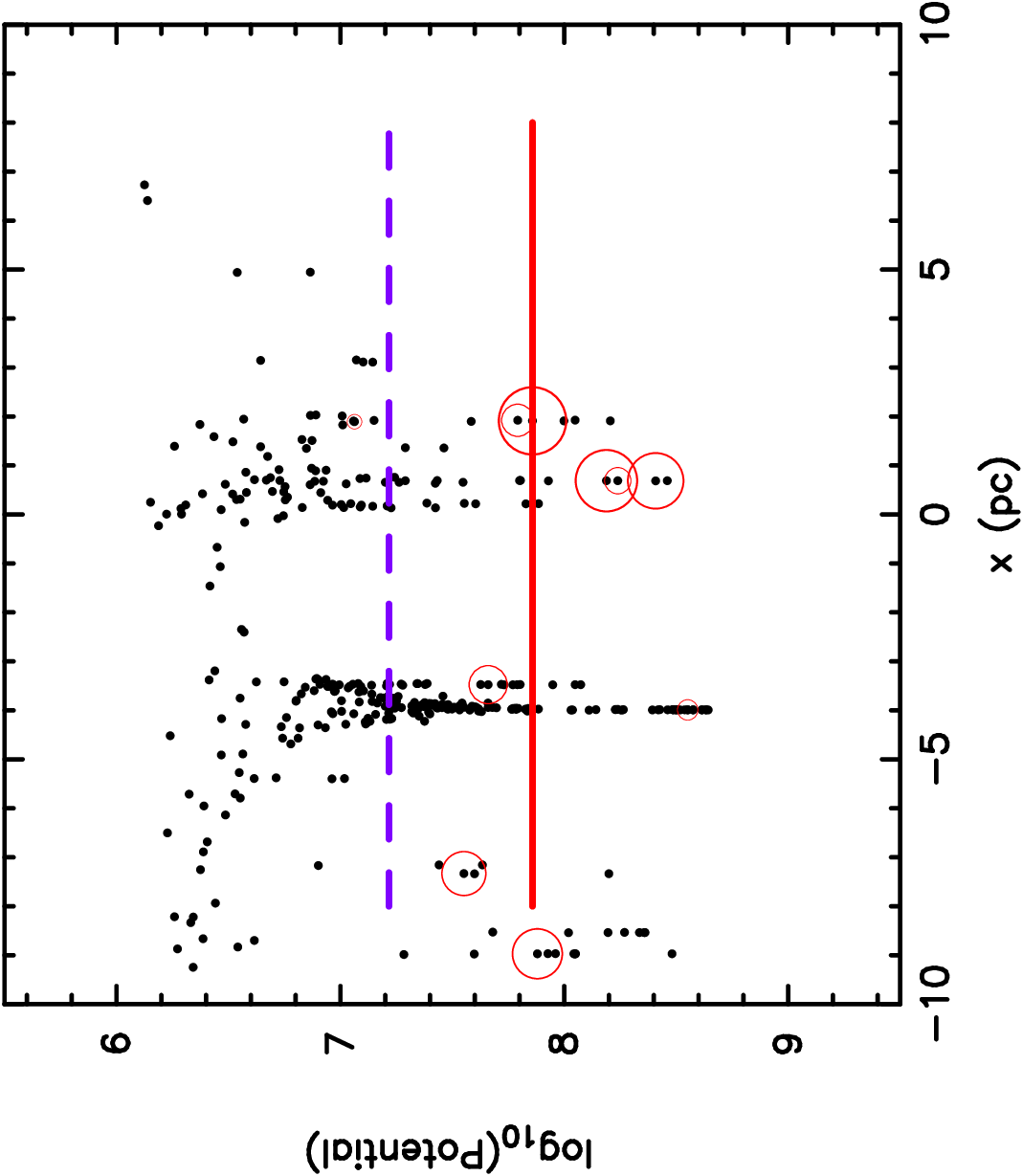}}}
\subfigure[]{\rotatebox{270}{\includegraphics[scale=0.21]{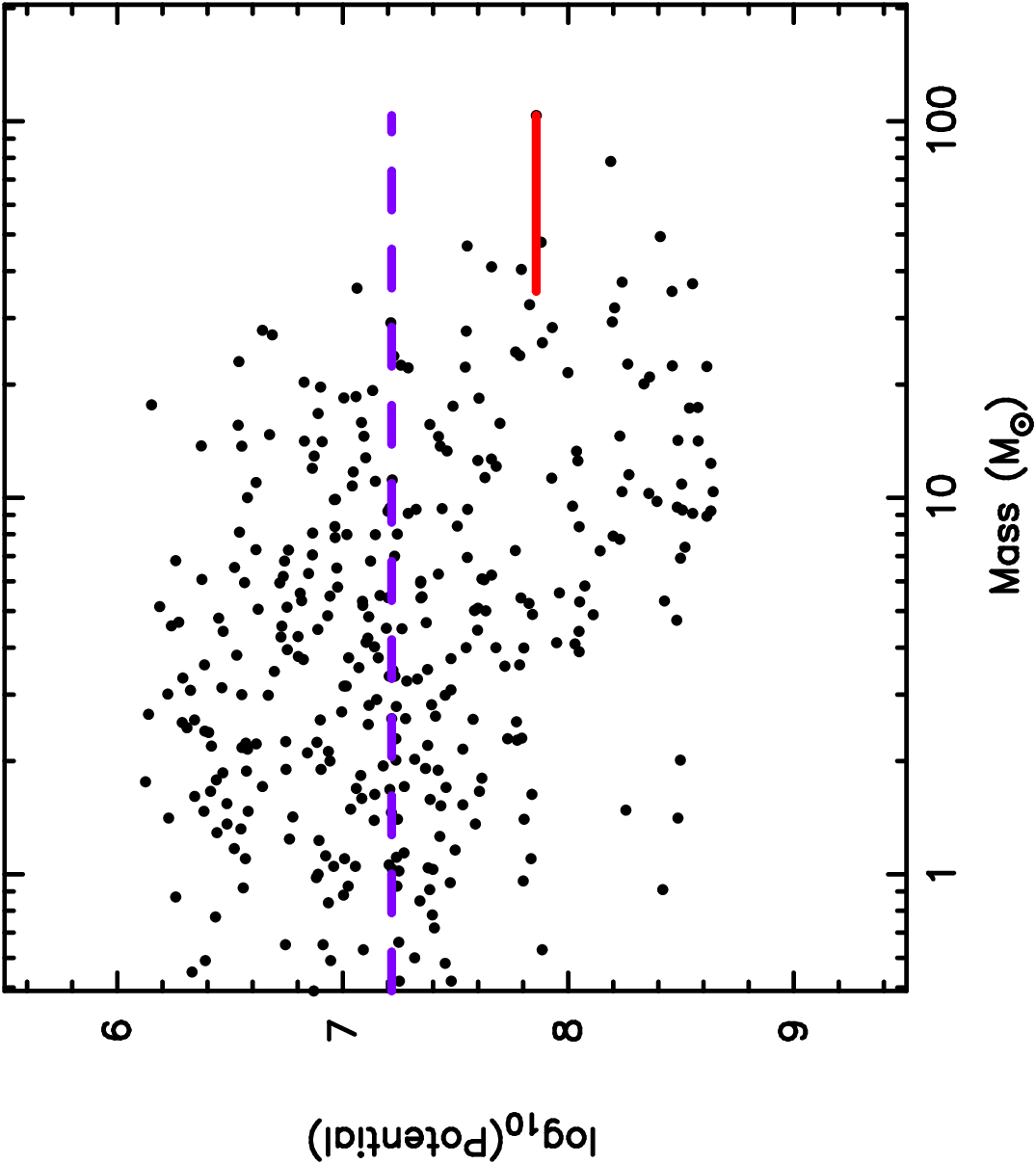}}}
\subfigure[]{\rotatebox{270}{\includegraphics[scale=0.21]{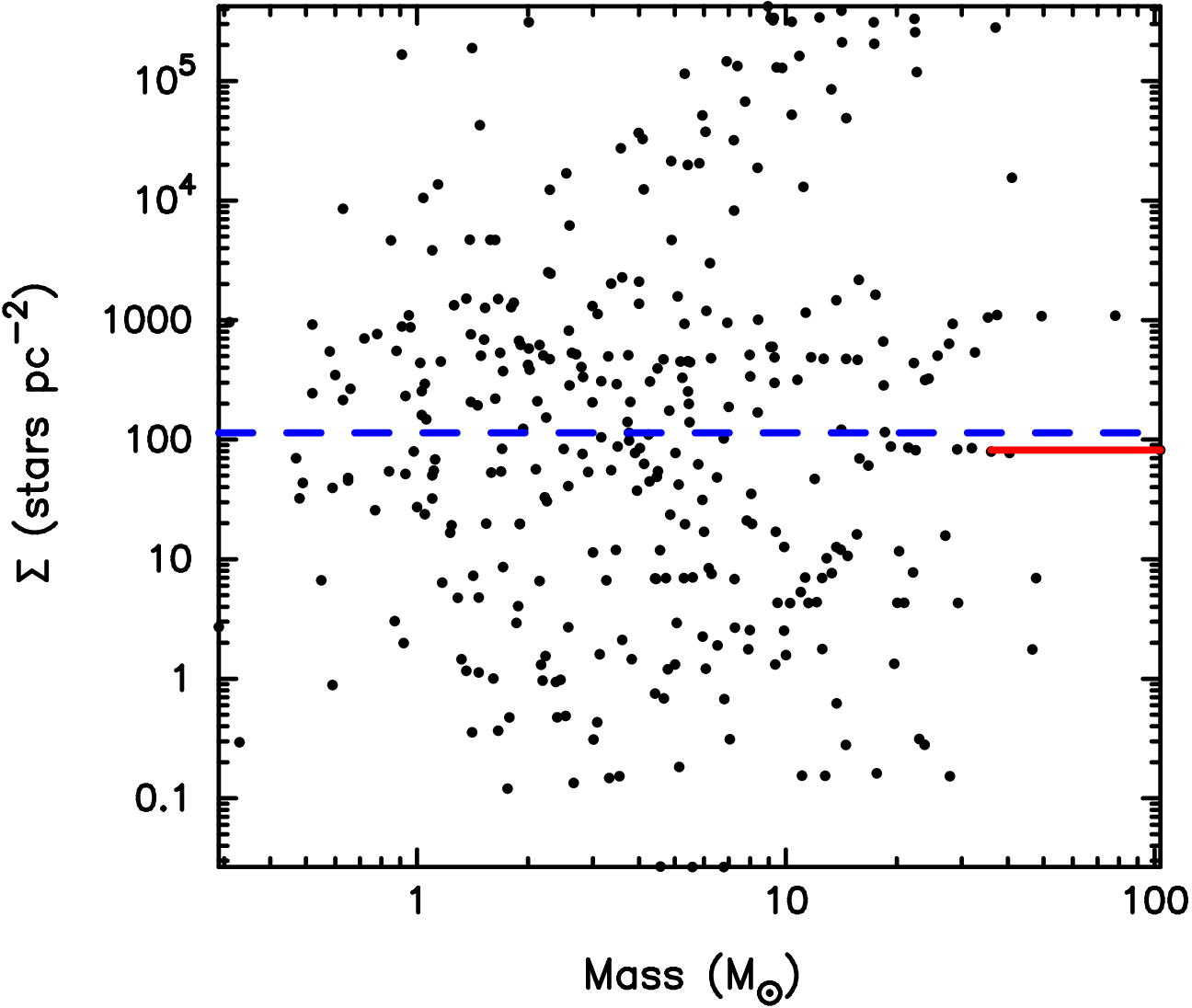}}}

\caption{UP Control   run.}
\label{UP202_sim_results_control}
  \end{center}
\end{figure*}

\begin{figure*}
  \begin{center}
\setlength{\subfigcapskip}{10pt}

\subfigure[]{\rotatebox{270}{\includegraphics[scale=0.21]{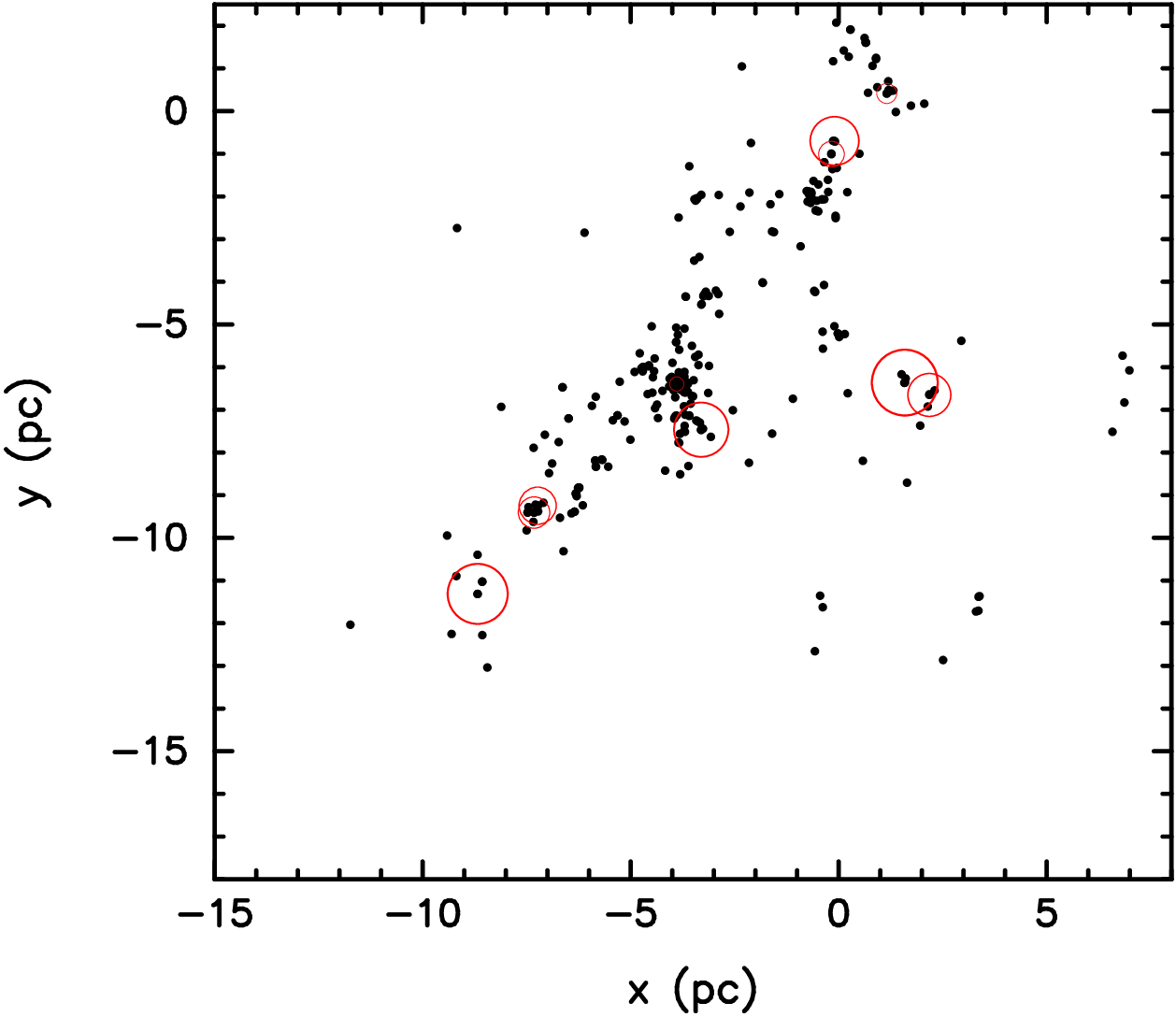}}}
\subfigure[]{\rotatebox{270}{\includegraphics[scale=0.21]{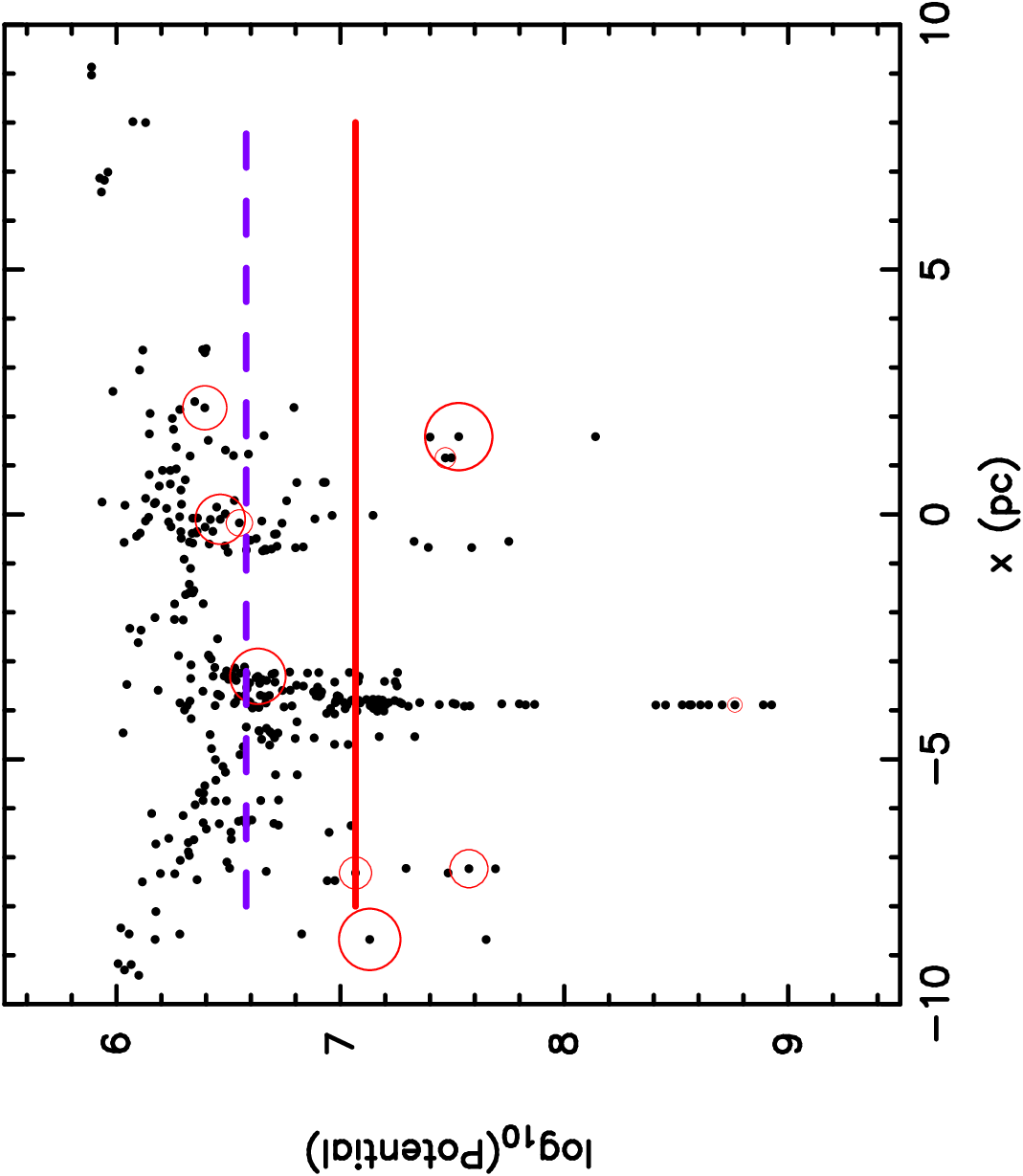}}}
\subfigure[]{\rotatebox{270}{\includegraphics[scale=0.21]{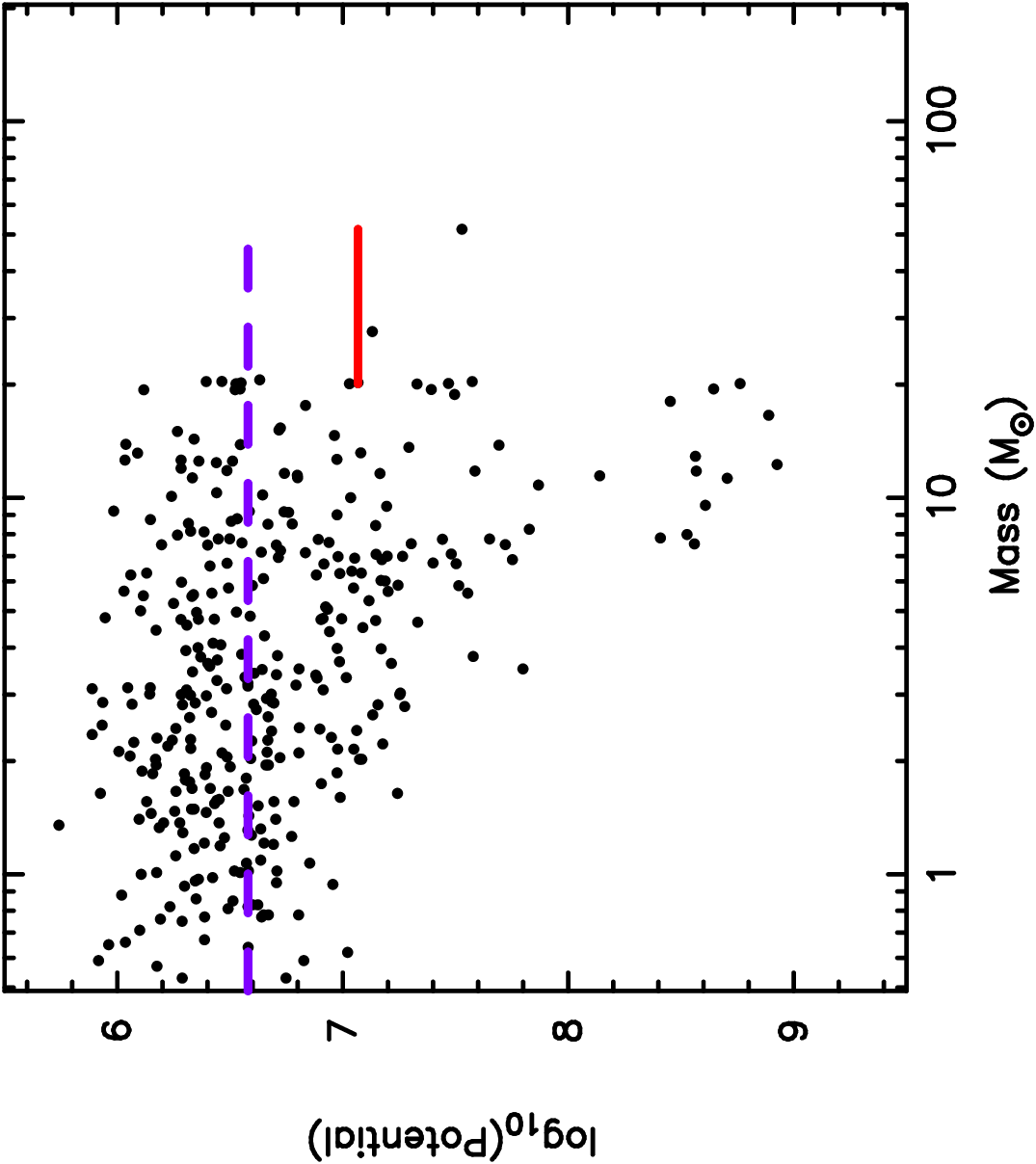}}}
\subfigure[]{\rotatebox{270}{\includegraphics[scale=0.21]{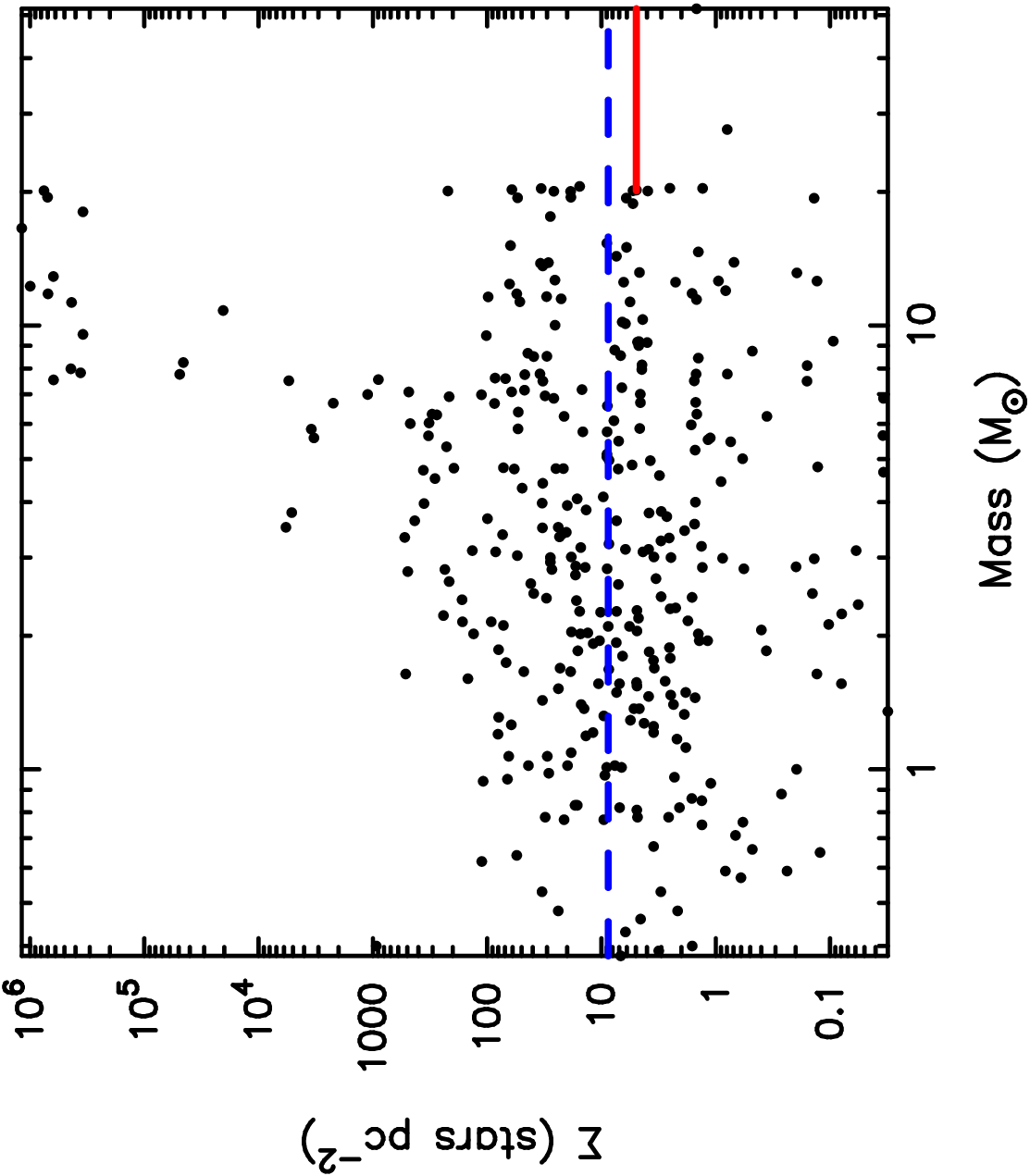}}}

\caption{UP Dual feedback   run.}
\label{UP200_sim_results_dual}
  \end{center}
\end{figure*}

\begin{figure*}
  \begin{center}
\setlength{\subfigcapskip}{10pt}

\subfigure[]{\rotatebox{270}{\includegraphics[scale=0.21]{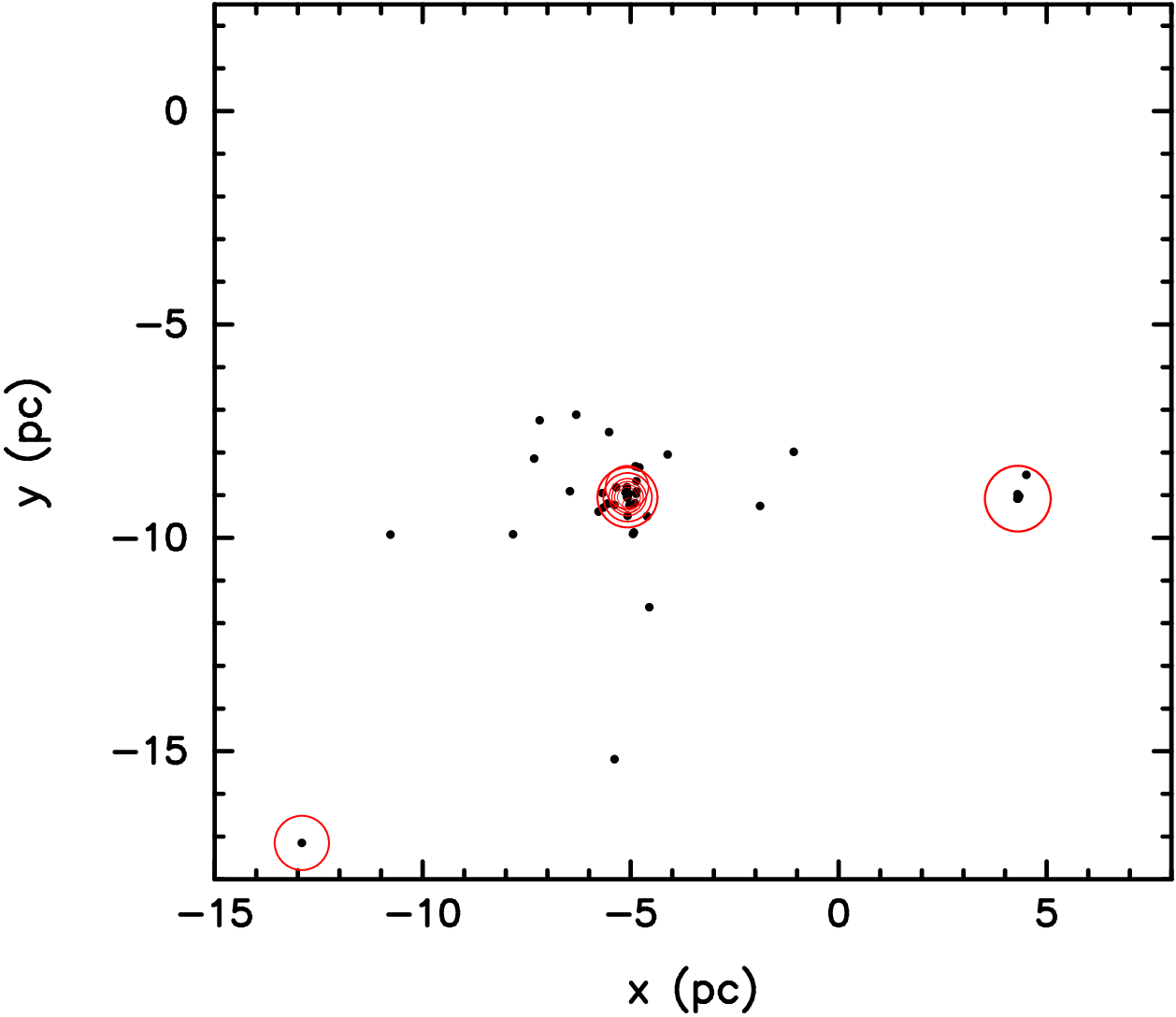}}}
\subfigure[]{\rotatebox{270}{\includegraphics[scale=0.21]{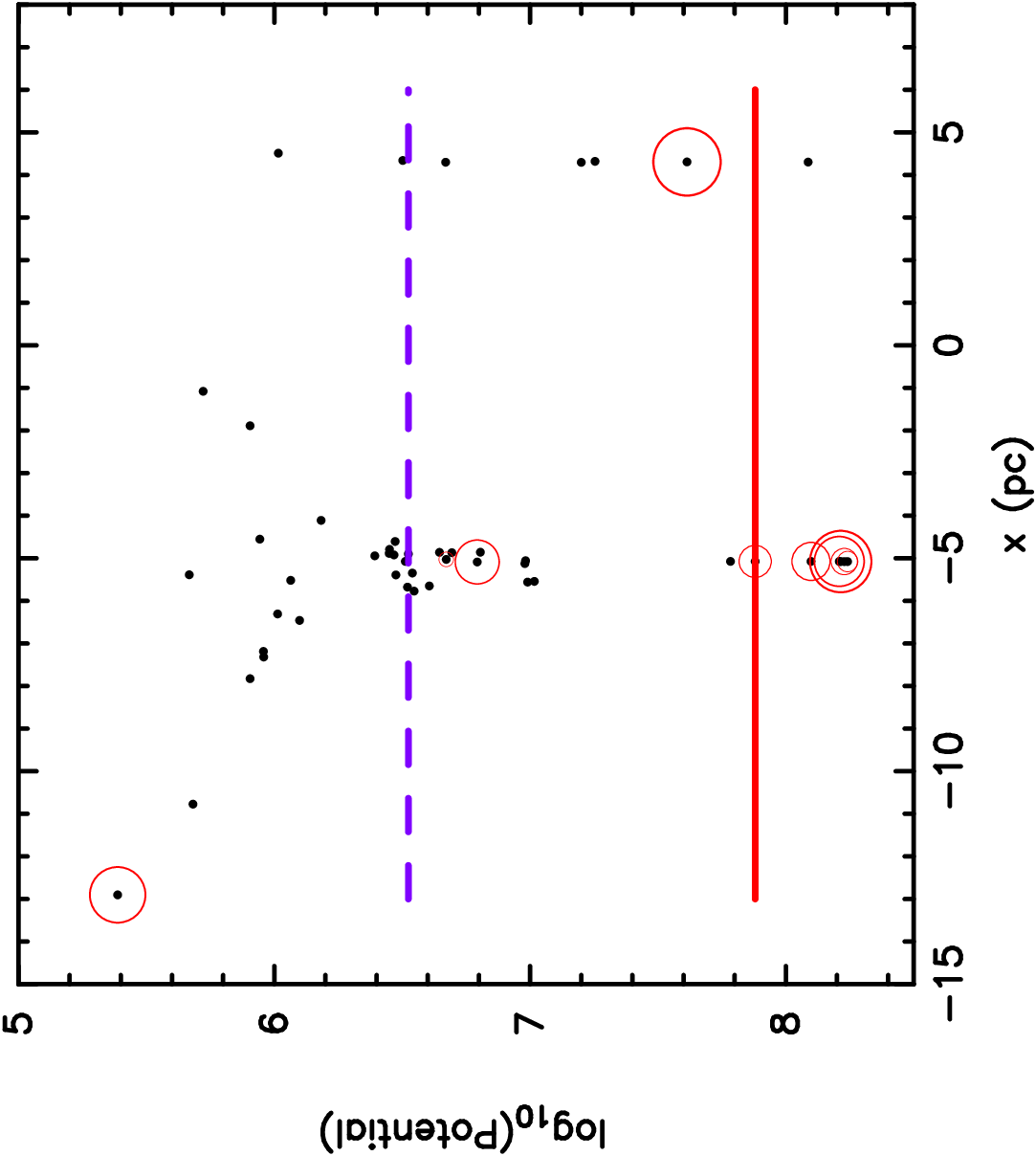}}}
\subfigure[]{\rotatebox{270}{\includegraphics[scale=0.21]{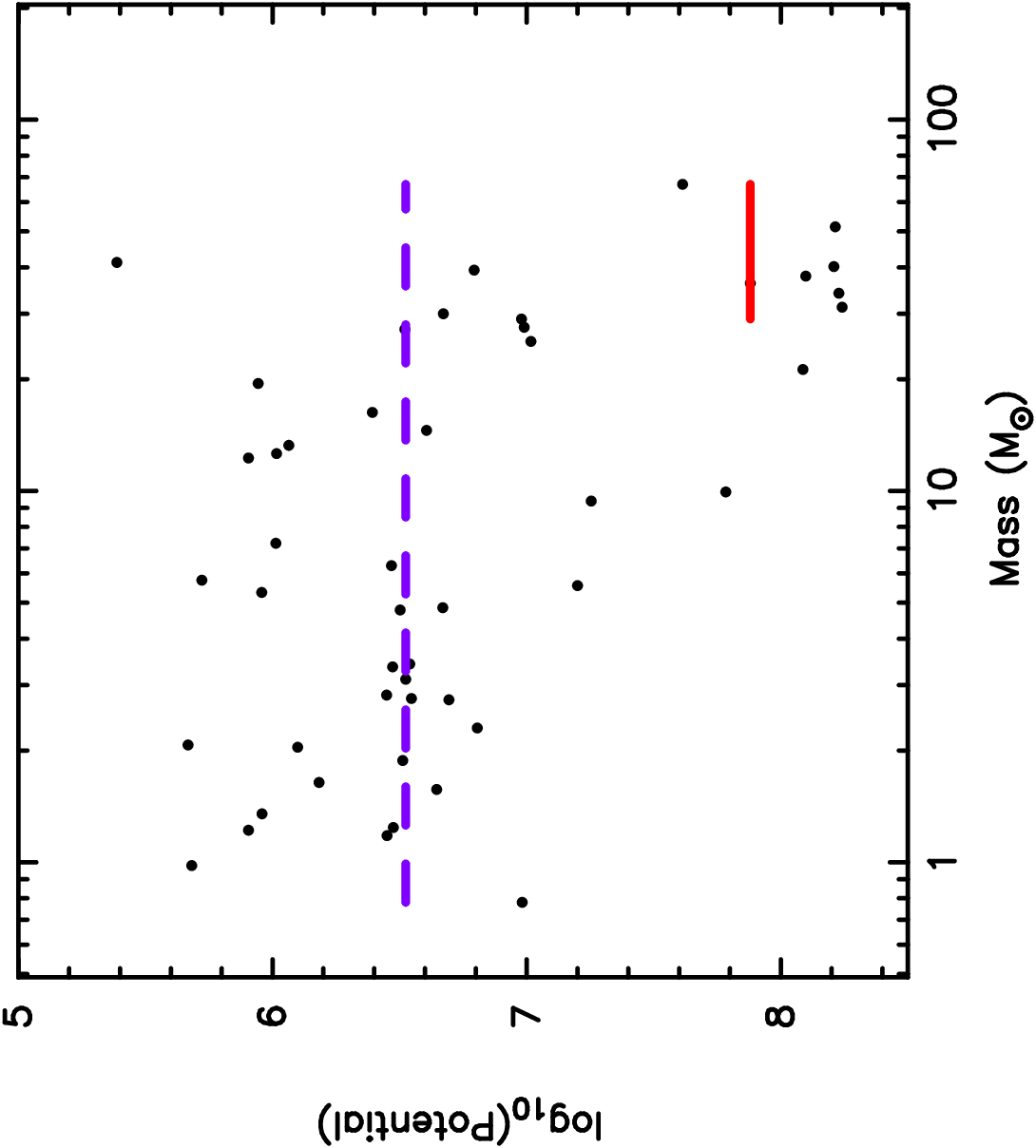}}}
\subfigure[]{\rotatebox{270}{\includegraphics[scale=0.21]{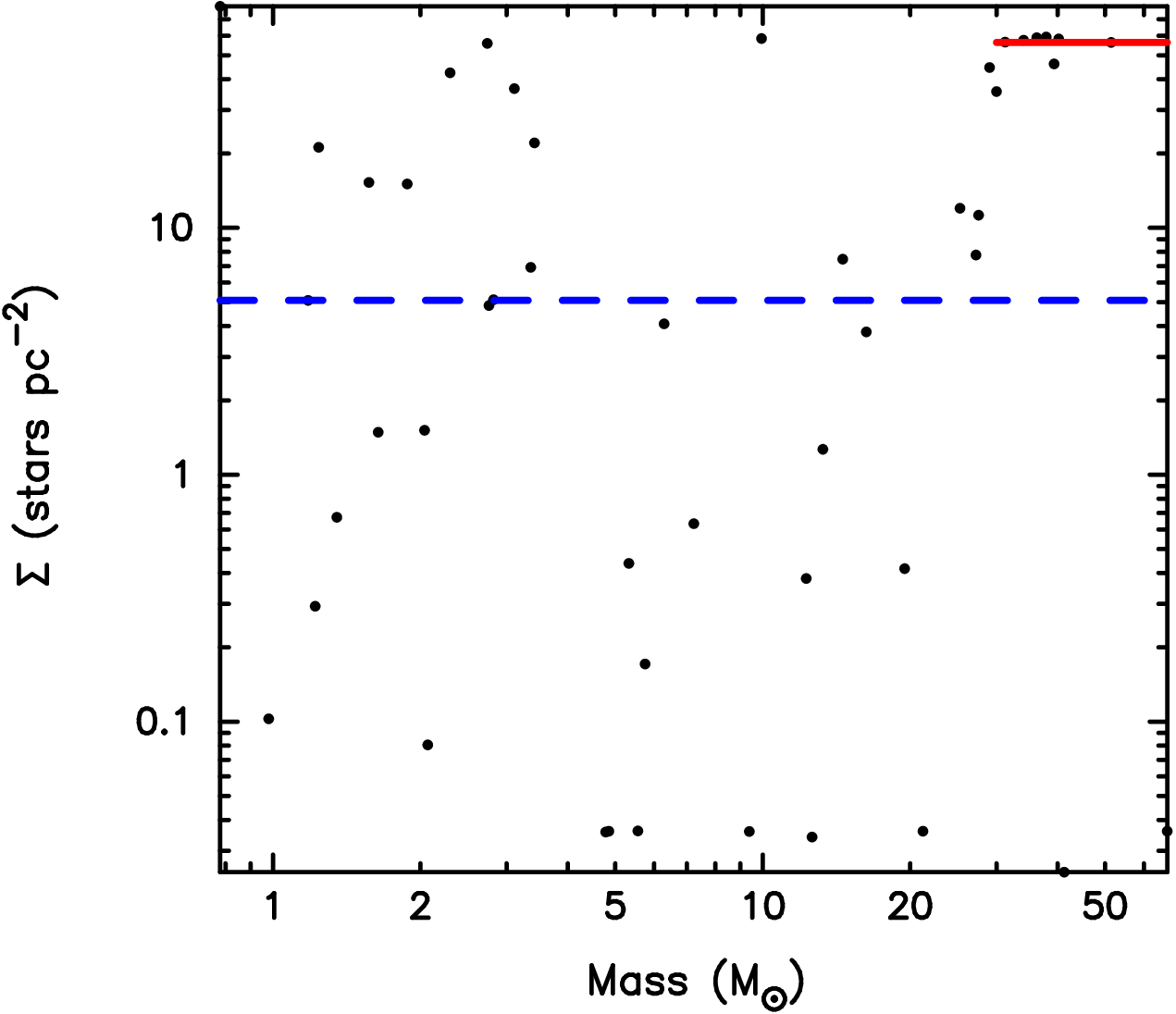}}}

\caption{UQ Control   run.}
\label{UQ202_sim_results_control}
  \end{center}
\end{figure*}

\begin{figure*}
  \begin{center}
\setlength{\subfigcapskip}{10pt}

\subfigure[]{\rotatebox{270}{\includegraphics[scale=0.21]{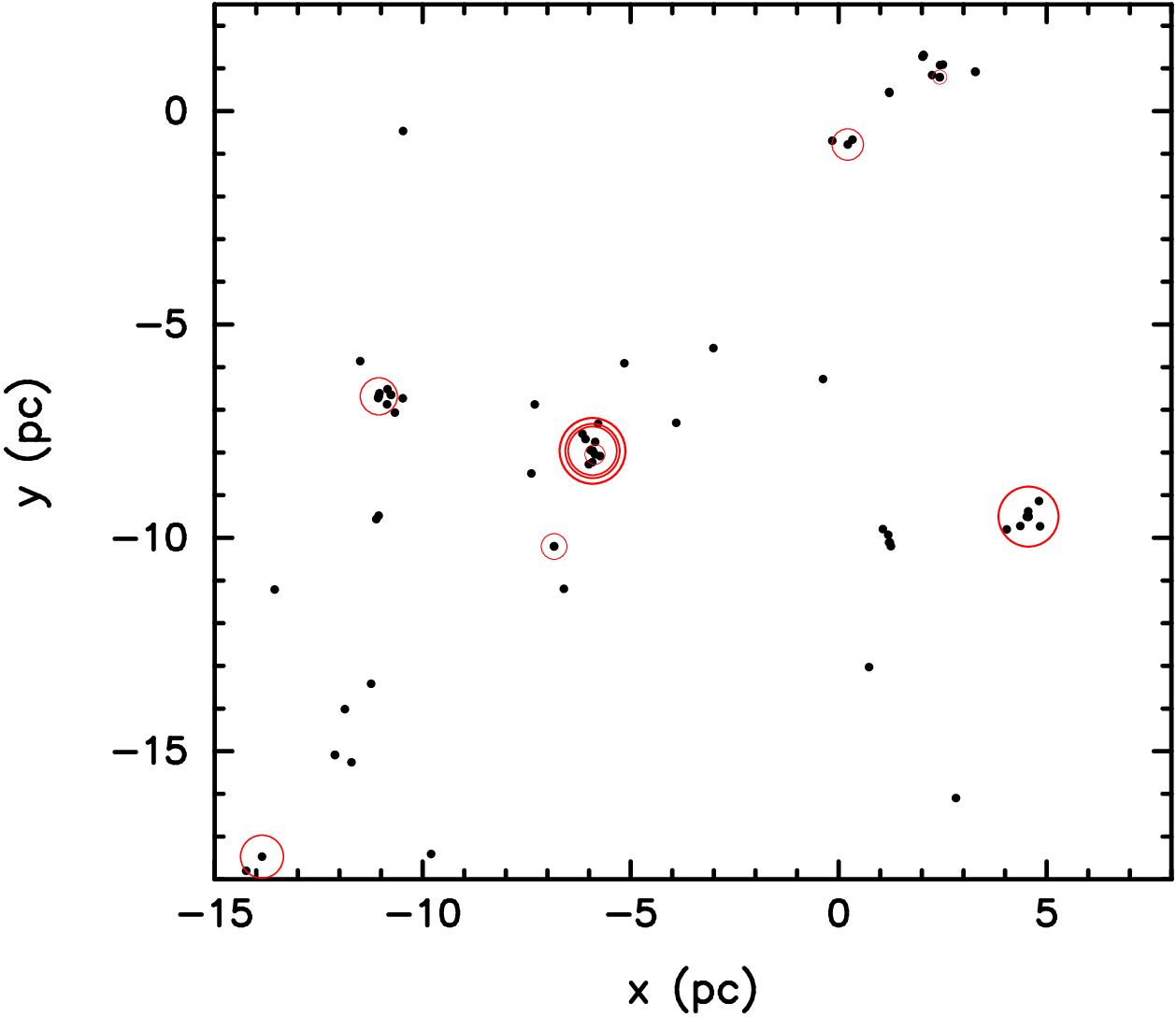}}}
\subfigure[]{\rotatebox{270}{\includegraphics[scale=0.21]{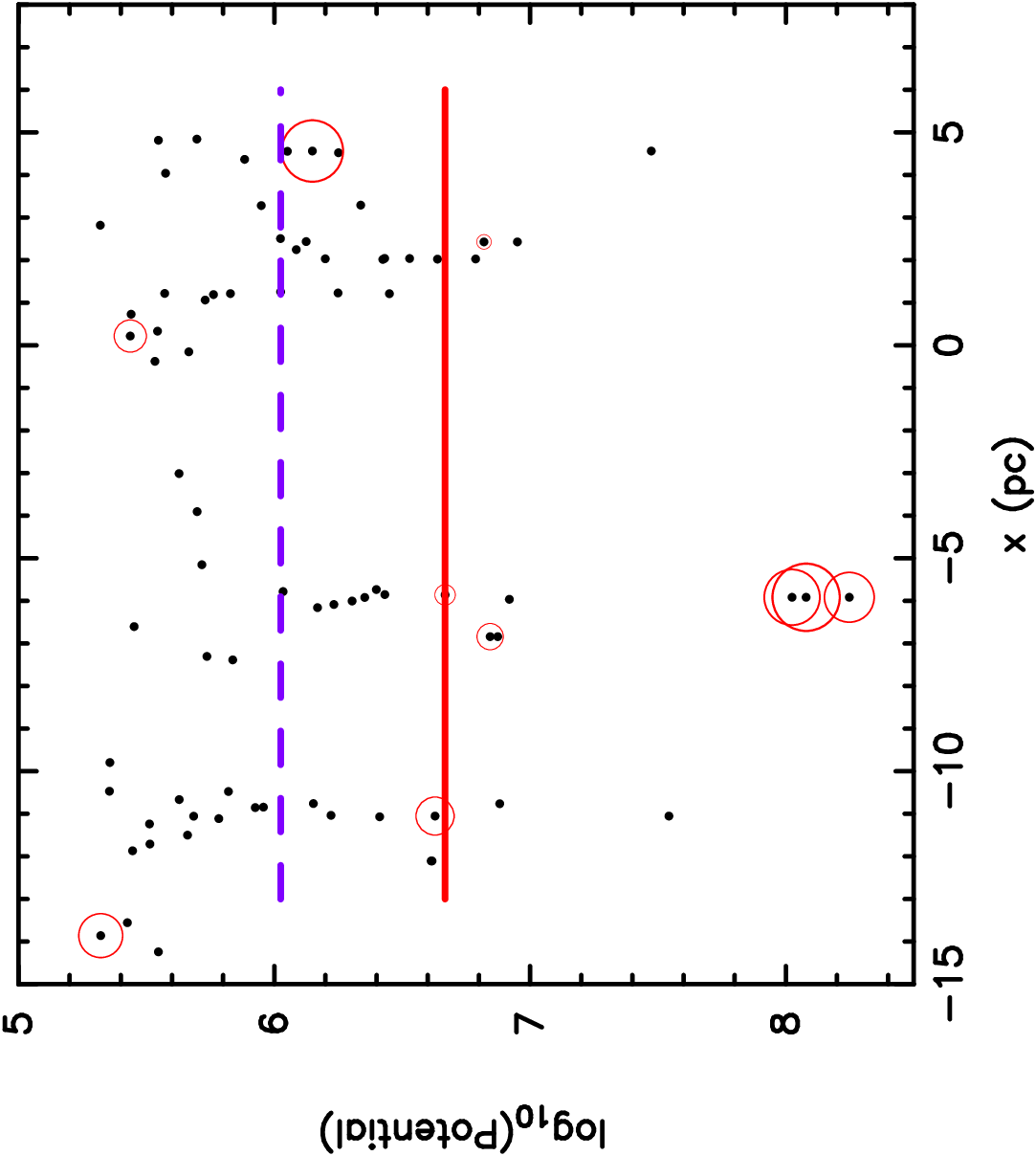}}}
\subfigure[]{\rotatebox{270}{\includegraphics[scale=0.21]{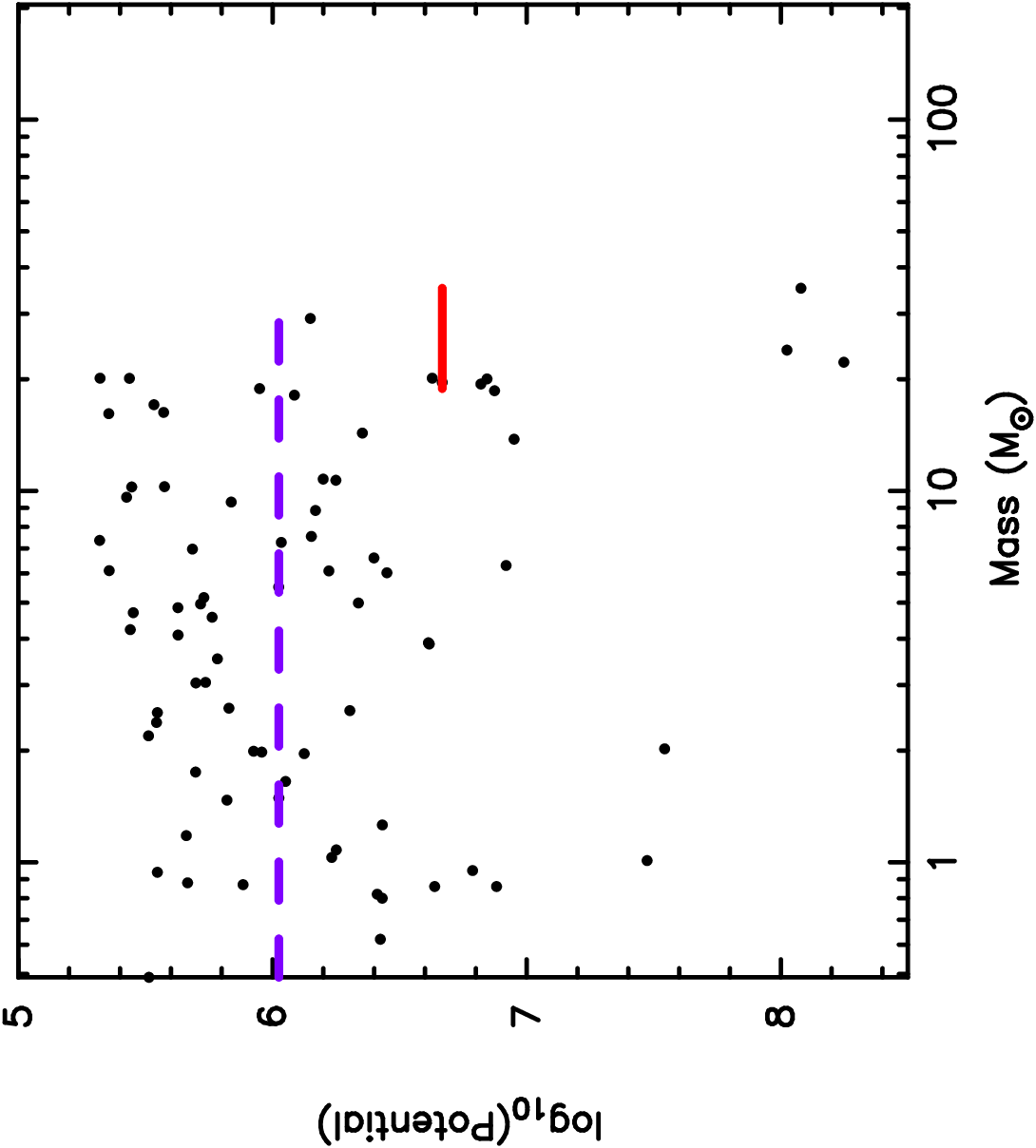}}}
\subfigure[]{\rotatebox{270}{\includegraphics[scale=0.21]{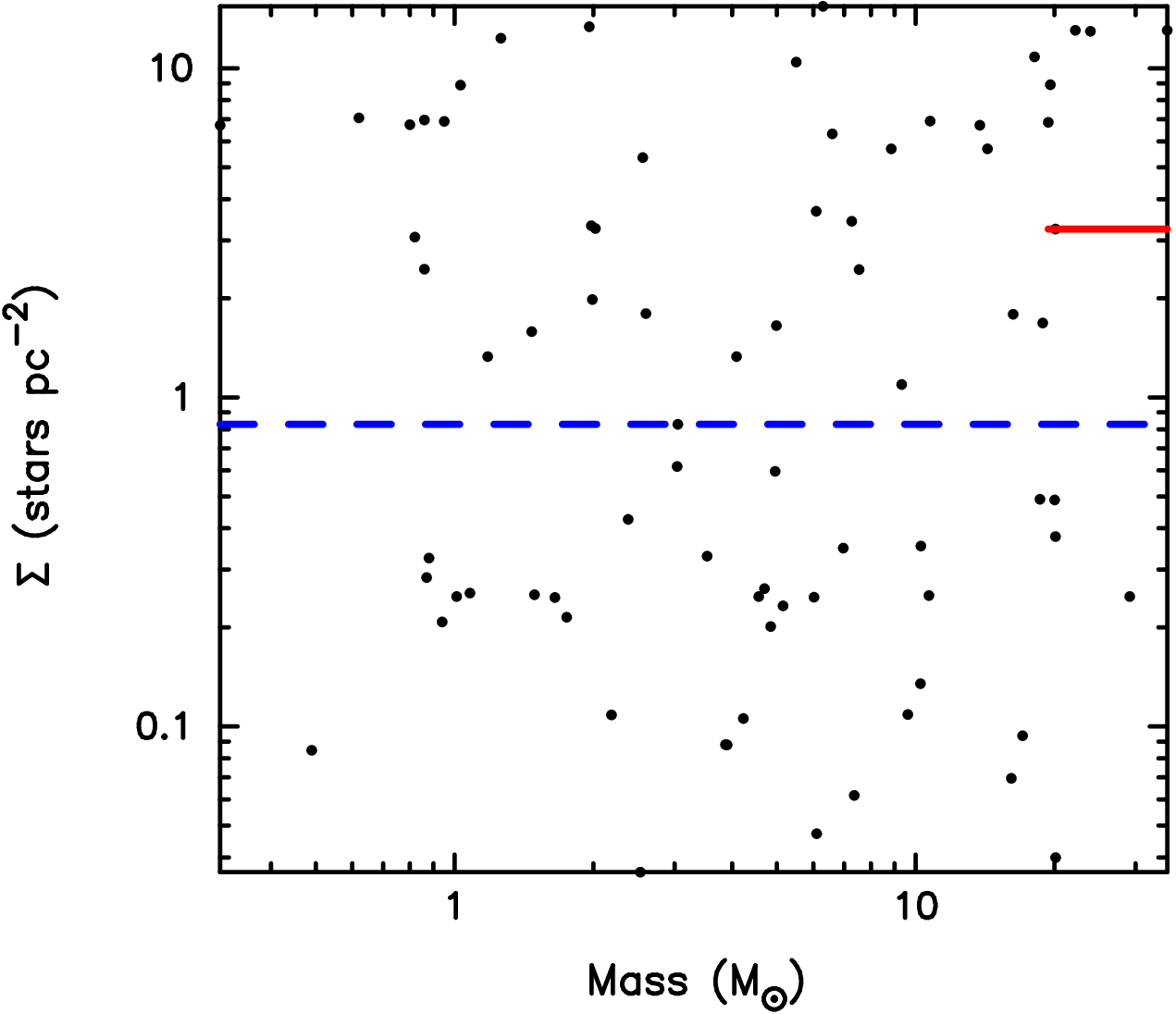}}}

\caption{UQ Dual feedback run.  }
\label{UQ202_sim_results_dual}
  \end{center}
\end{figure*}

\begin{figure*}
  \begin{center}
\setlength{\subfigcapskip}{10pt}

\subfigure[]{\rotatebox{270}{\includegraphics[scale=0.21]{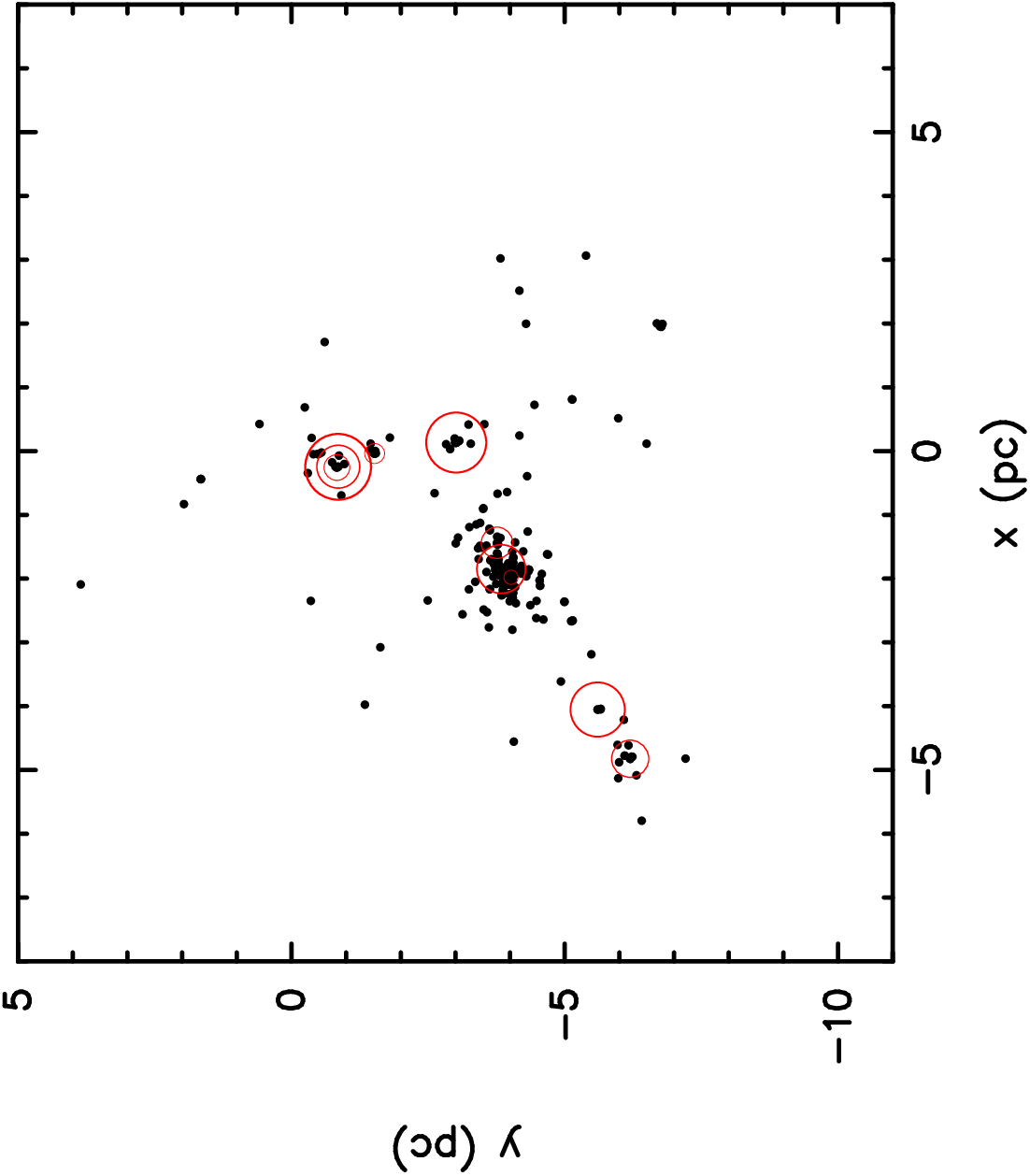}}}
\subfigure[]{\rotatebox{270}{\includegraphics[scale=0.21]{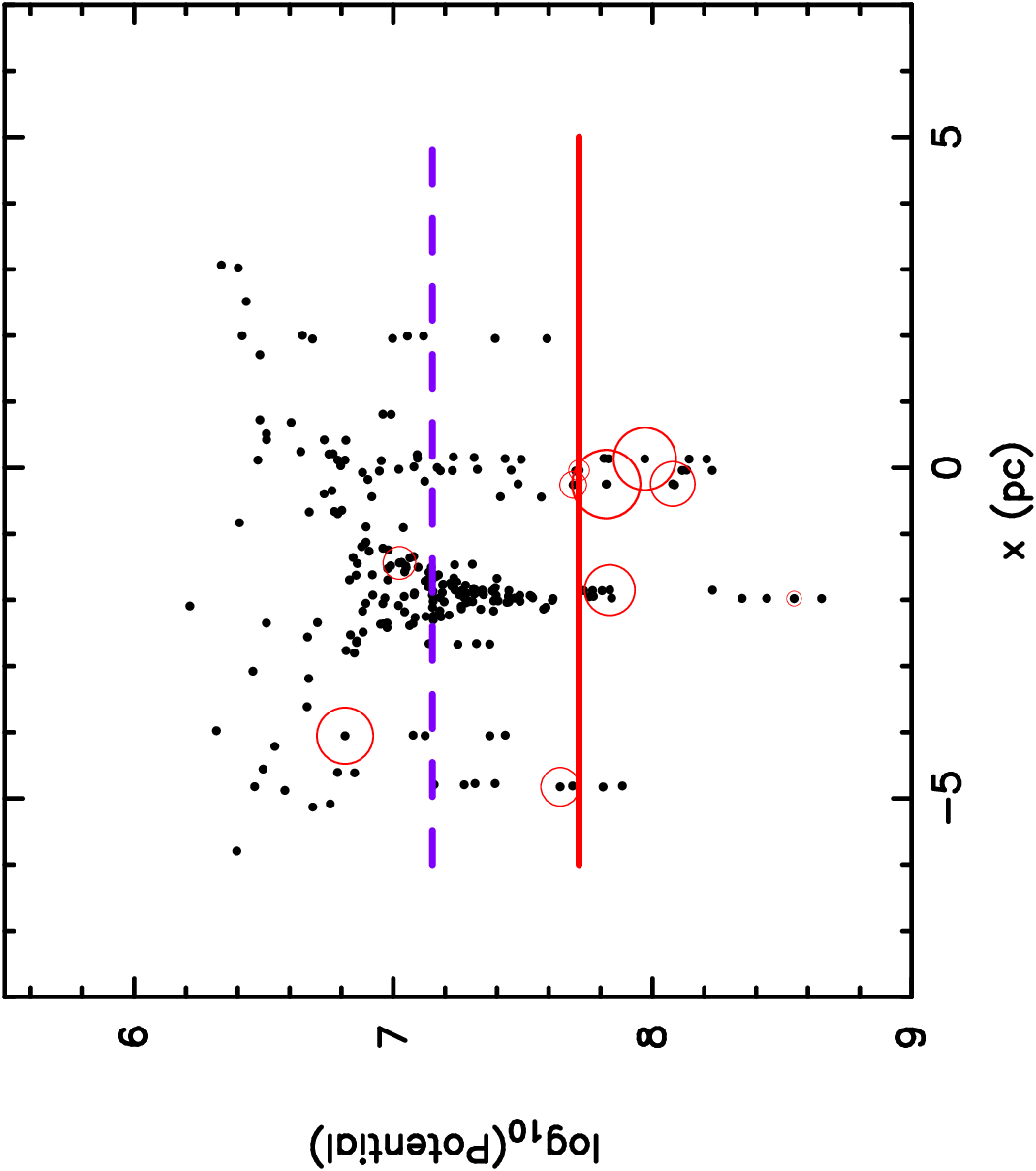}}}
\subfigure[]{\rotatebox{270}{\includegraphics[scale=0.21]{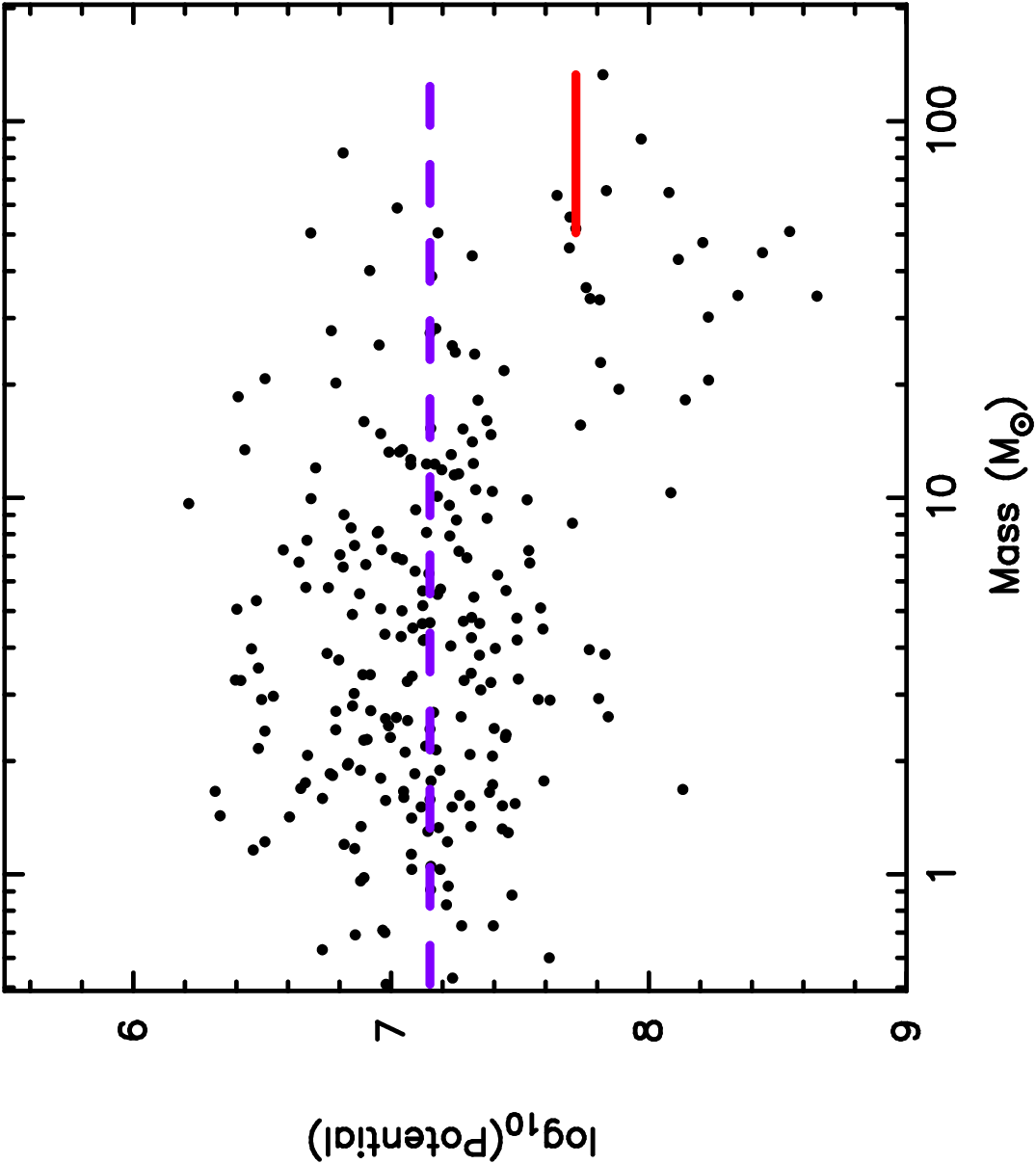}}}
\subfigure[]{\rotatebox{270}{\includegraphics[scale=0.21]{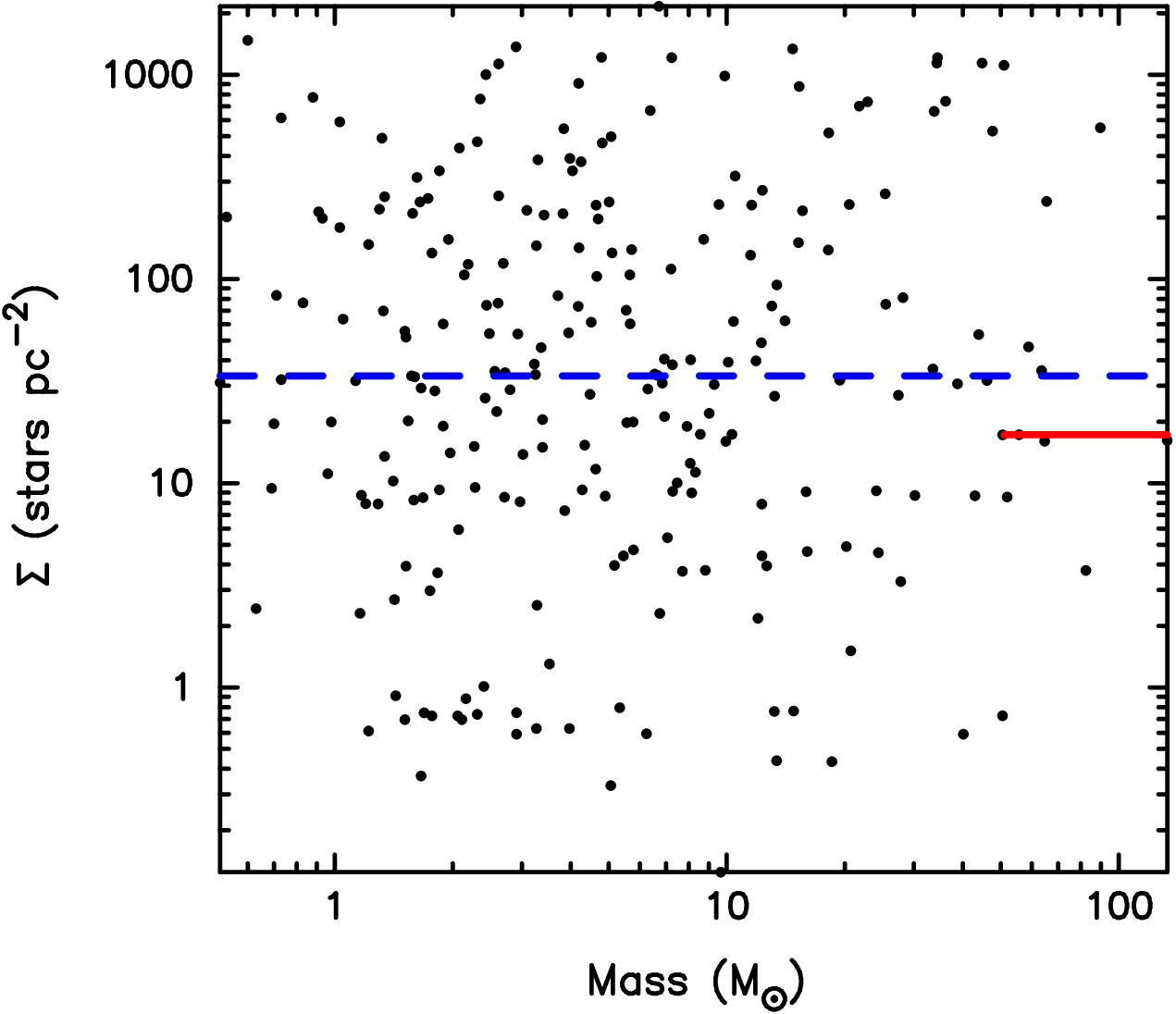}}}

\caption{R11O Control  run. }
\label{R11O_sim_results_control}
  \end{center}
\end{figure*}

\begin{figure*}
  \begin{center}
\setlength{\subfigcapskip}{10pt}

\subfigure[]{\rotatebox{270}{\includegraphics[scale=0.21]{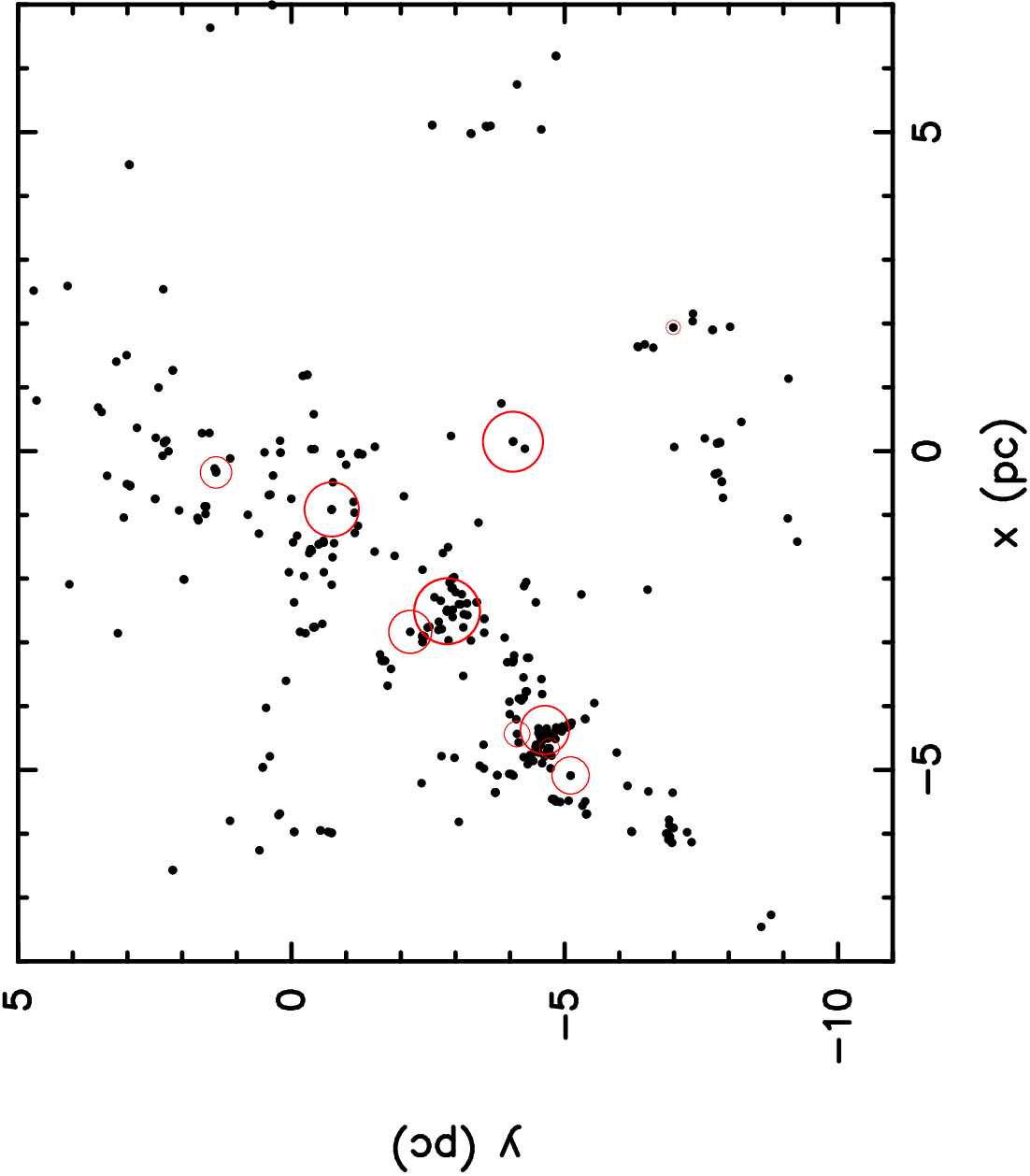}}}
\subfigure[]{\rotatebox{270}{\includegraphics[scale=0.21]{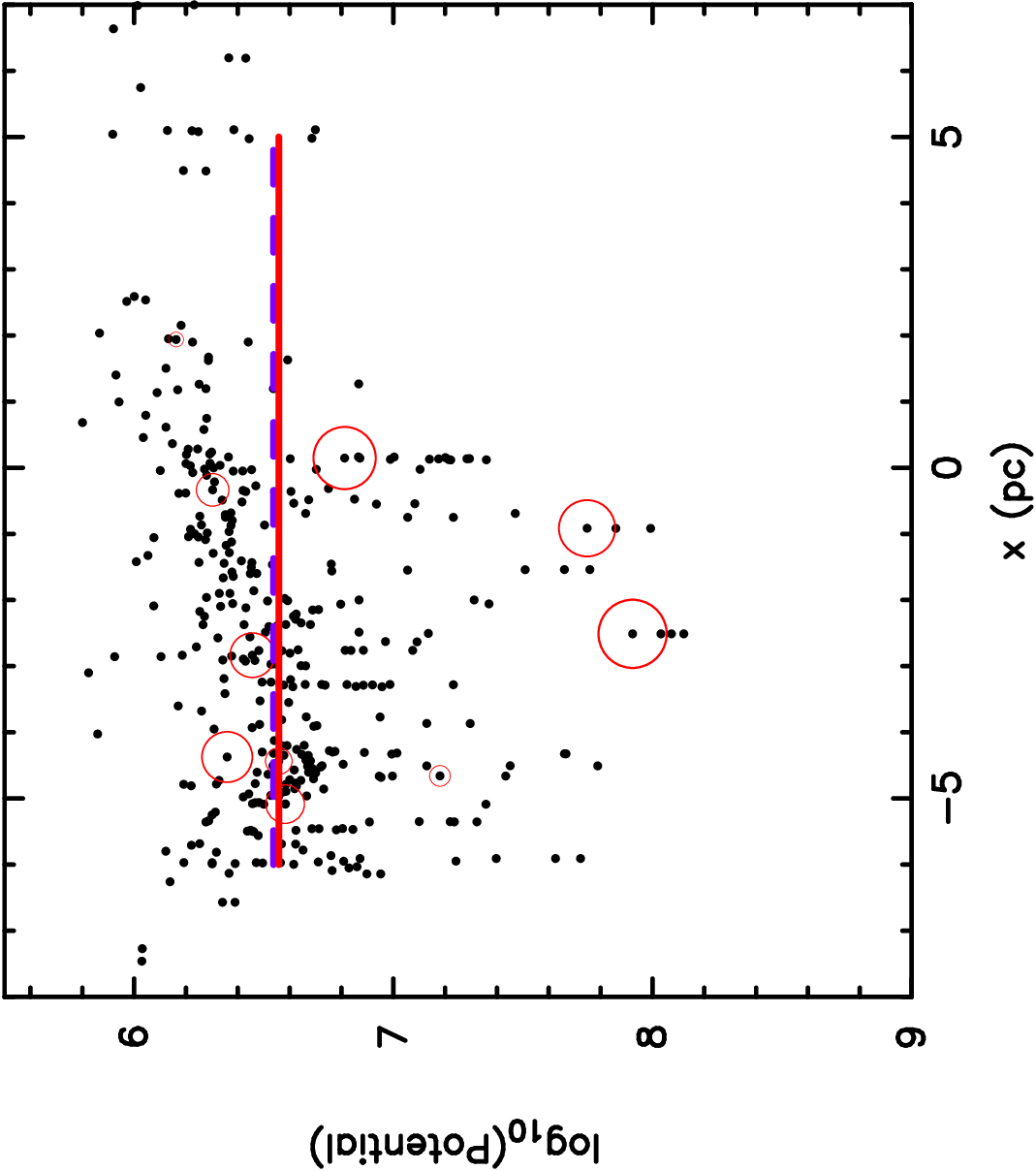}}}
\subfigure[]{\rotatebox{270}{\includegraphics[scale=0.21]{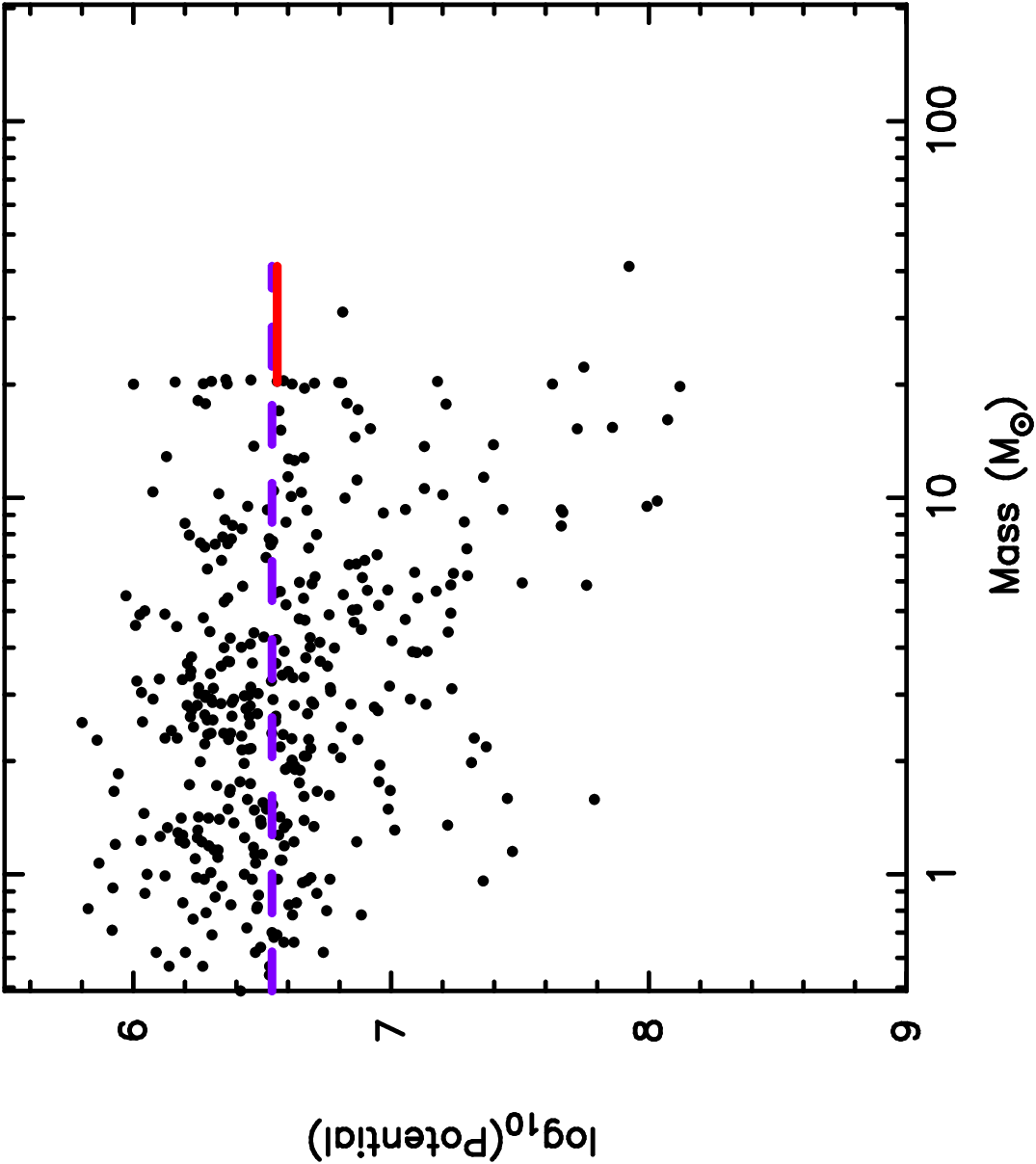}}}
\subfigure[]{\rotatebox{270}{\includegraphics[scale=0.21]{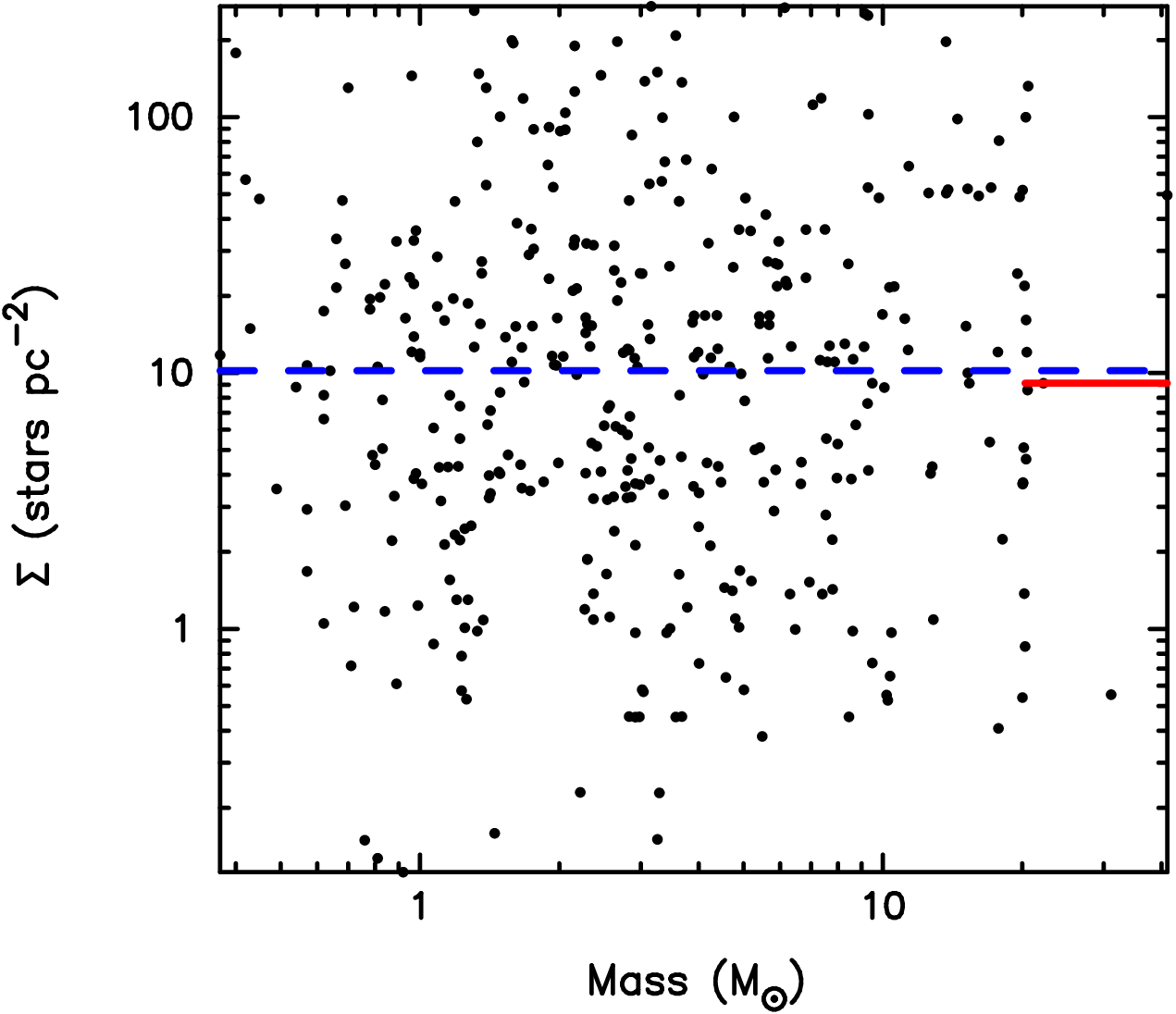}}}

\caption{R11O Dual feedback  run. }
\label{R11O_sim_results_dual}
  \end{center}
\end{figure*}

\begin{figure*}
  \begin{center}
\setlength{\subfigcapskip}{10pt}

\hspace*{-0.1cm}\subfigure[]{\rotatebox{270}{\includegraphics[scale=0.21]{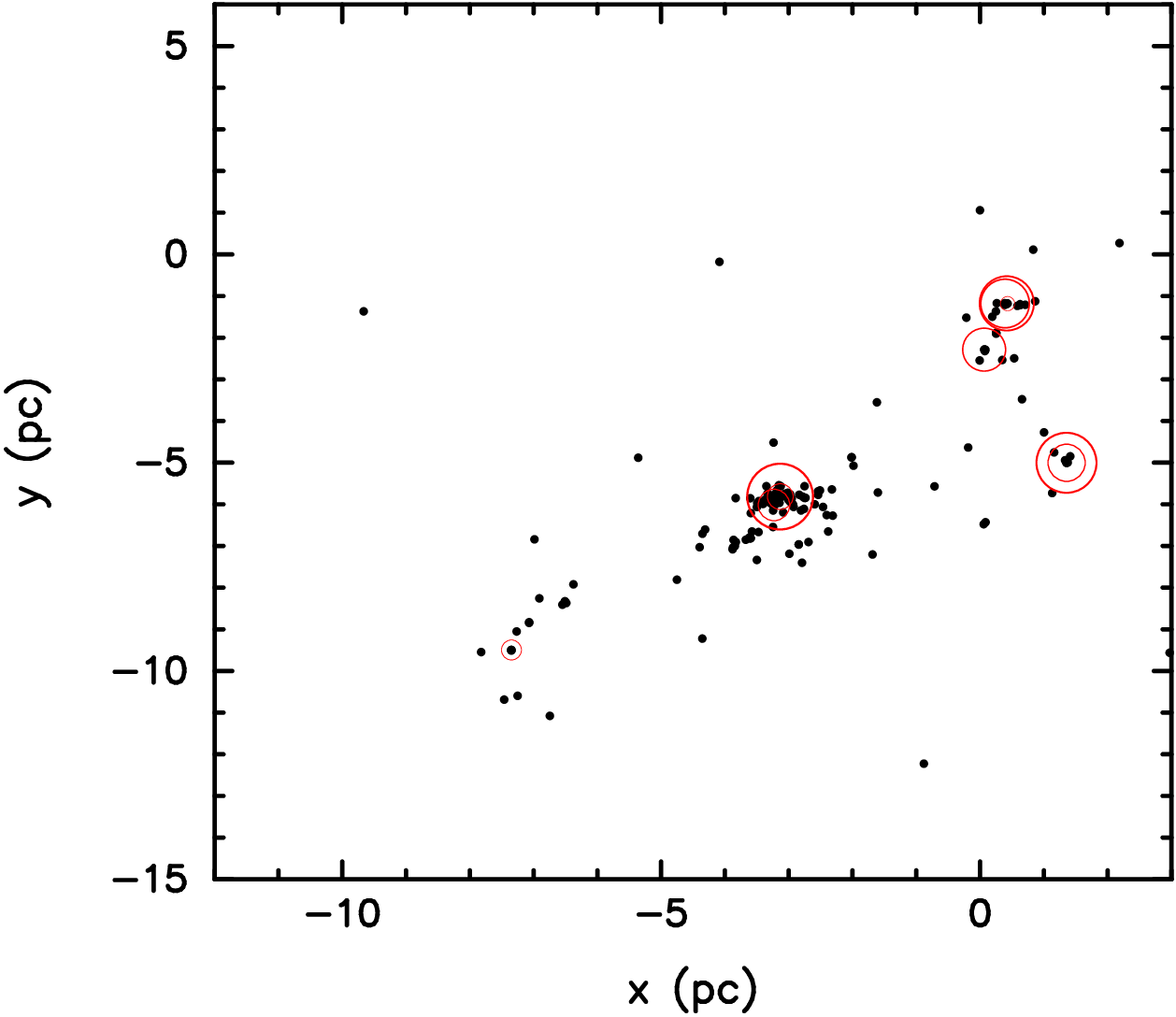}}}
\subfigure[]{\rotatebox{270}{\includegraphics[scale=0.21]{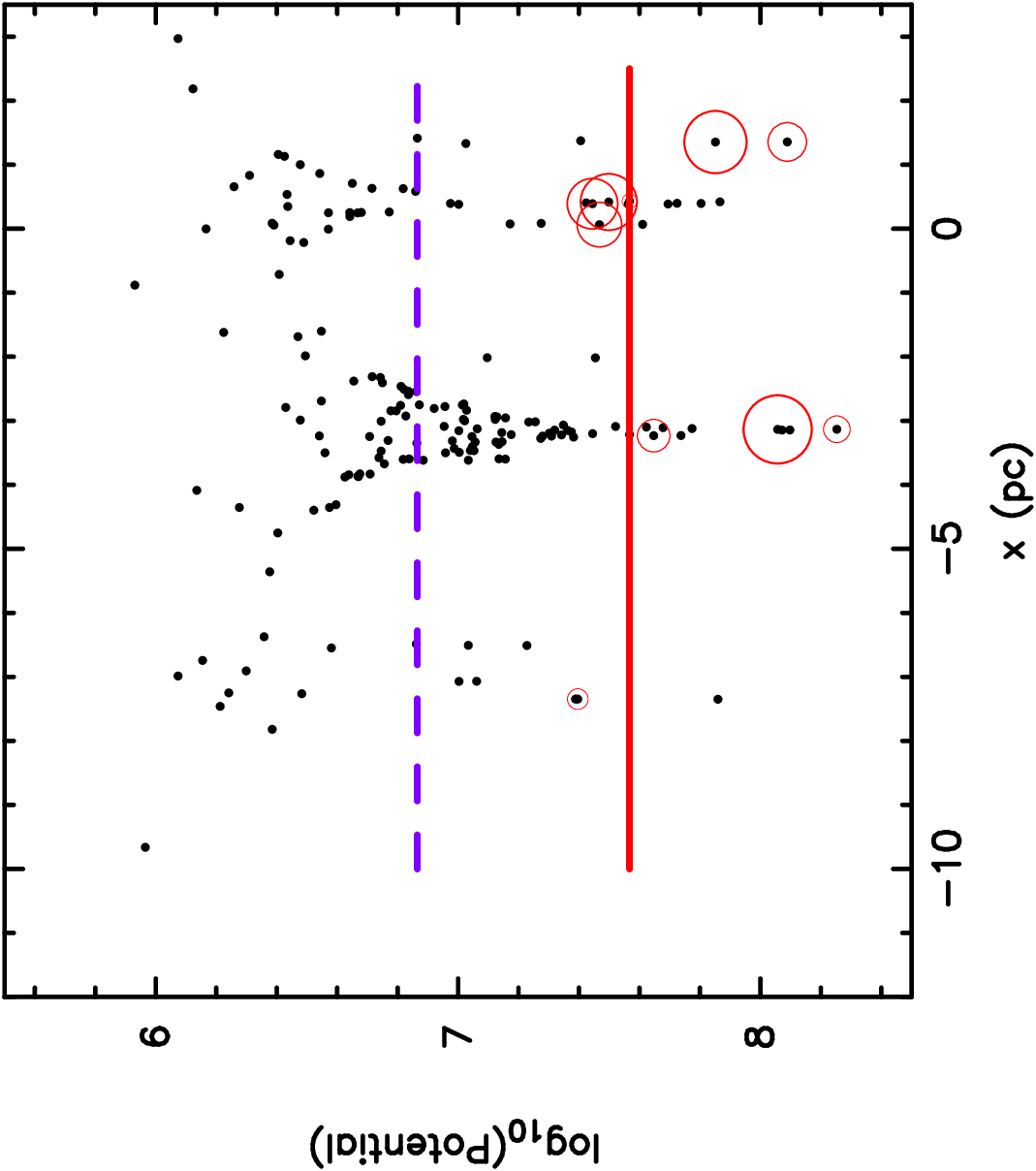}}}
\subfigure[]{\rotatebox{270}{\includegraphics[scale=0.21]{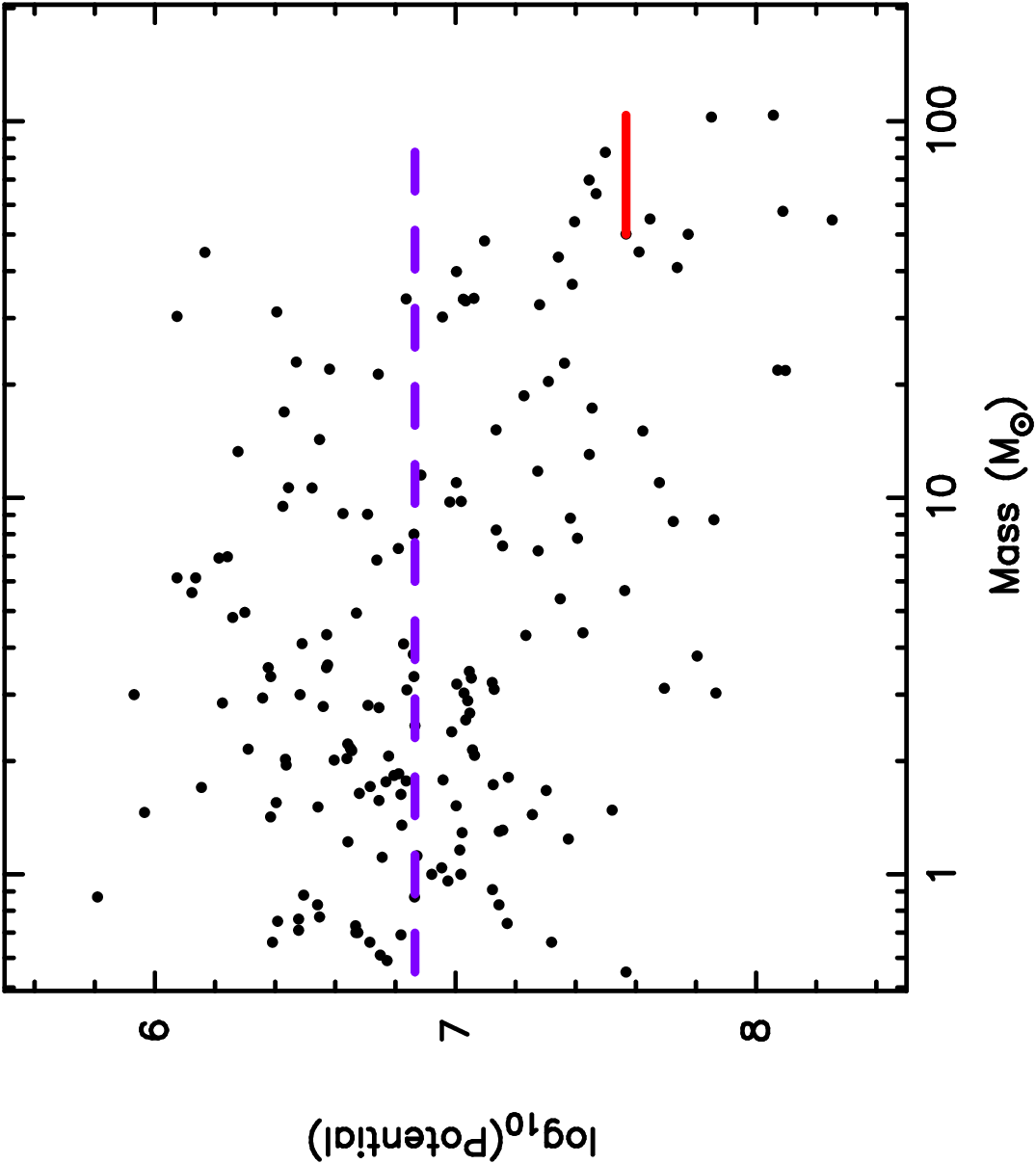}}}
\subfigure[]{\rotatebox{270}{\includegraphics[scale=0.21]{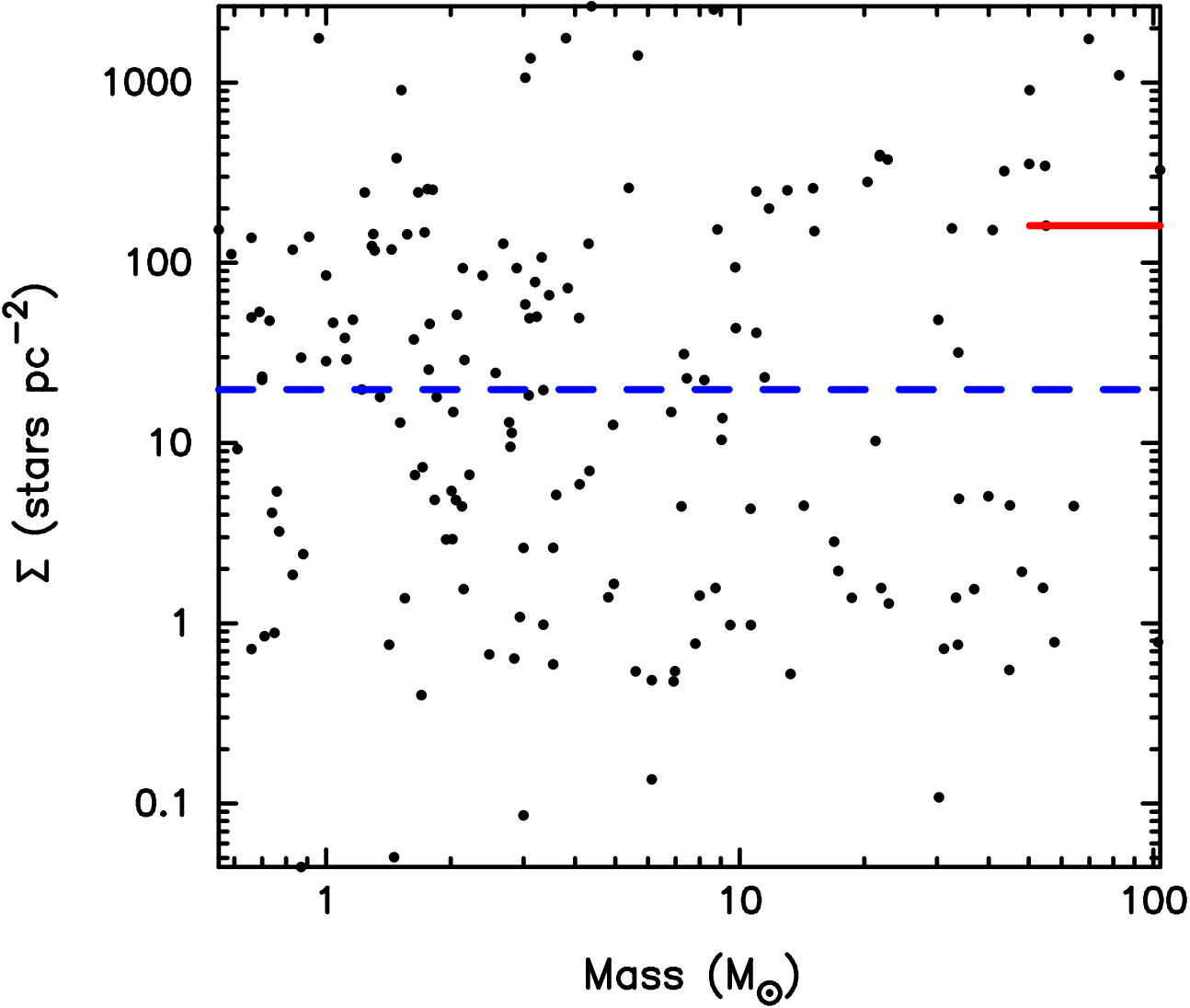}}}

\caption{R15L Control  run. }
\label{R15L_sim_results_control}
  \end{center}
\end{figure*}

\begin{figure*}
  \begin{center}
\setlength{\subfigcapskip}{10pt}

\subfigure[]{\rotatebox{270}{\includegraphics[scale=0.21]{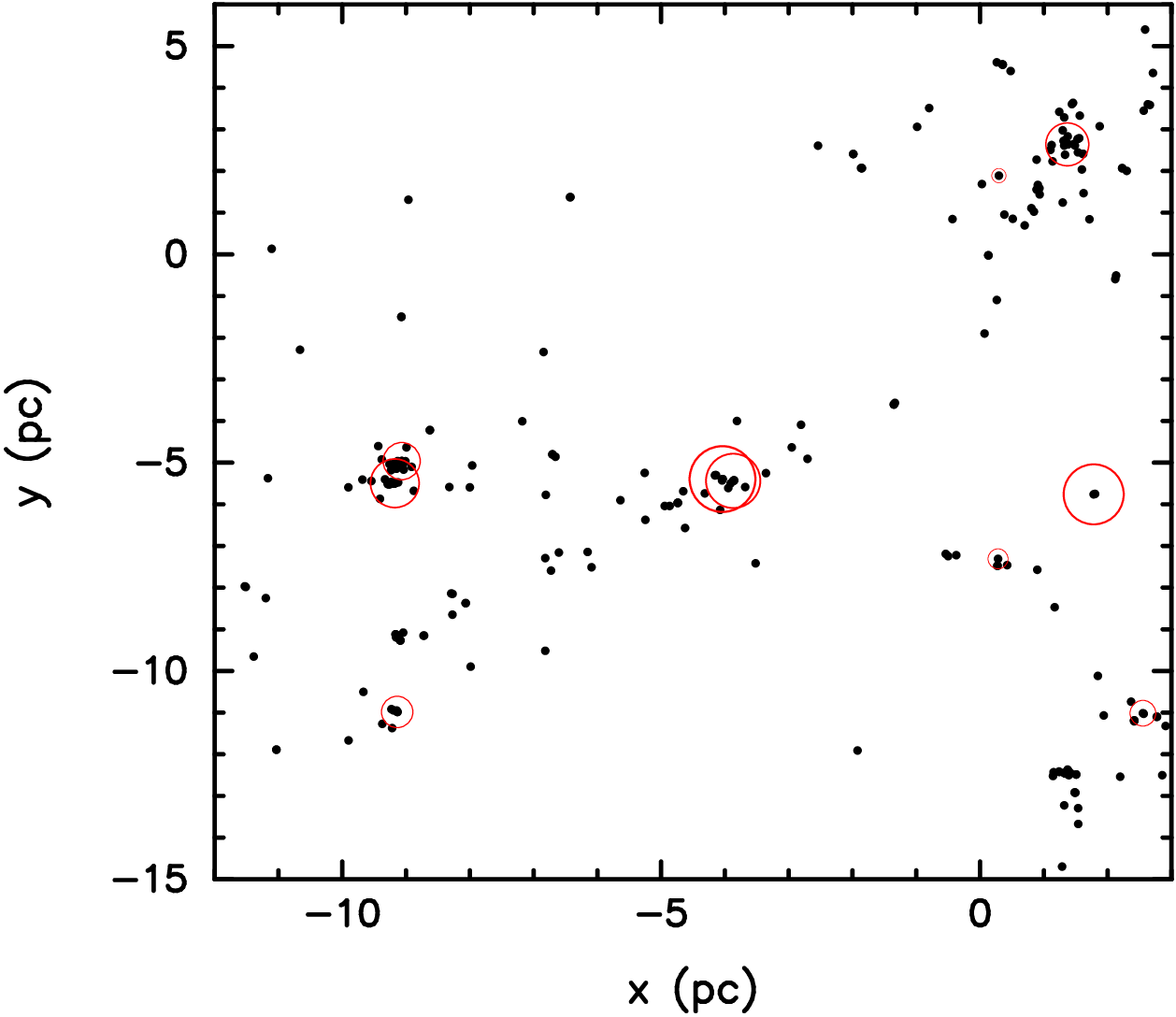}}}
\subfigure[]{\rotatebox{270}{\includegraphics[scale=0.21]{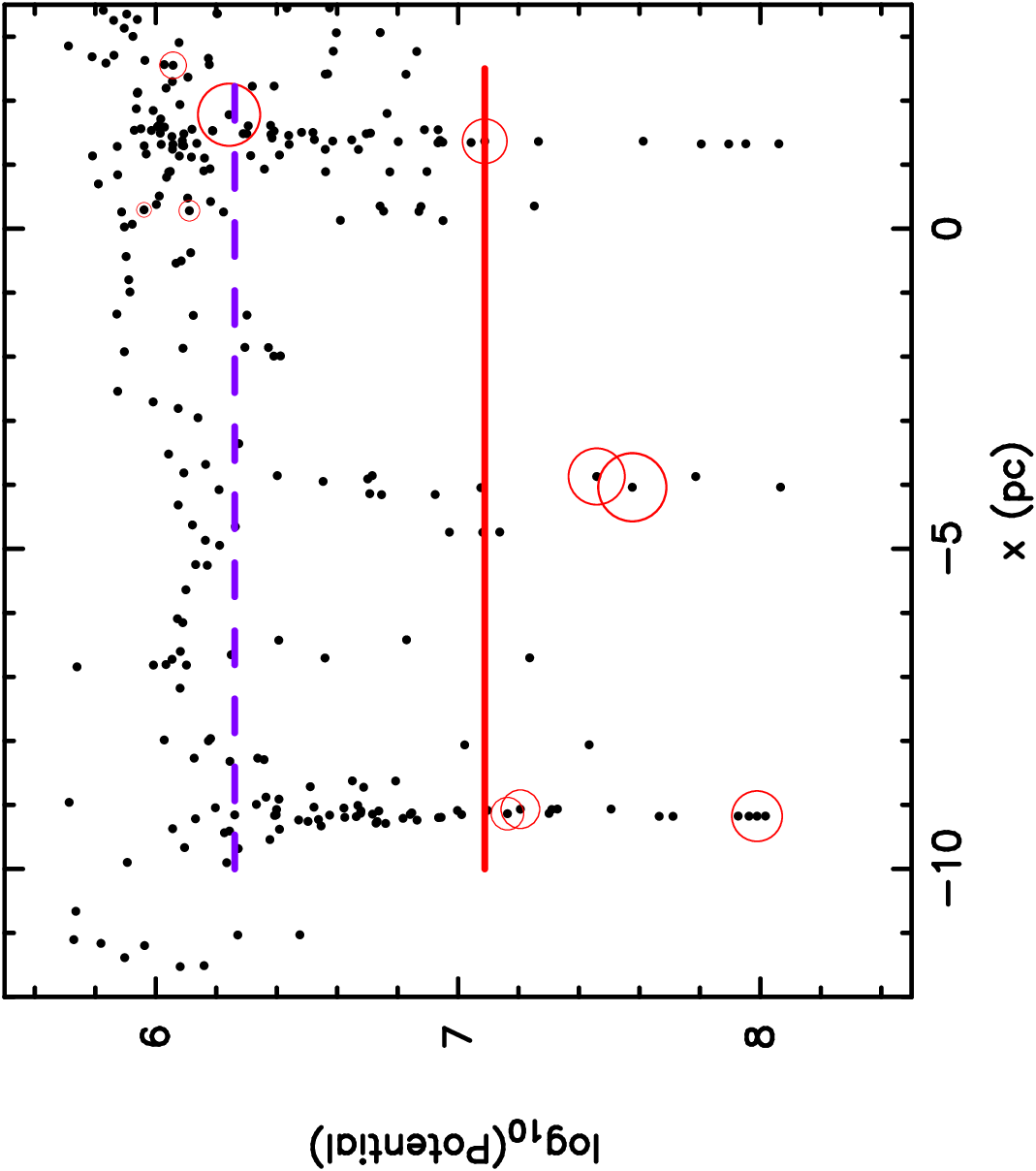}}}
\subfigure[]{\rotatebox{270}{\includegraphics[scale=0.21]{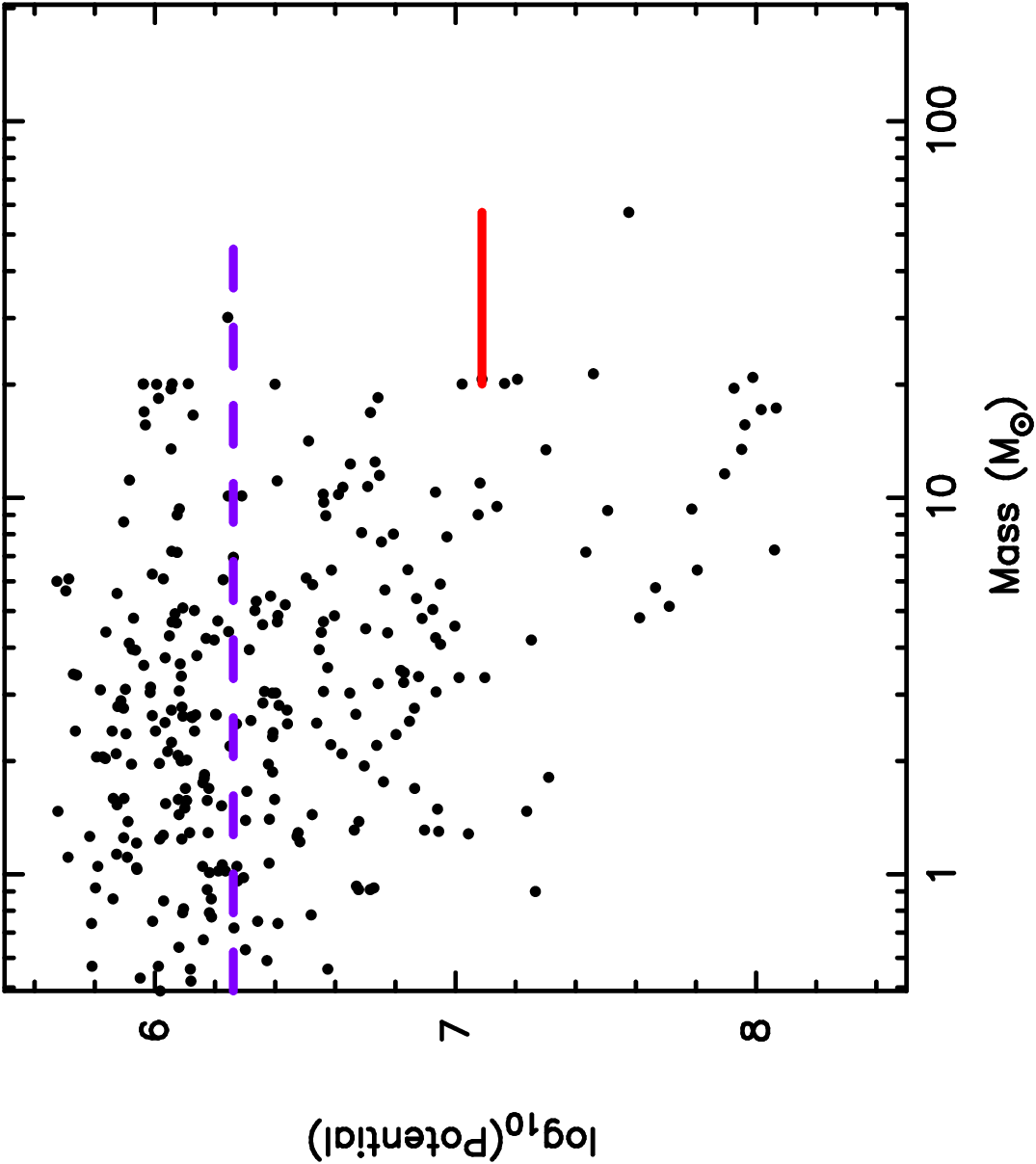}}}
\subfigure[]{\rotatebox{270}{\includegraphics[scale=0.21]{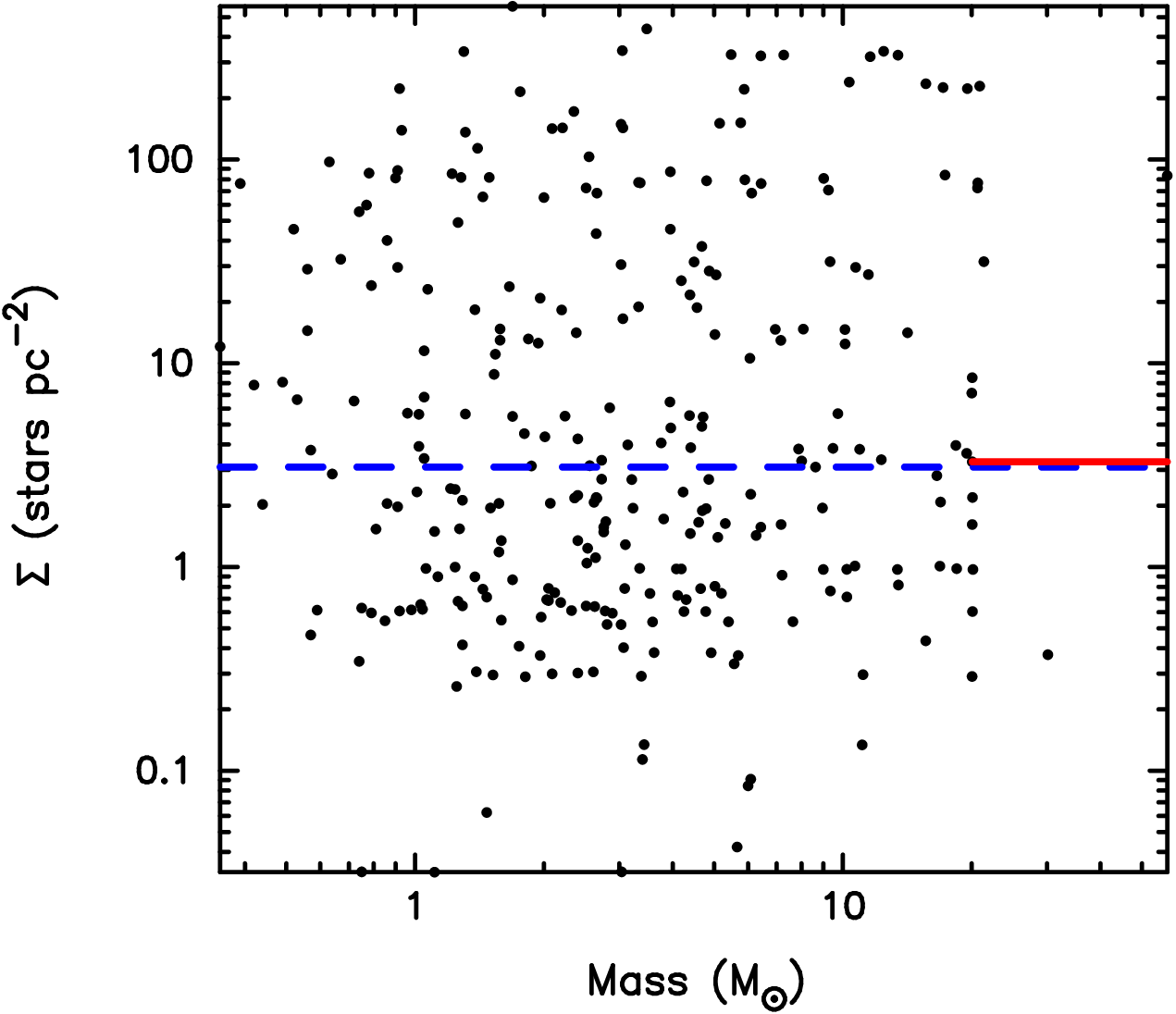}}}

\caption{R15L Dual feedback  run. }
\label{R15L_sim_results_dual}
  \end{center}
\end{figure*}

\begin{figure*}
  \begin{center}
\setlength{\subfigcapskip}{10pt}

\subfigure[]{\rotatebox{270}{\includegraphics[scale=0.21]{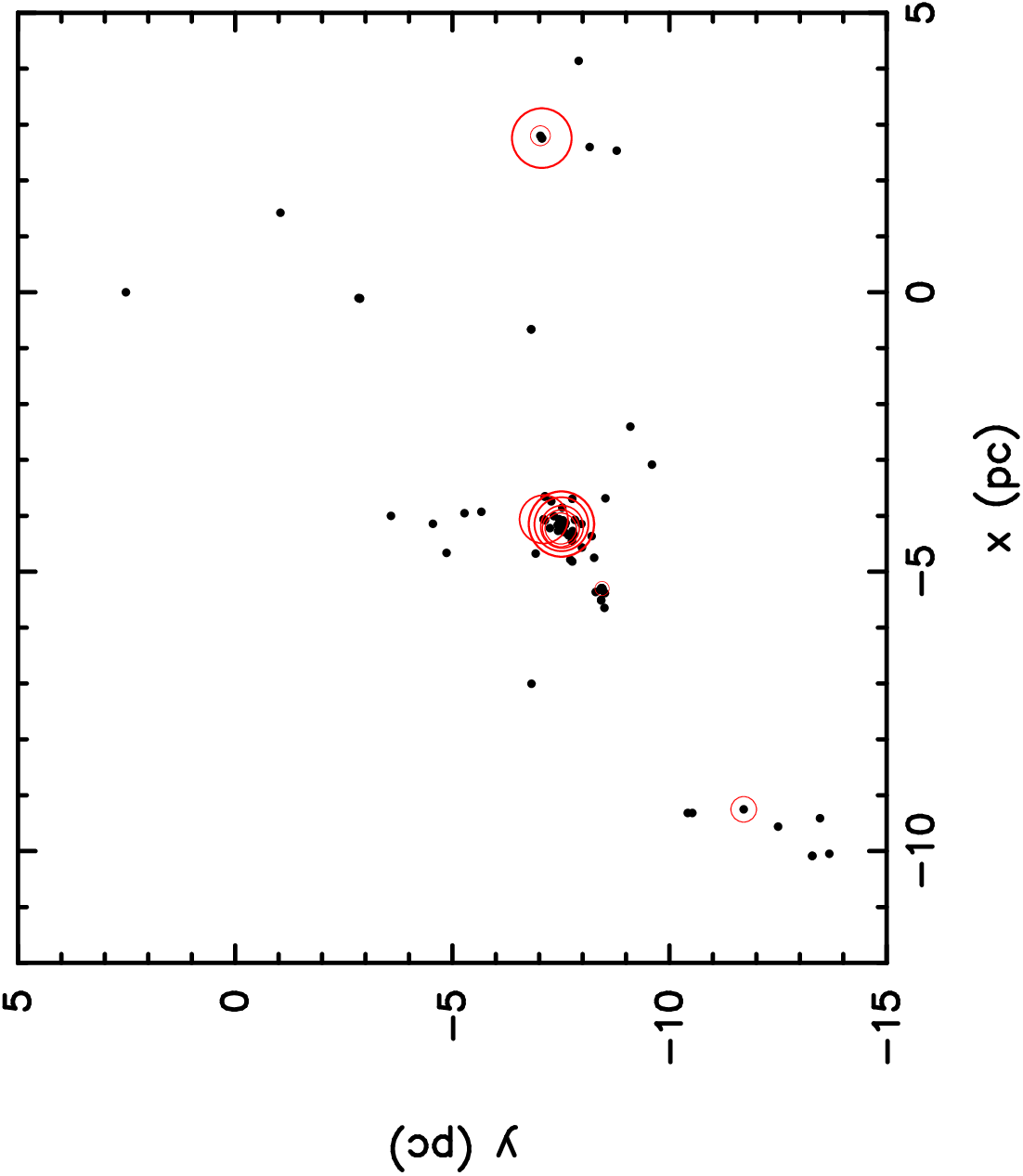}}}
\subfigure[]{\rotatebox{270}{\includegraphics[scale=0.21]{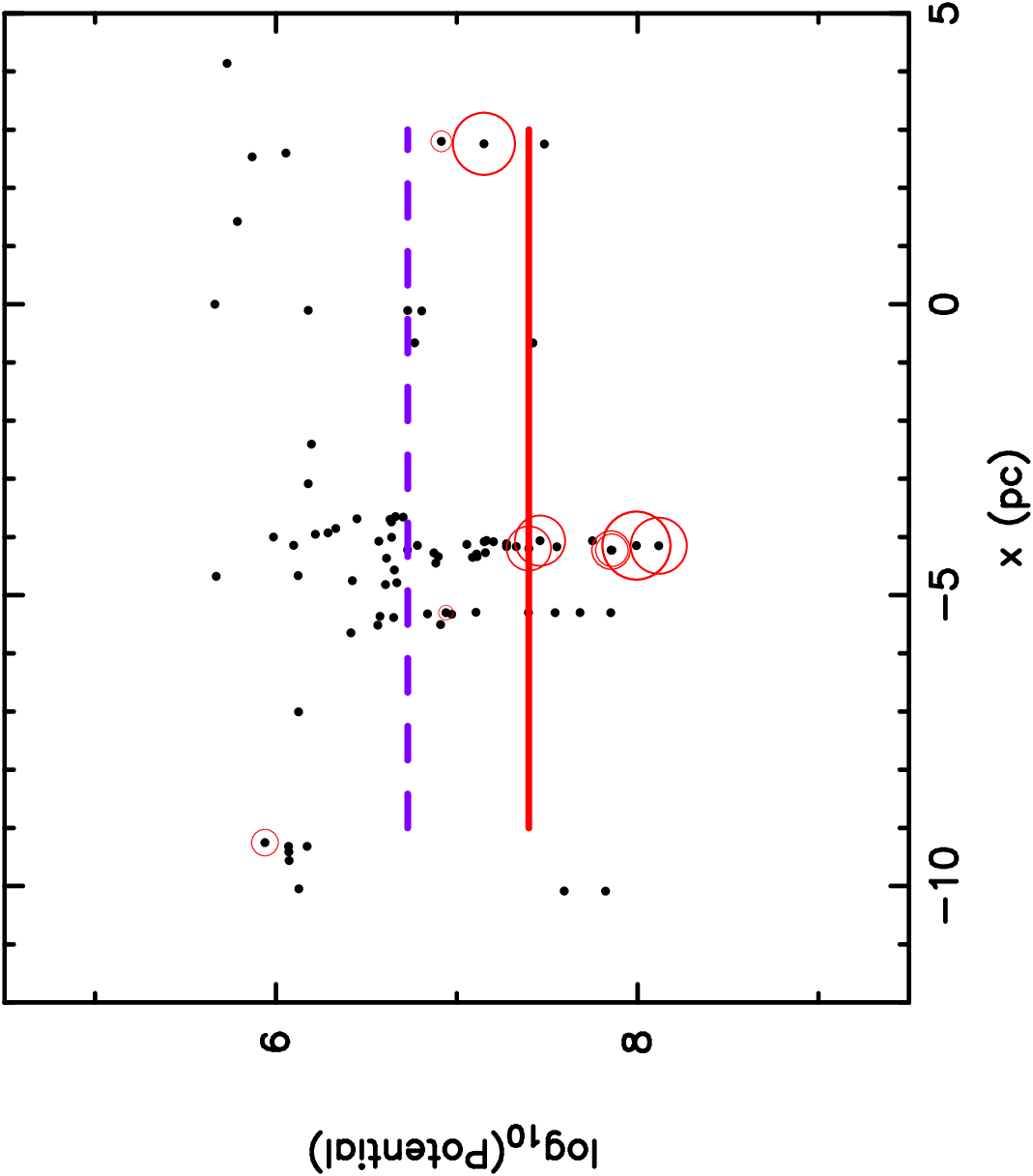}}}
\subfigure[]{\rotatebox{270}{\includegraphics[scale=0.21]{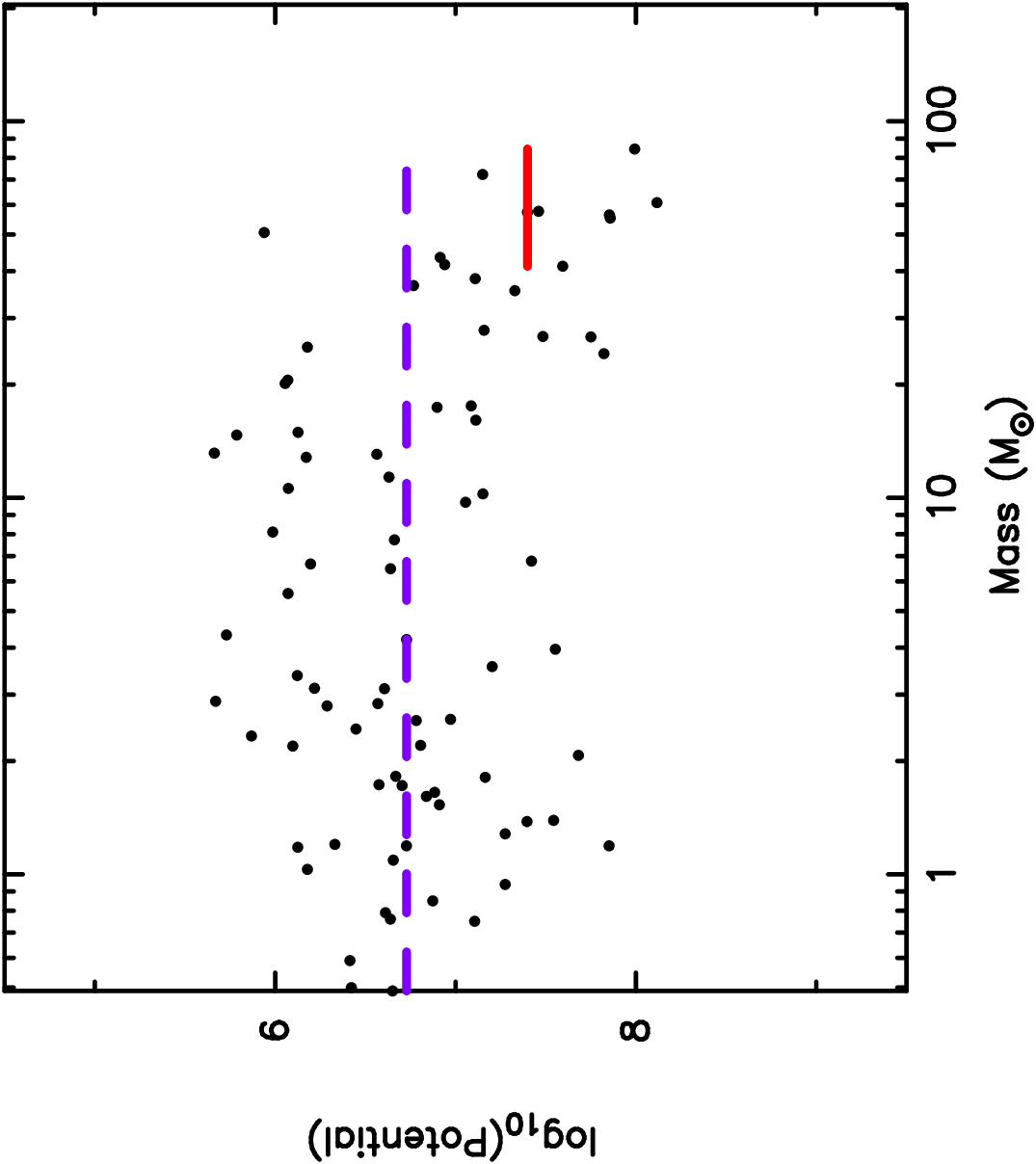}}}
\subfigure[]{\rotatebox{270}{\includegraphics[scale=0.21]{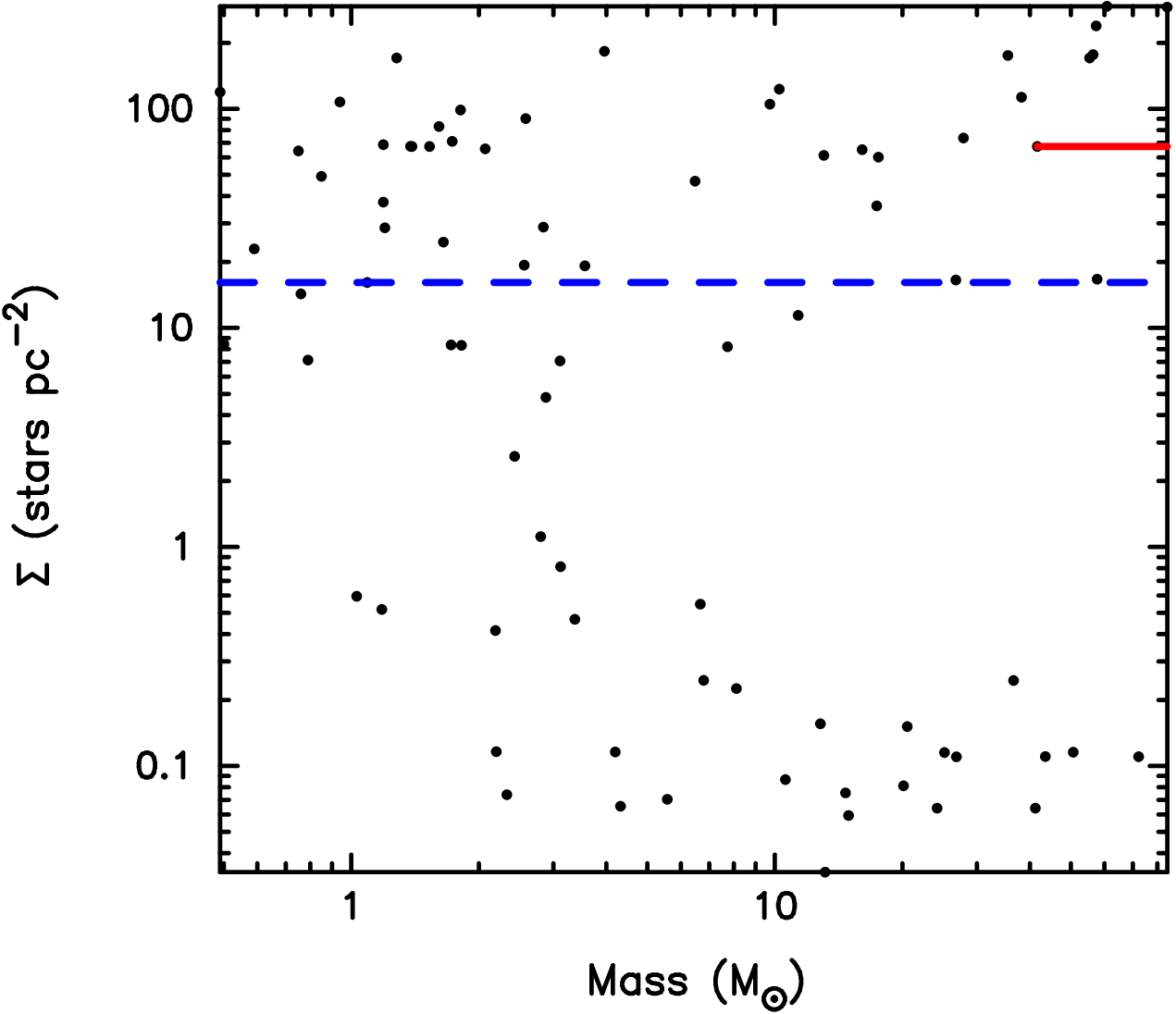}}}

\caption{R19S Control  run. }
\label{R19S_sim_results_control}
  \end{center}
\end{figure*}

\begin{figure*}
  \begin{center}
\setlength{\subfigcapskip}{10pt}

\subfigure[]{\rotatebox{270}{\includegraphics[scale=0.21]{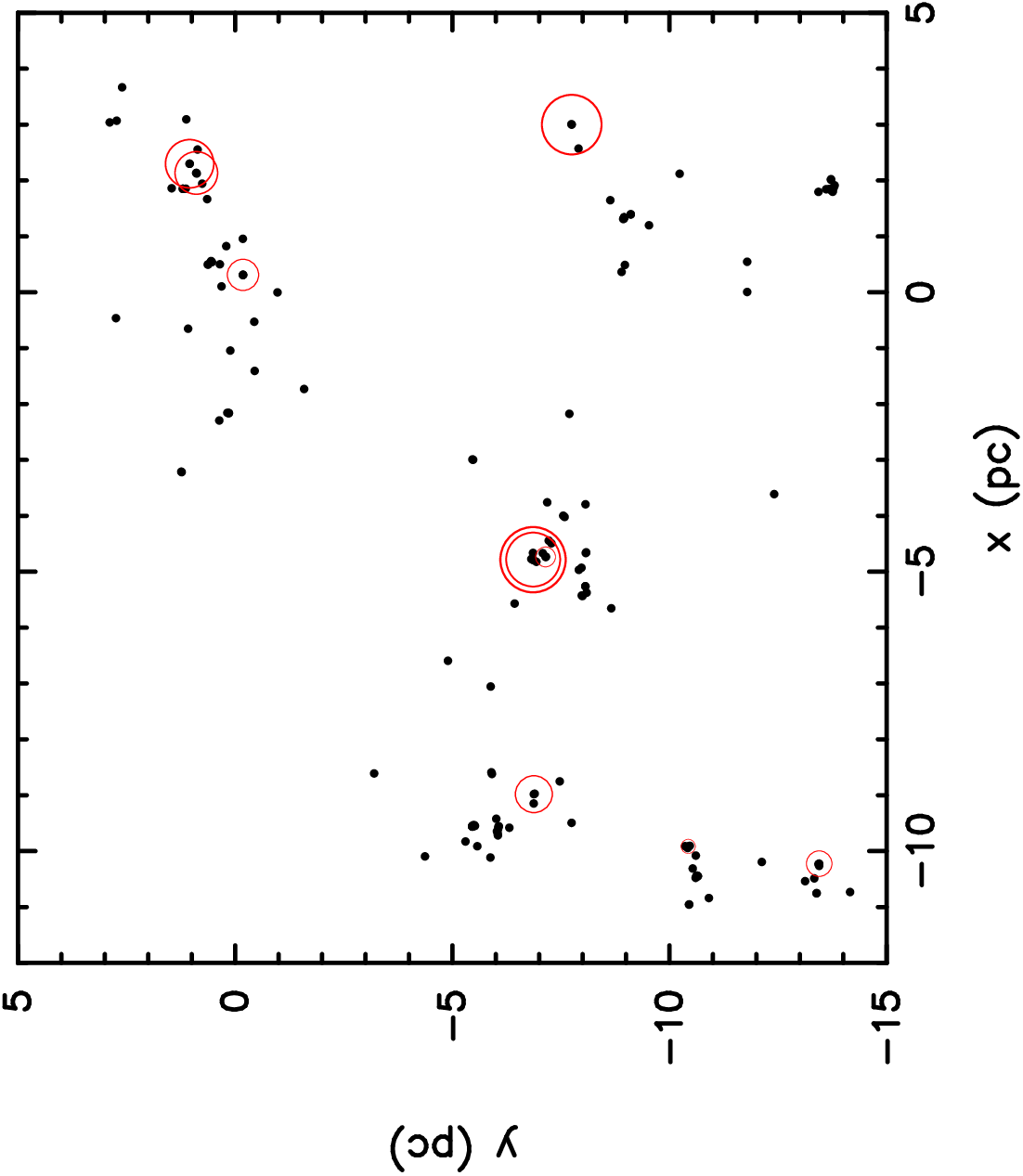}}}
\subfigure[]{\rotatebox{270}{\includegraphics[scale=0.21]{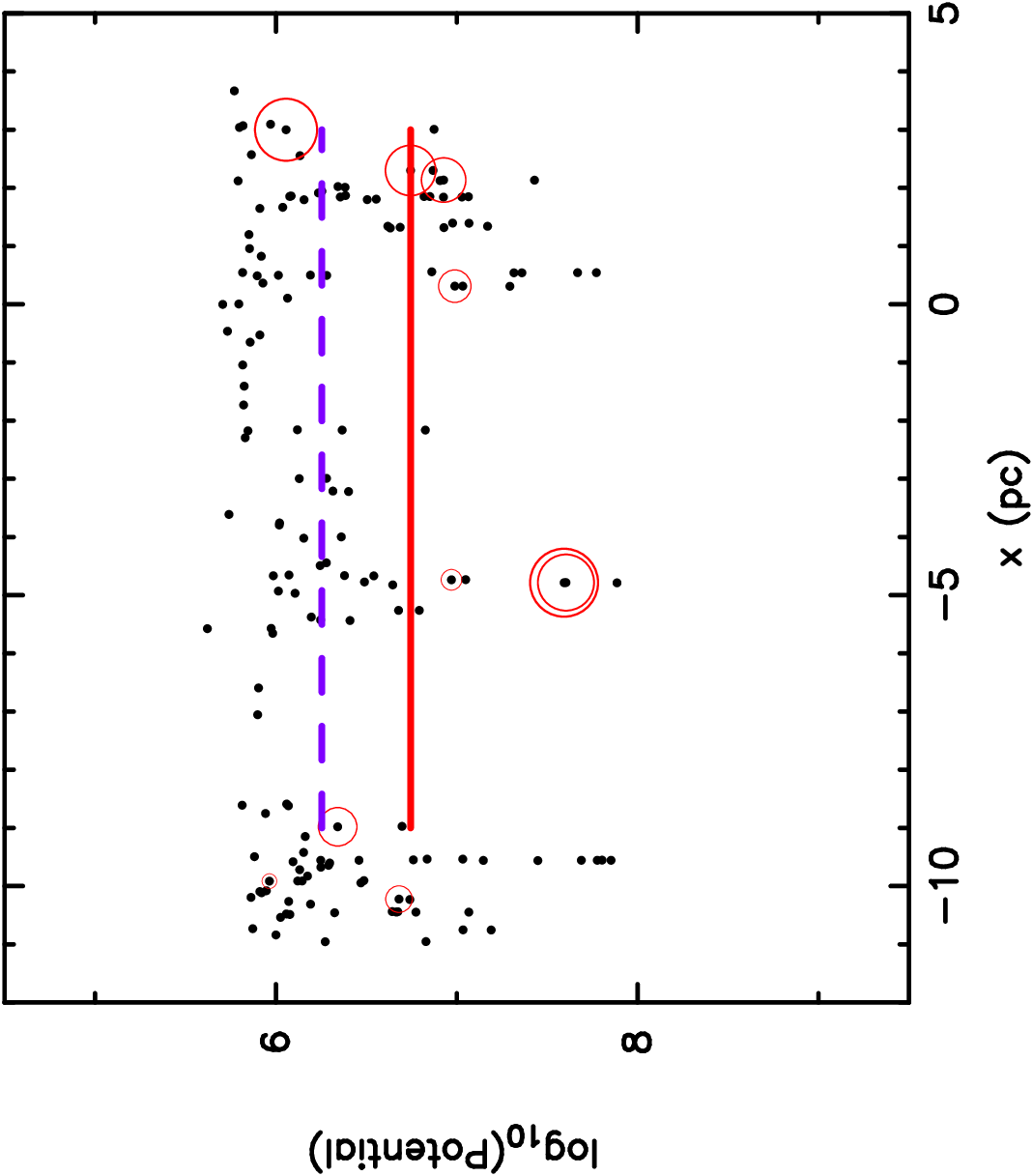}}}
\subfigure[]{\rotatebox{270}{\includegraphics[scale=0.21]{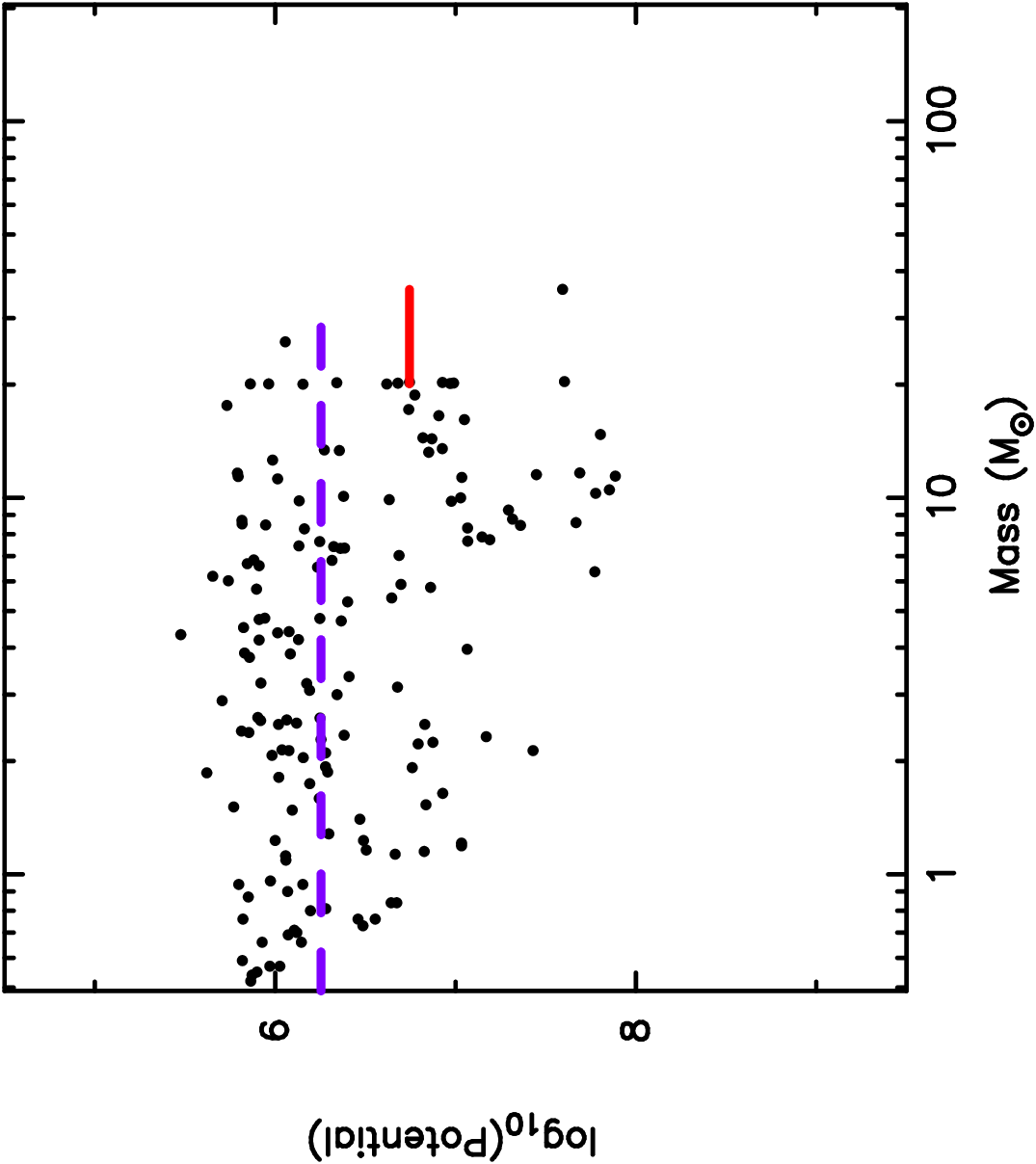}}}
\subfigure[]{\rotatebox{270}{\includegraphics[scale=0.21]{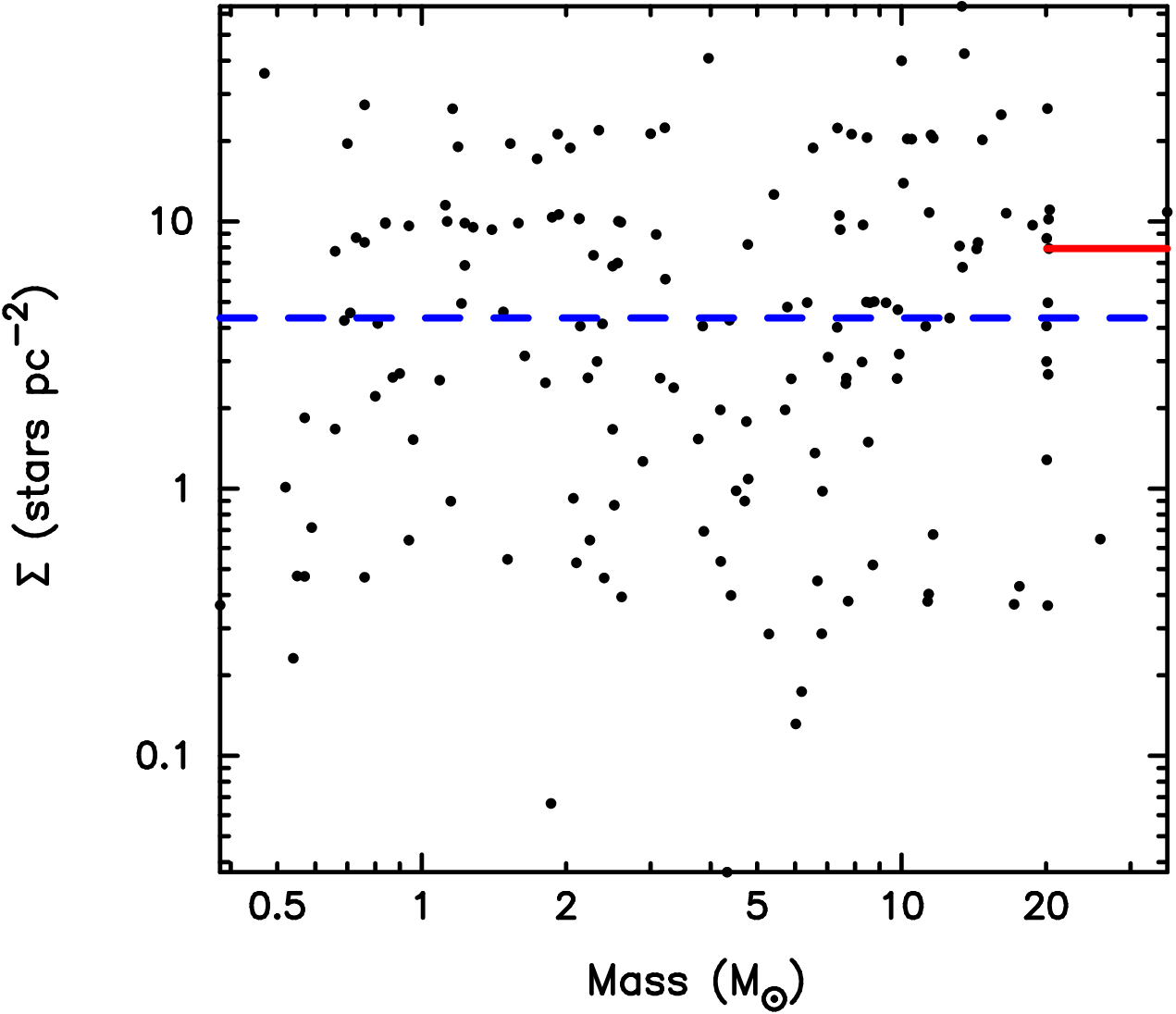}}}

\caption{R19S Dual feedback  run. }
\label{R19S_sim_results_dual}
  \end{center}
\end{figure*}

\begin{figure*}
  \begin{center}
\setlength{\subfigcapskip}{10pt}

\subfigure[]{\rotatebox{270}{\includegraphics[scale=0.21]{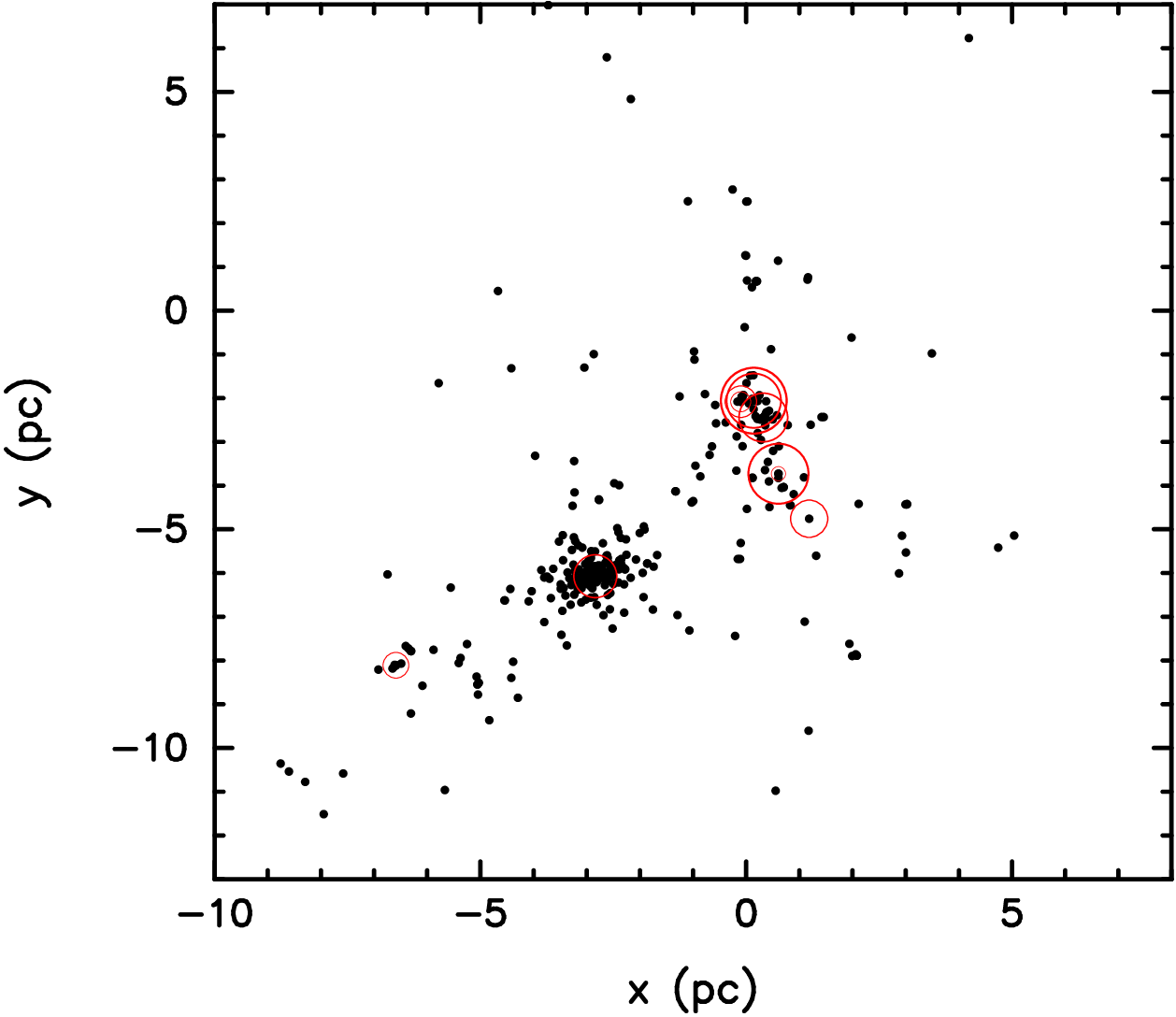}}}
\subfigure[]{\rotatebox{270}{\includegraphics[scale=0.21]{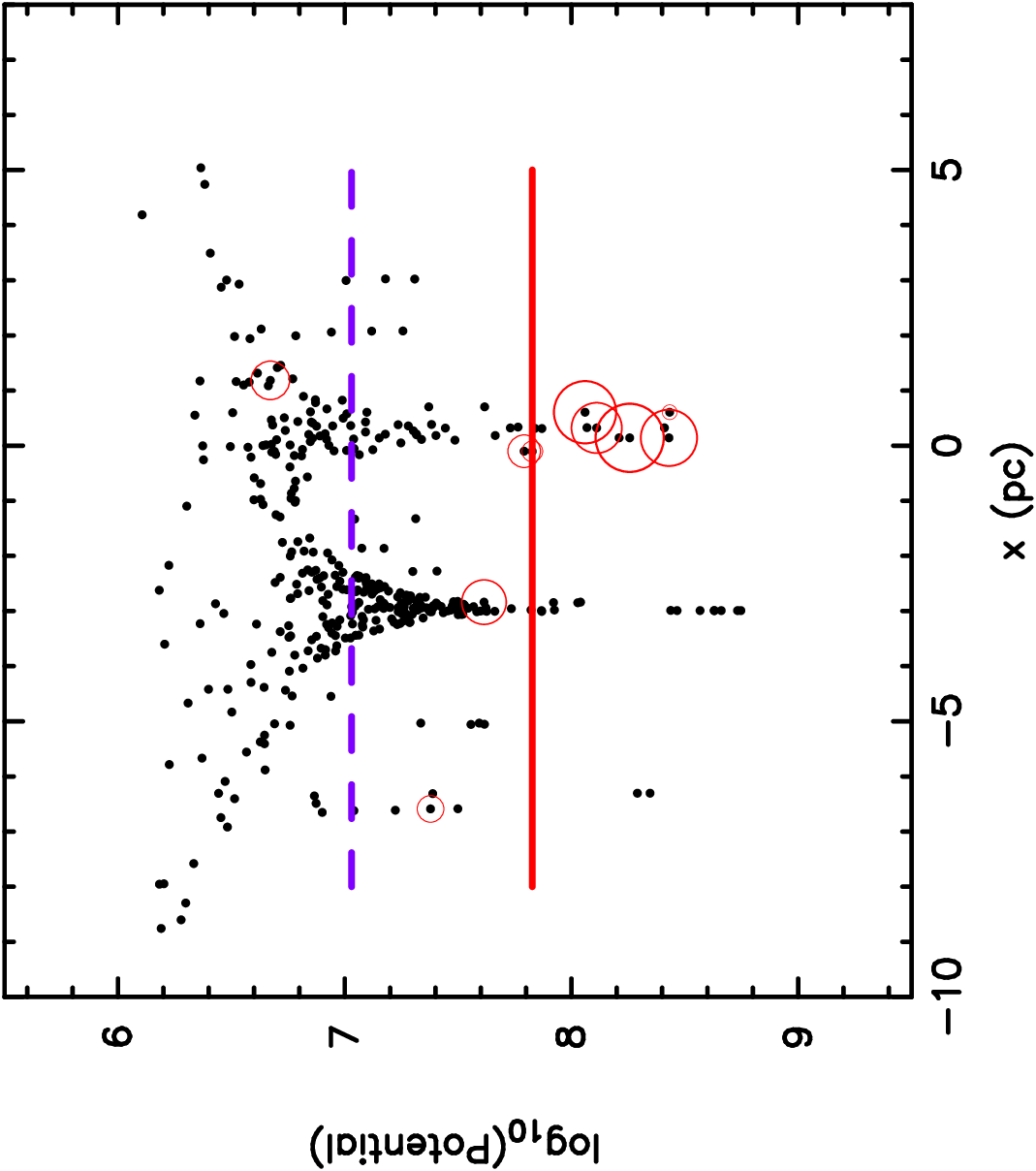}}}
\subfigure[]{\rotatebox{270}{\includegraphics[scale=0.21]{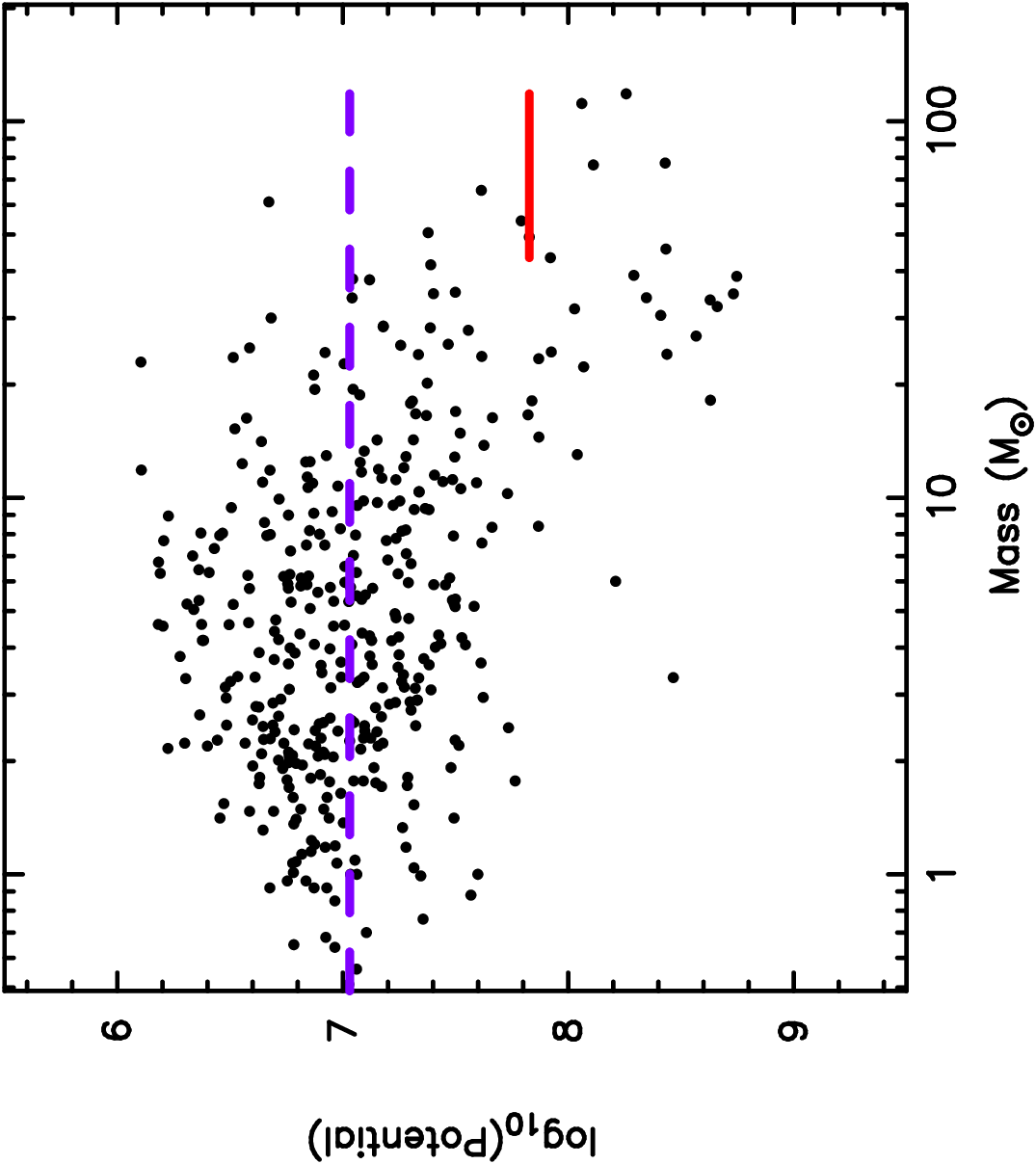}}}
\subfigure[]{\rotatebox{270}{\includegraphics[scale=0.21]{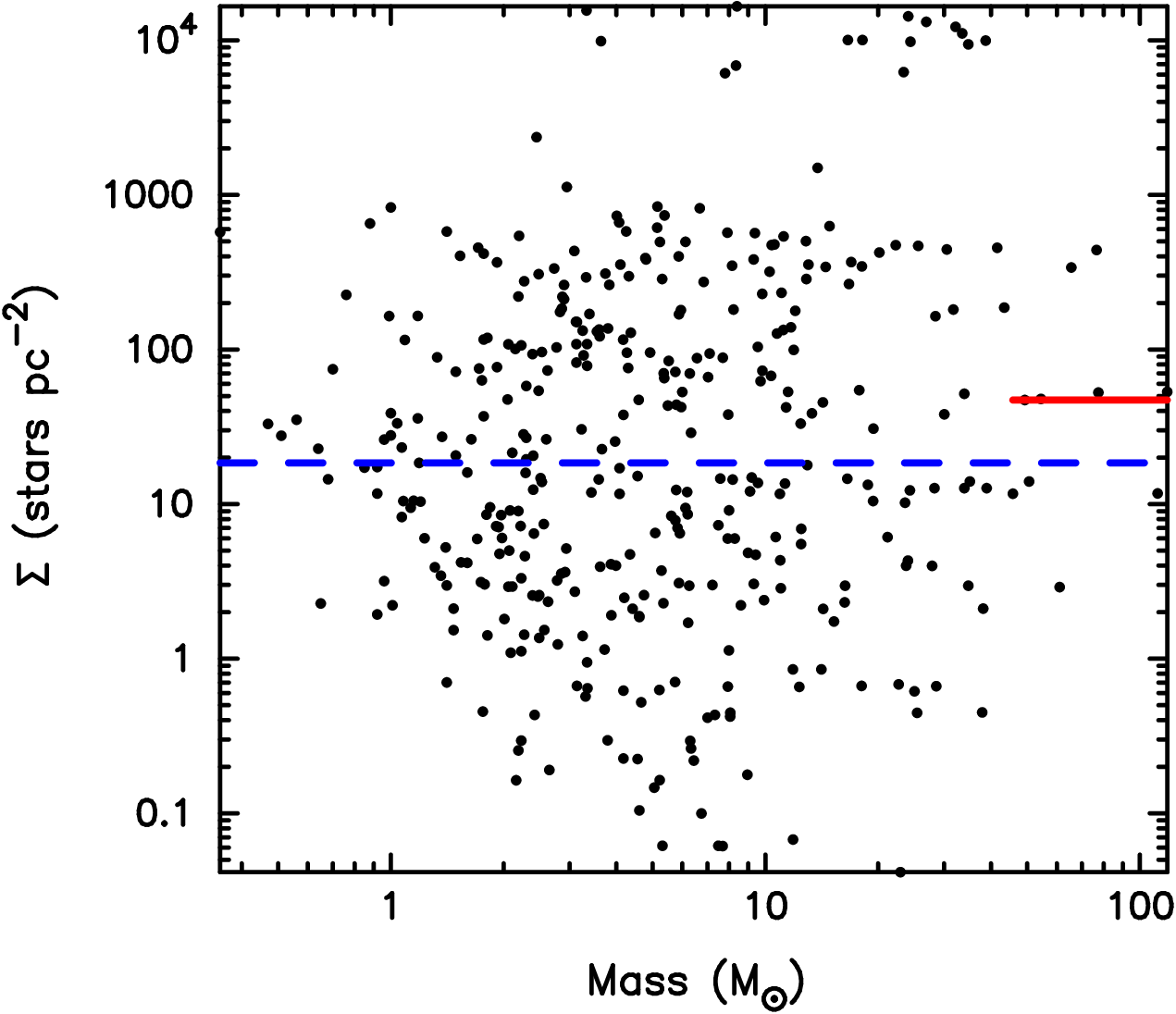}}}

\caption{R19T Control  run. }
\label{R19T_sim_results_control}
  \end{center}
\end{figure*}

\begin{figure*}
  \begin{center}
\setlength{\subfigcapskip}{10pt}

\subfigure[]{\rotatebox{270}{\includegraphics[scale=0.21]{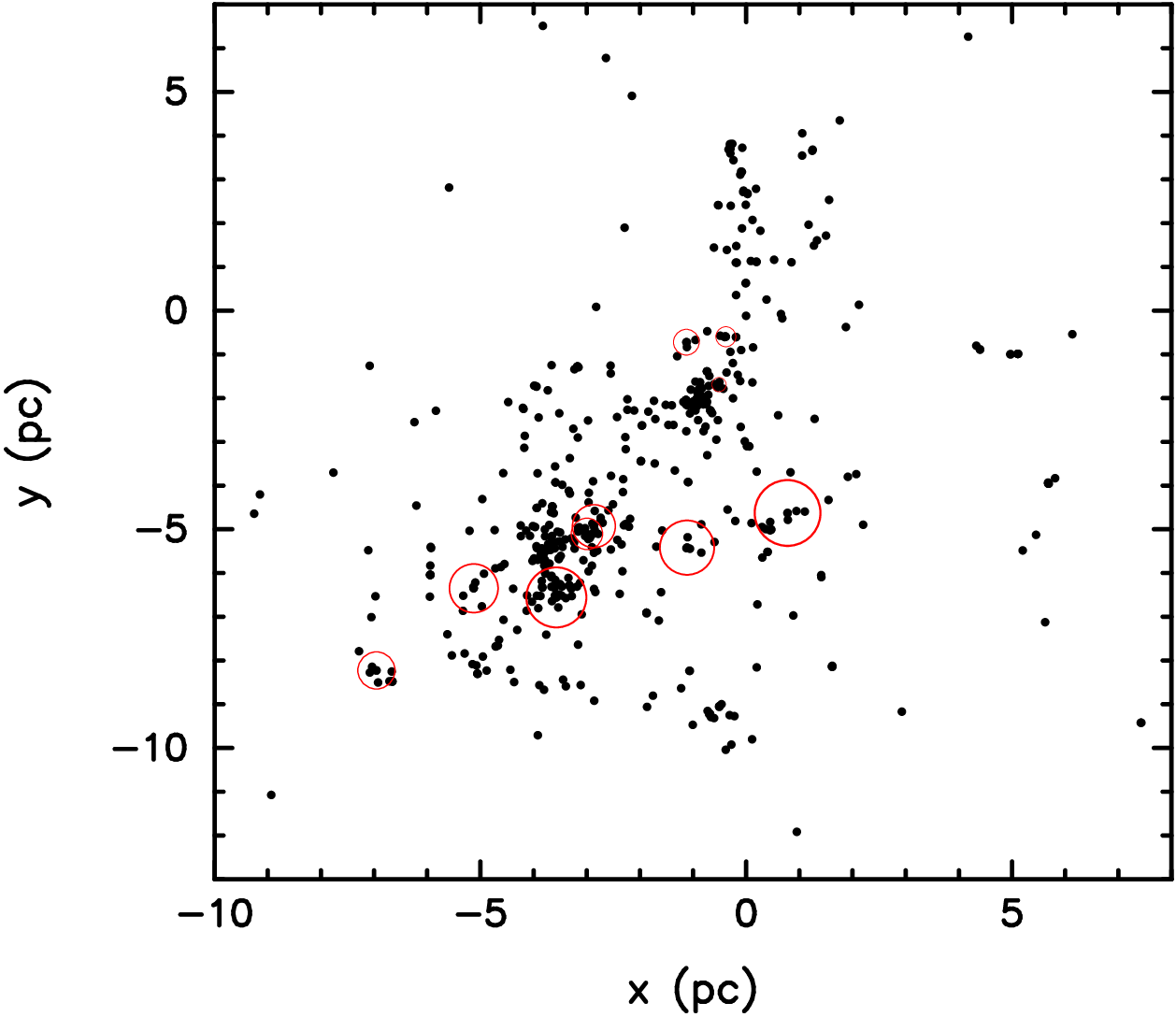}}}
\subfigure[]{\rotatebox{270}{\includegraphics[scale=0.21]{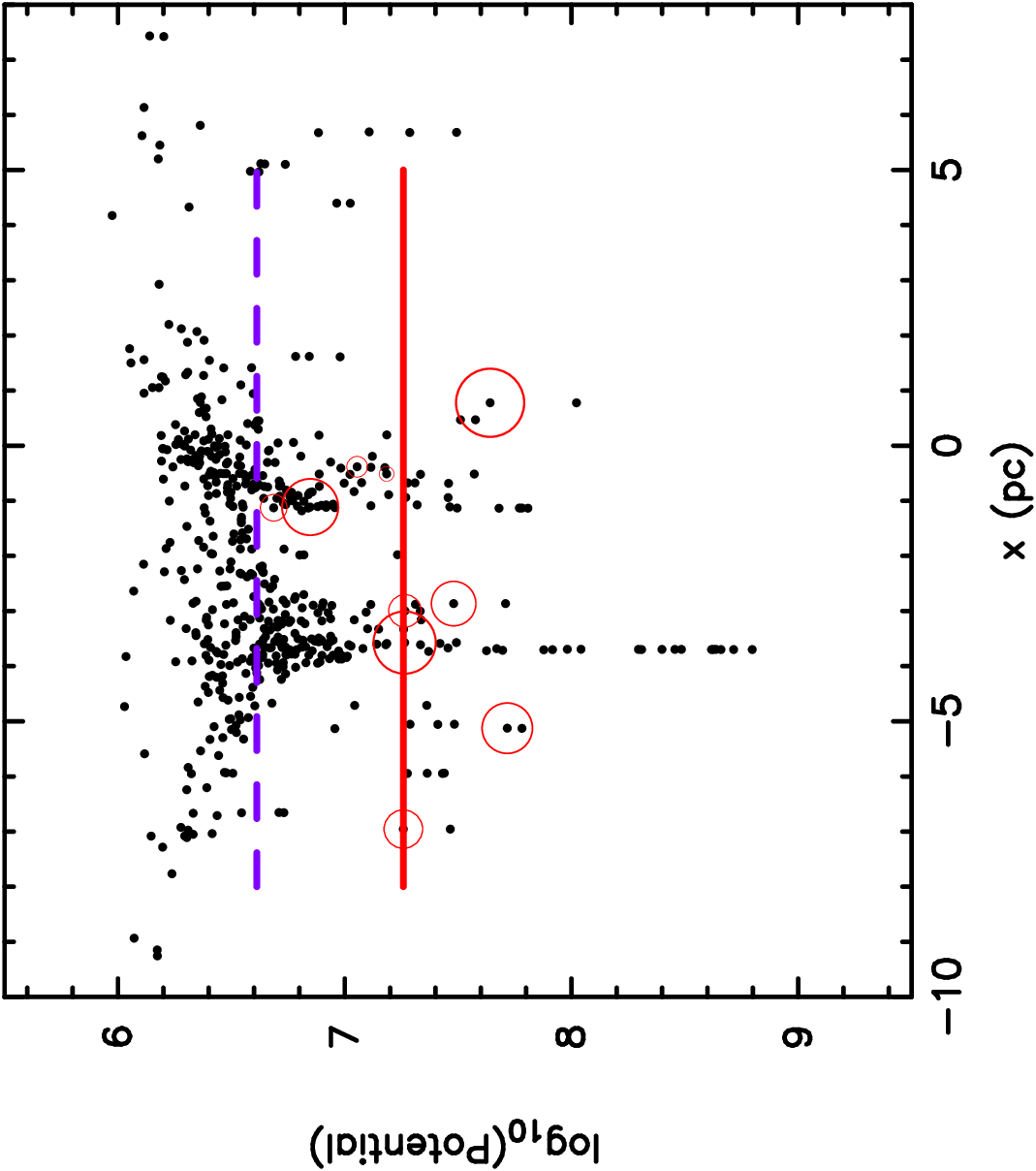}}}
\subfigure[]{\rotatebox{270}{\includegraphics[scale=0.21]{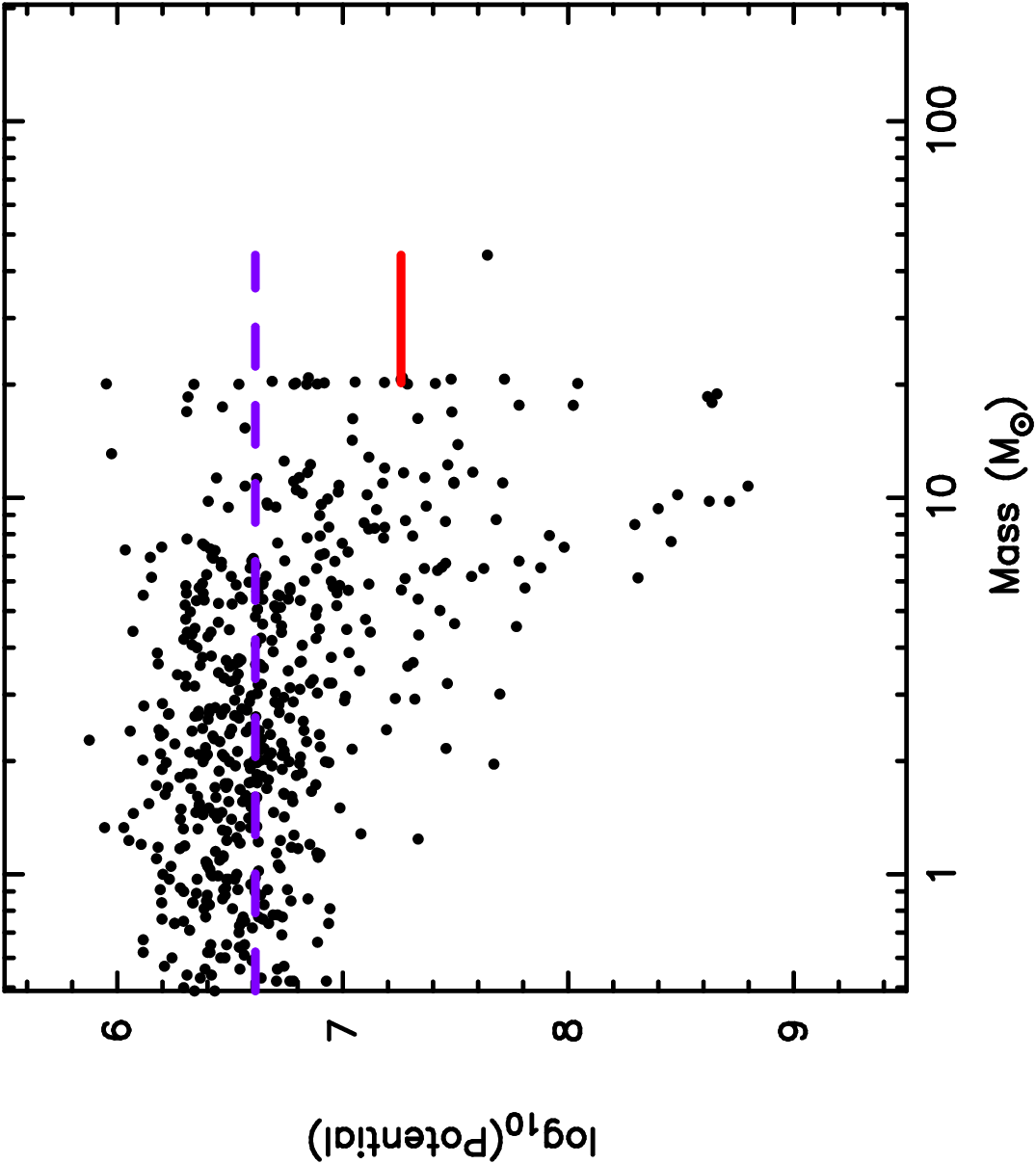}}}
\subfigure[]{\rotatebox{270}{\includegraphics[scale=0.21]{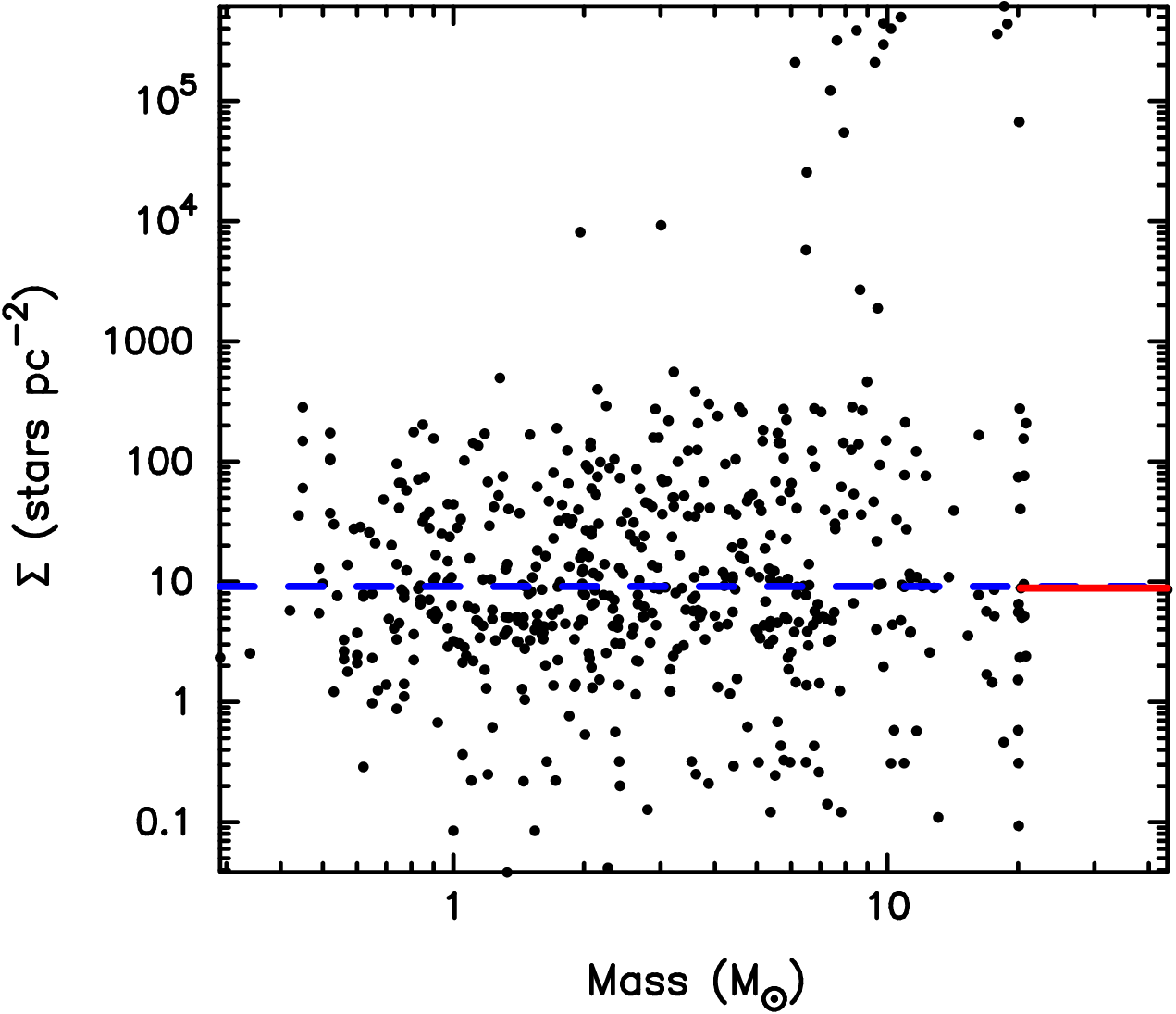}}}

\caption{R19T Dual feedback  run. }
\label{R19T_sim_results_dual}
  \end{center}
\end{figure*}

\label{lastpage}

\end{document}